\documentclass[11pt,a4paper]{article}
\usepackage{epsfig}
\usepackage[T1]{fontenc}    
\usepackage{graphics}
\usepackage{graphicx}
\usepackage{pstricks,pst-coil,pst-fill,pst-plot}
\usepackage[fleqn]{amsmath}    
\usepackage{amssymb}    
\usepackage{amsfonts}   
\usepackage{verbatim}   
\usepackage{mathrsfs}   
\usepackage{dsfont}
\usepackage{euscript}
\usepackage{yfonts}
\usepackage{enumerate}     
\usepackage{amsthm}         
\usepackage{txfonts}
\usepackage{marvosym}
\usepackage{stmaryrd}
\usepackage{vmargin}        
\usepackage{wasysym}		

\setmarginsrb{1.8cm}{2cm}{1.8cm}{2cm}{1cm}{1cm}{1cm}{1.6cm}
 \makeatletter
 \@addtoreset{equation}{section}
 \makeatother

\providecommand{\bysame}{\leavevmode\hbox to3em{\hrulefill}\thinspace}
\providecommand{\MR}{\relax\ifhmode\unskip\space\fi MR }

\providecommand{\href}[2]{#2}

\let\tend=\rightarrow

\long\def\symbolfootnote[#1]#2{\begingroup%
\def\thefootnote{\fnsymbol{footnote}}\footnote[#1]{#2}\endgroup}

\newtheorem{theorem}{Theorem}[section]
\newtheorem{prop}[theorem]{Proposition}
\newtheorem*{theorem*}{Theorem}

\newtheorem{defin}[theorem]{Definition}

\newtheorem{conj}[theorem]{Conjecture}

\newtheorem{lemme}[theorem]{Lemma}

\def\Proof{\medskip\noindent {\it Proof --- \ }}

\def\qed{\hfill\rule{2mm}{2mm}}

\newcommand\beq{\begin{equation}}
\newcommand\enq{\end{equation}}
\newcommand\bem{\begin{multline}}
\newcommand\enm{\end{multline}}

\def\beqa{\begin{eqnarray}}
\def\eeqa{\end{eqnarray}}
\def\ba{\begin{array}}
\def\ea{\end{array}}
\def\det{\operatorname{det}}

\newcommand{\f}[2]{{\ensuremath{%
    \mathchoice%
    {\dfrac{#1}{#2}}
    {\dfrac{#1}{#2}}
    {\frac{#1}{#2}}
    {\frac{#1}{#2}}
}}}
\newcommand{\tf}[2]{\ensuremath{#1/#2}}

\def\a{\alpha}

\def\be{\beta}
\def\ga{\gamma}
\def\Ga{\Gamma}

\def\de{\delta}

\def\De{\Delta}
\def\eps{\epsilon}
\def\veps{\varepsilon}
\def\la{\lambda}

\def\sg{\sigma}
\def\vsg{\varsigma}

\def\Ups{\Upsilon}
\def\ups{\upsilon}
\def\th{\theta}

\def\vth{\vartheta}

\def\Om{\Omega}
\def\om{\omega}
\def\vp{\varphi}

\newcommand{\mc}[1]{\ensuremath{\mathcal{#1}}}
\newcommand{\mf}[1]{\ensuremath{\mathfrak{#1}}}
\newcommand{\msc}[1]{\ensuremath{\mathscr{#1}}}

\newcommand{\bs}[1]{\ensuremath{\boldsymbol{#1}}}

\DeclareFontFamily{OT1}{pzc}{}
\DeclareFontShape{OT1}{pzc}{m}{it}{<-> s * [1.10] pzcmi7t}{}
\DeclareMathAlphabet{\mathpzc}{OT1}{pzc}{m}{it}

\def \i{ \mathrm i}

\newcommand{\ov}[1]{\ensuremath{\overline{#1}}}
\newcommand{\wt}[1]{\ensuremath{\widetilde{#1}}}
\newcommand{\wh}[1]{\ensuremath{\widehat{#1}}}

\newcommand{\Int}[2]{\ensuremath{\int\limits_{#1}^{#2}}}

\newcommand{\sul}[2]{\ensuremath{\sum\limits_{#1}^{#2}}}
\newcommand{\pl}[2]{\ensuremath{\prod\limits_{#1}^{#2}}}

\newcommand{\R}{\ensuremath{\mathbb{R}}}
\newcommand{\Cx}{\ensuremath{\mathbb{C}}}

\newcommand{\Dp}[1]{\ensuremath{\partial_{#1}}}

\newcommand{\limit}[2]{\ensuremath{\underset{#1 \tend #2}{\longrightarrow} }}

\newcommand{\ex}[1]{\ensuremath{\e{e}^{#1}}}

\newcommand{\op}[1]{ \boldsymbol{ \texttt{#1} } }

\newcommand{\norm}[1]{\ensuremath{  || #1 || }}

\newcommand{\dd}{\mathrm{d}}
\newcommand{\e}[1]{\ensuremath{\mathrm{#1}}}

\newcommand{\intff}[2]{\ensuremath{ [  #1 \,; #2 ] }}
\newcommand{\intfo}[2]{\ensuremath{ [  #1 \,; #2 [ }}
\newcommand{\intof}[2]{\ensuremath{ ]  #1 \,; #2 ] }}
\newcommand{\intoo}[2]{\ensuremath{ ]  #1 \,; #2 [ }}

\newcommand{\intn}[2]{\ensuremath{[\![ \, #1 \,;\, #2 \,]\!]}}

\begin{document}

\begin{center}
\begin{LARGE}
{\bf On singularities of dynamic response functions in the massless regime of the XXZ spin-1/2 chain}
\end{LARGE}

\vspace{1cm}

\vspace{4mm}
{\large Karol K. Kozlowski \footnote{e-mail: karol.kozlowski@ens-lyon.fr}}%
\\[1ex]
Univ Lyon, ENS de Lyon, Univ Claude Bernard Lyon 1, CNRS, Laboratoire de Physique, F-69342 Lyon, France \\[2.5ex]

\par 

\end{center}

\vspace{40pt}

\centerline{\bf Abstract} \vspace{1cm}
\parbox{12cm}{\small}

This work extracts, by means of an exact analysis, the singular behaviour of the dynamical response functions -the  Fourier transforms of dynamical two-point functions-
in the vicinity of the various excitation thresholds  in the massless regime of the XXZ spin-1/2 chain. The analysis yields the edge exponents and associated amplitudes 
which describe the local behaviour of the response function near a threshold. The singular behaviour is derived starting from first principle considerations:
the method of analysis \textit{does not rely, at any stage}, on 
some hypothetical correspondence with a field theory or other phenomenological approaches.  
The analysis builds on the massless form factor expansion for the response functions of the XXZ chain obtained recently by the author. 
It confirms the non-linear Luttinger based predictions relative to the power-law behaviour and of the associated edge exponents 
which arise in the vicinity of the dispersion relation of one massive excitation (hole, particle or bound state). 
In addition, the present analysis shows that, due to the lack of strict convexity of the particles dispersion relation
and due to the presence of slow velocity branches of the bound states, there exist excitation thresholds with a different structure of edge exponents. These origin from multi-particle/hole/bound state excitations
maximising the energy at fixed momentum.

\vspace{40pt}

\tableofcontents

\section{An outline of the problem and main results}

\subsection{The XXZ chain}

Due to the substantial progress which took place in experimental condensed matter physics,  one-dimensional models of quantum many body physics evolved from a status of purely theoretical toy-models 
of many body physics to 
concrete compounds exhibiting a genuine one-dimensional behaviour. Even more remarkably, there exist a plethora of compounds 
whose properties are grasped, within a very good precision, by one-dimensional quantum integrable Hamiltonians. 
The most prominent example is probably given by the XXZ spin-$\tf{1}{2}$ chain in an external longitudinal magnetic $h$. The  Hamiltonian of the model  takes the form 
\beq
\op{H} \, = \, J \sum_{a=1}^{L} \Big\{ \sigma^x_a \,\sigma^x_{a+1} +
  \sigma^y_a\,\sigma^y_{a+1} + \De  \,\sigma^z_a\,\sigma^z_{a+1}\Big\} \, - \, \f{h}{2} \sul{a=1}{L} \sg_{a}^{z} \; . 
\label{ecriture hamiltonien XXZ}
\enq
Here  $J>0$ represents the so-called exchange interaction, $\De$ is the anisotropy parameter,
$h>0$ the external magnetic field and $L \in 2\mathbb{N}$ corresponds to the number of sites. $\op{H}$ acts on the Hilbert space $\mf{h}_{XXZ}=\otimes_{a=1}^{L}\mf{h}_a$ with $\mf{h}_a \simeq \Cx^2$,  
 $\sg^{w}$, $w=x,y,z$, are the Pauli matrices and the operator $\sg_a^{w}$  acts as the Pauli matrix $\sg^{w}$
on $\mf{h}_a$ and as the identity on all the other spaces, \textit{viz}. 
\beq
\sg_a^{w} \; = \; \underbrace{ \e{id}\otimes \cdots \otimes \e{id} }_{ a-1 \; \e{times} } \otimes \; \sg^{w} \otimes \underbrace{ \e{id}\otimes \cdots \otimes \e{id} }_{ L-a \; \e{times} }  \;. 
\enq
 Finally, the model is subject to periodic boundary conditions, \textit{viz.} $\sg_{a+L}^{\ga}=\sg_{a}^{\ga}$.

\vspace{2mm}

Crystals such as  $\e{K}\e{Cu}\e{F}_3$ \cite{NaglerTennantCowleyPerringSatijaTestOfXXXDispersionRelationForKCuF3}  or 
$\e{Cu}(\e{C}_4\e{H}_4\e{N}_2)(\e{N}\e{O}_3)_2$  \cite{HammarStoneReichBroholmGibsonLandeeOshikawaCu(C4H4N2)(NO3)2AsXXXMagnetIdentified} have been identified
 to be well-grasped by the isotropic XXX Hamiltonian, \textit{viz}. the Hamiltonian $\op{H}$ given in \eqref{ecriture hamiltonien XXZ} when $\De$ is set to $1$.  
In its turn,  the behaviour of $\e{CsCoCl}_3$ has been found to be well-captured \cite{GoffTennantNaglerCsCoCl3IndetifiedasMassiveXXZChain} by the XXZ antiferromagnetic Heisenberg chain with $\De \simeq 10$
while  certain aspects of the  behaviour of the spin-ladder compound $(\e{C}_5\e{H}_{12}\e{N})_2\e{Cu}\e{Br}_4$ are well-described \cite{BinerBoehmCauxGudelHabichtKieferKramerLauchliMcMorrowMesotNormandRueggRonnowStahnThielemann(C5H12N)2CuBr4AsDeltaHalfXXZChain}
by an effective XXZ Hamiltonian with $\De=1/2$. 

Most experiments on the above and many other effectively one-dimensional materials measure the Fourier transforms of two-point correlation functions -the so-called dynamic response functions (DRF)- and typically rely on techniques 
such as Bragg  \cite{StamperetalMeasureStructureFactorbyBraggSpectroscopyonBEC} or photoemission spectroscopy or inelastic neutron scattering 
\cite{LakeTennantCauxBarthelSchollwockNaglerFrostKCUF3DSFComparisionFromExpMEasureAndABA,StoneReichBroholmLefmannRischelLandeeTurnbullVeryClearDSFMeasureForXXXMagnetCu(C4H4N2)(NO3)2}.
In fact, most experiments take place at rather low temperatures, what effectively means that they measure, with good accuracy, the zero-temperature DRF. 
In the case of the XXZ chain, the zero temperature DRFs\symbolfootnote[4]{The connectedness of the correlator allows one to regularise the convergence of the transforms at infinity.} take the form 
\beq
\msc{S}^{(\ga)}(k,\om) \, = \, \sul{ m \in \mathbb{Z} }{} \Int{ \R }{} \big< (\sg_1^{\ga})^{\dagger}\!(t) \,  \sg_{m+1}^{\ga} (0) \big>_{\e{c}} \cdot \ex{ \i(\om t - k m) } \dd t  \;. 
\label{definition dyn resp fct general}
\enq
Above, $\dagger$ stands for the Hermitian conjugation, and 
the integrand refers to the presumably existing\symbolfootnote[3]{The results of \cite{KitanineMailletTerrasElementaryBlocksPeriodicXXZ,KozProofOfDensityOfBetheRoots,YangYangXXZproofofBetheHypothesis} put together entail 
the existence of the limit at $t=0$ in that they provide a rigorous derivation of a well-defined multiple integral representation for the reduced density matrix of the chain. An appropriate trace thereof 
allows one to compute $\big<  (\sg_1^{\ga})^{\dagger}\!(0) \,  \sg_{m+1}^{\ga} (0) \big>_{\e{c}}$. Note that the existence of the limit at $t=0$ also follows from the general theory developed in \cite{RuelleRigorousResultsForStatisticalMechanics}.
It is also fairly easy to see that the limit \eqref{definition comme limite volume infini correlateurs connexes} exists for extracted subsequences in $L$.}  infinite volume limit of the connected  dynamical two-point function at zero temperature
\beq
\big< (\sg_1^{\ga})^{\dagger}\!(t) \, \sg_{m+1}^{\ga}(0) \big>_{\e{c}} \; = \;
\lim_{L \tend + \infty} \bigg\{  \Big( \Om, (\sg_1^{\ga})^{\dagger}\!(t) \,   \sg_{m+1}^{\ga} (0) ,\Om \Big)  \, - \, \Big|   \Big( \Om,   \sg_{1}^{\ga} ,\Om \Big)  \Big|^2   \bigg\} \;. 
\label{definition comme limite volume infini correlateurs connexes}
\enq
 Here $\Om$ stands for the model's ground state while the time and space evolution of a spin operator takes the form 
\beq
\sg^{\ga}_{m+1}(t)\, = \, \ex{\i m \op{P} + \i \op{H}  t} \cdot \sg^{\ga}_1 \cdot \ex{ -\i t \op{H} -\i m \op{P}} \;, 
\enq
where $\op{P}$ is the momentum operator and, hence, $\ex{\i  \op{P} } $ the translation operator by one-site.

\subsubsection{Singularities of response functions}

Taken that dynamic response functions are natural experimental observables, there is a clear demand to build effective and reliable theoretical 
tools allowing for their study, at least in some limiting regimes, and  providing a satisfactory explanation of the experimental observations. 
Typically, dynamic response functions in one-dimensional models are observed to exhibit a singular structure in the momentum $k$ -- frequency $\om$ plane.
Namely, at fixed momentum $k$, they exhibit a power-law behaviour $(\de \om)^{\mu}$ in $\de \om=\om-\mc{E}(k)$, this  in the vicinity of certain curves $(k,\mc{E}(k))$. 
The curves $k\mapsto \mc{E}(k)$ correspond to dispersion relations of the excitations that are at the root of generating the given non-analytic behaviour. 
The edge exponent $\mu$ governing a given singularity may be positive or negative. The range of possible values of the edge exponent $\mu$ strongly depends on whether the model is in a massive or massless phase 
and, in the latter case, on the universality class governing the massless regime. 
In fact, the singular structure of the DRF, and in particular the form taken by the edge exponents is deeply connected with the critical exponents driving the long-distance and large-time  power-law decay 
of the real space correlators.  In the massive case, one expects, unless some non-generic accident happens, that this decay is driven by Gaussian saddle-points, be it in one or several dimensions.  
 Thus, in the massive case, the edge exponents are expected to be of the form $-\tf{1}{2}+n$, $n\geq 0$ an integer, the typical behaviour being
either a square root divergence or a square root cusp in the vicinity of the dispersion curves  $k\mapsto \mc{E}(k)$. The situation appears to be much richer in a massless model precisely due to the existence 
of infinitely many zero energy excitations. The latter generate a non-trivial tower of critical exponents which give rise
to edge exponents $\mu$ that, generically, exhibit a dependence on the momentum $k$, can be positive or negative and which are, generically, non-rational.

Ideally, one would like to have at one's disposal tools allowing one to unravel the mentioned singularity structure of the DRFs for a generic, not necessarily integrable,
one-dimensional model at zero temperature. The approach should also provide accurate and explicit enough predictions.

\subsection{The main achievement of the work}

A reasonable path for achieving the goal described above appears to start by devising exact tools allowing one to fully describe the singularity structure of the DRF
in at least some instances of quantum integrable models; indeed, then,  one can hope to rely on the exact solvability of the model which provides one with numerous
additional algebraic properties allowing one to simplify the calculations. As will be discussed below, such calculations could have been carried out, at least in part, for some examples of quantum integrable models. 
However, what would be really useful for the purpose of unravelling a larger picture would be to construct tools and a framework of analysis  allowing one to stay as close as possible to objects
and pictures usually used in condensed matter physics. The success of such an approach could then allow, by extrapolating the features responsible for the emergence of singularities in an
integrable model, to devise an exact phenomenological approach allowing one to grasp the universal part of the structure of DRFs, at least, in certain classes of non-integrable models. 
 By exact phenomenological approach, I mean one being able to produce an exact and analysable to the end expression for the DRF in which the building blocks 
 will be given by specific to the model -but not explicit- functions and such that the part responsible for the singular behaviour of the DRF is 
captured by a universal structure common to all models belonging to the universality class of interest. 
In conjunction with the representations that were obtained in my previous work \cite{KozMasslessFFSeriesXXZ},  this is precisely the program that is achieved in this work. 

By starting from the massless form factor\symbolfootnote[4]{I remind that a "form factor" refers to a matrix element of some local operator taken between two Eigenstates of the model's
Hamiltonian. Such objects are well-defined in finite volume $L$ as it is the case for the XXZ Hamiltonian \eqref{ecriture hamiltonien XXZ}. See \cite{KitanineMailletTerrasFormfactorsperiodicXXZ}
where finite-size determinant representation for these objects have been obtained} based representation, which I  
obtained in \cite{KozMasslessFFSeriesXXZ} for the  zero temperature DRF of the massless XXZ spin-1/2 chain, I develop a method of rigorous analysis of the behaviour of 
each multiple integral present in the series. While the construction of the series that was carried out in \cite{KozMasslessFFSeriesXXZ} relies  on a certain amount of hypotheses that are yet to be proven to hold, 
the analysis of each multiple integral carried out in this work is rigorous.

This allows me to extract the singular behaviour of the DFRs for the XXZ chain and hence determine, through an exact approach, the value of edge exponents $\mu$, singularities curves 
$k \mapsto \mc{E}(k)$ and amplitudes characterising the singularities in the $(k,\om)$ plane. Doing so, allows me to:
\begin{itemize}

\item[{\bf i)}] test and confirm the predictions, issuing form the existing heuristic methods, in respect to the structure of the subset of the singularities associated with one particle/hole/bound state excitations;

\item[{\bf ii)}] fully analyse the effect of multi-particle/hole/bound state processes in the generation of the excitation thresholds. These thresholds take origin in that the velocity of the excitations is \textit{not} monotonously increasing
and, more importantly, in that particles, holes  or bound states may share same values of their velocities. 
Such multi-particle thresholds were, so far, mostly unaccounted for within the existing heuristic methods and not all of the effects present at such thresholds were
fully grasped.

\end{itemize}

I stress that this is the first \textit{ab inicio} calculation of the singularities of the response functions in the XXZ spin-$1/2$ chain, an interacting integrable model containing bound states.

An important point is that the approach developed in the present work is \textit{universal} in the sense given earlier. I argued in  \cite{KozMasslessFFSeriesXXZ} that the massless form factor series 
expansions of the zero temperature DRF
that I obtained for the massless regime of the XXZ spin-$1/2$ chain has, in fact, a universal form that should be shared by all models belonging to the Luttinger liquid universality class.
I refer to that paper for a more precise discussion of that fact. 
As a consequence, the techniques of analysis -up to trivial modifications- developed in this work will allow one to grasp the singular structure of DRFs in  models belonging to the Luttinger liquid universality class.
Of course, these will then only be phenomenological results since, for a general model, one does not have an explicit access to form factor densities of local operators 
or to dispersion relations of the elementary excitations. However, such an approach is not so uncommon in physics and, more importantly,
the approximations made to get the result are genuinely constructive and do not rely on this or that heuristics which, in concrete situations, might turn out to be 
complicated to verify or even simply to have an intuition of. Furthermore, the data (form factor densities, dispersion relations) on which the phenomenological approach builds can, in principle, 
be computed perturbatively in the vicinity of a free theory, at least on formal grounds \cite{CauxGlazmanImambekovShasiNonUniversalPrefactorsFromFormFactors}.

\subsection{The principal theorems}

On technical grounds, the main achievement of this work are the two theorems given below. These results allow one to grasp the small parameter asymptotic expansion   
of a class of integrals which, upon specialisation, correspond to the one arising in the series expansion for the dynamic response functions obtained in \cite{KozMasslessFFSeriesXXZ}.  
In order to state these theorems, I first need to introduce a specific class of smooth functions. This definition involves smooth functions on closed set,
see Definition \ref{Definition fct lisse sur ferme} for a precise characterisation of the concept.

\begin{defin}
\label{definition smooth class on K}
Given $K$ a compact subset $\R^n$ for some $n\in \mathbb{N}^*$, a function  $\msc{G}$ on $K\times \R^+\times \R^+$ is said to be in the smooth class of $K$ associated with functions $d_{\pm}$ 
 and constant $\tau \in \intoo{0}{1}$, if there exists a decomposition 
\bem
\msc{G}\big( \bs{x}, u , v  \big) \, = \, d_{+}(\bs{x}) \, d_{-}(\bs{x}) \, \msc{G}^{(1)}\big( \bs{x}  \big) \; + \;    d_{-}(\bs{x}) \,  \msc{G}^{(2)}\big( \bs{x}, u  \big) \cdot [u]^{1-\tau}   \\
\; + \;    d_{+}(\bs{x}) \,  \msc{G}^{(3)}\big( \bs{x}, v  \big) \cdot  [v]^{1-\tau} \; + \;   \msc{G}^{(4)}\big( \bs{x}, u , v \big) \cdot [u\,  v]^{1-\tau} \;, 
\label{ecriture decomposition smooth class K}
\end{multline}
where   $\msc{G}^{(1)}$ is smooth on $ K $, $\msc{G}^{(2)}, \msc{G}^{(3)}$ are smooth and bounded on $ K \times \R^+$, 
$\msc{G}^{(4)}$ is smooth and bounded on $ K \times \R^+\times \R^+$.

\vspace{2mm}

  These functions are such that, for any $(\bs{s},\ell_u,\ell_v) \in \mathbb{N}^{n} \times\mathbb{N}\times \mathbb{N}$, 
$s\in \intn{1}{4}$ and $\eps>0$
\begin{itemize}

 \item[Hi)] $ \pl{a=1}{ n }   \Dp{ x_a }^{s_a }    \, \cdot \,  \Dp{u}^{\ell_u} \, \Big\{ \msc{G}^{(s)}\big( \bs{x}, u  \big)  [u]^{1-\tau}  \Big\} \, = \, \e{O}\Big( [u]^{1-\tau-\ell_u} \Big) $ 
 uniformly in $\bs{x}\in K$, $u\in \intof{ 0 }{ \eps^{-1} }$ and for $s=2,3$;
 
 \item[Hii)] $ \pl{ a=1 }{ n }\Dp{ x_a }^{s_a }   \, \cdot \,  \Dp{u}^{\ell_u} \, \cdot \,  \Dp{v}^{\ell_v} \, \Big\{ \msc{G}^{(4)}\big( \bs{x}, u , v  \big)  [u v]^{1-\tau}  \Big\}
 \, = \, \e{O}\Big( [u]^{1-\tau-\ell_u} [v]^{1-\tau-\ell_v} \Big)$  uniformly in $\bs{x}\in K$, $(u,v)\in \intof{ 0 }{ \eps^{-1} }^2$.

\end{itemize}
\vspace{2mm}

Finally, if $n \geq 2$, the functions $\msc{G}^{(s)}$, $s\in \intn{1}{4}$, along with any of their partial derivatives, all vanish on $\Dp{}K$, $\Dp{}K \times \R^+$, $\Dp{}K \times \R^+\times \R^+$.

\end{defin}

\vspace{2mm}

The first theorem deals with the case of one-dimensional integrals.

\vspace{2mm}

\begin{theorem}
\label{Theorem Principal cas 1D}

 Let $a<b$ be two reals. Let $\mf{z}_{\pm}(\la)$ be two real-holomorphic functions in a neighbourhood of the interval $\msc{J}=\intff{a}{b}$, such that
\begin{itemize}
\item all the zeroes of $\mf{z}_{\pm}$ on $\msc{J}$ are simple;

\item[$\bullet$] $\mf{z}_{+}$ and $\mf{z}_{-}$ admit a unique common zero $\la_0 \in \e{Int}(\msc{J})$ that, furthermore, is such that $\mf{z}_{+}^{\prime}(\la_0) \not= \mf{z}_{-}^{\prime}(\la_0)$. 
\end{itemize}

 Let $\De_{\ups}$ be real analytic on $\e{Int}(\msc{J})$ and such that $\De_{\ups} \geq 0$. Let $\msc{G}$ be in the smooth class of $\msc{J}$
associated with the functions $\De_{\pm}$ and with a constant $\tau$. 
Then, for $\mf{x}\not=0$ and small enough,
\beq
\la \mapsto  \msc{G}\Big(\la, \wh{\mf{z}}_{+}(\la), \wh{\mf{z}}_{-}(\la)\Big) \cdot \pl{\ups= \pm }{} 
\Big\{ \Xi\big( \; \wh{\mf{z}}_{\ups}(\la) \big) \cdot \big[ \, \wh{\mf{z}}_{\ups}(\la) \big]^{ \De_{\ups}(\la)-1 } \Big\} \in L^{1}\big( \msc{J} \big)
\enq
where $\wh{\mf{z}}_{\pm}(\la)=\mf{z}_{\pm}(\la)+ \mf{x}$. Let $\mc{I}(\mf{x})$ denote the integral
\beq
\mc{I}(\mf{x})\, = \, \Int{ \msc{J} }{}   \msc{G}\Big(\la, \wh{\mf{z}}_{+}(\la), \wh{\mf{z}}_{-}(\la)\Big) \cdot \pl{\ups= \pm }{} 
\Big\{ \Xi\big( \; \wh{\mf{z}}_{\ups}(\la) \big) \cdot \big[ \, \wh{\mf{z}}_{\ups}(\la) \big]^{ \De_{\ups}(\la)-1 } \Big\} \cdot \dd \la \;. 
\enq

 \vspace{2mm}
Assume that $ \de_{\pm} \, = \,  \De_{\pm}(\la_0) >0$. 

\vspace{2mm}
 
 \noindent {\bf a) } If $\mf{z}_{+}^{\prime}(\la_0)\cdot \mf{z}_{-}^{\prime}(\la_0) <0$, then  $\mc{I}(\mf{x})$ admits the $\mf{x} \tend 0$ asymptotic expansion
\beq
\mc{I}(\mf{x})\, = \, \Xi\Big(   \mf{z}_{+}^{\prime}(\la_0) \cdot   \mf{X}  \Big) \cdot  \Bigg\{ 
\f{  \msc{G}^{(1)}(\la_0) \cdot \de_+ \de_- \cdot |  \mf{X}  |^{\de_+ + \de_- - 1}     }
{   |\, \mf{z}_{+}^{\prime}(\la_0) |^{\de_-} \cdot  | \, \mf{z}_{-}^{\prime}(\la_0)  |^{\de_+}       }
\cdot \f{ \Ga\big( \de_+ \big) \cdot \Ga\big(\de_- \big)  }{  \Ga\big( \de_++\de_-\big)  }
   \, + \, \e{O}\Big(   |\mf{x}|^{ \de_+ + \de_- - \tau} \Big)  \Bigg\}
 \, + \, f_{<}(\mf{x})
\enq
where 
\beq
 \mf{X}  \, = \,  \mf{x} \cdot  \big[ \mf{z}_{+}^{\prime}(\la_0)-\mf{z}_{-}^{\prime}(\la_0) \big]  \;, 
\enq
$\msc{G}^{(1)}$ is as appearing in \eqref{ecriture decomposition smooth class K}
and $f_{<}$ is a smooth function of $ \mf{x}$. Furthermore, if $\mf{z}_{\pm}$ have no zeroes on $\msc{J}$  other than $\la_0$, then $f_{<}=0$.  

 \vspace{2mm}
 
 \noindent {\bf b) }  If $\mf{z}_{+}^{\prime}(\la_0)\cdot \mf{z}_{-}^{\prime}(\la_0) >0$, then  $\mc{I}(\mf{x})$ admits the $\mf{x} \tend 0$ asymptotic expansion
\bem
\mc{I}(\mf{x})\, = \, 
\f{  \msc{G}^{(1)}(\la_0) \cdot  \de_+ \, \de_- \cdot  |  \mf{X}  |^{\de_+ + \de_- - 1}      }
{   |\, \mf{z}_{+}^{\prime}(\la_0) |^{\de_-}   \cdot | \, \mf{z}_{-}^{\prime}(\la_0)  |^{\de_+}       }
\cdot  \Ga\big( \de_+ \big) \cdot \Ga\big(\de_- \big) \cdot \Ga\big( 1- \de_+  - \de_- \big)   \\
\times \bigg\{ \Xi(\mf{x}) \tfrac{1}{\pi} \sin\big[ \pi \de_{\mf{p}} \big] \, + \,  \Xi(-\mf{x}) \tfrac{1}{\pi} \sin\big[ \pi \de_{-\mf{p}} \big]  \bigg\}
\, + \, \e{O}\Big(   |\mf{x}|^{ \de_+ + \de_- - \tau } \Big)
\; + \; f_{>}(\mf{x}) 
\end{multline}
where $ \mf{X} $ and $\de_{\pm}$ are as above,
\beq
\mf{p}\, = \, - \e{sgn} \big[ \mf{z}_{+}^{\prime}(\la_0) \big] \cdot \e{sgn} \big[ \mf{z}_{+}^{\prime}(\la_0)-\mf{z}_{-}^{\prime}(\la_0) \big]  
\enq
and $f_{>}$ is a smooth function of $ \mf{x}$.

\end{theorem}

The second theorem, deals with a multi-dimensional analogue of the integral given in \eqref{definition integrale 1D type Beta genralise}. Its statement demands to introduce a few notations and objects. One assumes to be given:
\begin{itemize}
\item a strictly positive real $\op{v} >0$;
\item a choice of signs $\zeta_r\in \{ \pm \}$;
\item  a collection of compact intervals $\msc{I}_{r}$, $r=1,\dots, \ell$ ;
\item smooth functions $\mf{u}_r$ on $\msc{I}_r$ such that $\mf{u}^{\prime}_r$ is strictly monotonous on $\msc{I}_r$, and such that 
\beq
\mf{u}^{\prime}_r(k) \not= \pm \op{v}  \qquad \e{for} \qquad k \in \e{Int}\big(\msc{I}_{r} \big) \;. 
\enq

\end{itemize}

The intervals $\msc{I}_r$ are such that they partition as 
\beq
\msc{I}_{r}\; = \; \msc{I}_{r}^{(\e{in})}\sqcup \msc{I}_{r}^{(\e{out})} \qquad \e{with} \qquad \msc{I}_{1}\; = \; \msc{I}_{1}^{(\e{in})}
\enq
so that 
\beq
\mf{u}_{r}^{\prime}\Big( \e{Int}\big(\msc{I}_{r}^{(\e{out})} \big) \Big) \, \cap  \, \mf{u}_{1}^{\prime}\Big( \e{Int}\big(\msc{I}_{1}^{(\e{in})} \big) \Big) \, = \, \emptyset
\qquad \e{and} \qquad 
\mf{u}_{r}^{\prime}\Big( \e{Int}\big(\msc{I}_{r}^{(\e{in})} \big) \Big) \, =  \, \mf{u}_{1}^{\prime}\Big( \e{Int}\big(\msc{I}_{1}^{(\e{in})} \big) \Big) \;. 
\enq

The  above ensures that there exist homeomorphisms 
\beq
t_r \, : \, \msc{I}_{1}^{(\e{in})} \tend \msc{I}_{r}^{(\e{in})} \qquad \e{such} \; \e{that} \qquad 
\mf{u}_1^{\prime}(k) \; = \; \mf{u}_r^{\prime}\big( t_r(k) \big) \;.
\enq

\vspace{2mm}

One defines the  macroscopic "momentum" and "energy" as 
\beq
\mc{P}(k) \; = \; \, \sul{r=1}{\ell }n_r \, \zeta_r  \, t_r(k)   \qquad \e{and} \qquad \mc{E}(k) \; = \; \sul{r=1}{\ell } n_r  \, \zeta_r \,  \mf{u}_r\big( t_r(k) \big)  \;\; ,  \qquad k\in  \msc{I}_{1} \; .
\enq
 It is assumed that $k\mapsto \mc{P}(k)$ is strictly monotonous on $\e{Int}(\msc{I}_1)$.

\begin{theorem}
\label{Theorem Principal}

Let $\msc{I}_{\e{tot}}= \msc{J}_1^{n_1}\times \cdots \times \msc{J}_{\ell}^{n_{\ell}}$ and 
$ \De_{\pm} $ be smooth positive functions on $ \msc{I}_{\e{tot}} $ admitting smooth square roots on $ \msc{I}_{\e{tot}} $.  
Let $\msc{G}$ be in the smooth class of $ \msc{I}_{\e{tot}} $ associated with the functions $\De_{\pm}$ and a constant $\tau \in \intoo{0}{1}$, 
\textit{c.f.} Definition \ref{definition smooth class on K}. 

Finally, let  
\beq
\mf{z}_{\ups}(\bs{p})\;=\; \mc{E}_0 \, -\,  \sul{ r=1}{\ell}\sul{a=1}{n_r}  \zeta_r \mf{u}_r\big( p_a^{(r)} \big)  \, +  \; \ups  \op{v} \, \bigg\{ \mc{P}_0 -  \sul{ r=1}{\ell}\sul{a=1}{n_r}  \zeta_r   p_a^{(r)}  \bigg\} \; , 
\quad \ups \in \{ \pm \} , 
\enq
with $\zeta_r \in \{\pm 1\}$ and where $(\mc{P}_0,\mc{E}_0) \in \R^2$.

\noindent Let $\mc{I}\big( \mf{x} \big)$ correspond to the multiple integral 
\beq
\mc{I}\big( \mf{x} \big) \, = \,\pl{r=1}{\ell}  \bigg\{ \Int{  \msc{I}_r^{n_r}   }{}  \dd \bs{p}^{(r)} \bigg\}  \; \msc{G}_{\e{tot}}(\bs{p})   
\qquad with \qquad 
\bs{p} \; = \; \big( \bs{p}^{(1)} , \dots ,  \bs{p}^{(\ell)} \, \big) \; \; , \;\;  \bs{p}^{(r)}  \in \msc{I}_r^{n_r} \;, 
\enq
where  
\beq
\msc{G}_{\e{tot}}(\bs{p})  \, = \, \msc{G}\Big( \bs{p},  \mf{z}_{+}(\bs{p})+ \mf{x} , \mf{z}_{-}(\bs{p})+ \mf{x}  \Big)  \cdot \pl{\ups=\pm }{} \bigg\{ \Xi\Big( \,  \mf{z}_{\ups}(\bs{p})+ \mf{x}  \Big)
\cdot \Big[ \,   \mf{z}_{\ups }(\bs{p})+ \mf{x}  \Big]^{ \De_{\ups}(\bs{p}) -1 }    \bigg\} \cdot \pl{r=1}{\ell} \pl{a<b}{n_r} \big( p_a^{(r)}-p_b^{(r)} \big)^2  \;. 
\enq
The type of $\mf{x}\tend 0$ asymptotic expansion of $\mc{I}(\mf{x})$ depends on the value of $(\mc{P}_0,\mc{E}_0)$.

\vspace{2mm} 

 {\bf a)} \texttt{The regular case.}

\vspace{2mm} 
 \noindent If the two conditions given below  hold 
\beq
\big( \mc{P}_0,\mc{E}_0 \big) \, \not\in \, \Big\{  \big( \mc{P}(k),\mc{E}(k) \big) \; : \; k \in \msc{I}_1   \Big\}  
\enq
and 
\beq
\underset{ \substack{ \a \in \Dp{}\msc{I}_{1}  \\ \ups=\pm}  }{\e{min}} \big| \mc{E}_0\,-\, \mc{E}(\a) +  \ups \op{v}\, (\mc{P}_0\,-\, \mc{P}(\a))\big| \; > \; 0
\label{ecriture condition positivite energie impulsion macroscopique thm principal}
\enq
then $\msc{G}_{\e{tot}} \in L^1\big( \msc{I}_{\e{tot}} \big)$ and  $\mc{I}\big( \mf{x} \big) $ is smooth in  $\mf{x}$, for $|\mf{x}|$ small enough.

\vspace{2mm} 
  {\bf b)}  \texttt{The singular case.} 

\vspace{2mm}  \noindent Let $k_0 \in \e{Int}(\msc{I}_1)$ 
\beq
\De_{\ups}^{(0)} \, = \, \De_{\ups}\big( \bs{t}(k_0) \big) \qquad and \qquad 
\vth \; = \; \f{1}{2} \sul{r=1}{\ell}n_r^2 \; - \; \f{3}{2}   \, + \, \De_{+}^{(0)} \, + \, \De_{-}^{(0)} \;, 
\enq
with 
\beq
\bs{t}(k_0)\, = \, \big( \bs{t}_1(k_0), \dots, \bs{t}_{\ell}(k_0) \big) \in \R^{\ov{\bs{n}}_{\ell}} \qquad with \qquad \bs{t}_r(k_0) \, = \, \big( t_r(k_0), \dots, t_r(k_0) \big) \in \R^{n_{r}} \;. 
\enq

If 
\beq
\big( \mc{P}_0,\mc{E}_0 \big) \, = \,   \big( \mc{P}(k_0),\mc{E}(k_0) \big)    \;, 
\quad \vth \not\in \mathbb{N} \;, \quad and \quad \De_{ \pm }^{(0)} >0
\enq
then $\msc{G}_{\e{tot}} \in L^1\big( \msc{I}_{\e{tot}} \big)$ and   $\mc{I}(\mf{x})$ admits the $\mf{x}\tend 0^+$ asymptotic expansion:
\bem
\mc{I}\big( \mf{x} \big)   \; = \; 
  \f{ \De_{ +}^{(0)} \, \De_{ - }^{(0)}\,  \mc{G}^{(1)}\big( \bs{t}(k_0) \big)  \cdot  \big(2 \op{v}  \big)^{\De_{ +}^{(0)} + \De_{-}^{(0)} - 1  } }
{ \sqrt{ | \mc{P}^{\prime}(k_0) | } \cdot  \pl{\ups=\pm }{} \big| \op{v} - \ups \mf{u}_1^{\prime}(k_0)  \big|^{ \De_{ \ups }^{(0)} }  }   
   \cdot \Ga\big( \De_{ +}^{(0)} \big)   \Ga\big( \De_{ -}^{(0)} \big) \Ga\big( - \vth   \big)  \cdot 
   \pl{r=1}{\ell} \Bigg\{   \f{ G(2+n_r) \cdot \big( 2\pi\big)^{\frac{ n_r - \de_{r,1} }{2} }  }{   \big|\mf{u}_{r}^{\prime\prime}(t_r(k_0))\big|^{ \frac{1}{2} ( n_r^2 - \de_{r,1} )  }  } \Bigg\}                \\ 
\times  
|\mf{x}|^{ \vth   }    \cdot \Bigg\{ \Xi(\mf{x}) \f{ \sin \big[ \pi  \nu_{+}\big] }{\pi}   \, + \,  \Xi(-\mf{x})\f{ \sin \big[ \pi  \nu_{-}\big] }{\pi}   \Bigg\}  
\, + \,     \mf{r}(\mf{x})  \, + \, \e{O} \Big( |\mf{x}|^{ \vth  + 1 -\tau  }   \Big)\;.
\end{multline}
Above $\mf{r}(\mf{x})$ is smooth in $\mf{x}$, for $|\mf{x}|$ small enough.
Finally, 
\beq
\nu_{ \pm } \; = \;  \f{1}{2}\sul{  \substack{ r=1 \, : \, \\ \veps_r= \mp 1} }{ \ell }  n_r^2 \; - \; \f{ 1 \mp \mf{s} }{ 4 }    
\; +  \hspace{-4mm}  \sul{ \substack{ \ups=\pm  \, : \, \\   \pm [ \op{v} - \ups \mf{u}_1^{\prime}(k_0) ]>0 } }{}  \hspace{-4mm} \De_{ \ups }^{(0)}
\enq
where $\mf{s}=-\e{sgn}\Big(  \frac{ \mc{P}^{\prime}(k_0) }{  \mf{u}_{1}^{\prime\prime}(k_0) }  \Big) $ and $\veps_r=-\zeta_r \e{sgn}\Big(\mf{u}^{\prime\prime}_{r}(t_r(k_0)) \Big) $\;.

\end{theorem}

\subsection{Outline of the paper}

This paper is organised as follows. This is the Introduction. Sub-section \ref{SousSSection Histoire de analyse fct rep dyn} to come  reviews the 
various developments  that took place in the analysis of the dynamic response functions of one-dimensional models. 
Section \ref{Section main results} contains a short review of the structure of excitations in the model followed by 
a discussion of the obtained results in the simple case of the singular structure of the at most two-particle/hole contribution to the longitudinal DRF  $\msc{S}^{(z)}(k,\om)$. 
Finally, this section closes on the description of the series of multiple integrals representation for $\msc{S}^{(\ga)}(k,\om)$
derived in  \cite{KozMasslessFFSeriesXXZ}.  In particular, I discuss the various properties enjoyed by the integrands of the multiple integrals building up the series. 
The singular behaviour of each multiple integral, \textit{viz}.  summands arising in the series, is then extracted, for the most typical excitations, in Section \ref{Section Edge singular behaviour des fcts spectrales}. 
All the technical details necessary for obtaining these results are relegated to several appendices. 
Appendix \ref{Appendix Fcts speciales} lists the main notations contained in this work. Appendix \ref{Appendix auxiliary theorems} recalls the statements of four theorems, 
the Morse lemma, the Weierstrass  and the Malgrange preparation theorems as well as the Whitney extension theorem. All these will 
be used in the core of the analysis developed in Appendices \ref{Appendix DA integrales unidimensionnelles} and \ref{Appendix DA integrales multidimensionnelles}. 
Appendix \ref{Appendix Observables XXZ} recalls the properties of interest of certain observables in the XXZ chain. 
Sub-appendix \ref{Appendix Lin Int Eqns Defs et al} recalls the linear integral equation based description of the observables 
in the XXZ chain. Sub-appendix \ref{Appendix Section phase oscillante dpdte de la vitesse} discusses the properties of the 
velocity of the particles and hole excitations that play an important role in the analysis. 
Appendices \ref{Appendix DA integrales unidimensionnelles} and \ref{Appendix DA integrales multidimensionnelles} are devoted to a detailed analysis of the 
asymptotic behaviour of auxiliary integrals whose understanding is necessary for obtaining the \textit{per se} singular behaviour of the DRF studied in Section 
 \ref{Section Edge singular behaviour des fcts spectrales}. 
Appendix \ref{Appendix DA integrales unidimensionnelles} is devoted to the analysis of the asymptotics of a generalisation of one-dimensional Euler $\be$-integrals while 
Appendix \ref{Appendix DA integrales multidimensionnelles} carries such an analysis relatively to a multi-dimensional generalisation of one-dimensional $\be$ integrals. 
The rigorous analysis developed in Appendices \ref{Appendix DA integrales unidimensionnelles} and \ref{Appendix DA integrales multidimensionnelles} constitutes
the main technical achievement of this work. Finally, Appendices \ref{Appendix Section Auxiliary results} and \ref{Appendix Section Asymptotics of model integral}
develop several technical results that are needed so as to carry out the analysis developed in Appendix \ref{Appendix DA integrales multidimensionnelles}.

\subsection{Some history of the analysis of dynamic response functions}
\label{SousSSection Histoire de analyse fct rep dyn}

\subsubsection{Heuristic approaches}

There is clearly little hope, for a generic one-dimensional model, to extract the singular structure of DRFs  by means of direct, \textit{ab inicio}, calculations. 
Still, over the years, there emerged various approximation techniques allowing one to analyse certain features of such a singular behaviour. 
In the massive case, the singularities of the DRFs appear to be controlled by Van Hove singularities and this completely catches the aforementioned behaviour. 
A whole lot more  attention was dedicated to the massless case where one expects a much richer behaviour and where no such simple explanation exists. 

To start with, one can argue that the equal-time long-distance asymptotics of the correlators in a massless model should be grasped by putting the model in 
correspondence with a Luttinger liquid \cite{LutherPeschelCriticalExponentsXXZZeroFieldLuttLiquid} or, more generally, with a conformal field theory (CFT) \cite{BeliavinPolyakovZalmolodchikovCFTin2DQFT}. 
Mappings of this kind are built by looking at the momentum and energy of the low-lying excited states above the 
ground state of the model \cite{BloteCardyNightingalePredictionL-1correctionsEnergyAscentralcharge,CardyConformalDimensionsFromLowLSpectrum} 
from where one can read-off the scaling dimensions of the operators on the CFT side which give access to the critical exponents arising in the equal-time long-distance 
asymptotic behaviour of the zero temperature correlation functions in the original model. In its turn this allows one to argue the 
behaviour of the Fourier transforms in the vicinity of the point $(k,\om)=(0,0)$.



The situation becomes much more involved if one would like to grasp, at least qualitatively, the behaviour of DRFs in the whole $(k,\om)$ plane. 
Indeed, then,  it becomes necessary to take into account certain of the non-linearities in the spectrum of the model's excitations. 
A first phenomenological description of the DRF's singularities in the $(k,\om)$ plane was argued by Beck, Bonner and Müller \cite{BeckBonnerMullerFenomenologicalFormDSFinXXX} in 1979. The approach 
was substantially developed one year later by these authors and Thomas \cite{BeckBonnerMullerThomasSpectralFctsXXXGeneralFeatures}
for the XXX Heisenberg spin-$1/2$ chain at zero magnetic field. These authors also proposed heuristic reasonings based on selection rules so as to predict 
some of the features of the DRF in the presence of a non-zero magnetic field.  
A substantial progress towards the setting of an operative phenomenological approach occurred, however, only in the mid '00.  In 2006, Glazmann, Kamenev, Khodas and Pustilnik \cite{GlazmanKamenevKhodasPustilnikNLLLTheoryAndSpectralFunctionsFremionsFirstAnalysis} 
managed to take into account the non-linearities in the dispersion relation of one-dimensional spinless fermions and argued, in the case of the density structure factor\symbolfootnote[3]{The latter corresponds to 
$\msc{S}^{(z)}(k,\om)$ in the case of the XXZ chain, \textit{c.f.} \eqref{definition dyn resp fct general}.}, 
the presence of a singular behaviour along single particle $k\mapsto \mf{e}_p(k)$ or hole $k\mapsto \mf{e}_h(k)$  excitations thresholds characterised by a non-trivial, \textit{viz}. 
differing from a half-integer, edge exponents $\mu$. 
Next year, the authors generalised their approach in \cite{GlazmanKamenevKhodasPustilnikNLLLTheoryAndSpectralFunctionsFremionsBetterStudy} so as to encompass other DRFs
and computed perturbatively the edge exponents arising in the $\de$-function Bose gas in \cite{GlazmanKamenevKhodasPustilnikDSFfor1DBosons}. 
More explicit results appeared later on. Pustilnik \cite{PustilnikDSFForCalogero} built on the expressions for the  exact form factors in the Calogero-Sutherland model
so as to unravel the singular behaviour of the density structure factor in that model. Further, building on the explicit expressions for the spectrum of excitations provided by the Bethe Ansatz,
Glazmann and Imambekov \cite{GlazmanImambekovComputationEdgeExpExact1DBose} proposed closed expressions for the edge exponents arising in the $\de$-function Bose gas,
while Cheianov and Pustilnik  \cite{CheianovPustilnikXXZLoweredgeNLLLEdgeExp} argued the expression for the edge exponents associated with the lower threshold -corresponding to one hole excitations- 
in the massless regime of the XXZ spin-1/2 chain. 
Such kinds of predictions for the edge exponents were generalised, in 2008-09 by Affleck, Pereira and White  \cite{AffleckPereiraWhiteEdgeSingInSpin1-2,AffleckPereiraWhiteSpectralFunctionsfor1DLatticeFermionsBoundStatesContributions}, 
to various other thresholds present in the XXZ chain.
In  \cite{GlazmanImambekovDvPMTCompletTheoryNNLL}, Glazmann and Imambekov advocated the manifestation of various universal behaviours in the amplitudes appearing in front of the power-law
behaviour $(\de \om)^{\mu}$ of the DRF, hence providing a firm ground to the so-called non-linear Luttinger liquid theory supposed to govern the edge singular behaviour of 
dynamic response functions in massless models. 
I refer to the review \cite{GlazmanImambekovSchmidtReviewOnNLLuttingerTheory} and references therein for a broader discussion of that approach.
Similarly to the case of the edge exponents, Caux, Imambekov, Panfil and Shashi \cite{CauxImambekovPanfilShashiHeuristicsPrefactorsForEdgeExpInIntModels}, 
by building on recent techniques pioneered in \cite{KozKitMailSlaTerEffectiveFormFactorsForXXZ,KozKitMailSlaTerThermoLimPartHoleFormFactorsForXXZ,SlavnovFormFactorsNLSE} and allowing one to study the large-volume behaviour of form factors
of local operators in quantum integrable models, argued the expressions of the amplitudes in front of the singular power-law behaviour of the DRF in the case of the XXZ spin-1/2 chain, the $\de$-function Bose gas and 
the Calogero-Sutherland model. 
One should also mention that, more recently, the lower thresholds present in DRFs of the spin-$\tf{1}{2}$ XXX Heisenberg chain where analysed,
by Campbell, Carmelo, Machado and Sacramento \cite{CampbellCarmeloMachadoSacramentoLowerTresholdsXXXLongAndTransDSFPseudoFermionDynTheor}, within the pseudofermion dynamical theory 
 and by taking the Bethe Ansatz issued input for the energies. 
 
 \subsubsection{Exact approaches}

The heuristic approaches described above appear quite powerful. 
 It is necessary to check and test the limits of applicability of the mentioned methods \textit{versus} results stemming from exact, \textit{ab inicio},
 calculations of DRF and the extraction of their singularities, carried out on quantum integrable models. 
Obtaining such exact results constituted a hard and long-standing problem, despite that numerous techniques of exact computations of correlation functions have been developed
after the invention of the algebraic Bethe Ansatz \cite{FaddeevSklyaninTakhtajanSineGordonFieldModel} on the one hand 
and of the vertex operator approach \cite{DavisFodaJimboMiwaNakayashikiDiagonalizationXXZinfiniteDelta>1} on the other hand. 

First results relative to DRFs appeared for free fermion equivalent models. The density structure factor, \textit{viz}. the longitudinal Fourier transform $\msc{S}^{(z)}(k,\om)$, 
of the XX chain was computed in a closed form by Beck, Bonner, M\"{u}ller and Thomas \cite{BeckBonnerMullerThomasSpectralFctsXXXGeneralFeatures} in  1981. 
The case of transverse response functions was much harder, even for the XX chain, due to the much more involved
structure of the transverse correlators. An analysis of the power-law divergencies in $\om$ for the transverse frequency Fourier transform $\int_{0}^{2\pi}\msc{S}^{(x)}(k,\om)\cdot \dd k $
of the XX chain was achieved in 1984 by M\"{u}ller and Shrock \cite{MullerShrockDynamicCorrFnctsTIandXXAsymptTimeAndFourier} by exploiting the 
connection between the associated two-point function and Painlevé transcendents.

The development of the vertex operator approach  \cite{DavisFodaJimboMiwaNakayashikiDiagonalizationXXZinfiniteDelta>1} in the 
mid '90s allowed for a substantial progress in the computation of the correlation functions in an interacting, \textit{viz}. 
away from the free fermion point, quantum integrable model, namely for the XXZ chain in its massive regime, \textit{i.e.} the Hamiltonian \eqref{ecriture hamiltonien XXZ} for $\De>1$, this in presence of a zero external magnetic field $h=0$. 
In 1995, Jimbo and Miwa \cite{JimboMiwaFormFactorsInMassiveXXZ} obtained 2n-fold multiple integral representations for the form factors of local operators
of the chain taken between the ground state and an excited state containing $2n$-spinon excitations.
Although initially obtained for the XXZ chain at $\De>1$, these integral representation admitted a regular $\De \tend 1^+$ limit, hence yielding the corresponding expressions for the XXX Heisenberg chain. 
The construction of integral representations for the form factors opened the possibility to estimate the DRF of the XXZ chain at $\De \geq 1$
by taking explicitly the space and time Fourier transforms of the form factor series. Doing so allowed Bougourzi, Karbach and M\"{u}ller \cite{BougourziKarbachMuller2SpinonDSFMassiveXXZFromVopExplicit}  
to obtain, in 1998, the two-spinon sector contribution $\msc{S}^{(x)}_2(k,\om)$ to the transverse dynamic response function  $\msc{S}^{(x)}(k,\om)$ in the massive regime of the XXZ chain. 
This analysis was  revisited and corrected by Caux, Mossel and Perez-Castillo \cite{CauxMosselPerezCastillo2SpinonDSFMassiveXXZReloaded} in 2008
what allowed them to explain the presence of an asymmetry in these DRF.
Relatively to singularities, the bottom line of these investigations is that  $\msc{S}^{(x)}_2(k,\om)$ exhibits square root cusps or singularities along two-spinon excitations thresholds -as expected from a DRF of a massive model-. 
The two-spinon contribution to $\msc{S}^{(x)}(k,\om)$ in the case of the XXX chain was computed by Bougourzi, Couture, Kacir 
 \cite{BougourziCoutureKacirAll2SpinonDSFforXXXFromVOpsColseFormulae} in 1996. Building on these results, Bougourzi, Fledderjohan, Karbach, M\"{u}ller and M\"{u}tter \cite{BougourziFledderjohanKarbachMullerMutterDynamicTwoSpinonStructureFactor}
have shown in 1997 that the two-spinon sector saturates \textit{ca}. 73\%  of the total intensity of $\msc{S}^{(x)}(k,\om)$. They carried as well  
a thorough analysis of the singularity structure of this DRF,  showing the presence of a square root cusp behaviour on the upper two-spinon treshold and 
a square root divergence on the lower-treshold (plus a logarithmic behaviour). 
Although the complexity of the integral representations for the higher than 2 spinon sector  form factors makes the computations more involved, 
Abada, Bougourzi, SiLakhal \cite{AbadaBougourziSiLakhalFourSpinonDSFXXXMultInt} and, later, Caux and Hagemans \cite{CauxHagemansDSFXXXFourSpinonContributionDeeperAnalysis} 
still managed to deal with the four spinon contributions to the XXX DRFs. 
Finally, in 2012, Caux, Konno, Sorrel and Weston \cite{CauxKonnoSorrelWestonFFofMasslessXXZfromXYZResults} managed to compute explicitly the two-spinon contribution to the XXZ chain directly in the massless regime 
and at $h=0$ by using earlier results of Lashkievich and Pugai \cite{LashkevichPugaiFormFactorsEightVertex} and later rewritings thereof. 
Here, much in the spirit of the results for the XXX case, the DRF were obtained by first starting from the integral representations for the two-spinon form factors in a massive model (the XYZ chain 
in this case) and then by taking an appropriate massless scaling limit thereof. Again, the analysis unraveled the presence of square root cusps or divergences, depending on the spinon tresholds. 
However, due to the much more involved structure on the XYZ chain side, no result exists so far for the higher than two spinon contribution to the form factor series in the massless 
regime of the XXZ chain at $h=0$.

 \subsubsection{Numerical and Bethe Ansatz based approaches}

All the exact results mentioned so far were obtained in a zero external magnetic field. The obtention of exact results in the presence of a non-zero  magnetic field
turned out to be much more involved. Nonetheless, it was possible to estimate the response functions numerically. 
First numerical plots of the longitudinal and transverse DRF in the XXX chain at $h\not=0$ were obtained by Karbach and M\"{u}ller
\cite{KarbachMullerDSFXXXSelectionRuleAndPlotsFromCBA} in 2000 and then by Biegel, Karbach and M\"{u}ller \cite{BiegelKarbachMullerDSFXXXNumericsFromFormCBAEigenvectors} in 2002.
The plots were obtained by means of a brute force numerical evaluation of the matrix elements of 
local operators which, in their turn, were computed by using the coordinate Bethe Ansatz representation for the Eigenfunctions of the chain. 
A qualitative and quantitative step forward of the numerical approach was enabled by the construction of determinant representations for the form factors
of local operators in the XXZ chain by Kitanine, Maillet and Terras \cite{KitanineMailletTerrasFormfactorsperiodicXXZ} in 1998. Using such representations which 
remarkably simplified the numerics, Biegel, Karbach and M\"{u}ller \cite{BiegelKarbachMullerDSFXXXLongItudandTransverseAtqPiPisur2VariousExcitationClasses} 
obtained in 2002 plots of the longitudinal and transverse response functions at fixed momentum $k \in \big\{ \pi, \tfrac{1}{2}  \pi \big\}$
for the XXX chain at finite magnetic field and by distinguishing the contributions of various classes of excitations. Again, for the same values of the momentum,
the two-spinon contribution to the longitudinal response function, at various values of the anisotropy $1>\De>0$ of the XXZ chain,
was evaluated numerically by Biegel, Karbach and M\"{u}ller \cite{BiegelKarbachMullerDSFXXZTransverseAtqPiPisur2VariousvaluesDelta} in 2003.
Then, Sato, Shiroishi and Takahashi \cite{SatoShiroishiTakahashiDSFLongitudXXZABA} obtained in 2004 plots at fixed momentum $k=\tf{\pi}{2}$ and energy-momentum plots of
the two-spinon contribution to the longitudinal response function at half-saturation field in the massless regime of the XXZ chain, this for various values of the anisotropy. 
In 2005, Caux and Maillet  \cite{CauxMailletDynamicalCorrFunctXXZinFieldPlots} and then Caux, Maillet and Hagemans \cite{CauxHagemansMailletDynamicalCorrFunctXXZinFieldPlots}
obtained $(k,\om)$ plots of the multi-particle and bound state contribution to the longitudinal $\msc{S}^{(z)}(k,\om)$ and transverse $\msc{S}^{(+)}(k,\om)$ DRF.
Similar numerics related to the $\msc{S}^{(-)}(k,\om)$ response function for the XXX chain were carried out by Kohno \cite{KohnoDynamDominantStringExcitationXXZChainVariousDSFXXXFiniteandZeroh} in 2009,
for various values of $h$. In particular, this work has shown that, for the $\msc{S}^{(-)}(k,\om)$ DRF, the two and three string bound states carry a certain non-negligible part of the spectral weight,
as opposed to the $\msc{S}^{(z)}(k,\om)$ and $\msc{S}^{(+)}(k,\om)$ response functions where most of the spectral weight is carried by particle-hole excitations. 
A similar type of numerical analysis was performed in 2006 for the DRFs of the $\de$-function Bose gas by Caux and Calabrese  \cite{CauxCalabreseDynamicalStructureFactoBoseGas} and in 2007 by
Caux, Calabrese and Slavnov \cite{CauxCalabreseSlavnovSpectralFunctionBoseGas}.

 \subsubsection{The restricted sum approach}
 
A breakthrough in the exact analysis of certain regimes of form factor expansions of two-point functions in the massless regime of 
the XXZ chain was achieved by Kitanine, Maillet, Slavnov, Terras and myself  \cite{KozKitMailSlaTerRestrictedSums} in 2011.  
In that work, we proposed a way to sum up the expansion of XXZ's static two-point correlation functions over the 
so-called  critical\symbolfootnote[2]{Expectation values of local operators taken between the ground state 
and the low-lying excited states exhibiting a conformal structure of their energies.} form factor. 
By heuristically arguing that only such form factors should contribute to the leading order of the large-distance 
asymptotic behaviour of the two-point functions in this chain, we have been able to compute the amplitude and critical exponent of the
leading term associated to every harmonic arising in the long-distance behaviour. 
Owing to the sole presence of particle-hole excitations in the  $\delta$-function Bose gas, we have extended \cite{KozKitMailSlaTerRestrictedSumsEdgeAndLongTime} in 2012
the above analysis so as to encompass the case of dynamic two-point functions of that model.  
We managed to extract, on the basis of first principle arguments,  the leading long-time and large-distance asymptotic behaviour 
of two-point functions while also providing the leading amplitude and critical exponent of every oscillating  harmonic 
(oscillating term at a given frequency and momentum) arising in the asymptotics. 
The method of analysis we employed also allowed us to investigate the singularity structure of the edge exponents for the 
dynamic response functions hence confirming, through an \textit{ab inicio} analysis, the predictions stemming from 
the non-linear Luttinger liquid approach. Although successful for that particular case, the analysis
left several open questions. In itself, the method used in \cite{KozKitMailSlaTerRestrictedSums,KozKitMailSlaTerRestrictedSumsEdgeAndLongTime}
only allows one to argue the various asymptotic regimes of the correlators 
(be it the long-distance/time or the edge singular behaviour of DRFs). In particular, it 
does not provide one with a way to write down a closed form for a massless form factor expansion in the thermodynamic limit and invokes certain heuristics in the
handlings of the asymptotic analysis. Furthermore, the applicability of the method to the case of integrable models containing bound states was open. 
These points were recently solved by myself in \cite{KozMasslessFFSeriesXXZ}. There I managed to circumvent the various problems
associated with defining form factor series expansions for massless models and constructed an explicit form factor expansion
representation for the dynamical two-point functions in the massless regime of the XXZ spin-$1/2$ chain at non-zero magnetic field. 
This representation was enough to take the Fourier transforms explicitly and led to a series of multiple integral representation
for the DRFs of the model. The series will be starting point for the analysis carried out in the present work.

The main goal of this paper is to provide a thorough analysis of the edge singularities in the dynamic response functions of the XXZ chain
at finite magnetic field and throughout the massless regime, this on the basis of first-principle based calculation: 
the work starts from the series of multiple integrals representation for the DRFs obtained, on the level of the microscopic model, in \cite{KozMasslessFFSeriesXXZ}. 
It then carries out rigorously -the well-definiteness and some of the properties of the representation obtained  in \cite{KozMasslessFFSeriesXXZ} being taken for granted- only those approximations that are 
consistent with the limiting regimes considered. 
As a consequence, the analysis carried out in this work does not relies, at any point of our calculations, upon some conjectural or heuristically argued correspondence with a
simplified effective model such as a CFT, a Luttinger liquid or its non-linear generalisation. 

Furthermore, although obtained for the massless regime of the XXZ chain, taken the "universal" nature of the massless form factor expansion based representation for the 
DRFs and that the analysis developed in this work solely uses this universal structure, the results will hold -provided one accepts the validity of the
phenomenological form of massless form factor expansions advocated in \cite{KozMasslessFFSeriesXXZ}- for any massless one-dimensional quantum Hamiltonian belonging to the Luttinger
liquid universality class.

\section{Main results}
\label{Section main results}

\subsection{The setting and some generalities on the model}

I shall focus on the so-called massless anti-ferromagnetic regime at positive magnetic field  which corresponds to 
$-1<\De <1$ and $h_{\e{c}}>h>0$, where the critical field $h_{\e{c}}$ takes the form $h_{\e{c}}=4J(1+\De)$. $h_{\e{c}}$ is the saturation field above which the model becomes
ferromagnetic. Then, it appears convenient to parametrise the anisotropy $\De$ introduced in \eqref{ecriture hamiltonien XXZ}
as 
\beq
\De= \cos(\zeta) \quad \e{with} \quad \zeta \in \intoo{0}{\pi} \, .
\label{ecriture reparametrisation anisotropie}
\enq
In the thermodynamic limit, the Bethe Ansatz analysis ensures that, for this range of parameters, the excited states above the ground state are built from
a pile up  of elementary dressed excitations of different types: holes and $r$-strings. 
For given value of $\zeta$, only certain values of $r$ are possible for the $r$-strings and it is convenient to collect these in the set $\mf{N} = \{ r_1, \dots, r_{ |\mf{N}|}  \}$. 
The set $\mf{N}$ is finite when $\tf{\zeta}{\pi}$ is rational and infinite otherwise \cite{TakahashiThermodynamics1DSolvModels}. 
Furthermore, independently of the value of $\zeta$, there always exists $1$-strings excitations (\textit{viz}. $r_1=1$). 
The $1$-string excitations correspond to so-called particle excitations. Among all possible $r$-string excitations, only the particles -\textit{i.e.} $1$-strings- may generate 
massless excitation, \textit{i.e.} carrying a zero energy. A given excited state will be made up of $n_h\in \mathbb{N}$ holes, $n_{r_k}\in \mathbb{N}$ $r_k$-strings and left/right Fermi boundary Umklapp excitations with deficiencies $\ell_{\pm}\in \mathbb{Z}$. 
These integers  satisfy  to the constraint 
\beq
 n_h= \sul{r \in \mf{N} }{} r n_r \, + \, \sul{\ups=\pm}{} \ell_{\ups} \;. 
\label{ecriture contraintes entiers trous et strings}
\enq
It is convenient to collect the integers labelling the number of excitations of each kind  into a single vector 
\beq
\bs{n} \, = \, \big(\ell_+, \ell_-  ; n_h, n_{r_1},\dots, n_{r_{|\mf{N}|}} \big)\;. 
\label{definition premiere apparition vecteur n}
\enq
Owing to the constraint \eqref{ecriture contraintes entiers trous et strings}, there are only finitely many non-zero entries in $\bs{n}$. 

$\bs{n}$ being fixed, the $n_h$ holes will carry momenta $t_1,\dots, t_{n_h}$ which take values in $\msc{I}_h=\intff{-p_F}{p_F}$, $p_{F} \in \intff{0}{\tf{\pi}{2}}$ being the Fermi momentum,
and the $n_{r}$ $r$-strings will carry momenta $k^{(r)}_1,\dots, k_{n_r}^{(r)}$ taking values in $\msc{I}_r=\intff{ p_-^{(r)} }{ p_+^{(r)} }$. 
I refer to Appendix \ref{Appendix Lin Int Eqns Defs et al} for more precise definitions of these intervals.  It appears convenient to gather the momenta carried by the various elementary excitations into the single vector 
\beq
\bs{\mf{K}} \, = \, \big(\ell_{+},\ell_{-} ; \bs{t}, \bs{k}^{(r_1)},\cdots , \bs{k}^{(r_{ |\mf{N}|})}  \big) \quad \e{with} \quad \bs{t} \in \msc{I}_h^{n_h} \quad \e{and} \quad \bs{k}^{(r )} \in \msc{I}_r^{n_r}  \;. 
\label{ecriture vecteur des impuslions des excitations diverses}
\enq
This notation should be understood as follows. If $n_h=0$, resp. $n_r=0$ with $r \in \mf{N}$, then the associated vectors $\bs{t}$, resp. $\bs{k}^{(r)}$, are to be read as 
$\emptyset$, meaning that there is simply no component of the hole or of this $r$-string momenta in $\bs{\mf{K}}$, since there are no excitations of this type 
in the given excited state. The use such a notation allows one to keep the precise track, on the level of the vector $\bs{\mf{K}}$, of the types of excitations which are present and those which are absent. 
I stress that formally  $\bs{\mf{K}}$ may contain infinitely many components with such a conventions, but only finitely many of them correspond to non-empty sets since, for fixed $\ell_{\pm}$
and $n_h$, there is only a finite number of integers $n_r$ that are non-zero. Hence  $\bs{\mf{K}}$
makes sense as an inductive limit. Furthermore, effectively speaking,  $\bs{\mf{K}}$ is built up from vector momenta $\bs{t}$, resp. $\bs{k}^{(r)}$ with $r \in \mf{N}$, 
such that $n_h\not=0$, resp. $n_r\not=0$.

A given excited state in a sector of relative spin $\op{s}_{\ga}$ above the ground state and associated with a vector momentum $\bs{\mf{K}}$ has a total excitation momentum 
\beq
\mc{P}(\bs{\mf{K}}) \, = \, \sul{ r \in  \mf{N} }{}  \sul{a=1}{n_r} k_a^{(r)}  \, + \, p_{F}\sul{\ups=\pm}{}  \ups \ell_{\ups} 
+ \pi  \op{s}_{\ga} - \sul{a=1}{n_h} t_a
\label{definition impulsion excitation}
\enq
and carries a total excitation energy
\beq
\mc{E}(\bs{\mf{K}}) \, = \, \sul{ r \in  \mf{N} }{}  \sul{a=1}{n_r} \mf{e}_r\big( k_a^{(r)} \big) \,- \,  \sul{a=1}{n_h} \mf{e}_1(t_a) \;. 
\label{definition energie excitation}
\enq
The functions $\mf{e}_a$ correspond to the dispersion relation of the various excitations, $\mf{e}_r$ for the $r$-strings, $-\mf{e}_1$ for the holes. They are defined as solutions to linear integral equations,
see Appendix \ref{Appendix Lin Int Eqns Defs et al}, equations \eqref{definition fct mathfrak e 1}-\eqref{definition fct mathfrak e r}  for more details. 
Again, by convention, sums that are subordinate to $n_r=0$ or $n_h=0$ are simply understood to be absent. 

The velocity of a given $r$-string excitation with momentum $k$ is defined as $\mf{v}_r(k)=\mf{e}_r^{\prime}(k)$. Moreover, $\mf{v}_1(k)$ gives the velocity, depending on the domain where $k$ evolves, of $1$-strings (particles) 
if $k \in \msc{I}_1$ or holes if $k \in \msc{I}_{h}$. 
Particles, holes, and more generally strings, may share the same value of their velocities. In particular, one can prove, \textit{c.f.} Proposition \ref{Proposition proprietes fondamentales de la vitesse des particules trous} 
in Appendix \ref{Appendix Section phase oscillante dpdte de la vitesse}, 
that for certain regimes of the model's parameters, that there exists an interval $\intff{K_m}{K_M}\subset \intoo{ p_-^{(1)} }{ p_+^{(1)}  }$ and a diffeomorphism 
\beq
\mf{t}:\intff{K_m}{K_M}\tend \msc{I}_h \quad  \e{such}\; \e{that} \quad  \mf{v}_1(k)=\mf{v}_1(\mf{t}(k)) \, .
\label{definition isomorphisme mf t}
\enq
It is conjectured that this property holds for any regime of the parameters and this is backed by an extensive numerical analysis.

\subsection{The behaviour of the longitudinal dynamic response function in the  two hole excitations in $\msc{S}^{(z)}(k,\om)$}
\label{SousSection description edge exponents}

\begin{figure}
\begin{center}
\includegraphics[width=.5\textwidth]{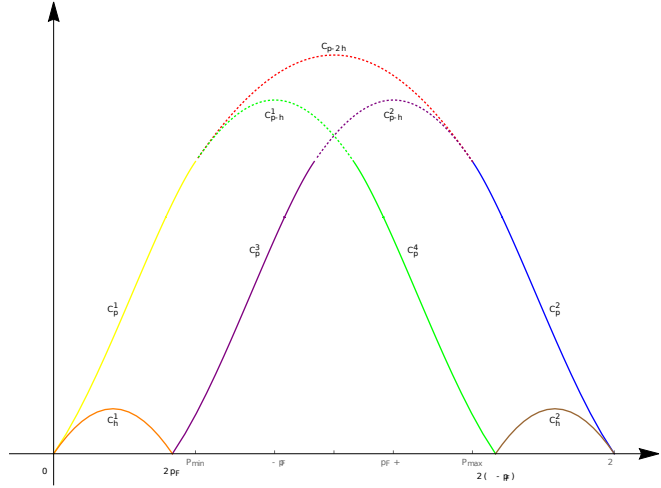}
\caption{\label{Figure relation dispersion 2 particules trous} Singularity curves issued from the sectors involving up to two particles, two-holes and no $r$-strings with $r\geq 2$ for $\Delta=0.57$ and 
in presence of a magnetic field $h$ which fixes the \textit{per} site magnetisation $\mf{m}=1-2D$ such that $D=0.21$.
Continuous curves correspond to one massive -hole or particle- excitation. Dotted curves correspond to a collective, coordinated, multi-particle-hole excitation. This excitation is such that all particles
and holes building it up have the same velocity.}
\end{center}
\end{figure}

 The excitations thresholds giving rise to singularities of the longitudinal dynamic response function $\msc{S}^{(z)}(k,\om)$ and built up from excitations containing
 at most two holes and/or two $1$-strings are depicted in Figure \ref{Figure relation dispersion 2 particules trous}. 
The curves   $\mc{C}_{h}^{(a)}$, $a=1,2$, resp.  $\mc{C}_{p}^{(b)}$, $b=1,\dots,4$, correspond to one hole, resp. particle, excitation above the ground state. 
The curves $\mc{C}_{p-h}^{(a)}$, $a=1,2$ correspond to a joint particle-hole excitation where the particle and hole both have the same velocity. Finally,  the curve $\mc{C}_{2p-h}$ is built up from a 
 two particle - one hole excitation, all having the same velocity. 
All the particles or holes building up the excitations in the curves depicted in Figure \ref{Figure relation dispersion 2 particules trous} are massive -\textit{viz}. carry a finite excitation energy- 
with the exception of the $\om=0$ line and of the junctures between the curves that are drawn in continuous and 
dotted lines. The present approach is unable to analyse the singularity structure at these points. Below, $0<\tau<1$ is arbitrary and can be taken as small as necessary. 

More precisely, the results established in Section \ref{Section Edge singular behaviour des fcts spectrales} entail that
\begin{itemize}

 \item $\mc{C}_h^{(1)}$ is realised as a one hole excitation with $\ell_+=1, \ell_-=0$. It takes the parametric form 
\beq
(\mc{P}_0,\mc{E}_0)= \big( p_F-t_0,-\mf{e}_1(t_0) \big) \quad \e{with} \quad  t_0\in \intoo{-p_F}{p_F}.
\enq
 Along this curve, the response function behaves as 
\beq
\msc{S}^{(z)}(\mc{P}_0,\mc{E}_0+\de \om) \; = \; \msc{S}^{(z)}_{h;\e{reg}}(\de \om) \; + \; \mc{A}^{(h)} \cdot (\de \om)^{\De^{(h)}} \cdot \Xi(\de\om)+\e{O}\Big( (\de\om)^{\De^{(h)}+1-\tau} \Big) \;. 
\enq
$\msc{S}^{(z)}_{h;\e{reg}}(\de \om)$ is smooth in $\de \om$  while the critical exponent takes the form $\De^{(h)}=\de_+^{(h)}+\de_-^{(h)}-1$. 
$\de_{\pm}^{(h)}$ are expressed in terms of the dressed phase, \textit{c.f.} \eqref{ecriture phase habillee dans rep impuslion}, as 
\beq
\de_+^{(h)} \, = \, \Big( \vp_1(p_F,t_0)-\vp_1(p_F,p_F)-1\Big)^2 \quad , \quad 
\de_-^{(h)}\, = \, \Big( \vp_1(-p_F,t_0)-\vp_1(-p_F,p_F)\Big)^2 \;. 
\enq
Finally, the amplitude $\mc{A}^{(h)}$ is closely related to the properly renormalised in the volume form factor squared $\msc{F}^{(z)}\big( \bs{\mf{K}}_{0}^{(h)} \big) $ of the operator $\sg^{z}$ taken between the ground state 
and the excited state associated with $\mc{C}^{(1)}_{p}$:
\beq
 \mc{A}^{(h)}  \; = \;  
\f{   (2\pi)^2  \cdot  \msc{F}^{(z)}\big( \bs{\mf{K}}_{0}^{(h)} \big)   }
{  \Ga\Big( \de_{+}^{(h)} +  \de_{-}^{(h)} \Big) \cdot \big[ \op{v}_F+\mf{v}_1(t_0) \big]^{ \de_{-}^{(h)} } \cdot  \big[ \op{v}_F-\mf{v}_1(t_0) \big]^{ \de_{+}^{(h)} }  } \;. 
\label{ecriture amplitude Ah}
\enq
The precise definition of  $ \msc{F}^{(z)}\big( \bs{\mf{K}}_{0}^{(h)} \big) $ is given in \eqref{definition facteur de forme convenablement renormalise} 
and  $\bs{\mf{K}}_{0}^{(h)}\, = \, \Big( \ell_{+}=1, \ell_{-}=0 ; \bs{t}= t_0 , \emptyset, \dots   \Big)$. 
All building blocks of  $\mc{A}^{(h)} $ other than the renormalised form factor $ \msc{F}^{(z)}\big( \bs{\mf{K}}_{0}^{(h)} \big) $ correspond to the universal part of the 
amplitude associated with this hole excitation branch. Finally, $\op{v}_F=\mf{v}_1(q)$ is the velocity of the excitations on the right Fermi boundary.

  \item $\mc{C}_p^{(1)}$ is realised as a one particle excitation with $\ell_+=-1, \ell_-=0$. It takes the parametric form 
\beq
(\mc{P}_0,\mc{E}_0)= \big( k_0-p_F,\mf{e}_1(k_0) \big) \quad \e{with} \quad k_0\in \intoo{p_F}{ K_m } \;. 
\enq
 Along this curve, the  response function behaves as 
\bem
\msc{S}^{(z)}(\mc{P}_0,\mc{E}_0+\de \om) \; = \; \msc{S}^{(z)}_{p;\e{reg}}(\de \om) \; + \; \mc{A}^{(p)} \cdot |\de \om|^{\De^{(p)}} 
\bigg\{ \Xi(\de \om) \f{\sin[ \pi \de_-^{(p)}] }{ \pi }  +  \Xi(-\de \om) \f{\sin [\pi \de_+^{(p)}] }{ \pi }  \bigg\} \\ 
+\e{O}\Big( (\de\om)^{\De^{(p)}+1-\tau} \Big) \;. 
\end{multline}
 $\msc{S}^{(z)}_{p;\e{reg}}(\de \om)$ is smooth in  $\de \om$. 
 The critical exponent takes the form $\De^{(p)}=\de_+^{(p)}+\de_-^{(p)}-1$ and $\de_{\pm}^{(p)}$ are expressed in terms of the dressed phase, \textit{c.f.} \eqref{ecriture phase habillee dans rep impuslion}, as 
\beq
\de_+^{(p)} \, = \, \Big( 1-\vp_1(p_F,k_0)+\vp_1(p_F,p_F)\Big)^2 \quad , \quad 
\de_-^{(p)}\, = \, \Big( \vp_1(-p_F,p_F)-\vp_1(-p_F,k_0)\Big)^2 \;. 
\enq
The amplitude $\mc{A}^{(p)}$ takes the form
\beq
\mc{A}^{(p)} \, = \, \f{ (2 \pi)^2   \cdot  \Ga\Big( 1 - \de_{+}^{(p)} -  \de_{-}^{(p)} \Big)    }
{   \big| \op{v}_F+\mf{v}_1\big( k_0 \big) \big|^{ \de_{-}^{(p)} } \cdot  \big| \op{v}_F-\mf{v}_1\big(  k_0\big) \big|^{ \de_{+}^{(p)} }   }   \cdot  \msc{F}^{(z)}\big( \bs{\mf{K}}^{(p)}_0\big) \;. 
\label{ecriture amplitue particule Ap}
\enq
 $ \msc{F}^{(z)}\big( \bs{\mf{K}}_{0}^{(p)} \big) $ has the same interpretation as given above, is defined in \eqref{definition facteur de forme convenablement renormalise} 
and  is parameterised by the vector momentum $\bs{\mf{K}}_{0}^{(p)}\, = \, \Big( \ell_{+}=-1, \ell_{-}=0 ; \bs{t}=\emptyset, \bs{k}^{(1)}= k_0, \emptyset ,\dots  \Big)$. 
All the other building blocks of  $\mc{A}^{(p)} $ correspond to the universal part of this particle branch amplitude.

  \item $\mc{C}_{ph}^{(1)}$ is realised as an excitation with $\ell_+=0, \ell_-=0$, and containing a particle and a hole,  both having the same velocity. It takes the parametric form 
\beq
(\mc{P}_0,\mc{E}_0)\, = \, \Big( k_0 - \mf{t}(k_0), \mf{e}_1(k_0) - \mf{e}_1(\mf{t}(k_0)) \Big) \quad \e{with} \quad k_0\in \intoo{ K_m }{K_M} \,  
\enq
and where $\mf{t}$ has been introduced in \eqref{definition isomorphisme mf t}. 
 Along this curve, the response function behaves as 
\bem
\msc{S}^{(z)}(\mc{P}_0,\mc{E}_0+\de \om) \; = \; \msc{S}^{(z)}_{ph;\e{reg}}(\de \om) \; + \; \mc{A}^{(ph)} \cdot |\de \om|^{\De^{(ph)}} 
\bigg\{ \Xi(\de \om) \f{\cos[ \pi  \De^{(ph)}] }{ \pi }  +  \Xi(-\de \om) \f{1 }{ \pi }  \bigg\} \\
+\e{O}\Big( (\de\om)^{ \De^{(ph)}+ 1 - \tau }  \Big) \;. 
\end{multline}
 $\msc{S}^{(z)}_{ph;\e{reg}}(\de \om)$ is smooth in  $\de \om$. The critical exponent takes the form $\De^{(ph)}=\de_+^{(ph)}+\de_-^{(ph)}-\tf{1}{2}$ and $\de_{\pm}^{(ph)}$
 are expressed in terms of the dressed phase, \textit{c.f.} \eqref{ecriture phase habillee dans rep impuslion}, as 
\beq
\de_+^{(ph)} \, = \, \Big( \vp_1(p_F,\mf{t}(k_0) ) - \vp_1(p_F,k_0) \Big)^2 \quad , \quad 
\de_-^{(ph)}\, = \, \Big( \vp_1(-p_F,\mf{t}(k_0)) - \vp_1(-p_F,k_0)\Big)^2 \;. 
\label{eqn edge + et - ph excitation}
\enq
Finally, the amplitude $\mc{A}^{(ph)}$ takes the form 
\beq
 \mc{A}^{(ph)}  \, = \,  
\f{ (2 \pi)^2   }{ \sqrt{ 1-\mf{t}^{\prime}(k_0) } } 
 \cdot \;\;  \bigg( \f{ 2 \pi }{ \mf{v}^{\prime}_1\big( \mf{t}(k_0 ) \big) }  \bigg)^{ \f{1}{2}} 
 \cdot  \f{  \Ga\Big( -\De^{(ph)}    \Big)    }
{   \big| \op{v}_F+\mf{v}_1\big( k_0  \big) \big|^{ \de_{-}^{(ph)}  }   \big| \op{v}_F-\mf{v}_1\big(  k_0 \big) \big|^{ \de_{+}^{(ph)}  }   }  
\cdot  \msc{F}^{(z)}\big( \bs{\mf{K}}^{(ph)}_0  \big) \;. 
\enq
 $ \msc{F}^{(z)}\big( \bs{\mf{K}}_{0}^{(ph)} \big) $ has the same interpretation  
and  $\bs{\mf{K}}_{0}^{(ph)}\, = \, \Big(\ell_{+}=0, \ell_{-}=0 ; \bs{t}=\mf{t}(k_0), \bs{k}^{(1)}=k_0, \emptyset,\dots   \Big)$. 
All the other building blocks of  $\mc{A}^{(ph)} $ correspond to the universal part of this equal velocity particle-hole branch amplitude.

\item $\mc{C}_{p}^{(4)}$ is realised as a one particle excitation with $\ell_+=0, \ell_-=-1$. It takes the parametric form 
\beq
(\mc{P}_0,\mc{E}_0)\, = \, \big( k_0 +p_F, \mf{e}_1(k_0)  \big) \quad \e{with} \quad  k_0\in \intoo{K_M}{2\pi - 3 p_{F} } \; .
\enq
Along this curve, the response function behaves as
\bem
\msc{S}^{(z)}(\mc{P}_0,\mc{E}_0+\de \om) \; = \; \msc{S}^{(z)}_{p;\e{reg}}(\de \om) \; + \; \mc{A}^{(p)} \cdot |\de \om|^{\De^{(p)}} 
\bigg\{ \Xi(\de \om) \f{\sin[ \pi \de_+^{(p)}] }{ \pi }  +  \Xi(-\de \om) \f{\sin [\pi \de_-^{(p)}] }{ \pi }  \bigg\}  \\ 
+\e{O}\Big( (\de\om)^{\De^{(p)}+1-\tau} \Big)  \;. 
\end{multline}
 $\msc{S}^{(z)}_{p;\e{reg}}(\de \om)$ is smooth in $\de \om$. The critical exponent takes the form $\De^{(p)}=\de_+^{(p)}+\de_-^{(p)}-1$ and $\de_{\pm}^{(p)}$ are expressed in terms of the dressed phase, \textit{c.f.} \eqref{ecriture phase habillee dans rep impuslion}, as 
\beqa
\de_+^{(p)} & = & \Big( \vp_1(p_F,p_F ) - \vp_1(p_F,k_0)  + \bs{1}_{I_{-}}(k_0) \e{sgn}(\pi-2\zeta)\mc{Z}(p_F)  \Big)^2  \;,     \\
\de_-^{(p)} & = & \Big( -1+\vp_1(-p_F,-p_F) - \vp_1(-p_F,k_0) + \bs{1}_{I_{-}}(k_0) \e{sgn}(\pi-2\zeta)\mc{Z}(p_F) \Big)^2 \;. 
\eeqa
Here, $I_-=\intoo{2\pi-2p_F \e{sgn}(\pi-2\zeta) - (\pi-\zeta)-(\pi-2\zeta)\tfrac{p_F}{\pi} }{ 2\pi-p_F -2p_F \e{sgn}(\pi-2\zeta) }$. Finally, the amplitude $\mc{A}^{(p)}$ takes the same
form as in \eqref{ecriture amplitue particule Ap}, with the constants appropriately substituted.

  \item $\mc{C}_{p2h}$ is realised as an excitation with $\ell_+=1, \ell_-=0$ that consists of one particle and two holes, all having the same velocity. It takes the parametric form 
\beq
(\mc{P}_0,\mc{E}_0)\, = \, \Big( k_0 - 2\mf{t}(k_0)+p_F, \mf{e}_1(k_0) - 2\mf{e}_1(\mf{t}(k_0)) \, \Big)  \quad \e{with} \quad  k_0\in \intoo{ K_m }{K_M} \,, 
\enq
and $\mf{t}$ as in \eqref{definition isomorphisme mf t}. 
For this parameterisation, $\mc{P}_0$ increases on the interval $\intff{K_m-p_F}{K_M+3p_F}$.
Along this curve, the response function has the singular structure
\beq
\msc{S}^{(z)}(\mc{P}_0,\mc{E}_0+\de \om) \; = \; \msc{S}^{(z)}_{p2h;\e{reg}}(\de \om) \; + \; \mc{A}^{(p2h)} \cdot (\de \om)^{\De^{(p2h)}} 
  \cdot \Xi(\de \om) \cdot  \f{\sin[ \pi \De^{(p2h)}] }{ \pi }  +\e{O}\Big( (\de\om)^{ \De^{(p2h)} +  1 -  \tau } \Big) \;. 
\enq
 The critical exponent takes the form $\De^{(p2h)}=\de_+^{(p2h)}+\de_-^{(p2h)}+1$ and $\de_{\pm}^{(p2h)}$ are expressed in terms of the dressed phase, \textit{c.f.} \eqref{ecriture phase habillee dans rep impuslion}, as 
\beqa
\de_+^{(p2h)} & = & \Big(-1 + 2 \vp_1(p_F,\mf{t}(k_0) ) - \vp_1(p_F,k_0) -  \vp_1(p_F,p_F)\Big)^2 \;,  \label{eqn edge + p2h excitation} \\
\de_-^{(p2h)} & = & \Big( 2\vp_1(-p_F,\mf{t}(k_0)) - \vp_1(-p_F,k_0) -  \vp_1(-p_F,p_F)\Big)^2 \;. 
\label{eqn edge - p2h excitation}
\eeqa
Finally, 
\beq
 \mc{A}^{(p2h)}  \, = \, 
\f{ - (2 \pi)^{3} }{  \sqrt{ 1-2 \mf{t}^{\prime}(k_0)  } }  \cdot \;\;    \bigg( \f{  1 }{ \mf{v}^{\prime}_1\big( \mf{t}(k_0 ) \big) }  \bigg)^2  \\
\cdot  \f{  \Ga\Big( -\De^{(p2h)}       \Big)    }
{   \big| \op{v}_F+\mf{v}_1\big( k_0  \big) \big|^{ \de_{-}^{(p2h)}  }   \big| \op{v}_F-\mf{v}_1\big(  k_0 \big) \big|^{ \de_{+}^{(p2h)}  }   }  
\cdot \msc{F}^{(z)}\big( \bs{\mf{K}}^{(p2h)}_0  \big) \;. 
\enq
 $ \msc{F}^{(z)}\big( \bs{\mf{K}}_{0}^{(p2h)} \big) $ has the same interpretation 
and  $\bs{\mf{K}}_{0}^{(p2h)}\, = \, \Big( \ell_{+}=1, \ell_{-}=0 ; \bs{t}=(\mf{t}(k_0),\mf{t}(k_0)), \bs{k}^{(1)}=k_0,\emptyset,\dots  \Big)$. 
All the other building blocks of  $\mc{A}^{(p2h)} $ correspond to the universal part of this equal velocity one particle two hole branch amplitude.

\end{itemize}

The curves appearing in Fig.~\ref{Figure relation dispersion 2 particules trous} are symmetric in respect to the $k=\pi$ axis.
This symmetry also applies relatively to the behaviour along these curves. Thus, the cases that were not listed above can be inferred by this symmetry operation. 
Also, one should observe that certain curves are realised as $2p_F$ or $2(\pi-p_F)$ translations of other curves. This is reminiscent of the possibility, in the model, to realise zero energy excitations carrying a 
non-zero discrete momentum which is an integer multiple of $2p_F$. $C_p^{(2)}$ is deduced from $\mc{C}^{(1)}_p$ by adding a particle on the right end of the Fermi zone and a hole on the left end what corresponds to 
$(\ell_+,\ell_-)=(-1,0)\hookrightarrow (\ell_+^{\prime},\ell_-^{\prime})=(0,1)$. This, however, changes the values of the critical exponents.

    The excitation thresholds $\mc{C}_{h}^{(a)}$, $a=1,2$ and  $\mc{C}_{p}^{(b)}$, $b=1,\dots,4$, along with the associated universal structure of the singular behaviour have been argued in the literature
by means of heuristic approaches: the non-linear Luttinger liquid \cite{CheianovPustilnikXXZLoweredgeNLLLEdgeExp} in what concerns $\mc{C}_h^{(a)}$, 
\cite{AffleckPereiraWhiteEdgeSingInSpin1-2,AffleckPereiraWhiteSpectralFunctionsfor1DLatticeFermionsBoundStatesContributions} relatively to $\mc{C}_{h}^{(a)}$,  $\mc{C}_{p}^{(b)}$ 
and the pseudofermion dynamic theory  \cite{CampbellCarmeloMachadoSacramentoLowerTresholdsXXXLongAndTransDSFPseudoFermionDynTheor} relatively to $\mc{C}_h^{(a)}$. 
 The present analysis does confirm these predictions on the basis of rigorous considerations. 
 
 The thresholds corresponding to the curves $\mc{C}_{ph}$ and $\mc{C}_{p2h}$ have never been discussed within the aforementioned approaches. 
These excitation thresholds are characterised  by a different structure of edge exponents as clearly appears in \eqref{eqn edge + et - ph excitation}
and \eqref{eqn edge + p2h excitation}-\eqref{eqn edge - p2h excitation}. On physical grounds, these thresholds issue from 
the presence of excited states built up from various excitations (particles, holes and/or r-strings), all having equal velocities. 
The singular structure of the dynamic response functions in the vicinity of multi-hole/r-string excitations, $r \in \mf{N}_{\e{st}}$ 
is discussed in Theorem \ref{Theorem DRF behaviour of multi string hole threshold}. Finally, although it is 
not detailed in the body of the paper, the structure of the behaviour of dynamic response functions in the vicinity of equal velocity multi-particle/hole/r-string thresholds 
can be readily worked out by appropriately adjusting the results of the main theorem established in this paper, 
Theorem \ref{Theorem Principal}.
One should mention, that solely the work \cite{AffleckPereiraWhiteSpectralFunctionsfor1DLatticeFermionsBoundStatesContributions} considered
the thresholds generated by a joint multi-particle massive excitation. In  \cite{AffleckPereiraWhiteSpectralFunctionsfor1DLatticeFermionsBoundStatesContributions},
the authors argued heuristically the expression for the edge exponents in the case of an equal velocity excitation built up from two holes and one two-string. 
They also asserted that the singularity is only one-sided. They did not discuss the form of the amplitude though. The present analysis recovers all these features 
and provides much more thorough information on the amplitude. The presence of one sided singularities does not hold, however, for generic $r$-string excitations.

\subsection{The series representation for the dynamic response functions}

Under certain assumptions, I have derived in \cite{KozMasslessFFSeriesXXZ} a series of multiple integral representation for the dynamic response functions of the 
XXZ spin-$1/2$ chain in the massless regime $-1<\De<1$ and at finite magnetic field $h_{\e{c}}>h>0$ . The derivation of the representation relied on the assumption that it is licit to exchange certain limits with summations, 
that the remainders were uniformly summable and that the resulting series was convergent. The rest of the handling were rigorous. I shall not discuss here further the rigour of the obtained series. 
\textit{In the present work, I shall take for granted the existence and well-definiteness of the series of multiple integrals representing the DRF}.
The justification of the exchange of limits procedures used in its derivations along with the convergence of the series is left for future investigations and will quite probably demand to 
invent new mathematical tools adapted for dealing with such questions.

 The  present work carries out \textit{a rigorous analysis of the singularity structure of each summands in the series} representing $\msc{S}^{(\ga)}(k,\om)$.
Developing a technique allowing one for a rigorous analysis of a class of multiple integrals containing, upon specialisations, the integrals of interest constitutes the main achievement of this work.  
The series of multiple integrals obtained in \cite{KozMasslessFFSeriesXXZ} takes the form

\beq
\msc{S}^{(\ga)}(k,\om) \; = \;     \sul{   \bs{n} \in  \mf{S}   }{}    \msc{S}^{(\ga)}_{\bs{n}}(k,\om) 
\label{ecriture series complete representant la DSF}
\enq
where the summation runs through  all the allowed choices of hole, $r$-string and Umklapp integers, all gathered in a single vector $\bs{n}$,  as in \eqref{definition premiere apparition vecteur n}, while 
\beq
\mf{S} \, = \, \Big\{ (\ell_{+},\ell_{-}; n_h,  n_{r_1},\dots, n_{ r_{ |\mf{N}|} })\; : \; \ell_{\pm}\in \mathbb{Z}\; , \; n_h, n_r \in \mathbb{N} \; \quad \e{and} \quad 
\; n_h= \sul{r \in \mf{N} }{} r n_r \, + \, \sul{\ups=\pm}{} \ell_{\ups} \Big\}  \;. 
\label{ecriture range summation dans series DRF}
\enq

A given summand  $\msc{S}^{(\ga)}_{\bs{n}}(k,\om) $ represents the contribution to the dynamic response function of all the excited states whose number of excitations of each type 
 is equal to the corresponding entry of the vector $\bs{n}$. It is given by the multidimensional integral 
\bem
\msc{S}^{(\ga)}_{\bs{n}}(k,\om)  \, = \, 
 \Int{  \big( \msc{J}_h^{(\eps)} \big)^{n_h}    }{}  \dd^{n_h}t  \cdot \pl{ r \in  \mf{N} }{} \; \bigg\{  \Int{ \big( \msc{J}_r^{(\eps)} \big)^{n_r} }{}   \dd^{n_r}k^{(r)}   \bigg\}  \cdot \;  \mc{F}^{(\ga)}\big( \bs{\mf{K}} \big) \\
  \times     \sul{s\in \mathbb{Z} }{} \pl{\ups= \pm }{} \bigg\{  \Xi\Big(\,  \wh{\mf{y}}_{\ups}\big( \bs{\mf{K}}; s\big)  \Big) \cdot \Big[ \, \wh{\mf{y}}_{\ups}\big( \bs{\mf{K}}; s\big)  \Big]^{ \De_{\ups}(\bs{\mf{K}}) -1 }  \bigg\}  
\cdot \bigg( 1 + \mf{r}\big( \bs{\mf{K}};s \big)  \bigg)  \;. 
\label{ecriture contrib excitation donnee facteur structure}
\end{multline}
Just as earlier on, by convention, if a hole $n_h$ or an $r$-string $n_r$ integer is zero, then the associated integration, and \textit{a fortiori} integration  variables, are simply absent.
The integration variables are collected in the vector $\bs{\mf{K}}$ that was introduced in \eqref{ecriture vecteur des impuslions des excitations diverses}. 
In the definition of this vector, it should be understood that, if $n_{r_k}=0$, than the corresponding vector $\bs{k}^{(r )}$ is simply absent. Thus, due to the summation constraint in \eqref{ecriture range summation dans series DRF}, 
for each $\bs{n}$, there are only finitely many $\bs{k}^{(r)}$ vectors present in $\bs{\mf{K}}$, \textit{c.f.} the discussion which followed after \eqref{ecriture vecteur des impuslions des excitations diverses}.

I now describe, in detail, the different building blocks of the multiple integral. 

\subsubsection*{$\bullet$ The integration domain and the regulator $\eps>0$}

The integration variables run  through the slightly deformed domains 
\beq
 \msc{J}_h^{(\eps)} \; = \; \intff{-p_F+\eps}{ p_F-\eps } \quad , \quad 
  \msc{J}_1^{(\eps)} \; = \; \intff{p_{-}^{(1)}+\eps}{ p_{+}^{(1)}-\eps }  
\label{definition domaines integrations eps deformees}
\enq
with $p_{\pm}^{(1)}$ being parameterised in terms of $\zeta$, introduced in \eqref{ecriture reparametrisation anisotropie},  as $p_{+}^{(1)} =   2\pi -p_F -2p_F\e{sgn}(\pi-2\zeta)$, $p_{-}^{(1)}  = p_F $
More generally,  $\msc{J}_r^{(\eps)}=\msc{I}_r=\intff{ p^{(r)}_- }{ p^{(r)}_+ }$ for any $r\geq 2$ and the explicit form for $p_{\pm}^{(r)}$ can be inferred from 
the content of Appendix \ref{Appendix Lin Int Eqns Defs et al}.

Recall that  massless excitations are realised by particles and/or holes whose momenta collapse, in the thermodynamic limit, on the left and right endpoint of the Fermi zone, \textit{i.e.}
the points $\pm p_F$ for the holes and the points $p_{\pm}^{(1)}$ for the particles. In their turn, the massive excitations 
carry a finite excitation energy in the thermodynamic limit. Thus, massive particles and/or holes have their momenta located uniformly away from the endpoints of the Fermi zone.  

The integral representation \eqref{ecriture contrib excitation donnee facteur structure} involves a small but otherwise arbitrary parameter $\eps>0$. 
The latter was introduced in \cite{KozMasslessFFSeriesXXZ} as a regulator defining a separating scale between the massive and massless particle and hole type excitations in the model.
The matter is that the contributions of the massless modes cannot be summed by means of a Lebesgue-measure based integral and demand a very different treatment. 
Their leading effect is already taken into account and manifests itself in the dependence on the functions $\wh{\mf{y}}_{\ups}\big( \bs{\mf{K}}; s\big)$, \textit{c.f.} \eqref{definition des fonctions hat y ups}. 
The $\eps$-dependence appears explicitly on the level of the domain of integration \eqref{definition domaines integrations eps deformees}, 
while the rest of $\eps$-dependence is contained in the remainder $\mf{r}\big( \bs{\mf{K}};s \big)$, \textit{c.f.} the later discussion. 
The whole series \eqref{ecriture series complete representant la DSF} does not depend on the regulator $\eps$. One cannot take the 
$\eps\tend 0^+$ individually in each multiple integral due to the presence of non-integrable singularities in the integrals and 
a non-uniformness in  of the control on the remainder in the $\eps\tend 0^+$ limit. However, one can always consider $\eps$ to be as small as necessary for the 
purpose of the analysis, as long as it remains fixed.

\subsubsection*{$\bullet$ The integrand}

The integrand in \eqref{ecriture contrib excitation donnee facteur structure} is built up from two contributions: the static part $\mc{F}^{(\ga)}\big( \bs{\mf{K}} \big)$,
and the dynamic part built up from the functions $\wh{\mf{y}}_{\pm}\big( \bs{\mf{K}} ; s\big) $ and $\De_{\pm}( \bs{\mf{K}} )$. 
Note that it is precisely the dynamic part that introduces singularities in the integrand and, as such, is the one responsible for the existence of an edge singular behaviour of the 
DRFs. 

\vspace{2mm}

  $\circledast$ {\bf The dynamic part}

\vspace{2mm}

The function 
\beq
\wh{\mf{y}}_{\ups}\big( \bs{\mf{K}}; s\big)  \; = \;  \om - \mc{E}(\bs{\mf{K}}) +  \ups \op{v}_{F} \big[ k - \mc{P}(\bs{\mf{K}})  + 2\pi s \big]    
\label{definition des fonctions hat y ups}
\enq
is the only building block of the integrand that depends on the momentum $k$ and the energy $\om$. 
Its expression involves the relative excitation momentum $\mc{P}(\bs{\mf{K}})$ and relative excitation energy  $\mc{E}(\bs{\mf{K}})$ which were defined, resp., 
in \eqref{definition impulsion excitation} and \eqref{definition energie excitation}. It also involves $\op{v}_F=\mf{v}_1(q)$, the velocity of the excitations on the Fermi boundary. 

The exponents  $\De_{\pm}( \bs{\mf{K}} ) \geq 0$ are smooth functions of $\bs{\mf{K}}$. Their explicit expression 
can be found in equation \eqref{definition shifted sfift function} of Appendix \ref{Appendix Lin Int Eqns Defs et al} and just above it.

The dynamic part is summed up over $s$ in \eqref{ecriture contrib excitation donnee facteur structure}. This summation is, in fact, finite. Indeed,  for fixed
$\bs{n} \in \mf{S}$, the functions $\mc{P}(\bs{\mf{K}})$ and $\mc{E}(\bs{\mf{K}})$ are bounded on the integration domain from below and above. Thus, for $(k,\om)$ belonging to 
any compact subset of $\R^2$, there will exist finitely many $s\in \mathbb{Z}$ such that both $ \wh{\mf{y}}_{\pm}\big( \bs{\mf{K}}; s\big) >0$. 
In fact, the summation over $s$ in \eqref{ecriture contrib excitation donnee facteur structure} simply translates the fact that the spectral function is a $2\pi$ periodic function of $k$, owing to the 
discrete nature of the XXZ chain.

\vspace{2mm}

$\circledast$ {\bf The static part}

\vspace{2mm}

The static part is  a smooth function of $\bs{\mf{K}}$, at least when the latter ranges through the integration domain given in \eqref{ecriture contrib excitation donnee facteur structure}. 
It is expressed as 
\beq
\mc{F}^{(\ga)}\big( \bs{\mf{K}} \big) \;=\; \f{ (2\pi)^2 \cdot \msc{F}^{(\ga)}\big( \bs{\mf{K}} \big)  \cdot    \big[ 2\op{v}_F \big]^{-\De_{+}  ( \bs{\mf{K}} ) -\De_{-} (\bs{\mf{K}} ) +1 } }
				    { n_h!    \cdot  \pl{ r \in  \mf{N} }{}    n_r!   \cdot  \Ga\Big(\De_{+} \big( \bs{\mf{K}} \big)  \Big)  \cdot  \Ga\Big(\De_{-} \big( \bs{\mf{K}} \big)   \Big)} \;. 
\enq
 $\msc{F}^{(\ga)}\big( \bs{\mf{K}} \big)$ corresponds to the properly renormalised in the volume, thermodynamic limit of the form factor squared 
 of the spin operator $\sg_1^{\ga}$ taken between the ground state $\Om$ and the state $  \Ups_{\bs{\mf{K}}}  $ which is the 
 Eigenstate of $\op{H}$ satisfying to the constraints:
\begin{itemize}
\item[ {\bf i) } ]  in the thermodynamic limit, $ \Ups_{\bs{\mf{K}}}  $ is parameterised in terms of elementary excitations whose momenta are gathered in the vector $\bs{\mf{K}}$; 
\item[ {\bf ii) } ] $ \Ups_{\bs{\mf{K}}}  $  has the lowest possible, compatible with {\bf i)}, relative excitation energy above the ground state in finite volume $L$. 
\end{itemize}
This properly normalised form factor reads 
\beq
\msc{F}^{(\ga)}\big( \bs{\mf{K}} \big) \; = \; \lim_{L\tend +\infty} \bigg\{ \, \Big(\f{ L }{ 2\pi} \Big)^{ \tau(\bs{\mf{K}}) } \cdot \Big| \Big(\Ups_{\bs{\mf{K}}} ,  \sg_1^{\ga} \,  \Om \Big) \Big|^2  \bigg\}
\quad \e{with} \quad 
 \tau(\bs{\mf{K}}) \; = \; \De_{+}  ( \bs{\mf{K}} ) + \De_{-} (\bs{\mf{K}} ) + n_h + \sul{r \in \mf{N} }{} r n_r \;. 
\label{definition facteur de forme convenablement renormalise}
\enq
The explicit expression for $\msc{F}^{(\ga)}\big( \bs{\mf{K}} \big)$ can be found in \cite{KozProofOfAsymptoticsofFormFactorsXXZBoundStates}.

\vspace{2mm}

$\circledast$ {\bf The remainder}

\vspace{2mm}

Finally, $\mf{r}(\bs{\mf{K}};s)$ is a remainder term. It is controlled as 
\beq
\mf{r}(\bs{\mf{K}};s) \; = \; \e{O}\Big(  \sum_{ \ups = \pm } \big|\, \wh{\mf{y}}_{\ups}\big( \bs{\mf{K}}; s\big) \big|^{1-\tau}   \Big)
\enq
and this estimation is uniform throughout the integration domain. The parameter $\tf{1}{2}>\tau>0$ is arbitrary provided that it is taken small enough.  
The control on the remainder is also differentiable in respect to the parameters $(\om,k)$, in the sense of Definition \ref{defintion reste differentiable}. 
However, the control on the remainder  is \textit{not} uniform in respect to $\eps \tend 0^+$.

In fact, one expects that the optimal control on $\mf{r}(\bs{\mf{K}};s)$ is provided by the sharper bound 
\beq
\mf{r}(\bs{\mf{K}};s) =  \e{O}\Big(  \sum_{ \ups = \pm } \big|\, \wh{\mf{y}}_{\ups}\big( \bs{\mf{K}}; s\big) \big|\ln \big|\, \wh{\mf{y}}_{\ups}\big( \bs{\mf{K}}; s\big) \big|   \Big) \, .
\enq

\vspace{2mm}

$\circledast$ {\bf An additional property of the integrand }

\vspace{2mm}

As argued in \cite{KozLongDistanceLargeTimeXXZ}, the series \eqref{ecriture series complete representant la DSF} taken as a whole 
has a built-in mechanism which enforces the complete cancellation between the contributions of an immediate vicinity of the boundaries of integration 
\beq
\Dp{} \Big\{ \big( \msc{J}_h^{(\eps)} \big)^{n_h} \times_{r \in \mf{N} }^{}  \big( \msc{J}_r^{(\eps)} \big)^{n_r} \Big\}
\label{definition bord domaine integration}
\enq
arising in each of the multiple integrals \eqref{ecriture contrib excitation donnee facteur structure}. This cancellation property effectively results in  
that the form factor density $\msc{F}^{(\ga)}\big( \bs{\mf{K}} \big)$ can be considered as a function vanishing smoothly on the boundary \eqref{definition bord domaine integration}.

\section{The edge singular behaviour of dynamic response functions}
 \label{Section Edge singular behaviour des fcts spectrales}
 
 This section gathers various theorems capturing the singular behaviour of the dynamic response function issuing from various excitation sectors in the model's spectrum.  
The statements follows from an application of the general theorems proven throughout Appendix \ref{Appendix DA integrales unidimensionnelles} and \ref{Appendix DA integrales multidimensionnelles}.
All the theorems stated below, take as a hypothesis the smooth vanishing of the integrands on the boundary of integration which was discussed above. 
The precise and rigorous establishing of this property, beyond the arguments given in  \cite{KozLongDistanceLargeTimeXXZ}, is left for further study. 
Also, some of these rely on the properties stated in Conjecture \ref{Conjecture  diffeo liee a la vitesse}, which can be proven in certain cases of the 
coupling constants $\De$ and $h$, \textit{c.f.} Appendix \ref{Appendix Section phase oscillante dpdte de la vitesse}.

\subsection{The one free rapidity sector}

In this subsection, I extract the  singular behaviour of the dynamical response functions associated with one massive excitation, namely an excitation consisting either of one hole or one particle far from the Fermi boundaries, or 
one $r$-string with $r\in \mf{N}\setminus \{1\}$. Such an excitation can be accompanied by any value of the left or right Umklapp integers $\ell_{\pm}$ that are compatible with
the constraint \eqref{ecriture contraintes entiers trous et strings}.

\subsubsection{The one-hole contributions}

For the present purpose, it is convenient to parameterise the momentum-energy $(k,\om)$ combination as 
\beq
k=\mc{P}_0 \; \; \e{where} \;\; \mc{P}_0 \; = \;  \pi  \op{s}_{\ga} \, + \, p_{F}\sul{\ups=\pm}{}  \ups \ell_{\ups}  - t_0 -2\pi s_{0}
\quad \e{and}  \quad  \; \om=\de \om + \mc{E}_0  
\label{ecriture parametrisation impulsion energie hole spectral fct} 
\enq
where $t_0 \in \intfo{-\pi}{\pi}$, $s_0 \in \mathbb{Z}$, and $\ell_{\pm}$ are subject to the constraint $\sul{\ups=\pm}{}\ell_{\ups}=1$. 

The one-hole DRF takes the form 
\bem
\msc{S}^{(\ga)}_{\bs{n}_h}\big(\mc{P}_0,\mc{E}_0 + \de \om\big) \; = \; \Int{  \msc{J}_h^{(\eps)}    }{}  \dd t \;    \mc{F}^{(\ga)}\big( \bs{\mf{K}}^{(h)} \big)
 \cdot \sul{s\in \mathbb{Z} }{}\pl{\ups= \pm }{} \bigg\{   \Xi\Big( \de \om + \mf{y}_{\ups}^{(h)}\big( t ;s \big)  \Big) \cdot \Big[ \de \om + \mf{y}_{\ups}^{(h)}\big(  t ;s \big)  \Big]^{ \De_{\ups}(\bs{\mf{K}}^{(h)} ) -1 }  \bigg\}   
 \\
 \times   
\Big( 1+\mf{r}\big( \bs{\mf{K}}^{(h)}; s \big)   \Big) \;. 
\label{ecriture TF S un trou}
\end{multline}
Here, I have set 
\beq
\bs{n}_{h}\, = \, (\ell_+,\ell_-; n_h=1, 0,\dots  ) 
\quad \e{and} \quad
\bs{\mf{K}}^{(h)} \, = \, \Big( \ell_{+}, \ell_{-} ; \bs{t}= t   , \emptyset, \dots  \Big)  \;. 
\enq
 As a consequence, there  
are no $\bs{k}^{(r)}$ vectors present in $\bs{\mf{K}}^{(h)}$.  
The vector  $\bs{\mf{K}}^{(h)}$ only involves the momentum $t$ of one hole excitation and the Umklapp integers. The $\mf{y}^{(h)}_{\ups}$ functions appearing in \eqref{ecriture TF S un trou} take the form 
\beq
 \mf{y}_{\ups}^{(h)}\big( t ; s \big) \; = \;   \mf{e}_1(t) + \mc{E}_0 \, + \, \ups \op{v}_{F}\big[ t -  t_0 +2\pi (s-s_0) \big]  \;. 
\enq
Finally, the remainder satisfies 
\beq
\mf{r}\big( \bs{\mf{K}}^{(h)}; s \big) \, = \, \e{O}\Big(   \sum_{ \ups = \pm } \big|\, \de \om + \mf{y}_{\ups}^{(h)}\big( t ;s \big) \big|^{1-\tau}   \Big) \;, 
\enq
and the control is differentiable in the sense of Definition \ref{defintion reste differentiable}.

\begin{theorem}
\label{Theorem exictation a un trou} 
Assume that 
\begin{itemize}
\item[{\bf i)} ] $t_0 \not\in \intff{-p_F}{p_F}$, \textit{i.e.} does not belong to the range of available momenta for a hole, in which case $\mc{E}_0$ can take any value;
\item[ {\bf ii)} ] $t_0 \in \intff{-p_F}{p_F}$ is within the range of momenta of a hole excitation and that the subsidiary condition holds $\mc{E}_0\not= -\mf{e}_{1}(t_0)$ .
\end{itemize}
Then $\msc{S}^{(\ga)}_{\bs{n}_h}\big(\mc{P}_0,\mc{E}_0+ \de \om\big)$ is smooth in  $\de \om$ belonging to a neighbourhood of $0$. 

\vspace{2mm}

Assume that 
\beq
t_0\in \intoo{-p_F}{p_F} \; ,  \quad \mc{E}_0=-\mf{e}_1(t_0) \quad and \quad  \de_{\ups}^{(h)} \, = \,  \De_{\ups}( \bs{\mf{K}}_0^{(h)} ) >0
\enq
where  $\bs{\mf{K}}_{0}^{(h)}\, = \, \Big(\ell_{+}, \ell_{-}; \bs{t}= t_0 , \emptyset ,\dots    \Big)$.

Then, one has the $\de \om \tend 0$ asymptotic expansion  
\bem
\msc{S}^{(\ga)}_{\bs{n}_h}\big(\mc{P}_0,\mc{E}_0+ \de \om\big) \; = \;  \f{  \Xi(\de \om) \cdot  \big( \de \om \big)^{ \de_{+}^{(h)} +  \de_{-}^{(h)} - 1 }   }{ \Ga\Big( \de_{+}^{(h)} +  \de_{-}^{(h)} \Big) }
\f{   (2\pi)^2  \cdot  \msc{F}^{(\ga)}\big( \bs{\mf{K}}_{0}^{(h)} \big)   }{  \big[ \op{v}_F+\mf{v}_1(t_0) \big]^{ \de_{-}^{(h)} } \cdot  \big[ \op{v}_F-\mf{v}_1(t_0) \big]^{ \de_{+}^{(h)} }  } \\
 +\;     \e{O}\Big( |\de \om|^{ \de_{+}^{(h)} +  \de_{-}^{(h)} -\tau}   \Big) 
\; + \; \msc{S}^{(\ga)}_{h;\e{reg}}(\de\om)  \;. 
\label{ecriture comportement singulier DRF a un trou}
\end{multline}
The function $\msc{S}^{(\ga)}_{h;\e{reg}}(\de\om)$ appearing above is smooth in the neighbourhood of the origin. 
 
\end{theorem}

I recall that $ \mf{v}_1$ appearing in \eqref{ecriture comportement singulier DRF a un trou} is defined as 
\beq
 \mf{v}_1(t) \, = \,    \mf{e}^{\prime}_1(t)  \;. 
\label{rappel definition vitesse trous}
\enq
Further, one should observe that since $\De_{\ups}$ is an analytic function of the rapidity $t_0$ and that $\De_{\ups} \geq 0$ by construction, the constraint of the theorem is always satisfied 
for a generic choice of parameters.

\Proof

Consider the contribution to $\msc{S}^{(\ga)}_{\bs{n}_h}\big(\mc{P}_0,\mc{E}_0+ \de \om\big)$ stemming from the integrals in \eqref{ecriture TF S un trou} associated with 
picking $s\not=s_0$. Then, the functions 
$ \mf{y}_{\pm}^{(h)}\big( t ; s \big)$ cannot share a common zero on $\msc{J}^{(0)}_h$. Assume the contrary. Then, denoting this zero as $t^{\prime} \in \intoo{-p_F}{p_F}$ one would have that 
\beq
0 \, = \, \mf{y}_{+}^{(h)}(t^{\prime};s) \, - \,  \mf{y}_{-}^{(h)}(t^{\prime};s) \, = \, 2\op{v}_F \big(t^{\prime} - t_0 +2\pi (s-s_0) \big) \;. 
\label{ecriture equation zero joint de z hole} 
\enq
However, $|t^{\prime} - t_0|\leq p_F+\pi < 2\pi$, hence producing  a contradiction. Observe that one has
\beq
 \Dp{t} \mf{y}_{\ups}^{(h)}  \big( t ; s  \big) \; = \;   \mf{v}_1(t) + \ups \op{v}_F   \not= 0
\label{ecriture derivee y ups h}
\enq
with $\mf{v}_1$ as defined in \eqref{rappel definition vitesse trous}. As discussed in Appendix \ref{Appendix Section phase oscillante dpdte de la vitesse}, 
one has $ | \mf{v}_1(t) | < \op{v}_F$  for $t \in \intoo{-p_F}{p_F}$ what ensures that $\mf{y}_{\ups}^{(h)}(t ;s)$ has at most one zero on 
$\msc{J}_{h}^{(\eps)}$ and that the latter is simple. A straightforward application of Lemma \ref{Lemme integrale type beta reguliere} then ensures that integrals subordinate to $s\not=s_0$
only produce smooth functions of $\de \om$ in the neighbourhood of $0$.

It remains to focus on the $s=s_0$ case. First, consider the situation subordinate to the cases {\bf i)} and {\bf ii)}. 
If one is in case {\bf i)}, then due to \eqref{ecriture equation zero joint de z hole} the functions $\mf{y}_{\pm}^{(h)}(t ;s_0)$ cannot share a zero on 
$\msc{J}_h^{(\eps)}$. In case {\bf ii)}, \eqref{ecriture equation zero joint de z hole} would impose that a common zero $t^{\prime}$ necessarily coincides with $t_0$. The latter would then impose that 
$0= \mf{y}_{+}^{(h)}(t_0;s) \, = \, \mc{E}_0 + \mf{e}_1(t_0) $ what leads to a contradiction.
Thus, since in both cases the functions $\mf{y}_{\ups}^{(h)}$ do not share a common zero on $\msc{J}_{h}^{(\eps)}$, one has, by Lemma \ref{Lemme integrale type beta reguliere}, that
$\msc{S}^{(\ga^{\prime} \ga)}_{\bs{n}_h}\big(\mc{P}_0,\mc{E}_0 + \de \om\big) $ is smooth in $\de \om$ around $0$. 

Finally, I focus on the last case $t_0 \in \intoo{-p_F}{p_F}$ with  $\mc{E}_0=-\mf{e}_1(t_0)$. Since $t_0 \in \intoo{-p_F}{p_F}$, one can invoke the freedom of choosing the regulator $\eps>0$ 
so that $t_0 \in \e{Int}(\msc{J}_h^{(\eps)})$.  
$t_0$ is clearly a common zero to $t \mapsto \mf{y}_{\ups}^{(h)} \big( t ; s_0  \big)$. It is the only one on $\msc{J}_h^{(\eps)}$ owing to \eqref{ecriture equation zero joint de z hole}. 
Furthermore, due to \eqref{ecriture derivee y ups h}, one has 
$ \Dp{t} \mf{y}_{+}^{(h)} \big( t_0 ; s\big)  \cdot \Dp{t} \mf{y}_{-}^{(h)} \big( t_0 ;s \big) <0$ and $\Dp{t} \mf{y}_{+}^{(h)} \big( t_0 ; s\big)  - \Dp{t} \mf{y}_{-}^{(h)} \big( t_0 ;s \big)=2\op{v}_F\not= 0$.

All is set so as to apply Theorem \ref{Proposition integrale principale spectral function} and thus, in the $\de \om \tend 0$ limit, one indeed gets \eqref{ecriture comportement singulier DRF a un trou}. 
  \qed

\subsubsection{The one $r$-string contributions}

For the purpose of discussing the contribution of one $r$-string excitations to the DRF, it appears convenient to parametrise the momentum-energy $(k,\om)$ variables of the response function as
\beq
k=\mc{P}_0 \; \; \e{where} \;\; \mc{P}_0 \; = \;  \pi  \op{s}_{\ga} \, + \, p_{F}\sul{\ups=\pm}{}  \ups \ell_{\ups}  + k^{(r)}_0 -2\pi s_0
\quad \e{and} \quad  \; \om=\de \om + \mc{E}_0 \;. 
\label{ecriture parametrisation impulsion energie r string spectral fct} 
\enq
Above, $k^{(r)}_0 \in  \msc{I}_r$ while the Umklapp integers are subject to the constraints $\sul{\ups=\pm}{} \ell_{\ups}  \, = \, -r$ and $s_0 \in \mathbb{Z}$. 

The associated one $r$-string, $r \in \mf{N}$, DRF takes the form
\bem
\msc{S}^{(\ga)}_{\bs{n}_r}\big(\mc{P}_0,\mc{E}_0+ \de \om\big)\; = \; \Int{  \msc{J}_r    }{}  \dd k^{(r)} \;    \mc{F}^{(\ga)}\big( \bs{\mf{K}}^{(r)} \big)
 \cdot \pl{\ups= \pm }{} \bigg\{   \Xi\Big( \de \om + \mf{y}_{\ups}^{(r)}\big( k^{(r)}  ; s   \big)  \Big) \cdot \Big[ \de \om + \mf{y}_{\ups}^{(r)}\big(  k^{(r)}  ; s  \big)  \Big]^{ \De_{\ups}(\bs{\mf{K}}^{(r)}) -1 }  \bigg\}   
 \\
 \times   
\Big( 1+ \mf{r}\big( \bs{\mf{K}}^{(r)};s \big)   \Big) \;. 
\label{ecriture TF S un r string}
\end{multline}
Here, 
\beq
\left\{ \ba{ccc} 
\bs{n}_{r} &  = & (\ell_+,\ell_-;  n_h=0, n_{1}=0,\dots,0,  n_r=1,0,\dots )  \vspace{2mm}\\
\bs{\mf{K}}^{(r)} & = & \Big(\ell_{+},\ell_{-}; \bs{t}=\emptyset,\bs{k}^{(1)}=\emptyset,\dots, \emptyset, \bs{k}^{(r)}=k^{(r)} , \emptyset,\dots    \Big)   \ea \right. 
\enq
where the notation means that the only rapidity that is present in $\bs{\mf{K}}^{(r)}$ is the rapidity $k^{(r)}$ of one $r$-string 
while, Umklapp integers being set apart, the only non-zero integer in $\bs{n}_r$ is the one counting the $r$-string excitations, and it is set to one. 

Also \eqref{ecriture TF S un r string} involves the functions 
\beq
 \mf{y}_{\ups}^{(r)}\big( k^{(r)} ; s \big) \; = \;     \mc{E}_0   \, - \, \mf{e}_r\big( k^{(r)} \big) \, + \,  \ups \op{v}_{F}\big[ k_0^{(r)} -   k^{(r)} +2\pi (s-s_0)  \big]   \;. 
\enq
Finally, the remainder satisfies 
\beq
\mf{r}\big( \bs{\mf{K}}^{(r)}; s \big) \, = \, \e{O}\Big(   \sum_{ \ups = \pm } \big|\, \de \om + \mf{y}_{\ups}^{(r)}\big( k^{(r)} ;s \big) \big|^{1-\tau}   \Big) \;, 
\enq
and the control on $\mf{r}\big( \bs{\mf{K}}^{(r)}; s \big)$ it is differentiable in the sense of Definition \ref{defintion reste differentiable}.

\begin{theorem}
\label{Theorem exictation a une r corde} 
 Let $k_0^{(r)}(s)=k_0^{(r)}+2\pi(s-s_0)$. Assume that 
\begin{itemize}
\item[{\bf i)} ] $k_0^{(r)}(s) \not\in \msc{I}_r$, this for any $s$, in which case $\mc{E}_0$ can take any value;
\item[ {\bf ii)} ] $k_0^{(r)}(s) \in \msc{I}_r$, at least for one $s$ and that, for any such $s$, one has $\mc{E}_0\not= \mf{e}_r\big( k_0^{(r)}(s) \big)$ .
\end{itemize}
Then $\msc{S}^{(\ga^{\prime} \ga)}_{\bs{n}_r}\big(\mc{P}_0,\mc{E}_0+ \de \om\big)$ is smooth in  $\de \om$ belonging to a neighbourhood of $0$. 

\vspace{5mm}

\noindent Assume that 
\begin{itemize}
 
 \item $k_0^{(r)}(s) \in \e{Int}(\msc{I}_r)$ for at least for one $s$,
 \item $\mc{E}_0=\mf{e}_r\big( k^{(r)}_0 + 2\pi (s-s_0) \big)$ for the same value of $s$. 
 \item $\de_{\ups}^{(r)}(s) \, = \, \De_{\ups}\Big( \bs{\mf{K}}^{(r)}_0(s) \Big) >0$, where 
\beq
 \bs{\mf{K}}_{0}^{(r)}(s)\, = \,  \Big(\ell_{+},\ell_{-}; \bs{t}=\emptyset,\bs{k}^{(1)}=\emptyset,\dots, \emptyset, \bs{k}^{(r)}=k^{(r)}_0(s) , \emptyset,\dots    \Big)    \;.
\enq

\end{itemize}

\vspace{3mm}
 \noindent {$\blacklozenge$ Case 1 : \quad If $| \mf{v}_r\big(k_0^{(r)}(s) \big) | > \op{v}_F$}
\vspace{3mm}

 \noindent then, agreeing upon 
\beq
\eta(s) \, = \,- \e{sgn}\Big\{ \mf{v}_r\Big( k_0^{(r)}(s) \Big)  \Big\}   
\enq
one has the asymptotic expansion 
\bem
\msc{S}^{(\ga)}_{\bs{n}_r}\big(\mc{P}_0,\mc{E}_0+ \de \om\big) \; = \;   \sul{ \substack{ s\; : \; \mc{E}_{0}=\mf{e}_{r}(k_0^{(r)}(s)) \\ k_0^{(r)}(s) \in \msc{I}_r } }{} 
\f{ (2 \pi)^2  \cdot  \msc{F}^{(\ga)}\big( \bs{\mf{K}}^{(r)}_0(s) \big) \cdot  \Ga\Big( 1 - \de_{+}^{(r)}(s) -  \de_{-}^{(r)}(s) \Big)    }
{   \Big| \op{v}_F+\mf{v}_r\Big( k^{(r)}_0(s) \Big) \Big|^{ \de_{-}^{(r)}(s) } \cdot  \Big| \op{v}_F-\mf{v}_r\Big(  k^{(r)}_0(s) \Big) \Big|^{ \de_{+}^{(r)}(s) }   } \cdot 
\big| \de \om \big|^{ \de_{+}^{(r)}(s) +  \de_{-}^{(r)}(s) - 1 }  \\
\times \Bigg\{  \Xi(\de \om)  \f{ \sin\big[ \pi \de_{\eta(s)}^{(r)}(s)  \big] }{ \pi } \; + \;  \Xi(-\de \om)  \f{ \sin\big[ \pi \de_{-\eta(s)}^{(r)}(s)  \big] }{ \pi } \Bigg\}
 \quad  + \; \msc{S}^{(\ga)}_{r;\e{reg}}(\de\om)  \; + \;   \e{O}\Big(  |\de \om|^{\de_{+}^{(r)}(s) +  \de_{-}^{(r)}(s) -\tau} \Big)    \;. 
\end{multline}
 $\msc{S}^{(\ga)}_{r;\e{reg}}(\de\om)$ is smooth in the neighbourhood of the origin, and 
\beq
 \mf{v}_r(t) \, = \,    \mf{e}^{\prime}_r(t)   \;. 
\label{rappel definition vitesse  r cordes}
\enq

\vspace{3mm}
 \noindent {$\blacklozenge$  Case 2 :  \quad  If $| \mf{v}_r\big(k_0^{(r)}(s) \big) | < \op{v}_F$}
\vspace{3mm}

\noindent then, under the same conventions as above 
\bem
\msc{S}^{(\ga)}_{\bs{n}_r}\big(\mc{P}_0,\mc{E}_0+ \de \om\big) \; = \;   \sul{ \substack{ s\; : \; \mc{E}_{0}=\mf{e}_{r}(k_0^{(r)}(s)) \\ k_0^{(r)}(s) \in \msc{I}_r } }{} 
\f{ (2 \pi)^2  \cdot  \msc{F}^{(\ga)}\big( \bs{\mf{K}}^{(r)}_0(s) \big) }{  \Ga\Big( \de_{+}^{(r)}(s) +  \de_{-}^{(r)}(s) \Big)    }
\cdot \f{  \big| \de \om \big|^{ \de_{+}^{(r)}(s) +  \de_{-}^{(r)}(s) - 1 }   \Xi(\de \om)   }
{   \big| \op{v}_F+\mf{v}_r\Big( k^{(r)}_0(s) \Big) \big|^{ \de_{-}^{(r)}(s) } \cdot  \big| \op{v}_F-\mf{v}_r\Big(  k^{(r)}_0(s) \Big) \big|^{ \de_{+}^{(r)}(s) }   }  \\
\; + \; \msc{S}^{(\ga)}_{r;\e{reg}}(\de\om)  \; + \;   \e{O}\Big(  |\de \om|^{\de_{+}^{(r)}(s) +  \de_{-}^{(r)}(s) -\tau} \Big)    \;. 
\end{multline}

\end{theorem}

\Proof 

The analysis is quite similar to the one-hole excitation case. 
Cases ${\bf i)}$ and ${\bf ii)}$ are dealt with by means of Lemma \ref{Lemme integrale type beta reguliere}.

It remains to focus on the case where $k_0^{(r)}(s) \in \e{Int}(\msc{I}_r)$. Again, if $r=1$, then one adjusts $\eps>0$ so that $k_0^{(r)}(s) \in \e{Int}(\msc{J}_r^{ (\eps) }) $.
By hypothesis, one has that $\mc{E}_0=\mf{e}_r\big( k_0^{(r)}(s) \big)$, for at least one $s$. 
Then, one treats each integral subordinate to a value of $s$ separately. Only integrals subordinate to values of $s$ such that 
$k_0^{(r)}(s) \in \e{Int}(\msc{I}_r^{(\eps)})$  and $\mc{E}_0=\mf{e}_r\big( k_0^{(r)}(s) \big)$ will give rise to a non-smooth
behaviour when  $\de \om \tend 0$. For any such value of $s$, just as for the one-hole case, one concludes that $k_0^{(r)}(s)$ is the only simultaneous zero of $\mf{y}_{\pm}^{(r)}(\cdot; s)$ on $\msc{J}_r^{(\eps)}$
and that 
\beq
 \Dp{k^{(r)}} \mf{y}_{\ups}^{(r)}  \big( k^{(r)} ; s \big) \; = \;   - \big[  \mf{v}_r( k^{(r)} ) \, + \, \ups \op{v}_F  \big] \;. 
\enq
\textit{A priori}, and this is supported by numerical investigations, \textit{c.f.} Appendix \ref{Appendix Section phase oscillante dpdte de la vitesse}, 
$|\mf{v}_r|$ may or may not be smaller than $\op{v}_F$, namely depending on the choice of the anisotropy $\zeta$, the values of the magnetic field $h$ -and hence the endpoint of the 
Fermi zone-, the value of  $r \in \mf{N}$ and, finally, the value of $k_0^{(r)}(s)$ both situations may occur, namely  
\beq
| \mf{v}_r\big( k_0^{(r)}(s) \big) | < \op{v}_F \qquad \e{or} \qquad  | \mf{v}_r\big( k_0^{(r)}(s) \big) | > \op{v}_F \, .
\enq

\vspace{3mm}

\noindent In case 1 listed in the statement, \textit{viz}. $| \mf{v}_r\big(k_0^{(r)}(s) \big) | > \op{v}_F$, one observes that 
\beq
 \Dp{k^{(r)}} \mf{y}_{+}^{(r)}  \big( k_0^{(r)}(s) ; s \big)  \cdot  \Dp{k^{(r)}} \mf{y}_{-}^{(r)}   \big( k_0^{(r)}(s) ;  s   \big)  \; > \; 0
\enq
and that 
\beq
  -\e{sgn}\Big\{ \Dp{k^{(r)}}  \mf{y}_{+}^{(r)}   \big( k_0^{(r)} (s) ; s \big) 
\cdot \big[  \Dp{k^{(r)}}  \mf{y}_{+}^{(r)}  \big( k_0^{(r)}(s) ; s \big) \, - \,   \Dp{k^{(r)}} \mf{y}_{-}^{(r)} \big( k_0^{(r)}(s) ; s \big)  \big]  \Big\}
\; = \; - \e{sgn}\Big\{ \mf{v}_r\Big( k_0^{(r)}(s) \Big)  \Big\} \;. 
\enq
This is all that is needed so as to apply Theorem \ref{Proposition integrale principale spectral function} to the situation of interest. 
 
\vspace{3mm}
 
Finally, when $| \mf{v}_r\big(k_0^{(r)}(s) \big) | < \op{v}_F$ the analysis parallels the one exposed for the contribution of the one hole excitation sector. The details are left to the interested reader. \qed

\subsection{The multi-hole/$r$-string excitation sector}

Below, I will consider the contribution to the DRF issuing form the sector containing multiple hole and multiple $r$-strings all having the same value of $r \in \mf{N}$. 
 Although it will be not discussed here, the case of multiple hole and various numbers $r$-strings can be treated analogously and
leads to a similar structure of singularities. Likewise, one may derive the behaviour in the sector built up only from multi-particle excitations.
Such results may be easily extracted from the main structural theorem governing the asymptotics of the 
class of multiple integrals of interest to the analysis of dynamic response functions, Theorem \ref{Theorem Principal}
which is established in the Appendix.

\subsubsection{Excitations built up from holes and, possibly, particles}

For the purpose of the present section, it is convenient to parametrise the momentum-energy $(k,\om)$ combination as 
\beq
k=\mc{P}_0 \; \; \e{where} \;\; \mc{P}_0 \; = \;  \pi  \op{s}_{\ga} \, + \, p_{F}\sul{\ups=\pm}{}  \ups \ell_{\ups}  + \mf{q}_0 -2\pi s_{0}
\quad \e{and}  \quad  \; \om=\de \om + \mc{E}_0  
\enq
where $\ell_{\pm}$ are subject to the constraint $\sul{\ups=\pm}{}\ell_{\ups}=n_h-n_p$. The integers $n_p, n_h$ are assumed to satisfy  
\beq
n_h \geq 1 \quad \e{and} \quad  n_p+n_h \geq 2 \; . 
\enq
In this case of interest, the contribution of these types of excitations to the dynamical response function takes the form 
\bem
\msc{S}^{(\ga)}_{\bs{n}_{hp}}\big(\mc{P}_0,\mc{E}_0 + \de \om\big) \; = \; \Int{ \big( \msc{J}_h^{(\eps)}\big)^{n_h}    }{} \hspace{-2mm}  \dd^{n_h} \bs{t} \;  \Int{ \big( \msc{J}_1^{(\eps)}\big)^{n_p}    }{}  \hspace{-2mm} \dd^{n_p} \bs{k}
\; \; \mc{F}^{(\ga)}\big( \bs{\mf{K}}^{(hp)} \big) \\
 \times \sul{s\in \mathbb{Z} }{}\pl{\ups= \pm }{} 
 \bigg\{   \Xi\Big( \de \om + \mf{y}_{\ups}^{(hp)}\big(  \bs{\mf{K}}^{(hp)}  ;s \big)  \Big) \cdot \Big[ \de \om + \mf{y}_{\ups}^{(hp)}\big(  \bs{\mf{K}}^{(hp)}   ;s \big)  \Big]^{ \De_{\ups}(\bs{\mf{K}}^{(hp)} ) -1 }  \bigg\}   
\cdot \Big( 1+\mf{r}\big( \bs{\mf{K}}^{(hp)}; s \big)  \Big) \;. 
\end{multline}
Here, I have set 
\beq
\bs{n}_{hp}\, = \, (\ell_+,\ell_-;n_h, n_{1}=n_p, 0,\dots ) 
\quad \e{and} \quad
\bs{\mf{K}}^{(hp)} \, = \, \Big( \ell_{+}, \ell_{-} ; \bs{t} , \bs{k}^{(1)}=\bs{k},\emptyset,\dots   \Big)   
\enq
and agree upon 
\beq
\mf{y}_{\ups}^{(hp)}\big(  \bs{\mf{K}}^{(hp)}   ;s \big) \, = \, \mc{E}_{0}\, - \, \sul{a=1}{n_p} \mf{e}_1(k_a) \, + \, \sul{a=1}{n_h} \mf{e}_1(t_a) 
\, + \, \ups \op{v}_F \Big( \mf{q}_0   - \, \sul{a=1}{n_p} k_a \, + \, \sul{a=1}{n_h} t_a   + 2\pi (s-s_0) \Big) \;. 
\enq

Also, recall, \textit{c.f.} Appendix \ref{Appendix Section phase oscillante dpdte de la vitesse}, that for a reduced range of the model's parameters and as a conjecture more generally, 
there exists a strictly decreasing diffeomorphism $\mf{t}:\intff{K_m}{K_M}\tend \intff{-p_F}{p_F}$ such that 
$\mf{v}_1(k)=\mf{v}_1( \mf{t}(k))$. Set $\mf{p}=\mf{t}^{-1}$ for its inverse. 
Further, given  $t \in  \intff{ - p_F }{ p_F }$, let 
\beq
\mc{P}(t) \, = \, n_p \mf{p}(t) - n_h t \quad \e{and} \quad 
\mc{E}(t) \, = \, n_p \mf{e}_1( \mf{p}(t) ) - n_h \mf{e}_1(t)  \;. 
\enq
Finally, the remainder satisfies 
\beq
\mf{r}\big( \bs{\mf{K}}^{(hp)}; s \big) \, = \,  \e{O}\Big(  \sum_{ \ups = \pm } \big|\, \de \om + \mf{y}_{\ups}^{(hp)}\big( \bs{\mf{K}}^{(hp)}  ;s \big) \big|^{1-\tau }  \Big)  \;, 
\enq
and the control on $\mf{r}\big( \bs{\mf{K}}^{(hp)}; s \big) $  is differentiable in the sense of Definition \ref{defintion reste differentiable}.

\begin{theorem}
\label{Theorem DRF behaviour of multi particle hole threshold}

Let $\mf{q}_0(s)=\mf{q}_0+2\pi(s-s_0)$ and assume that $n_p+n_h \geq 2$ with $n_h \geq 1$. Also, assume that Conjecture \ref{Conjecture  diffeo liee a la vitesse} holds if the 
set of parameters of the model does not enter into the specifications of Theorem \ref{Proposition proprietes fondamentales de la vitesse des particules trous}. 

\vspace{2mm}

If, for any $s \in \mathbb{Z}$, 
\beq
\big( \mf{q}_0(s) , \mc{E}_0 \big) \not\in \Big\{ \big(\mc{P}(t), \mc{E}(t) \big) \; : \; t \in \intff{-p_F}{p_F} \Big\} \;, 
\enq
%
%
%
%
%
%
%
%
%
%
%
then $\msc{S}^{(\ga)}_{\bs{n}_{hp}}\big(\mc{P}_0,\mc{E}_0 + \de \om\big)$ is a smooth function in $\de \om$ belonging to a neighbourhood of the origin. 

\vspace{3mm}

\noindent Assume that, for at least one $s \in \mathbb{Z}$ 
\beq
\big( \mf{q}_0(s) , \mc{E}_0 \big) \, = \,  \Big( \mc{P}(t_0(s)), \mc{E}(t_0(s)) \Big) \quad for \; a  \quad  t_0(s) \in \intoo{-p_F}{p_F} \;, 
\label{definition rapidite k0 de s dans DRF multi points}
\enq
and that for any such $s$ it holds $\de_{\ups}^{(hp)}(s)>0$ where 
\beq
\de_{\ups}^{(hp)}(s) \, = \, \De_{\ups}( \bs{\mf{K}}^{(hp)}_0(s))   \quad with \quad 
\bs{\mf{K}}_{0}^{(hp)}\, = \,\Big( \ell_{+},\ell_{-} ; \bs{t_0(s)},\bs{\mf{p}( t_0(s) ) } , \emptyset ,\dots  \Big)   \;. 
\enq
There  $ \bs{ t_0(s) } \, = \,  \Big( t_0(s) ,\cdots, t_0(s)  \Big)\in \R^{n_h}$ and $ \bs{\mf{p}( t_0(s) ) } = \Big( \mf{p}( t_0(s) ),\cdots, \mf{p}( t_0(s) ) \Big)\in \R^{n_p}$.

Then, the multi particle-hole spectral function has the $\de \om \tend 0$ asymptotic expansion
\bem
\msc{S}^{(\ga)}_{\bs{n}_{hp}}\big(\mc{P}_0,\mc{E}_0 + \de \om\big)\; = \; \msc{S}^{(\ga)}_{\bs{n}_{hp};\e{reg}}(\de\om)  \; + \;  
   \hspace{-3mm}  \sul{  s\; : \; \exists t_0(s) }{}
%
\f{  \msc{F}^{(\ga)}\big( \bs{\mf{K}}^{(hp)}_0(s) \big) }{ \sqrt{ |\mc{P}^{\prime}(t_0(s)) | } } \cdot 
\Bigg( \f{-1 }{ \mf{v}^{\prime}_1\Big(  \mf{p}(t_0(s)) \Big) }  \Bigg)^{ \f{n_p^2}{2}} \hspace{-2mm} \cdot \;\;  \Bigg( \f{  1 }{ \mf{v}^{\prime}_1\Big( t_0(s) \Big) }  \Bigg)^{ \f{n_h^2-1}{2}} \\
\times G(n_p+1) \, G(n_h+1) \cdot  \f{ ( \sqrt{2 \pi} )^{3+n_p+n_h }   \Ga\Big( -\de_{+}^{(hp)}(s) -  \de_{-}^{(hp)}(s) -\tfrac{n_p^2+n_h^2 -3}{2}    \Big)    }
{   \big| \op{v}_F+\mf{v}_1\Big( t_0(s) \Big) \big|^{ \de_{-}^{(hp)}(s) }   \big| \op{v}_F-\mf{v}_1\Big(  t_0(s) \Big) \big|^{ \de_{+}^{(hp)}(s) }   }  
\cdot |\de \om |^{ \de_{+}^{(hp)}(s) + \de_{-}^{(hp)}(s) +\tfrac{n_p^2+n_h^2-3}{2}}  \\
\bigg\{    \Xi(\de \om) \f{1}{\pi} \sin \Big( \pi \Big[  \de_{+}^{(hp)}(s) + \de_{-}^{(hp)}(s)  \Big] \Big)     +     \Xi(- \de \om)    \f{1}{\pi} \sin \Big(\pi \f{n_p^2+n_h^2-1}{2} \Big)    \bigg\}
\; + \;   \e{O}\Big(   |\de \om |^{ \de_{+}^{(hp)}(s) + \de_{-}^{(hp)}(s) + \tfrac{n_p^2+n_h^2-1}{2} + \tau } \Big)     \;. 
\label{ecriture DA singulier de S pour multi part trou}
\end{multline}
$ \msc{S}^{(\ga)}_{\bs{n}_{hp};\e{reg}}(\de\om)$ is smooth in the neighbourhood of the origin. 

Finally, the summation over $s$ in \eqref{ecriture DA singulier de S pour multi part trou} runs through all solutions $t_0(s)$ to \eqref{definition rapidite k0 de s dans DRF multi points}. 
This summation contains at most two terms and, for generic parameters, it only contains one term.  

\end{theorem}

\Proof 

This is a direct consequence of Theorem \ref{Theorem Principal}. In order to identify quantities with the notations of that theorem, 
one should identify the quantities given in Section \ref{Appendix DA integrales multidimensionnelles} of the appendix:
\beq
\ell=2\; , \quad \msc{I}_1 \, = \, \msc{J}_h^{(\eps)} \;, \quad \msc{I}_2 \, = \, \msc{J}_1^{(\eps)} \;, \quad (n_1,n_2)=(n_h,n_p) 
\enq
in what concerns the intervals. Further, 
\beq
\big(\bs{p}^{(1)}, \bs{p}^{(2)} \big) \, = \, \big(\bs{t} , \bs{k} \big)  \; , \quad \big(\mf{u}_1, \mf{u}_{2} \big) \, = \, \big(\mf{e}_1 , \mf{e}_1 \big) \;, \quad 
\big( \zeta_1,\zeta_2 \big) \, = \, (-1,1) \;. 
\enq
From there one infers that one has 
\beq
\mc{P}^{\prime}(t) \, = \, -\big( n_h - n_p\mf{p}^{\prime}(t) \big)<0\quad \e{on} \quad  \intoo{-p_F}{p_F}
\enq
since $\mf{p}$ is strictly decreasing.  Also, it follows directly from the definition of $\mf{p}$ that 
\beq
\mc{E}^{\prime}(t) \, = \, \mf{v}_1(t) \cdot \mc{P}^{\prime}(t) 
\enq
so that $t\mapsto \mc{E}(t)$ is strictly increasing on $\intoo{-p_F}{0}$ and strictly decreasing on $\intoo{0}{p_F}$. 
Furthermore, one has that 
\beq
\Dp{t}\Big\{ \mc{E}_0-\mc{E}(t) \pm \op{v}_F\big( \mf{q}_0(s)-\mc{P}(t) \big)   \Big\} \; = \; -\mc{P}^{\prime}(t) \cdot \Big( \mf{v}_1(t)\pm \op{v}_F \Big) \not=0
\label{calcul derive grosse fct Z ups cas multiparticlues}
\enq
on $\intoo{-p_F}{p_F}$.

I first focus on the regular case, namely when, for any $s \in \mathbb{Z}$,
\beq
\big( \mf{q}_0(s) , \mc{E}_0 \big) \not\in \Big\{ \big(\mc{P}(t), \mc{E}(t) \big) \; : \; t \in \intff{-p_F}{p_F} \Big\} \;. 
\enq
Then observe that \eqref{calcul derive grosse fct Z ups cas multiparticlues} implies that $t\mapsto \mc{E}_0-\mc{E}(t) \pm \op{v}_F\big( \mf{q}_0(s)-\mc{P}(t) \big)$ are both strictly monotonous 
on $\msc{J}_h^{(\eps)}$. Thus, should one of these functions vanish on $\Dp{}\msc{J}_h^{(\eps)}$, it is enough to slightly change $\eps>0$, which is a free parameter in the problem (as long as it is small 
and strictly positive) so as to have a non-vanishing function. Thus, automatically, condition \eqref{ecriture condition positivite energie impulsion macroscopique thm principal} stated
in Theorem \ref{Theorem Principal} is satisfied. Then, the results of that theorem ensures the smooth behaviour 
of $\de\om \mapsto \msc{S}^{(\ga)}_{\bs{n}_{hp}}\big(\mc{P}_0,\mc{E}_0 + \de \om\big)$ around $0$. 

\vspace{3mm}

In the case when there exist at least one $s \in \mathbb{Z}$ such that $\big( \mf{q}_0(s) , \mc{E}_0 \big) \, = \,  \Big( \mc{P}(t_0(s)), \mc{E}(t_0(s)) \Big) $ for a $t_0(s) \in \intoo{-p_F}{p_F}$, 
one needs to identify additional constants. First, however, one fixes $\eps$ such that $t_0(s) \in \e{Int}(\msc{I}_h^{(\eps)})$ for any $s$ compatible with the mentioned constraint. 
One should also observe that the variations of  $t\mapsto \mc{E}(t)$ and  $t\mapsto \mc{P}(t)$ on $\intoo{-p_F}{p_F}$ entail that 
there may at most exist two different $s$ such that the previous equality holds. It is also evident that, in the generic case, only one such $s$ will exist.

Since $\mf{u}_1^{\prime\prime}(t)=\mf{v}_1^{\prime}(t)>0$ on $\intoo{-p_F}{p_F}$, while  $\mf{u}_{2}^{\prime\prime}( \mf{p}(t) ) = \tf{ \mf{v}_1^{\prime}(t) }{ \mf{p}^{\prime}(t) } <0$ on $\intoo{-p_F}{p_F}$, 
it follows that 
\beq
\left\{ \ba{cccccc} \veps_1 & =  - \zeta_1 \e{sgn}\Big( \mf{u}_1^{\prime\prime}(t_0(s)) \Big) & = &  \e{sgn}\Big( \mf{v}_1^{\prime}(t_0(s)) \Big)  & =1  \vspace{2mm} \\
\veps_2 & =   - \zeta_2 \e{sgn}\Big( \mf{u}_2^{\prime\prime}\big[ \mf{p}(t_0(s)) \big] \Big) &  =  &  -\e{sgn}\Big( \mf{v}_1^{\prime}\big[ \mf{p}(t_0(s)) \big] \Big) & =-1  \ea \right. 
\enq
and $\mf{s}= -\e{sgn} \bigg(  \f{\mc{P}^{\prime}(t_0) }{ \mf{u}_1^{\prime\prime}(t_0(s))  } \bigg) > 0 $\;. 
All parameters being identified, it remains to apply the results of Theorem \ref{Theorem Principal}
to each $s \in \mathbb{Z}$ such that a $t_0(s)$ exists.

Finally, the $n_h \geq 2 $ and $n_p=0$ case is treated along much the same lines. The results boils down to \eqref{ecriture DA singulier de S pour multi part trou} 
with $n_p$ being set to $0$. \qed



\subsubsection{Excitations built up from holes and a fixed $r$-string species}

Below, $r\in \mf{N}_{\e{st}}$ is assumed to be fixed. 
For the purpose of the present section, it is convenient to parametrise the momentum-energy $(k,\om)$ combination as 
\beq
k=\mc{P}_0 \; \; \e{where} \;\; \mc{P}_0 \; = \;  \pi  \op{s}_{\ga} \, + \, p_{F}\sul{\ups=\pm}{}  \ups \ell_{\ups}  + \mf{q}_0 -2\pi s_{0}
\quad \e{and}  \quad  \; \om=\de \om + \mc{E}_0  
\enq
where $\ell_{\pm}$ are subject to the constraint $\sul{\ups=\pm}{}\ell_{\ups}=n_h-n_{\e{st}}$. The integers $n_{\e{st}}, n_h$ are assumed to satisfy  
\beq
n_h \geq 1 \quad \e{and} \quad  n_{\e{st}} + n_h \geq 2 \; . 
\enq
In this case of interest, the contribution of these types of excitations to the dynamical response function takes the form 
\bem
\msc{S}^{(\ga)}_{\bs{n}_{hr}}\big(\mc{P}_0,\mc{E}_0 + \de \om\big) \; = \;
\Int{ \big( \msc{J}_h^{(\eps)}\big)^{n_h}    }{} \hspace{-2mm}  \dd^{n_h} \bs{t} \;  \Int{  \msc{J}_r^{ n_{\e{st}} }    }{}  \hspace{-2mm} \dd^{ n_{\e{st}} } \bs{k}
\; \; \mc{F}^{(\ga)}\big( \bs{\mf{K}}^{(hr)} \big) \\
 \times \sul{s\in \mathbb{Z} }{}\pl{\ups= \pm }{} 
 \bigg\{   \Xi\Big( \de \om + \mf{y}_{\ups}^{(hr)}\big(  \bs{\mf{K}}^{(hr)}  ;s \big)  \Big) \cdot \Big[ \de \om + \mf{y}_{\ups}^{(hr)}\big(  \bs{\mf{K}}^{(hr)}   ;s \big)  \Big]^{ \De_{\ups}(\bs{\mf{K}}^{(hr)} ) -1 }  \bigg\}   
\cdot \Big( 1+\mf{r}\big( \bs{\mf{K}}^{(hr)}; s \big)  \Big) \;. 
\end{multline}
Here, I have set 
\beq
\bs{n}_{hr}\, = \, (\ell_+,\ell_-;n_h, 0,\dots, 0, n_{r}=n_{\e{st}}, 0,\dots ) 
\quad \e{and} \quad
\bs{\mf{K}}^{(hr)} \, = \, \Big( \ell_{+}, \ell_{-} ; \bs{t} ,\emptyset , \dots, \emptyset,  \bs{k}^{(r)}=\bs{k},\emptyset,\dots,  \Big)   
\enq
and agree upon 
\beq
\mf{y}_{\ups}^{(hr)}\big(  \bs{\mf{K}}^{(hr)}   ;s \big) \, = \, \mc{E}_{0}\, - \, \sul{a=1}{ n_{\e{st}} } \mf{e}_r(k_a) \, + \, \sul{a=1}{n_h} \mf{e}_1(t_a) 
\, + \, \ups \op{v}_F \Big( \mf{q}_0   - \, \sul{a=1}{ n_{\e{st}} } k_a \, + \, \sul{a=1}{n_h} t_a   + 2\pi (s-s_0) \Big)
\enq

Also, recall, \textit{c.f.} Appendix \ref{Appendix Section phase oscillante dpdte de la vitesse}, that for a reduced range of the model's parameters and as a conjecture more generally, 
there exists a diffeomorphism $\mf{h}^{(r)}:\intff{-p_F}{p_F} \tend \intff{ K_m^{(r)} }{  K_M^{(r)} }$ such that 
$\mf{v}_1(t)=\mf{v}_r( \mf{h}^{(r)}(t))$. 
Further, given  $t \in  \intff{ - p_F }{ p_F }$, let 
\beq
\mc{P}(t) \, = \, n_{\e{st}} \mf{h}^{(r)}(t) - n_h t \quad \e{and} \quad 
\mc{E}(t) \, = \, n_{\e{st}} \mf{e}_r( \mf{h}^{(r)}(t) ) - n_h \mf{e}_1(t)  \;. 
\enq
Finally, the remainder satisfies 
\beq
\mf{r}\big( \bs{\mf{K}}^{(hr)}; s \big) \, = \,  \e{O}\Big(  \sum_{ \ups = \pm } \big|\, \de \om + \mf{y}_{\ups}^{(hr)}\big( \bs{\mf{K}}^{(hr)}  ;s \big) \big|^{1-\tau }  \Big)  \;, 
\enq
and the control on $\mf{r}\big( \bs{\mf{K}}^{(hr)}; s \big) $ is differentiable in the sense of Definition \ref{defintion reste differentiable}.

\begin{theorem}
\label{Theorem DRF behaviour of multi string hole threshold}

Let $\mf{q}_0(s)=\mf{q}_0+2\pi(s-s_0)$ and assume that $n_{\e{st}}+n_h \geq 2$ with $n_h \geq 1$. Also, assume that Conjecture \ref{Conjecture  diffeo liee a la vitesse} holds if the 
set of parameters of the model does not enter into the specifications of Theorem \ref{Proposition proprietes fondamentales de la vitesse des particules trous}. 
 Finally, assume that $t \mapsto \mc{P}(t)$ is a diffeomorphism on $\intff{-p_F}{p_F}$. 

\vspace{2mm}

If, for any $s \in \mathbb{Z}$, 
\beq
\big( \mf{q}_0(s) , \mc{E}_0 \big) \not\in \Big\{ \big(\mc{P}(t), \mc{E}(t) \big) \; : \; t \in \intff{-p_F}{p_F} \Big\} \;, 
\enq
%
%
%
%
%
%
%
%
%
%
%
then, for $\eps>0$ small enough, $\msc{S}^{(\ga)}_{\bs{n}_{hr}}\big(\mc{P}_0,\mc{E}_0 + \de \om\big)$ is a smooth function in $\de \om$ belonging to a neighbourhood of the origin. 

\vspace{3mm}

\noindent Assume that, for at least one $s \in \mathbb{Z}$, 
\beq
\big( \mf{q}_0(s) , \mc{E}_0 \big) \, = \,  \Big( \mc{P}(t_0(s)), \mc{E}(t_0(s)) \Big) \quad for \; a  \quad  t_0(s) \in \intoo{-p_F}{p_F} \;, 
\label{definition rapidite k0 de s dans DRF multi hole r string}
\enq
and that, for any such $s$ it holds $\de_{\ups}^{(hr)}(s)$ with  
\beq
\de_{\ups}^{(hr)}(s) \, = \, \De_{\ups}( \bs{\mf{K}}^{(hr)}_0(s))   \quad and \quad 
\bs{\mf{K}}_{0}^{(hr)}\, = \,\Big( \ell_{+},\ell_{-} ; \bs{t_0(s)},\emptyset, \dots, \emptyset, \underset{ r-\e{string} \, \e{position} }{ \bs{\mf{h}^{(r)}( t_0(s) ) } } , \emptyset ,\dots  \Big)   \;. 
\enq
There  $ \bs{ t_0(s) } \, = \,  \Big( t_0(s) ,\cdots, t_0(s)  \Big)\in \R^{n_h}$ and $ \bs{\mf{h}^{(r)}( t_0(s) ) } = \Big( \mf{h}^{(r)}( t_0(s) ),\cdots, \mf{h}^{(r)}( t_0(s) ) \Big)\in \R^{ n_{\e{st}} }$. 

Then, the multi r-string-hole spectral function has the $\de \om \tend 0$ asymptotic expansion
\bem
\msc{S}^{(\ga)}_{\bs{n}_{hr}}\big(\mc{P}_0,\mc{E}_0 + \de \om\big)\; = \; \msc{S}^{(\ga)}_{\bs{n}_{hr};\e{reg}}(\de\om)  \; + \;  
   \hspace{-3mm}  \sul{  s\; : \; \exists t_0(s) }{}
%
\f{  \msc{F}^{(\ga)}\big( \bs{\mf{K}}^{(hr)}_0(s) \big) }{ \sqrt{ |\mc{P}^{\prime}(t_0(s)) | } } \cdot 
\Bigg( \tfrac{ 1 }{ \big| \mf{v}^{\prime}_1\big(  \mf{h}^{(r)}(t_0(s)) \big) \big|  }  \Bigg)^{ \f{n_{\e{st}}^2}{2}} \hspace{-2mm} \cdot \;\;  \Bigg( \tfrac{  1 }{ \mf{v}^{\prime}_1\big( t_0(s) \big) }  \Bigg)^{ \f{n_h^2-1}{2}} \\
\times G( n_{\e{st}} + 1) \, G(n_h+1) \cdot  \f{ ( \sqrt{2 \pi} )^{3+n_{\e{st}}+n_h }   \Ga\Big( -\de_{+}^{(hr)}(s) -  \de_{-}^{(hr)}(s) -\tfrac{n_{\e{st}}^2+n_h^2 -3}{2}    \Big)    }
{   \big| \op{v}_F+\mf{v}_1\Big( t_0(s) \Big) \big|^{ \de_{-}^{(hr)}(s) }   \big| \op{v}_F-\mf{v}_1\Big(  t_0(s) \Big) \big|^{ \de_{+}^{(hr)}(s) }   }  
\cdot |\de \om |^{ \de_{+}^{(hr)}(s) + \de_{-}^{(hr)}(s) +\tfrac{n_{\e{st}}^2+n_h^2-3}{2}}  \\
\bigg\{    \Xi(\de \om) \f{1}{\pi} \sin \Big( \pi   \nu_+^{(hr)}(s)   \Big)     +     \Xi(- \de \om)    \f{1}{\pi} \sin \Big(\pi  \nu_-^{(hr)}(s)  \Big)    \bigg\}
\; + \;   \e{O}\Big(   |\de \om |^{ \de_{+}^{(hr)}(s) + \de_{-}^{(hr)}(s) + \tfrac{n_{\e{st}}^2+n_h^2-1}{2} + \tau  } \Big)     \;. 
\label{ecriture DA singulier de S pour multi hole r string}
\end{multline}
$ \msc{S}^{(\ga)}_{\bs{n}_{hr};\e{reg}}(\de\om)$ is smooth in the neighbourhood of the origin.
Further, one has
\beq
\nu_+^{(hr)}(s)  \; = \; \de_{+}^{(hr)}(s) + \de_{-}^{(hr)}(s)+\frac{1}{2}\de_{-,\sg_r}  n_{\e{st}}^2-\f{1-\mf{s}_r}{2}
\quad , \qquad 
\nu_-^{(hr)}(s)  \; = \;  \frac{1}{2}\de_{+,\sg_r}  n_{\e{st}}^2 \, + \,  \f{ 1}{2} n_h^2  + \f{1-\mf{s}_r}{2}
\enq
where 
\beq
\sg_r \, = \,  \e{sgn}\Big[ \mf{v}^{\prime}_1\Big(  \mf{h}^{(r)}(t_0(s)) \Big) \Big]  \qquad and \qquad \mf{s}_r \; = \; - \e{sgn}\Big[ \mc{P}^{\prime}(t_0(s)) \Big] \;. 
\enq

Finally, the summation over $s$ in \eqref{ecriture DA singulier de S pour multi hole r string} runs through all solutions $t_0(s)$ to \eqref{definition rapidite k0 de s dans DRF multi hole r string}. 
This summation contains at most two terms and, for generic parameters, it only contains one term.  

\end{theorem}

 The proof follows closely the case of the multi-hole multi-particle sector, so that I omit the details.

\section{Conclusion}

This work developed a technique allowing one to extract, on rigorous grounds, the asymptotic behaviour in certain parameters of a family of multiple integrals. These results are detailed
in Sections \ref{Appendix DA integrales unidimensionnelles}, \ref{Appendix DA integrales multidimensionnelles} of the appendix. The multiple integrals studied in these sections, upon specialisation, contain the multiple
integrals which define the coefficients of the series giving the 
massless form factor expansion issued representation for the DRF in the XXZ chain that was derived in \cite{KozMasslessFFSeriesXXZ}.
Hence, the analysis I developed allowed, upon relying on additional properties that were argued in \cite{KozLongDistanceLargeTimeXXZ,KozMasslessFFSeriesXXZ}, 
to give a precise characterisation of the singular behaviour in the $(k,\om)$ plane of the series' coefficients. 
 In doing so, this work provides a test and confirmation of the predictions, 
issuing from the non-linear Luttinger liquid approach, for some of the singularities of the DRF, namely those
issuing from a one massive excitation process. Furthermore, the work showed the existence of other singularity lines: the ones issuing from multi-particle/hole/r-string excitations and which correspond to 
configurations of the various momenta that maximise the multi-excitation energy at fixed momentum. 
Such multi-species singularity curves generate structurally different edge exponents and universality constants. The edge exponents associated with one such 
"mixed" excitation were discussed in \cite{AffleckPereiraWhiteSpectralFunctionsfor1DLatticeFermionsBoundStatesContributions}, but all the other cases were not
considered in the literature. Furthermore, the work \cite{AffleckPereiraWhiteSpectralFunctionsfor1DLatticeFermionsBoundStatesContributions} only focused on the exponents
and so did not provide any expression for the universal part of the amplitude. Thus, the expression for the universal part of the 
amplitude is new.

\section*{Acknowledgment}

K.K.K. acknowledges support from  CNRS and ENS de Lyon. The author is indebted to J.-S. Caux, F. Göhmann, J.M. Maillet, G. Niccoli for stimulating discussions
on various aspects of the project.

\appendix

\section*{Appendix}

\section{Main notations}
\label{Appendix Fcts speciales}

\subsection*{Sets}

\begin{itemize}
\item Given a set $A$, $\e{Int}(A)$ stands for its interior,  $\ov{A}$ for its closure and $\Dp{}A$ for its boundary.
\item Given a finite set $A$, $|A|$ stands for its cardinal. 
\item $\mathbb{N}=\{0,1, 2, \dots\}$, $\R^{+}=\intoo{0}{+\infty}$, $\R^{*}=\R\setminus \{0\}$.
\item $\intn{1}{n}=\{1,\dots, n\}$ and $\mf{S}_{n}$ stands for the permutation group of $\intn{1}{n}$.  
\item $\sqcup$ refers to the disjoint union of sets. 
\item $\de_{a,b}$ stands for the Kronecker symbol: $\de_{a,b}=1$ if $a=b$ and $\de_{a,b}=0$ otherwise. 
\item Given $\ell$ integers $n_1,\dots, n_{\ell}$, it is understood that 
\beq
\ov{\bs{n}}_{\ell} \; = \; \sul{ r = 1 }{ \ell } n_{r} \;. 
\label{definition bs ne ell}
\enq

\end{itemize}

\subsection*{Vectors and related objects}

\begin{itemize}

\item $^{\op{t}}$ stands for the transposition of a matrix or vector, depending on the context. 
\item $\op{I}_{n}$ stands for the identity matrix on $\R^n$, and it will sometimes also be denoted as $\e{id}$. 
\item Vectors are denoted in bold, \textit{viz}. $ \bs{x} \in \R^n$ corresponds to the vector $\bs{x}=(x_1,\dots,x_n)$. The dimensionality of the vector is 
always undercurrent by the context. 
\item If the vector space has a natural Cartesian product structure $\pl{r=1}{\ell} \R^{n_{r}}$, then any vector $\bs{x}$ is represented as
\beq
\bs{x} \; = \; \big( \bs{x}^{(1)} , \dots ,  \bs{x}^{(\ell)} \, \big)\qquad \e{with} \qquad \bs{x}^{(r)}\, = \, \big( x_1^{(r)}, \cdots, x_{n_r}^{(r)} \big) \in \R^{n_r} \;. 
\label{notation vecteur dans produit cartesien}
\enq
\item Vectors with omitted coordinates are denoted as :
\beq
\bs{x}^{(r)}_{[a]} \, = \, \big( x_1^{(r)} , \dots, x_{a-1}^{(r)},x_{a+1}^{(r)},\dots , x_{n_r}^{(r)}  \,  \big) \quad \e{and} \quad 
\bs{x}_{[r,a]} \, = \,  \Big( \bs{x}^{(1)} , \dots , \bs{x}^{(r)}_{[a]},   \dots , \bs{x}^{(\ell)} \big) \;. 
\label{notation vecteurs reduits}
\enq
\item Given $\bs{\a}, \bs{\be} \in  \pl{r=1}{\ell} \mathbb{N}^{n_{r}}$, it is understood that 
\beq
| \bs{\a} | \, = \, \sul{r=1}{ \ell }  \sul{a=1}{n_{r}} \a_{a}^{(r)} \qquad \e{and} \qquad 
\bs{\a} \geq  \bs{\be}  \quad \Longleftrightarrow \quad \forall (r,a) \quad \a_{a}^{(r)} \geq \be_{a}^{(r)} \;. 
\label{notation norme et ordre partiel sur entiers vecteurs}
\enq
\item Given $\bs{\a}  \in  \pl{r=1}{\ell} \mathbb{N}^{n_{r}}$ and $\bs{x} \in \pl{r=1}{\ell} \R^{n_{r}}$, one has
\beq
\bs{x}^{ \bs{\a} } \, = \, \pl{r=1}{\ell}\pl{a=1}{n_r} \big[ x_{a}^{(r)} \big]^{ \a_{a}^{(r)} } \;. 
\label{notation exposant polynomial vectoriel}
\enq
\end{itemize}

\subsection*{Functions}

\begin{itemize}
\item Given a set $A$, $\bs{1}_{A}$ stands for the indicator function of $A$. 
\item  $\Xi$ refers to the Heaviside step function, \textit{viz}. $\Xi=\bs{1}_{\R^+}$.
\item $\Ga$ refers to the Gamma function which allows one to express the Euler $\be$-integral as
\beq
\Int{0}{1} t^{x-1} (1-t)^{y-1} \cdot \dd t  \, = \, \f{ \Ga(x) \Ga(y) }{  \Ga(x + y) } \;. 
\label{ecriture formule integrale pour fct beta}
\enq
\item $G$ stands for the Barnes \cite{BarnesDoubleGaFctn1,BarnesDoubleGaFctn2} function. 
\item The Gaudin-Mehta integral is expressed in terms of the Barnes function as:

\beq
\Int{ \R^n }{}  \dd \bs{y}  \; \ex{-  (\bs{y},\bs{y})} \pl{a<b}{n} (y_a-y_b)^2 \; =\;  \Big( \f{1}{2} \Big)^{\f{1}{2} n^2 } \big( 2\pi \big)^{ \f{n}{2} } G(2+n) \;. 
\label{formule integrale Gaudin-Mehta}
\enq
\item Given  $S \subset \R^{n} $ measurable and a function $f:S \tend \R$,
\beq
 \norm{f}_{L^{\infty}(S)} \, = \, \e{supess}\Big\{|f(\bs{x})| \, : \,  \bs{x}\in S   \Big\} \;. 
\enq
\item Given $g:U\times V \tend W$, with $U\subset \R^n$, $V\subset \R^m$ and $W\subset \R^o$ the totally even part of a function 
in respect to a set of variables is defined as 
\beq
\Big[  g(\bs{z}, \bs{v}) \Big]_{ \bs{z}-\e{even} } \; = \; \f{1}{ 2^{n} } \sul{ \substack{ \eps_a=\pm \\  a=1,\dots, n }  }{ }  g(\bs{z}^{(\eps)}, \bs{v})
\qquad \e{with} \qquad 
\bs{z}^{(\eps)}\; = \; \big(\eps_1 z_1, \dots, \eps_d z_d \big) \;. 
\label{definition even part of a function}
\enq

\item Given a smooth function $f:U\tend V$ between two open subsets $U\subset \R^n$ and $V\subset \R^m$, $D_{\bs{x}}^{(k)}f$ denotes the $k^{\e{th}}$-order differential of $f$
at the point $\bs{x}\in U$. When $k=1$, it is simply denoted as $D_{\bs{x}}f$. 

\end{itemize}

\begin{defin}
\label{defintion reste differentiable}
 
 Given smooth functions $f,g$ on an open neighbourhood of a point $\bs{y}  \in \R^n$, one says that a $\e{O}$-remainder relation $f=\e{O}(g)$ when $\bs{x} \tend \bs{y}$ is 
differentiable if, for each $\bs{\ell}=(\ell_1,\dots, \ell_n) \in \mathbb{N}^n$ there exists a smooth function $\psi_{\bs{\ell}}$ in the vicinity of $\bs{y}$ and a constant $C_{\bs{\ell}}>0$
such that 
\beq
\pl{a=1}{n} \Dp{x_a}^{\ell_a} \cdot f(\bs{x}) \;  \leq  \; C_{\bs{\ell}} \cdot \pl{a=1}{n} \Dp{x_a}^{\ell_a} \cdot [\psi_{\bs{\ell}} g \big](\bs{x}) 
\label{Defintion reste differentiable}
\enq
on some open neighbourhood of $\bs{y}$.

\end{defin}

 Note that the use of $\psi_{\bs{\ell}}$ in this definition allows one to encompass a situation when $g$ does not depend explicitly on some of the variables.






\section{Auxiliary theorems}
\label{Appendix auxiliary theorems}

The proof of theorems \ref{Theorem Morse Lemma}, \ref{Theorem Weierstrass preparation theorem} and \ref{Theorem Malgrange preparation theorem} 
can be found in \cite{GolubitskyGuilleminStableMappingAndSings}. The proof of Theorem \ref{Theoreme extension de Whitney} can be found in \cite{BierstoneWhitneyExtensionTheoremAndOtherDiffFctsProperties}.

\begin{theorem}{\bf Morse Lemma}
 \label{Theorem Morse Lemma}

Let $f : U\tend \R $ be a smooth  function on an open set $U\subset \R^n$. Let $\bs{p} \in U$ be a non-degenerate critical point of $f$. 
Let $\op{M}$ be the matrix associated with the bilinear form $ D_{\bf{p}}^2f$:
\beq
\big( \bs{v}, \op{M} \bs{w} \big) \, = \,  D_{\bf{p}}^2f\big( \bs{v}, \bs{w} \big)  \; .
\enq
Then, there exists an open neighbourhood $V_0$ of $\bs{0}\in \R^n$ and a smooth diffeomorphism  onto $g: V_0 \mapsto U_0 \subset U$ such that 
\begin{itemize}
 
 \item  $\bs{0}\in V_0$ and $g(\bs{0})=\bs{p} \in U_0$;
 
 \item $f\circ g (\bs{x}) \, = \, (\bs{x},  \op{M} \bs{x})$  on $V_0$. 
 
\end{itemize}
 Here $(\cdot,\cdot)$ is the canonical scalar product on $\R^n$.
%
%
%
%
%
%
%
%
 
\end{theorem}

\begin{theorem}{\bf Weierstrass preparation theorem}
 \label{Theorem Weierstrass preparation theorem}
 
 Let $f$ be a holomorphic function on an open  set $V \subset \Cx^n$. Let $\bs{y}\in V$ and $d \in \mathbb{N}$ be such that 
\beq
\big( \Dp{z_n}^{k}f \big) (\bs{y}) \, =  \, 0 \quad  for \quad k=0,\dots, d-1, \quad and \quad 
\big( \Dp{z_n}^{d}f\big) (\bs{y}) \, \not= \, 0 \;. 
\enq
Then there exists 
\begin{itemize}
\item  open neighbourhoods $U_0 \subset \Cx^{n-1}$ of $\bs{y}_{[n]}=(y_1,\dots,y_{n-1})$ and $W_0 \subset \Cx $ of $y_n$
such that $V_0=U_0\times W_0 \subset V$; 
\item a holomorphic, non-vanishing, function h on $V_0$;
\item a Weierstrass polynomial
\beq
\mc{W}(\bs{z}) \, = \, (z_n-y_n)^d \,  + \, \sul{k=0}{d-1}   (z_n-y_n)^k \,  a_k(\bs{z}_{[n]}) \qquad  with   \quad  
\bs{z}_{[n]}\, = \, (z_1,\dots,z_{n-1})\; ,
\enq
and $a_k$, $k \in \intn{0}{d-1}$  ,  being holomorphic functions on $U_0$ satisfying  $a_k(\bs{y}_{[n]})=0$; 
 \end{itemize}
such that one has the factorisation
\beq
f = \mc{W} \cdot h \quad  on  \quad V_0=U_0\times W_0 \, .
\enq

\end{theorem}

\begin{theorem}{\bf Malgrange preparation theorem}
 \label{Theorem Malgrange preparation theorem}
 
 Let $f$ be a smooth function on an open  set $V \subset \R^n$. Let $\bs{y}\in V$ and $d \in \mathbb{N}$ be such that 
\beq
\big( \Dp{z_n}^{k}f\big)(\bs{y}) \, =  \, 0 \quad  for \quad k=0,\dots, d-1, \quad and \quad 
\big( \Dp{z_n}^{d}f \big)(\bs{y}) \, \not= \, 0 \;. 
\enq
Then there exists 
\begin{itemize}
\item  open neighbourhoods $U_0 \subset \R^{n-1}$ of $\bs{y}_{[n]}=(y_1,\dots,y_{n-1})$ and $W_0 \subset \R $ of $y_n$
such that $V_0=U_0\times W_0 \subset V$; 
\item a smooth, non-vanishing, function h on $V_0$;
\item a Weierstrass polynomial
\beq
\mc{W}(\bs{x}) \, = \, (x_n-y_n)^d \,  + \, \sul{k=0}{d-1}  (x_n-y_n)^k \, a_k(\bs{x}_{[n]}) \qquad  with   \quad  
\bs{x}_{[n]}\, = \, (x_1,\dots,x_{n-1})\; ,
\enq
and the $a_k$'s all being smooth on $U_0$ and such that $a_k(\bs{y}_{[n]})=0 $;
 \end{itemize}
such that one has the factorisation
\beq
f = \mc{W} \cdot h \quad  on  \quad  V_0=U_0\times W_0 \, .
\enq
\end{theorem}

\begin{defin}
\label{Definition fct lisse sur ferme}
 
 Let $F$ be a closed set in $\R^n$ such that $F \; = \; \ov{\e{Int}(F)}$. A function $f$ is said to be a smooth function on $F$ if
\begin{itemize}
 
 \item $f$ is smooth on $\e{Int}(F)$; 
 
 \item for any $\bs{k} \in \mathbb{N}^n$, $f^{(\bs{k})}\equiv \pl{a=1}{N}\Dp{x_a}^{k_a} f$ extends continuously to $F$;
 
 \item for any $\bs{a} \in \Dp{} F$,  $f$ admits an all order Taylor series expansion, \textit{viz}. for any $m\geq 0$ it holds 
\beq
f(\bs{x}) \; = \; \sul{ \substack{  \bs{k} \in \mathbb{N}^n  \; : \; \\   |\bs{k}| \leq m  }  }{} f^{(\bs{k})}(a) \big(\bs{x}-\bs{a} \big)^{\bs{k}} \; + \, R_a^{m}[f](x)
\qquad \e{with} \qquad 
R_a^{m}[f](x) \, = \, \e{o}\Big( ||\bs{x}-\bs{a} ||^m \Big)  \;. 
\enq
\end{itemize}

\end{defin}

This definition of smoothness can be stated, in greated generality, in the language of jets where it translates itself in the jet associated to a given function being a Whitney field.

\begin{theorem}{\bf Whitney extension theorem}
\label{Theoreme extension de Whitney}

Let $U\subset \R^n$ be open and $X \subset U$ be closed in $\R^n$. Any $f$ smooth on $X$ admits a smooth extension into a function $f_{\mf{e}}$ to $U$, with $f_{\mf{e}}^{(\bs{k})}=f^{(\bs{k})}$ on $X$.

\end{theorem}






\section{Observables in the infinite XXZ chain}
\label{Appendix Observables XXZ}

\subsection{Solutions to linear integral equations}
\label{Appendix Lin Int Eqns Defs et al}

The observables describing the thermodynamic limit of the spin-$1/2$ XXZ chain are characterised by means of a collection of functions 
solving linear integral equations. These equations are driven by an operator $\op{K}_{\eta,Q}$ on $L^2(\intff{-Q}{Q})$ characterised by the integral kernel $K(\la,\mu)=K(\la-\mu\mid \eta)$ with 
\beq
K(\la\mid \eta )  \, = \,   \f{ \sin(2\eta)   }{ 2\pi \sinh(\la + \i \eta)  \sinh(\la  - \i \eta)  }  \;. 
\label{ecriture fonction K de lambda et eta}
\enq

To introduce all of the functions of interest to this work, one starts by defining the $Q$-dependent dressed energy which allows one to construct the Fermi zone of the model. 
It is defined as the solution to the linear integral equation
\beq
\veps(\la\mid Q) \, + \, \Int{-Q}{Q} K\big(\la-\mu\mid \zeta \big) \, \veps(\la\mid Q)  \cdot \dd \mu \; = \;  h - 4 \pi J \sin(\zeta) K\big( \la \mid \tfrac{1}{2}\zeta \big)  \;. 
\label{definition energie habille et energie nue}
\enq
Note that the unique solvability of \eqref{definition energie habille et energie nue} follows from $\op{K}_{\zeta,Q}$ having its spectral radius $<1$. 

The endpoint of the Fermi zone is defined as the unique \cite{KozDugaveGohmannThermoFunctionsZeroTXXZMassless} positive solution $q$ to $\veps(q\mid q)=0$. 
Then, the function $\veps_1(\la)\equiv \veps(\la\mid q)$ corresponds to the dressed energy of the particle-hole excitations of the model. 
The functions 
\beq
\veps_r(\la)\; = \; r h - 4 \pi J \sin(\zeta) K\big( \la \mid \tfrac{r}{2}\zeta \big)   \, -\, \Int{-q}{q} K_{r}\big(\la-\mu \big) \veps_1(\mu)  \cdot \dd \mu 
\label{definition r energi habille}
\enq
with 
\beq
K_{r}(\la) \,  = \,  K\Big(\la \mid \tfrac{1}{2} \zeta(r+1) \Big) \, + \,  K\Big(\la \mid \tfrac{1}{2} \zeta(r-1) \Big)
\enq
correspond to the dressed energies of the $r$-bound state excitations.
For any $0<\zeta<\tf{\pi}{2}$ and under some additional constraints for $\tf{\pi}{2}< \zeta < \pi$, one can show \cite{KozProofOfStringSolutionsBetheeqnsXXZ}
that $\veps_{r}(\la+\i\de  \tf{\pi}{2}) > c_r>0$ for any $\la\in \R$, and $\de \in \{0, 1\}$. 
However, this lower bound should hold throughout the whole massless regime $0<\zeta<\pi$, irrespectively of some additional constraints. 
This property has been checked to hold by  numerical study of the solutions to \eqref{definition r energi habille}, \textit{c.f.} \cite{KozProofOfStringSolutionsBetheeqnsXXZ}. 

In order to introduce the dressed momenta of the $r$-bound states and of the particle-hole excitations, I first need to 
define the $r$-bound state bare phases $\theta_r$ :
\beq
\theta_r(\la) \, = \, 2\pi \Int{ \Ga_{\la}  }{}  K_r(\mu-0^+ ) \cdot \dd \mu \quad \e{for} \;\;  r \geq 2
\quad \e{and} \qquad \theta_1(\la) \, = \, \theta(\la\mid \zeta)
\enq
with 
\beq
\theta(\la\mid \eta) \, = \, 2\pi \Int{ \Ga_{\la}  }{}  K(\mu-0^+\mid \eta ) \cdot \dd \mu   \;. 
\enq

The contour of integration corresponds to the union of two segments $ \Ga_{\la} \; = \; \intff{ 0 }{ \i \Im(\la) }\cup \intff{\i \Im(\la) }{ \la } $ and the $-0^+$ prescription indicates that the poles 
of the integrand at $\pm \i \eta +\i \pi \mathbb{Z}$ should be avoided from the left.

Then, the function  
\bem
p_r(\la)\; = \; \theta\big(\la\mid \tfrac{r}{2}\zeta \big)  \, -\, \Int{-q}{q} \theta_{r}\big(\la-\mu \big) p^{\prime}_1(\mu)  \cdot \f{ \dd \mu }{2\pi} \\
\, + \, \pi \ell_r(\zeta)-p_{F}m_r(\zeta)
-2p_{F} \sul{ \sg=\pm }{} \big(1\, - \, \de_{\sg,-}\de_{r,1} \big) \e{sgn}\Big( 1- \tfrac{2}{\pi} \cdot \wh{\tfrac{r+\sg}{2}\zeta} \Big) \cdot \bs{1}_{ \mc{A}_{r,\sg} } (\la)\;, 
\label{definition r moment habille}
\end{multline}
extended by $\i \pi$-periodicity to $\Cx$, corresponds to the dressed momentum of the $r$-bound states. Above, I have introduced
\beq
 \ell_r(\zeta)=1-r+2 \lfloor \tfrac{r \zeta}{2\pi} \rfloor  \qquad \e{and} \qquad
   m_r(\zeta)=2-r - \de_{r,1} + 2 \sul{ \ups = \pm }{} \lfloor \zeta \tfrac{r + \ups   }{2\pi} \rfloor  \;. 
\enq
Furthermore, I agree upon  
\beq
\wh{\eta} \, = \, \eta  - \pi \lfloor \tfrac{ \eta }{ \pi } \rfloor \quad \e{and} \qquad \mc{A}_{r,\sg} \, = \, 
\Big\{ \la \in \Cx \; : \; \tfrac{\pi}{2}\geq |\Im(\la) | \geq \e{min}\big(   \wh{\tfrac{r+\sg}{2}\zeta} ,  \pi- \wh{\tfrac{r+\sg}{2}\zeta} \big)  \Big\}  \;. 
\enq
In order to obtain $p_r$, one should first solve the linear integro-differential equation for $p_1$ and then 
use $p_1$ to define $p_r$ by \eqref{definition r moment habille}.  $p_1$ corresponds to the dressed momentum of the particle-hole excitations and  $p_F=p_1(q)$ corresponds to the Fermi momentum. 
$1$-strings have their rapidities $\la \in \big\{  \R \setminus \intff{-q}{q} \big\} \cup \big\{ \R+\i\tf{\pi}{2}\big\}$ while $r\geq 2$ strings, $r \in \mf{N}\setminus \{1\}$, 
have their rapidities $\la \in \R+\i\de_{r}\tf{\pi}{2}$ for a $\de_r=0$ or $1$, depending on the value of $r$ and $\zeta$. See, \textit{e.g.} \cite{TakahashiThermodynamics1DSolvModels}
for more details on the string parities. 

One can show \cite{KozProofOfStringSolutionsBetheeqnsXXZ} under similar conditions on $\zeta$ as for the dressed energy that, for any $\la \in \R$, 
\beq
\big| p^{\prime}_{r}\big(\la  +\i \de_{r} \tfrac{\pi}{2} \big) \big| \; > \; 0 \quad \e{when} \quad r \in \mf{N}\setminus \{1\}
\qquad \e{and} \qquad 
\e{min}\Big( p^{\prime}_{1}\big(\la\big) \, , \, - p^{\prime}_{1}\big(\la  + \i \tfrac{\pi}{2} \big)   \Big) \; > \; 0   \;. 
\label{ecriture equation positivite pr prime}
\enq
Again, a numerical investigation indicates that \eqref{ecriture equation positivite pr prime} does, in fact, hold irrespectively of the value of $\zeta$.

It is convenient to introduce a piecewise shifted deformation of $p_1$:
\beq
\wh{p}_1(\la) \; = \; p_1(\la) \, -\, 2p_F \e{sgn}\big(\pi-2\zeta \big) \bs{1}_{\intoo{-\infty}{-q}}(\la) +2\pi \bs{1}_{\intoo{-\infty}{-q}\cup \{\R+\i\tfrac{\pi}{2}\}}(\la)
\enq
which is a diffeomorphism from the oriented concatenation of sets 
\beq
\intfo{q}{+\infty}\cup \big\{-\R+\i\tfrac{\pi}{2}\big\}\cup\intof{-\infty}{-q}  \quad 
\e{onto} \quad  \intff{p_F}{2\pi -p_F-\, 2p_F \e{sgn}\big(\pi-2\zeta \big)} \;. 
\enq
The image of  $\big\{-\R+\i\tfrac{\pi}{2}\big\}\cup\big\{ \R \setminus \intff{-q}{q} \big\}$ under $\wh{p}_1$ defines the range $ \msc{I}_1 = \intff{ p_-^{(1)} }{ p_+^{(1)} }$
with $p_-^{(1)}=p_F$ and $p_+^{(1)}=2\pi -p_F-\, 2p_F \e{sgn}\big(\pi-2\zeta \big)$, where the particles' momenta evolve. Likewise, the image  of $\R+\i \de_r \tf{\pi}{2}$
under $p_r$ defines the range $ \msc{I}_r = \intff{ p_-^{(r)} }{ p_+^{(r)} }$ where the $r$-string momenta evolve. $ p_{\pm}^{(r)}$ can be 
readily computed by taking the $\la -\i \de_r \tf{\pi}{2}\tend \pm \infty$ limits in \eqref{definition r moment habille}. However,
since their explicit values do not play a role, we do not provide them here. 

\vspace{3mm}

The dressed energies of the excitations in the momentum representation are defined as:
\beqa
\mf{e}_{1}(k) & = & \veps_1\circ \wh{p}_1^{\, -1}(k) \quad \e{for} \quad k \in \msc{I}_1
\label{definition fct mathfrak e 1} \\
\mf{e}_{r}(k) & = & \veps_r\circ p_r^{-1}(k) \quad \e{for} \quad k \in \msc{I}_r \quad \e{and} \quad r \in \mf{N}\setminus \{1\} \;. 
\label{definition fct mathfrak e r}
\eeqa

The $r$-bound dressed phase is defined as the solution to 
\beq
\phi_{r}(\la,\mu) \, = \, \f{ 1  }{ 2 \pi }  \theta_{r}\big( \la -\mu  \big) \, - \, \Int{-q}{q} K(\la-\nu)\,  \phi_{r}(\nu, \mu ) \cdot \dd \nu  \; + \; \f{m_{r}(\zeta)}{2}
\label{definition dressed phase}
\enq
and the dressed charge solves
\beq
Z(\la)\, + \,  \Int{-q}{q} K(\la-\mu)\,  Z(\mu ) \cdot \dd \mu   \, = \,  1 \;. 
\label{definition dressed charge}
\enq
The dressed charge is related to the dressed phase by the below identities \cite{KorepinSlavnovNonlinearIdentityScattPhase}:
\beq
\phi_1(\la,q) \, - \,   \phi_1(\la,-q) \, + \, 1 \; = \; Z(\la) \quad \e{and} \quad 1+\phi_1(q,q) - \phi_1(-q,q) \, = \, \f{1}{ Z(q) }  \;. 
\label{ecriture identites entre phase et charge habilles} 
\enq

Similarly to the dressed energy in the momentum representation, it is convenient to introduce the momentum representation of the $r$-bound dressed phase:
\beq
\vp_{r}(s,k) \, = \, \phi_{r}\Big(\wh{p}_1^{\, -1}(s), \wh{p}_r^{-1}(k)  \Big) \quad \e{for} \quad s \in \intff{ -p_F }{ 2\pi - p_F - 2 p_F\e{sgn}(\pi-2\zeta) }
\quad \e{and} \quad 
k \in \msc{I}_r \,.
\label{ecriture phase habillee dans rep impuslion}
\enq
Here, one should understand that $\wh{p}_r=p_r$ if $r\geq 2$. Also, one sets $\mc{Z}=Z\circ \wh{p}_1^{\,-1}$. 

Then, the exponents $\De_{\pm}( \bs{\mf{K}} )$ governing the dynamic part of the DRF are expressed as $ \De_{\pm}( \bs{\mf{K}} )=\vth_{\ups}^2( \bs{\mf{K}} ) $
where 
\bem
\vth_{\ups}( \bs{\mf{K}} ) \, = \, \, - \, \ups \ell_{\ups} \, + \, \tfrac{ 1 }{ 2 } \op{s}_{\ga}  \mc{Z}(  p_F ) \, + \,   \sul{ a=1 }{ n_h } \vp_1( \ups p_F , t_{a} )   
\; - \; \sul{  r \in \mf{N}  }{  } \sul{ a=1 }{ n_r } \vp_{r}(\ups p_F ,k_a^{(r)} ) \\
\; - \hspace{-1mm} \sul{ \ups^{\prime} \in \{ \pm \} }{}\ell_{ \ups^{\prime} }  \vp_1(\ups p_F , \ups^{\prime} p_F \, ) 
+ \, \e{sgn}(\pi-2\zeta) \cdot  n_1(\bs{\mf{K}}) \mc{Z}(p_F)    \;. 
\label{definition shifted sfift function}
\end{multline}
Here $ n_1(\bs{\mf{K}})  \, = \, \# \Big\{ k_a^{(1)}\; : \; \wh{p}_1^{\, -1}(  k_a^{(1)} ) \in \intoo{-\infty}{-q}  \Big\}$.

\subsection{The velocity of individual excitations}
\label{Appendix Section phase oscillante dpdte de la vitesse}

The velocity $\mf{v}_{r}$ of an $r$-string excitation if $r \in \mf{N}_{\e{st}}$ and of a particle/hole excitation if $r=1$ is defined by
\beq
\mf{v}_r(k) \, = \;  \mf{e}_{r}^{\prime}(k) \;.  \qquad  \e{In} \; \e{particular} \qquad  \op{v}_{F} =  \mf{e}^{\prime}_1(p_F)
\enq
is the Fermi velocity, namely the velocity of a particle or of a hole excitation located directly on the right edge of the Fermi zone $\intff{-p_F}{p_F}$ in the momentum representation. 
$\mf{v}_1$ is defined, originally, on 
\beq
\intff{-p_F}{2\pi-p_F-2p_F \e{sgn}(\pi-2\zeta) }
\enq
and it is easy to see that it extends to a $2\pi- 2p_F \e{sgn}(\pi-2\zeta)$ periodic function 
on $\R$. 

Furthemore, $\mf{v}_1$ enjoys the symmetry
\beq
\mf{v}_1\big( k  \big) \; = \; -\mf{v}_1\big(2\pi-2p_F \e{sgn}(\pi-2\zeta) -k \big) \;. 
\enq
These properties follow easily from its definition.

Also $\mf{v}_1$ is a continuous function on $\R$ that is piecewise smooth. The points where smoothness may fail correspond to the two momenta $\wh{p}_{1}^{\,-1}\big( \pm \infty \big) \, = \, \wh{p}_{1}^{\,-1}\big( \pm \infty + \i\tf{\pi}{2} \big)$.

One can easily prove for $p_F$ small enough, or for $\zeta$ belonging to a sufficiently small open neighbourhood of $\tf{\pi}{2}$, the below proposition 
characterising some of the properties of $\mf{v}_1$.

\begin{prop}
\label{Proposition proprietes fondamentales de la vitesse des particules trous}

There exists $p_F^{(0)}$ and $\de \zeta^{(0)}$ such that, if either of the two bounds holds 
\beq
0\leq p_F<p_F^{(0)} \qquad  or \qquad   \big| \zeta - \tf{\pi}{2} \big| < \de \zeta^{(0)} \; ,
\enq
then 
\begin{itemize}

 \item $|\mf{v}_1|<\op{v}_F$ on $\intoo{-p_F}{p_F}$;
 
 \item   there exists $ P_m \in \intoo{ p_F }{ \pi - 2 p_F \e{sgn}(\pi-2\zeta) }$ such that $\mf{v}_1$ is strictly increasing on 
\beq
\intoo{ -p_F }{ P_m  } \cup \intoo{ P_M }{ 2\pi-p_F-2p_F \e{sgn}(\pi-2\zeta) }
\enq
with $P_M= 2\pi - 2 p_F \e{sgn}(\pi-2\zeta) - P_m $, and strictly decreasing on $\intoo{P_m  } { P_M }$;

\item there exists an interval 
\beq
\intoo{K_m}{K_M} \subset \intoo{p_F}{2\pi-p_F-2p_F \e{sgn}(\pi-2\zeta) }  \qquad with  \qquad K_M \, = \, 2\pi \, - \,  2p_F \e{sgn}(\pi-2\zeta) \,- \, K_m 
\label{definition intervalle de egalite vitesses particule trou}
\enq
such that 
\beq
\big| \mf{v}_1(k) \big|  < \op{v}_F \;\; \e{for} \;\;  k \in \intoo{K_m}{K_M}
\enq
and
\beq
\big| \mf{v}_1(k) \big|  > \op{v}_F \;\;  \e{for} \;\; k \in \intoo{p_F}{2\pi-p_F-2p_F \e{sgn}(\pi-2\zeta)} \setminus \intff{K_m}{K_M} \, ; 
\enq

\item there exists a strictly decreasing homeomorphism
\beq
\mf{t}\; : \; \intff{K_m}{K_M} \mapsto \intff{-p_F}{p_F} \quad such \; that \quad \mf{t}(k) \;is \, the\, unique\, solution \, to \quad 
\mf{v}_{1}(k)=\mf{v}_1\big( \mf{t}(k) \big) 
\label{ecriture propriete fonction t} 
\enq
with $k \in \intff{K_m}{K_M}$ and $\mf{t}(k) \in \intff{-p_F}{p_F}$. The map $\mf{t}$ is smooth on $\intoo{K_m}{K_M}$;

\item there exists a strictly decreasing homeomorphisms $\mf{p}_L$, $\mf{p}_R$
\beq
\mf{p}_{L}\; : \; \intff{p_F}{P_m} \mapsto \intff{P_m}{K_m} \;\; , \qquad \mf{p}_{R}\; : \; \intff{P_M}{2\pi-p_F-2p_F\e{sgn}(\pi-2\zeta) } \mapsto \intff{K_M}{P_M}, 
\enq
such that $\mf{p}_{L}(k)$, resp. $\mf{p}_{R}(k)$, is  the unique solution  to $\mf{v}_{1}(k)=\mf{v}_1\big( \mf{p}_{L}(k) \big)$, 
resp. $\mf{v}_{1}(k)=\mf{v}_1\big( \mf{p}_{R}(k) \big)$, on their respective range. The maps $\mf{p}_{R,L}$ are smooth on the interior of their domains.

\item $k\mapsto \mf{v}_{r}$ is a diffeomorphism from $\msc{J}_r=\intoo{-p_-^{(r)} }{ p_+^{(r)} }$ onto $\intoo{-\op{v}^{(r)} }{ \op{v}^{(r)} }$
with $ \op{v}^{(r)} > \op{v}_F$. Furthermore, there exists $K_m^{(r)}, K_M^{(r)} \in \msc{J}_r$ and a diffeomorphism 
\beq
\mf{h}^{(r)} \; : \; \intff{-p_F}{p_F} \tend \intff{ K_m^{(r)} }{ K_M^{(r)} } \qquad  such  \,  that  \qquad 
\mf{v}_1(t) \, = \, \mf{v}_r \Big(\mf{h}^{(r)}( t ) \Big) \;.  
\enq

\end{itemize}

\end{prop}

\begin{figure}[t]
 
$\ba{cc}
\includegraphics[width=.4\textwidth]{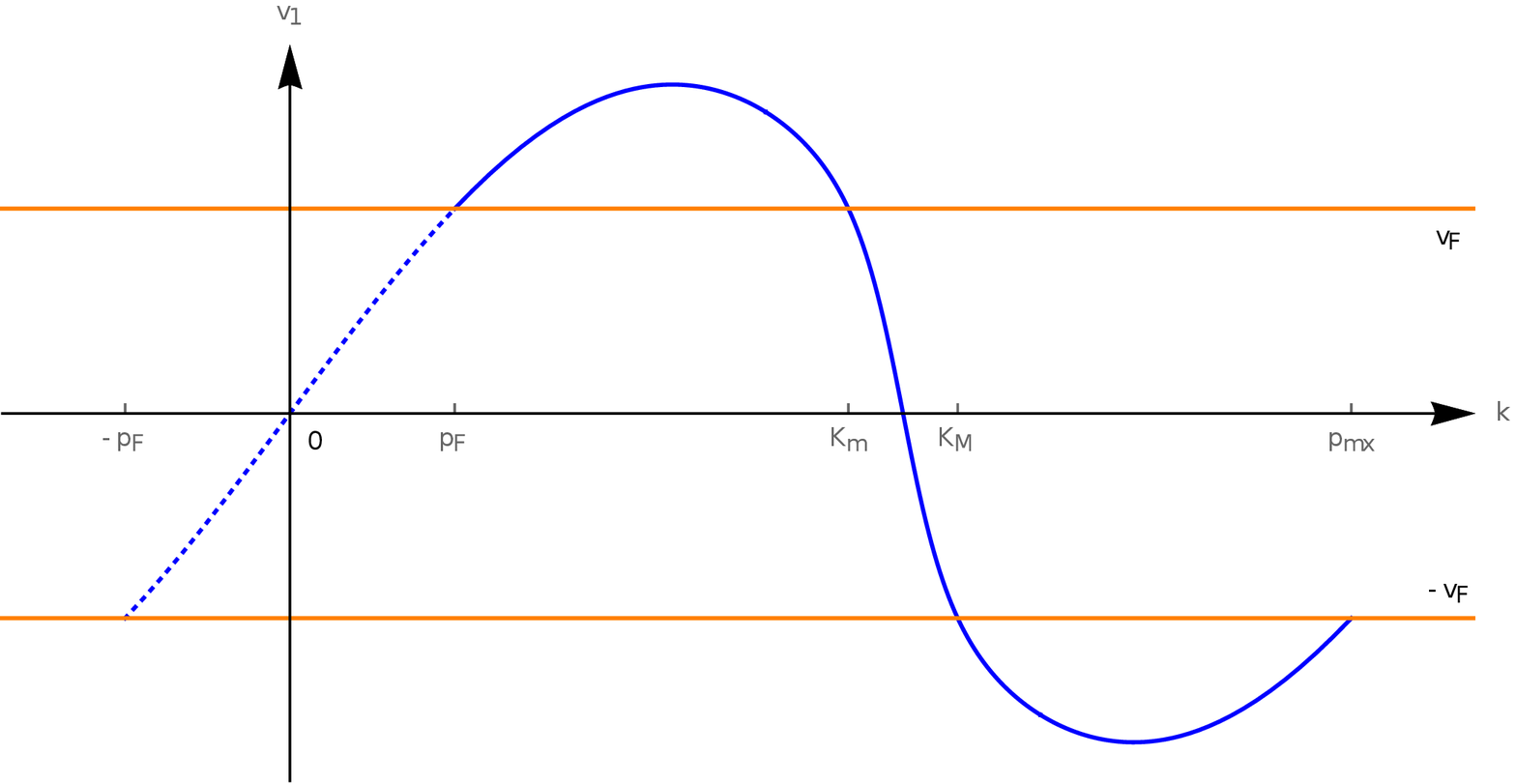}
& 
\includegraphics[width=.4\textwidth]{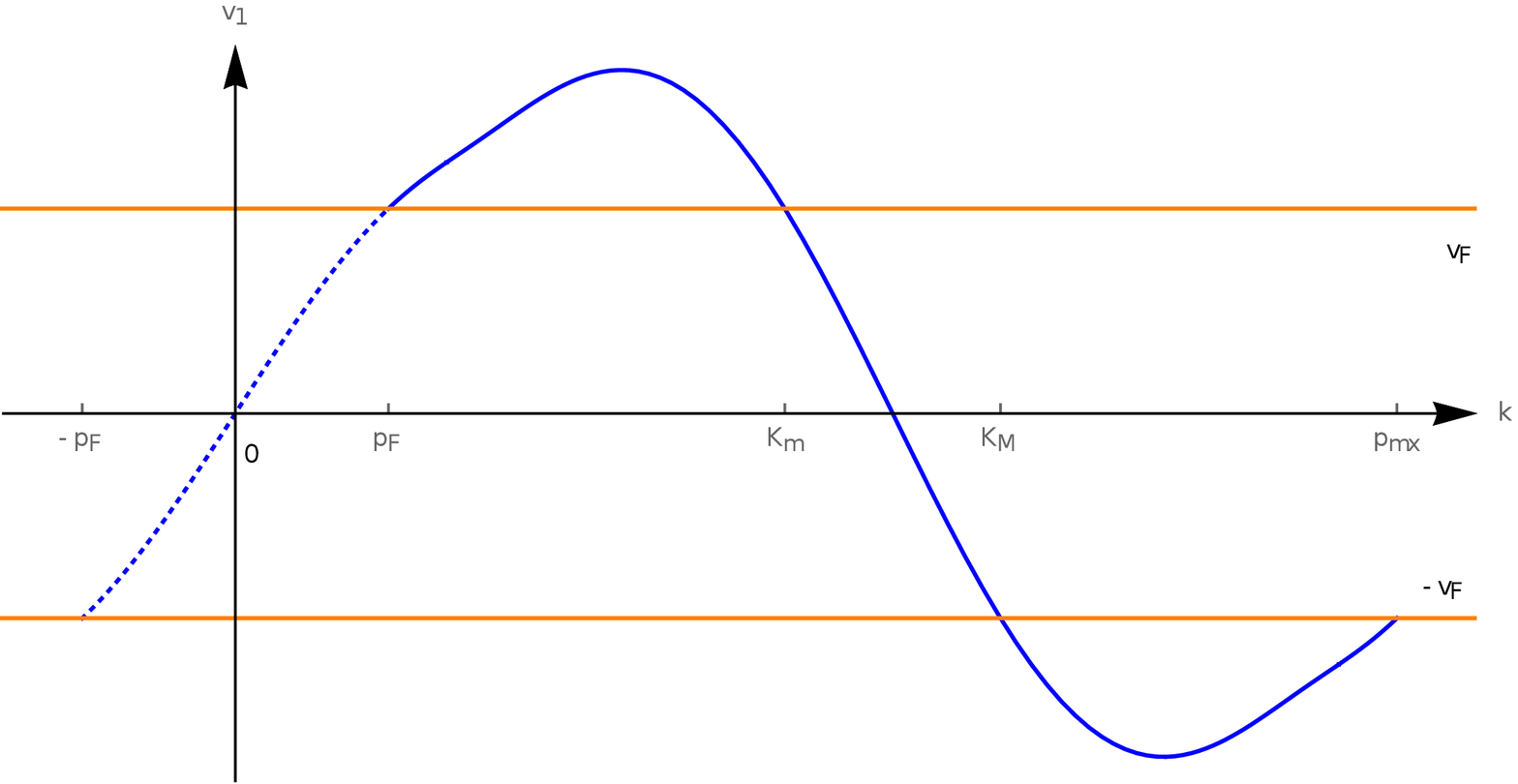}
\ea$
\caption{\label{Figure Vitesse particule trous} Velocity $\mf{v}_1$ plotted for $\Delta=0.57$ and magnetic field $h$ such that the per-site magnetisation $\mf{m}=1-2D$
is parameterised by $D=0.21$ (\textit{lhs}) and for $\Delta=-0.60$ and $D=0.30$ for the \textit{rhs}.}
 
 \end{figure}

In fact, one can check by means of numerical analysis (\textit{c.f.} Fig.\ref{Figure Vitesse particule trous}) that the properties listed 
in Proposition \ref{Proposition proprietes fondamentales de la vitesse des particules trous} above hold true for any $p_F\in \intff{0}{\tf{\pi}{2}}$ and 
$\zeta \in \intoo{0}{\pi}$. Thus the conjecture:
\begin{conj}
\label{Conjecture  diffeo liee a la vitesse}
 The conclusions of Proposition \ref{Proposition proprietes fondamentales de la vitesse des particules trous} hold true irrespectively of the values of $\zeta \in \intoo{0}{\pi}$ or 
$p_F\in \intff{0}{\tf{\pi}{2}}$.  
 
\end{conj}

\section{Asymptotics of a one-dimensional $\be$-like integral}
\label{Appendix DA integrales unidimensionnelles}

In this appendix, I establish the main theorem in the one-dimensional case, \textit{viz}. Theorem \ref{Theorem Principal cas 1D}. 
 For convenience, I recall the statement of the theorem below and then expose the proof. The latter is based on two auxiliary lemmata  
 which are discussed in a separate section.

\subsection{The structural theorem in the one-dimensional case}

\begin{theorem}
\label{Proposition integrale principale spectral function}

 Let $a<b$ be two reals. Let $\mf{z}_{\pm}(\la)$ be two real-holomorphic functions in a neighbourhood of the interval $\msc{J}=\intff{a}{b}$, such that
\begin{itemize}
\item all the zeroes of $\mf{z}_{\pm}$ on $\msc{J}$ are simple;

\item[$\bullet$] $\mf{z}_{+}$ and $\mf{z}_{-}$ admit a unique common zero $\la_0 \in \e{Int}(\msc{J})$ that, furthermore, is such that $\mf{z}_{+}^{\prime}(\la_0) \not= \mf{z}_{-}^{\prime}(\la_0)$. 
\end{itemize}

 Let $\De_{\ups}$ be real analytic on $\e{Int}(\msc{J})$ and such that $\De_{\ups} \geq 0$. Let $\msc{G}$ be in the smooth class of $\msc{J}$
associated with the functions $\De_{\pm}$ and with a constant $\tau$, c.f. Definition \ref{definition smooth class on K}. 
Then, for $\mf{x}\not=0$ and small enough,
\beq
\la \mapsto  \msc{G}\Big(\la, \wh{\mf{z}}_{+}(\la), \wh{\mf{z}}_{-}(\la)\Big) \cdot \pl{\ups= \pm }{} 
\Big\{ \Xi\big( \; \wh{\mf{z}}_{\ups}(\la) \big) \cdot \big[ \, \wh{\mf{z}}_{\ups}(\la) \big]^{ \De_{\ups}(\la)-1 } \Big\} \in L^{1}\big( \msc{J} \big)
\enq
where $\wh{\mf{z}}_{\pm}(\la)=\mf{z}_{\pm}(\la)+ \mf{x}$. Let $\mc{I}(\mf{x})$ denote the integral
\beq
\mc{I}(\mf{x})\, = \, \Int{ \msc{J} }{}   \msc{G}\Big(\la, \wh{\mf{z}}_{+}(\la), \wh{\mf{z}}_{-}(\la)\Big) \cdot \pl{\ups= \pm }{} 
\Big\{ \Xi\big( \; \wh{\mf{z}}_{\ups}(\la) \big) \cdot \big[ \, \wh{\mf{z}}_{\ups}(\la) \big]^{ \De_{\ups}(\la)-1 } \Big\} \cdot \dd \la \;. 
\label{definition integrale 1D type Beta genralise}
\enq

 \vspace{2mm}
Assume that $ \de_{\pm} \, = \,  \De_{\pm}(\la_0) >0$. 

\vspace{2mm}
 
 \noindent {\bf a) } If $\mf{z}_{+}^{\prime}(\la_0)\cdot \mf{z}_{-}^{\prime}(\la_0) <0$, then  $\mc{I}(\mf{x})$ admits the $\mf{x} \tend 0$ asymptotic expansion
\beq
\mc{I}(\mf{x})\, = \, \Xi\Big(   \mf{z}_{+}^{\prime}(\la_0) \cdot   \mf{X}  \Big) \cdot  \Bigg\{ 
\f{  \msc{G}^{(1)}(\la_0) \cdot \de_+ \de_- \cdot |  \mf{X}  |^{\de_+ + \de_- - 1}     }
{   |\, \mf{z}_{+}^{\prime}(\la_0) |^{\de_-} \cdot  | \, \mf{z}_{-}^{\prime}(\la_0)  |^{\de_+}       }
\cdot \f{ \Ga\big( \de_+ \big) \cdot \Ga\big(\de_- \big)  }{  \Ga\big( \de_++\de_-\big)  }
   \, + \, \e{O}\Big(   |\mf{x}|^{ \de_+ + \de_- - \tau} \Big)  \Bigg\}
 \, + \, f_{<}(\mf{x})
\label{ecriture Da integrale I de omege cas zeros entrelaces}
\enq
where 
\beq
 \mf{X}  \, = \,  \mf{x} \cdot  \big[ \mf{z}_{+}^{\prime}(\la_0)-\mf{z}_{-}^{\prime}(\la_0) \big]  \;, 
\label{defintion tau et de pm petits}
\enq
$\msc{G}^{(1)}$ is as appearing in \eqref{ecriture decomposition smooth class K}
and $f_{<}$ is a smooth function of $ \mf{x}$. Furthermore, if $\mf{z}_{\pm}$ have no zeroes on $\msc{J}$  other than $\la_0$, then $f_{<}=0$.  

 \vspace{2mm}
 
 \noindent {\bf b) }  If $\mf{z}_{+}^{\prime}(\la_0)\cdot \mf{z}_{-}^{\prime}(\la_0) >0$, then  $\mc{I}(\mf{x})$ admits the $\mf{x} \tend 0$ asymptotic expansion
\bem
\mc{I}(\mf{x})\, = \, 
\f{  \msc{G}^{(1)}(\la_0) \cdot  \de_+ \, \de_-  \cdot  |  \mf{X}  |^{\de_+ + \de_- - 1}     }
{   |\, \mf{z}_{+}^{\prime}(\la_0) |^{\de_-}  \cdot | \, \mf{z}_{-}^{\prime}(\la_0)  |^{\de_+}       }
\cdot  \Ga\big( \de_+ \big) \cdot \Ga\big(\de_- \big) \cdot \Ga\big( 1- \de_+  - \de_- \big)   \\
\times \bigg\{ \Xi(\mf{x}) \tfrac{1}{\pi} \sin\big[ \pi \de_{\mf{p}} \big] \, + \,  \Xi(-\mf{x}) \tfrac{1}{\pi} \sin\big[ \pi \de_{-\mf{p}} \big]  \bigg\}
\, + \, \e{O}\Big(   |\mf{x}|^{ \de_+ + \de_- - \tau } \Big)
\; + \; f_{>}(\mf{x}) 
\end{multline}
where $ \mf{X} $ and $\de_{\pm}$ are as above,
\beq
\mf{p}\, = \, - \e{sgn} \big[ \mf{z}_{+}^{\prime}(\la_0) \big] \cdot \e{sgn} \big[ \mf{z}_{+}^{\prime}(\la_0)-\mf{z}_{-}^{\prime}(\la_0) \big]  
\label{definition parametre eta integrale type fct beta}
\enq
and $f_{>}$ is a smooth function of $ \mf{x}$.

\end{theorem}

\Proof 

The hypothesis on $\mf{z}_{\pm}(\la)$ ensure that these functions have a holomorphic inverse in a neighbourhood of any of their zeroes. 
As a consequence, any zero of $\wh{\mf{z}}_{\pm}$ is holomorphic in $\mf{x}$ small enough. Thus, the integral can be decomposed as $\mc{I}(\mf{x})=\sul{k=1}{n}\mc{I}_{k}(\mf{x})$, 
where 
\beq
\mc{I}_k(\mf{x})\, = \, \Int{ a_k(\mf{x}) }{ b_k(\mf{x}) }    \msc{G}\Big(\la, \wh{\mf{z}}_{+}(\la), \wh{\mf{z}}_{-}(\la)\Big) \cdot \pl{\ups= \pm }{} \Big\{ \big[ \, \wh{\mf{z}}_{\ups}(\la) \big]^{ \De_{\ups}(\la)-1 } \Big\} \cdot \dd \la \;. 
\label{ecriture integrale Ik}
\enq
The endpoints $ a_k(\mf{x})$, $k \geq 2$, and  $b_k(\mf{x})$, $k \leq n-1$, all correspond to a zero of $\wh{\mf{z}}_{+}$ or $\wh{\mf{z}}_{-}$. 
Furthermore,  if $a_1(\mf{x})>a$ and/or $b_n(\mf{x})<b$, then these also correspond to a zero of $\wh{\mf{z}}_{+}$ or $\wh{\mf{z}}_{-}$. 
However, it may be that  $a_1(\mf{x})=a$ and/or $b_n(\mf{x})=b$, where I remind that $\msc{J}=\intff{a}{b}$. Then $a_1(\mf{x})=a$ and/or $b_n(\mf{x})=b$ may or may not correspond to 
zeroes of $\wh{\mf{z}}_{\ups}$.

The fact that $\msc{G}$ belongs to the smooth class of $\msc{J}$ with functions $\De_{\pm}$ and a constant $\tau$ ensures that the integrals $\mc{I}_k(\mf{x})$ are well-defined.
Indeed, problems with the $L^{1}$-nature of its integrand could, in principle, arise if some zero of $\wh{\mf{z}}_{\ups}$ coincides with a zero of $\De_{\ups}$. 
However, observe that due to the smooth class property and the hypotheses stated above (also \textit{c.f.} equation \eqref{decomposition zero mu pm}), the zeroes of $\wh{\mf{z}}_{+}$
and $\wh{\mf{z}}_{-}$ are all distinct and simple, at least provided that $\mf{x}$ is small enough. Furthermore, one has the decomposition 
\bem
 \msc{G}\Big(\la, \wh{\mf{z}}_{+}(\la), \wh{\mf{z}}_{-}(\la)\Big) \cdot \pl{\ups= \pm }{} \Big\{ \big[ \, \wh{\mf{z}}_{\ups}(\la) \big]^{ \De_{\ups}(\la)-1 } \Big\} \; = \; 
\msc{G}^{(1)}\big(\la\big) \cdot \pl{\ups= \pm }{} \Big\{ \De_{\ups}(\la)  \big[ \, \wh{\mf{z}}_{\ups}(\la) \big]^{ \De_{\ups}(\la)-1 } \Big\}  \\
\; + \; \msc{G}^{(2)}\Big(\la, \wh{\mf{z}}_{+}(\la) \Big) \cdot \De_{-}(\la) \,  \big[ \, \wh{\mf{z}}_{-}(\la) \big]^{ \De_{-}(\la)-1 }  \cdot   \big[ \, \wh{\mf{z}}_{+}(\la) \big]^{ \De_{+}(\la)-\tau } 
\; + \; \msc{G}^{(3)}\Big(\la, \wh{\mf{z}}_{-}(\la) \Big) \cdot \De_{+}(\la) \,  \big[ \, \wh{\mf{z}}_{+}(\la) \big]^{ \De_{+}(\la)-1 }  \cdot   \big[ \, \wh{\mf{z}}_{-}(\la) \big]^{ \De_{-}(\la)-\tau }  \\
\; + \; \msc{G}^{(4)}\Big(\la,  \wh{\mf{z}}_{+}(\la),  \wh{\mf{z}}_{-}(\la) \Big)   \cdot \pl{\ups= \pm }{} \Big\{  \big[ \, \wh{\mf{z}}_{\ups}(\la) \big]^{ \De_{\ups}(\la)-\tau } \Big\}  \;.  
\label{ecriture decomposition integrande Ik}
\end{multline}
By the above, $\wh{\mf{z}}_{\ups}$ vanishes linearly at its zeroes. $\De_{\ups}$ being holomorphic, it vanishes at least linearly at its zeroes. 
These two properties ensure the $L^{1}\big( \intff{ a_k(\mf{x}) }{ b_k(\mf{x}) }  \big)$ nature of the integrand in \eqref{ecriture integrale Ik}.

\vspace{2mm}

In the following, I denote by  $\mu_{\pm}(\mf{x})$ the zeroes of $\wh{\mf{z}}_{\pm}(\la)$ such that $\mu_{\pm}(0)=\la_0$. If neither $a_{k}(\mf{x})$ nor $b_{k}(\mf{x})$ coincides with $\mu_{\pm}(\mf{x})$,
then the endpoint  $a_{k}(0)$, resp. $b_{k}(0)$,  is at most a simple
zero of one of the functions $\mf{z}_{\pm}$, but not of both. The latter is a direct consequence of the assumed properties of the functions $\mf{z}_{\pm}$.
Hence, $\mc{I}_k(\mf{x})$ corresponds to the class of integrals 
studied in Lemma \ref{Lemme integrale type beta reguliere} and, as such, is smooth in $\mf{x}$ small enough. Its contribution is thus included
in one of the functions $f_{<}(\mf{x})$ or $f_{>}(\mf{x})$, depending on the case of interest.

\vspace{2mm}

It thus remains to focus on the integral containing, as one of its endpoints, the zero $\mu_{\pm}(\mf{x})$. For convenience, denote 
this integral by  $\mc{J}_{\la_0}(\mf{x})$. 

\vspace{2mm}

As already stated, the zeroes $\mu_{\pm}(\mf{x})$ are analytic functions of $\mf{x}$, at least for $\mf{x}$ small enough. 
Furthermore, it is readily checked that 
\beq
\mu_{\pm}(\mf{x}) \; = \; \la_0 \, - \, \f{  \mf{x} }{  \mf{z}_{\pm}^{\prime}(\la_0) }  \; + \; \e{O}\big( \mf{x}^2\big) \;. 
\label{decomposition zero mu pm}
\enq
This ensures that $\mu_{+}(\mf{x})\not= \mu_{-}(\mf{x})$ for $\mf{x}$ small enough. Being holomorphic,  $\wh{\mf{z}}_{\ups}(\la)$ admits the factorisation:
\beq
\wh{\mf{z}}_{\ups}(\la) \, = \,  \big( \la-\mu_{\ups}(\mf{x}) \big) \cdot h_{\ups}(\la ,  \mf{x} )  \qquad \e{with} \qquad h_{\ups} \Big(  \mu_{\ups}(\mf{x}) , \mf{x}  \Big) \; = \; \mf{z}_{\ups}^{\prime} \Big( \mu_{\ups}(\mf{x}) \Big)  \;. 
\label{ecriture factorisation fct z pm hat}
\enq
By the Weierstrass preparation theorem, \textit{c.f.} Theorem \ref{Theorem Weierstrass preparation theorem}, $h_{\ups}$ is holomorphic in $\la$ and $\mf{x}$, at least for $|\mf{x}|$ small enough.

In order to proceed further, one has to distinguish between the cases ${\bf a)}$ and ${\bf b)}$ outlined in the statement of the theorem.

 \vspace{2mm}

\noindent {\bf $\bullet$ Case a): $\mf{z}_{+}^{\prime}(\la_0)\cdot \mf{z}_{-}^{\prime}(\la_0) <0$ }

 \vspace{2mm}

Let $ \vsg= \e{sgn}\big[  \mf{z}_{+}^{\prime}(\la_0)  \big] $. Then, 
\beq
\wh{\mf{z}}_{+}(\la_0)  >0  \; \; \e{on}  \; \;    \vsg \intoo{ \mu_{+}(\mf{x}) }{ \nu_{+}^{(\vsg)}(\mf{x}) }  \quad \e{while}\quad 
 \wh{\mf{z}}_{-}(\la_0)  >0 \; \;  \e{on}   \;\;  \vsg \intoo{ \nu_{-}^{(\vsg)}(\mf{x})   }{ \mu_{-}(\mf{x}) } 
\label{ecriture tableau positivite fcts z pm}
\enq
where $\nu_{\pm}^{(\vsg)}(\mf{x})$ is the closest zero of $\wh{\mf{z}}_{\pm}$ to $\mu_{\pm}(\mf{x})$ such that 
the function satisfies to the above properties\symbolfootnote[4]{In case there are no more zeroes, one should simply take $\nu_{+}^{(+)}(\mf{x})=b$ and $\nu_{+}^{(-)}(\mf{x})=a$ or 
$\nu_{-}^{(+)}(\mf{x})=a$ or $\nu_{-}^{(-)}(\mf{x})=b$ depending on the situation, where I remind that $\msc{J} \, = \, \intff{a}{b}$.}. 
The $\vsg$ pre-factor in front of the intervals means that the interval is always oriented from the smallest to the largest element.  

  One can convince oneself that 
\beq
\nu_{\pm}^{(\vsg)}(\mf{x}) \, = \, \mu_{\pm}(\mf{x}) \pm \vsg \de \nu_{\pm}^{(\vsg)}(\mf{x)} \quad  \e{where} \quad 
   \de \nu_{\pm}^{(\vsg)}(\mf{x)}   \, > \, C
\enq
for some $\mf{x}$-independent constant $C>0$. 

The above means that the neighbourhood of $\la_0$ will produce non-vanishing contributions to $\mc{J}_{\la_0}(\mf{x})$ only if $\mu_{\vsg}(\mf{x}) < \mu_{-\vsg}(\mf{x})$. Provided this inequality holds,
 the integration in $\mc{J}_{\la_0}(\mf{x})$ runs through the interval $\intff{  \mu_{\vsg}(\mf{x})   }{  \mu_{-\vsg}(\mf{x})  }$. 

Since one has
\beq
\mu_{-\vsg}(\mf{x}) - \mu_{\vsg}(\mf{x}) \, = \, \mf{x} \cdot \f{       \mf{z}_{\vsg}^{\prime}(\la_0) - \mf{z}_{-\vsg}^{\prime}(\la_0)   }{   - \mf{z}_{+}^{\prime}(\la_0)\cdot \mf{z}_{-}^{\prime}(\la_0)  }
\Big( 1 \, + \,  \e{O}\big( \mf{x} \big) \Big) \; , 
\enq
the condition $\mu_{\vsg}(\mf{x}) < \mu_{-\vsg}(\mf{x})$ can be recast, for $|\mf{x}|$ small enough, as $\mf{z}_{+}^{\prime}(\la_0) \cdot  \mf{X} >0 $ where $\mf{X}$ is as defined in \eqref{defintion tau et de pm petits}.
Thence, upon inserting the factorisation \eqref{ecriture factorisation fct z pm hat}
into $\mc{J}_{\la_0}(\mf{x})$, the integral can be recast, for $|\mf{x}|$ small enough, as
\beq
\mc{J}_{\la_0}(\mf{x}) \, = \,  \Xi\Big(   \mf{z}_{+}^{\prime}(\la_0) \cdot   \mf{X}  \Big) \cdot \Int{  \mu_{\vsg}(\mf{x}) }{  \mu_{-\vsg}(\mf{x}) } 
\msc{H}(\la) \cdot \pl{\ups = \pm }{} \Big\{ \ups \vsg \big[\la -\mu_{\ups}(\mf{x}) \big] \Big\}^{\De_{\ups}(\la)-1}  \cdot \dd \la
\enq
with 
\beq
 \msc{H}(\la) \; = \; \msc{G}\Big(\la, \wh{\mf{z}}_{+}(\la), \wh{\mf{z}}_{-}(\la)\Big)
 \cdot \pl{\ups=\pm}{} \Big\{ \ups \vsg h_{\ups}( \la ,  \mf{x} ) \Big\}^{\De_{\ups}(\la)-1} \;. 
\enq
The representation \eqref{ecriture decomposition integrande Ik}, the properties of the functions $\msc{G}^{(k)}$ and the fact that the $\msc{G}$ independent-part of the integrand has constant sign, all lead together to the decomposition
\beq
\mc{J}_{\la_0}(\mf{x}) \, = \, \check{\mc{J}}\big[ H,\De_{+},\De_{-}\big](\mf{x}) \; + \; \sul{\ups=\pm}{} \e{O}\Big( \check{\mc{J}}\big[1,\De_{+} + (1-\tau)\de_{\ups,+} \, ,\De_{-} + (1-\tau) \de_{\ups,-}\big](\mf{x}) \Big)
\enq
where $\de_{a,b}$ stands for the Kronecker symbol. Here
\beq
 \check{\mc{J}}\big[H,\De_{+},\De_{-}\big](\mf{x}) \, = \,  \Xi\Big(   \mf{z}_{+}^{\prime}(\la_0) \cdot   \mf{X}  \Big) \cdot \Int{  \mu_{\vsg}(\mf{x}) }{  \mu_{-\vsg}(\mf{x}) } 
H(\la) \cdot \pl{\ups = \pm }{} \Big\{ \ups \vsg \big[\la -\mu_{\ups}(\mf{x}) \big] \Big\}^{\De_{\ups}(\la)-1}  \cdot \dd \la
\enq
with 
\beq
H(\la) \; = \;  \De_{+}(\la) \, \De_{-}(\la) \cdot \msc{G}^{(1)}(\la)     
 \cdot \pl{\ups=\pm}{} \Big\{ \ups \vsg h_{\ups}( \la ,  \mf{x} ) \Big\}^{\De_{\ups}(\la)-1} \;. 
\enq
Then, the change of variables
\beq
t \; = \; \f{ \la - \mu_{\vsg}(\mf{x})  }{  \mu_{-\vsg}(\mf{x}) -  \mu_{\vsg}(\mf{x}) }
\enq
recasts the integral as
\beq
 \check{\mc{J}}\big[H,\De_{+},\De_{-}\big](\mf{x}) \, = \,  \Xi\Big(   \mf{z}_{+}^{\prime}(\la_0) \cdot    \mf{X} \Big) \cdot 
\Int{ 0 }{ 1 }  t^{ \wt{\De}_{\vsg}(t)-1} \cdot \big( 1 - t \big)^{ \wt{\De}_{-\vsg}(t)-1} \cdot \wt{H}(t) \cdot \dd t 
\enq
where 
\beq
 \wt{H}(t) \; = \; H\Big( \mu_{\vsg}(\mf{x}) +t \cdot \big[\mu_{-\vsg}(\mf{x}) -  \mu_{\vsg}(\mf{x})\big]   \Big) \cdot 
 \Big( \mu_{-\vsg}(\mf{x}) -  \mu_{\vsg}(\mf{x})   \Big)^{ \wt{\De}_+(t)+\wt{\De}_-(t) - 1}
\enq
and 
\beq
 \wt{\De}_{\ups}(t) \; = \; \De_{\ups}\Big(  \mu_{\vsg}(\mf{x}) +t \cdot \big[\mu_{-\vsg}(\mf{x}) -  \mu_{\vsg}(\mf{x})\big]   \Big) \,.
\enq
Being smooth, all functions have an expansions in $\mf{x}$ that is uniform in $t \in \intff{0}{1}$. This fact ensures that the leading asymptotic expansion of the 
integral is obtained by setting the argument $t$ of all functions to $0$, leading to 
\beq
 \check{\mc{J}}\big[H,\De_{+},\De_{-}\big](\mf{x}) \, = \,  \Xi\Big(   \mf{z}_{+}^{\prime}(\la_0) \cdot    \mf{X} \Big) \cdot \wt{H}(0) \cdot 
\f{ \Ga\Big(\, \wt{\De}_{\vsg}(0) \Big) \cdot \Ga\Big( \, \wt{\De}_{-\vsg}(0) \Big) }{ \Ga\Big( \, \wt{\De}_{\vsg}(0) + \wt{\De}_{-\vsg}(0) \Big)  }
\cdot \Big( 1 \, + \, \e{O}\big( \mf{x} \ln \mf{x} \big) \Big) \;. 
\enq
Note that the $\e{O}\big( \mf{x} \ln \mf{x} \big) $ remainder issues from the expansion of the exponents in $\wt{H}(t)$. 
One can simplify the formula further. One has $ \wt{\De}_{\ups}(0)= \de_{\ups} + \e{O}\big( \mf{x} \big)  $ with $\de_{\pm}$
as in \eqref{defintion tau et de pm petits} as well as  
\beq
\wt{H}(0) \; = \;\f{   \de_{+} \de_{-} \cdot \msc{G}^{(1)}(\la_0) \cdot  |  \mf{X} |^{\de_+ + \de_- - 1}     }
{   |\, \mf{z}_{+}^{\prime}(\la_0) |^{\de_-} \cdot  | \, \mf{z}_{-}^{\prime}(\la_0)  |^{\de_+}       }   \, + \, \e{O}\Big( |\mf{x}|^{\de_+ + \de_- } \ln |\mf{x}|  \Big) 
\enq
what allows one to conclude regarding to \eqref{ecriture Da integrale I de omege cas zeros entrelaces}.


\vspace{2mm}

\noindent {\bf $\bullet$ Case b): $\mf{z}_{+}^{\prime}(\la_0)\cdot \mf{z}_{-}^{\prime}(\la_0) > 0$ }


\vspace{2mm}
Still agreeing upon $ \vsg= \e{sgn}\big[  \mf{z}_{+}^{\prime}(\la_0)  \big] $, and keeping the same definition of $\nu_{\pm}^{(\vsg)}(\mf{x})$, one gets that 
\beq
\wh{\mf{z}}_{\pm }(\la_0)  >0  \; \; \e{on}  \; \;    \vsg \intoo{ \mu_{\pm}(\mf{x}) }{   \nu_{\pm}^{(\vsg)}(\mf{x}) } \;. 
\enq
Here, as earlier, the $\vsg$ prefactor indicates that the interval is oriented from its smallest to its largest element. 
It is easy to  convince oneself that, in the present case of interest,  
\beq
\nu_{\pm}^{(\vsg)}(\mf{x}) \, = \, \mu_{\pm}(\mf{x})  + \vsg \de \nu_{\pm}^{(\vsg)}(\mf{x)} \quad  \e{where} \quad 
 \de \nu_{\pm}^{(\vsg)}(\mf{x)}  >C
\enq
for some $\mf{x}$-independent constant $C>0$. 
Thus, after imposing the positivity constraints and using that $|\mf{x}|$ is small, the integral $\mc{J}_{\la_0}(\mf{x})$ runs through $\vsg \intff{ b_{\vsg} }{ c_{\vsg} }$ where 
\beq
b_{\vsg} \, = \,  \vsg \, \e{max}\Big( \vsg  \mu_{+}(\mf{x}) , \vsg  \mu_{-}(\mf{x}) \Big)  \qquad \e{and} \qquad 
c_{\vsg} \, =  \,   \vsg \, \e{min}\Big(  \vsg \nu_{+}^{(\vsg)}(\mf{x}) ,\vsg \nu_{-}^{(\vsg)}(\mf{x}) \Big)\;. 
\enq
The integral of interest can then be decomposed as $\mc{J}_{\la_0}(\mf{x} ) \, = \,\mc{J}_{\la_0}^{(1)}(\mf{x} )  \, + \,  \mc{J}_{\la_0}^{(2)}(\mf{x} )$
\beq
\mc{J}_{\la_0}^{(1)}(\mf{x} )\; = \; \vsg \hspace{-2mm} \Int{ b_{\vsg} } { b_{\vsg} + \vsg \de } \hspace{-2mm} 
\msc{G}\Big(\la, \wh{\mf{z}}_{+}(\la), \wh{\mf{z}}_{-}(\la)\Big) \pl{\ups= \pm }{}  \big[ \, \wh{\mf{z}}_{\ups}(\la) \big]^{ \De_{\ups}(\la)-1 }\hspace{-1mm}  \cdot \dd \la 
\enq
and
\beq
\mc{J}_{\la_0}^{(2)}(\mf{x} )\; = \; \vsg \hspace{-2mm} \Int{ b_{\vsg} + \vsg \de }{ c_{\vsg} } \hspace{-2mm}   \msc{G}\Big(\la, \wh{\mf{z}}_{+}(\la), \wh{\mf{z}}_{-}(\la)\Big) 
\pl{\ups= \pm }{}   \big[ \, \wh{\mf{z}}_{\ups}(\la) \big]^{ \De_{\ups}(\la)-1 } \hspace{-1mm}  \cdot \dd \la  \; , 
\enq
where $\de>0$ is taken small enough. 

$\mc{J}_{\la_0}^{(2)}(\mf{x} )$ is a smooth function of $\mf{x}$. This can be seen as follows. If $ c_{\vsg} \in \Dp{}\msc{J}$ and if the endpoints of $\msc{J}$ are not 
zeroes of $\mf{z}_{\pm}$, then the integrand in  $\mc{J}_{\la_0}^{(2)}(\mf{x})$ can be expanded into 
powers of $\mf{x}$ owing to 
\beq
\e{inf}\Big\{ \mf{z}_{\pm}(s) \; : \; s \in \intff{ b_{\vsg} + \vsg \de }{ c_{\vsg} }  \Big\}>C^{\prime} \; ,
\enq
for some $C^{\prime}>0$. 
This entails the claim. Otherwise, $ c_{\vsg} $ coincides with a zero $\wh{\mf{z}}_{\pm}$. One treats the part of the integral corresponding to an integration over a domain
uniformly away from $c_{\vsg}$, exactly as in the first case. Then, the neighbourhood of $c_{\vsg}$  can  be treated as in Lemma \ref{Lemme integrale type beta reguliere}, and the claim follows.  

\vspace{2mm}
It thus remains to focus on $\mc{J}_{\la_0}^{(1)}(\mf{x} )$. Recalling the definition \eqref{definition parametre eta integrale type fct beta} of $\mf{p}$, one can readily check that, for $\mf{x}$
small enough,
\beq
b_{\vsg} \, =  \,  \mu_{-\e{sgn}(\mf{x}) \mf{p}}(\mf{x}) \qquad \e{and} \qquad 
\vsg \, \e{min}\Big( \vsg  \mu_{+}(\mf{x}) , \vsg  \mu_{-}(\mf{x}) \Big) \, = \,  \mu_{ \e{sgn}(\mf{x}) \mf{p}}(\mf{x})  \;.  
\enq
After the change of variables $\la=b_{\vsg} + \vsg t$, by using the factorisation \eqref{ecriture factorisation fct z pm hat} and setting 
\beq
a_{\vsg} \; = \; \vsg \Big( \mu_{-\e{sgn}(\mf{x}) \mf{p} }(\mf{x}) \, - \,  \mu_{ \e{sgn}(\mf{x}) \mf{p} }(\mf{x})  \Big) \, \geq 0 \, , 
\enq
one gets that 
\beq
\mc{J}_{\la_0}^{(1)}(\mf{x} ) \; = \; \Int{ 0 }{ \de } \msc{H}\Big( t, t u(t),(t+a_{\vsg}) v(t) \Big) \cdot t^{A(t)} \cdot (t+a_{\vsg})^{ B(t) } \cdot \dd t \;. 
\enq
Above, we have set 
\beq
A(t) \, = \, \De_{-\e{sgn}(\mf{x}) \mf{p} }(b_{\vsg} +\vsg t)-1 \; ,\quad  B(t) \, = \, \De_{\e{sgn}(\mf{x}) \mf{p}}(b_{\vsg} +\vsg t)-1 \;
\enq
and
\beq
 u(t) \, = \, \vsg   h_{-\e{sgn}(\mf{x}) \mf{p}} (b_{\vsg} +\vsg t ,  \mf{x}) \; , \quad 
 v(t) \, = \, \vsg   h_{ \e{sgn}(\mf{x}) \mf{p}} (b_{\vsg} +\vsg t ,  \mf{x}) \; . 
\enq
Finally,  
\beq
\msc{H}(t, x, y ) \, = \, \msc{G}\Big(  b_{\vsg} +\vsg t, x , y\Big) 
\cdot \pl{\ups=\pm}{} \Big\{ \vsg h_{\ups}(b_{\vsg} +\vsg t ,  \mf{x}) \Big\}^{ \De_{\ups }(b_{\vsg} +\vsg t)-1 } \;. 
\enq
The properties of $\msc{G}$ entail that 
\beq
\msc{H}(t, x, y ) \, = \, H(t) \; + \; \e{O}\Big( x^{1- \tau} \, + \,   y^{1- \tau}\Big)
\enq
with a differentiable remainder in the sense of Definition \ref{Defintion reste differentiable} and where 
\beq
H(t) \, = \,  \Big(\De_{+} \De_{-} \msc{G}^{(1)} \Big)\big(  b_{\vsg} +\vsg t \big) \cdot \pl{\ups=\pm}{} \Big\{ \vsg h_{\ups}(b_{\vsg} +\vsg t ,  \mf{x}) \Big\}^{ \De_{\ups }(b_{\vsg} +\vsg t)-1 } \;. 
\enq

One is now in position to apply the result of Lemma \ref{Lemme integrale beta auxiliaire} given below. This yields that 
\bem
\mc{J}^{(1)}_{\la_0}(\mf{x}) \; = \;   -   \tfrac{ H(0) }{ \pi } \big( a_{\vsg} \big)^{ 1+A(0)+B(0) } \sin[\pi B(0)]    
\cdot \Ga\Big(1+A(0)\Big) \Ga\Big(1+B(0)\Big)\Ga\Big(-1-A(0)-B(0)\Big) \\
 \, + \, \e{O}\Big( a_{\vsg}^{2+A(0)+B(0)} \cdot \ln a_{\vsg}  \Big) \; + \; r(a_{\vsg})  \;, 
\end{multline}
where $r$ is some smooth function. At this stage, it remains to use that 
\beq
\De_{\ups}( b_{\vsg} ) = \de_{\ups}+\e{O}(\mf{x})  \qquad \e{and} \qquad 
a_{\vsg} =  |\mf{X}| \cdot  \big[ \mf{z}^{\prime}_{+}(\la_0) \cdot  \mf{z}^{\prime}_{-}(\la_0) \big]^{-1}  \cdot \big( 1 + \e{O}(\mf{x}) \big)
\enq
with $\mf{X}$ as given in \eqref{defintion tau et de pm petits}, so  as to conclude. \qed




\subsection{Auxiliary lemmata}

\begin{lemme}
\label{Lemme integrale beta auxiliaire}

Let $1> \de>0$ be fixed and $f(t),A(t), B(t)$ be smooth real valued functions on $\intff{0}{\de}$ admitting the expansion around zero 
\beq
f(t) \, = \, f_0 \, + \, \e{O}(t) \; , 
\qquad A(t) \, = \, a_0 \, + \, \e{O}(t)   \; , 
\qquad B(t) \, = \, b_0 \, + \, \e{O}(t) \;,
\enq
where $a_0>-1$ and $b_0>-1$ are such that $a_0+b_0 \not\in \mathbb{N}$. Further, let $\msc{F}$ be smooth on $\intff{0}{\de}\times \R^+\times \R^+$ and such that, for $x,y$ bounded
\beq
\msc{F}\big(t; x, y \big) \; = \; f(t)   \, + \, \e{O}\big( x^{\a}+y^{\a} \big)   \quad for\; some \quad 0<\a<1
\enq
with a differentiable remainder, \textit{c.f.} Definition \ref{defintion reste differentiable}. Let $u,v$ be smooth on $\intff{0}{\de}$ and such that $u(t),v(t)>0$. 
Then, the integral 
\beq
\mc{J}[\msc{F},A,B]( \mf{x} ) \;= \; \Int{0}{\de} \msc{F}\big(t; t \, u(t) , \,  (t+\mf{x})\,  v(t) \big)  \cdot t^{ A(t) } \cdot (t+  \mf{x} )^{ B(t) } \cdot \dd t 
\enq
has the $ \mf{x} \tend 0^+$ asymptotic expansion
\bem
\mc{J}[\msc{F}, A, B](  \mf{x} ) \;= \; -f_0 \tfrac{ \sin [ \pi b_0 ] }{ \pi } \Ga\Big( 1+a_0 \Big)\Ga\Big( 1+b_0 \Big)\Ga\Big( -1-a_0-b_0 \Big) \cdot  \mf{x}^{ 1+a_0+b_0 } \\
\, + \, r( \mf{x}) \; + \; \e{O}\Big(  \mf{x}^{1 + a_0 + b_0  + \a }    \Big) \;. 
\end{multline}
where  the function $r$ is  smooth in $\mf{x}$, and does depend on $\de, A, B$. 

\end{lemme}

\Proof 

Observe that $\mf{x} \mapsto \mc{J}[\msc{F}, A, B](  \mf{x} )$ is smooth in $\mf{x} \in \R^+$ and that the $\mf{x}$ derivatives are obtained by differentiating under the integral. 
Let $n\in \mathbb{N}$ be such that 
\beq
-2 < 1+a_0+b_0-n <-1 \;. 
\label{ecriture choix entier n}
\enq
Observe that the hypotheses on the differentiability of the remainder ensures that 
\beq
\Dp{\mf{x}}^{n} \Big\{ \msc{F}\big(t; t \, u(t) , \,  (t+\mf{x})\,  v(t) \big)   \cdot (t+  \mf{x} )^{ B(t) } \Big\} \; = \; \wt{f}(t) (t+  \mf{x} )^{  \wt{B}(t) } \, + \, \e{O}\Big(  (t+  \mf{x} )^{  \wt{B}(t) +\a } \Big) 
\, + \, \e{O}\Big( t^{\a} (t+  \mf{x} )^{  \wt{B}(t) } \Big)
\enq
with
\beq
\wt{f}(t)\, = \, f(t) \cdot \f{  \Ga\big( B(t)+1 \big)  }{  \Ga\big( 1 + B(t) - n \big) } \qquad \e{and} \qquad \wt{B}(t) \, = \, B(t) -n \;. 
\enq
Furthermore, being smooth functions on $\intff{0}{\de}$, one has that 
\beq
\wt{f}(t) \, = \, \sul{k=0}{p} \wt{f}_k t^k \, + \, \e{O}(t^{p+1}) 
\qquad A(t) \, = \, \sul{k=0}{p} a_k t^k \, + \, \e{O}(t^{p+1})
\qquad \wt{B}(t) \, = \,  \sul{k=0}{p} \wt{b}_k t^k  \, + \, \e{O}(t^{p+1}) \;. 
\enq
From there and the fact that $u(t), v(t)>c$ for some $c>0$, one readily deduces that 
\beq
\wt{f}(t) \cdot t^{A(t) } \cdot (t+ \mf{x})^{ \wt{B}(t) } \; = \; t^{ a_0 } \cdot (t+ \mf{x})^{ \wt{b}_0 }  \cdot \bigg( \wt{f}_0 \, + \, \e{O}\Big(  t \cdot \e{max}\big\{  | \ln t | ,   |\ln (t  +  \mf{x})|, 1    \big\}  \Big) \bigg) \;. 
\enq
Then, since $\ln t$, $\ln(t+ \mf{x})$ have constant sign on $\intff{0}{\de}$ provided that $|\mf{x}|$ is small enough, straightforward bounds  lead to 
\bem
\Dp{ \mf{x} }^{n} \Big\{\,  \mc{J}[\msc{F}, A, B](  \mf{x} ) \, \Big\} \;= \; \wt{f}_0 \, \mc{T}(a_0,\wt{b}_0) 
\; + \; \e{O}\bigg( \,  \big| \Dp{a} \mc{T}(a, b) \big| \, + \, \big| \Dp{b}\mc{T}(a,b) \big| \, + \, \big|  \mc{T}(a,b) \big| \, \bigg)_{ \big| \substack{ a=a_0+1  \\ b=\wt{b}_0 } }  \\
\; + \; \e{O}\bigg(  \big|  \mc{T}(\a + a_0,\wt{b}_0) \big| \, + \,   \big|  \mc{T}( a_0,\wt{b}_0+\a) \big|  \bigg) 
\label{ecriture dvpmt derivee nieme integrale J 1D}
\end{multline}
where 
\beq
\mc{T}(a,b) \; = \; \Int{0}{ \de } t^{ a} \cdot (t+ \mf{x})^{b} \cdot \dd t \; = \;  \mf{x}^{ a + b + 1} \Int{0}{ \tf{\de}{ \mf{x}} } t^{ a } \cdot (t+1)^{ b } \cdot \dd t \;. 
\enq
Note that, if need be, one may always slightly decrease the value of $\a$ so that $\a+a_0+b_0\not \in \mathbb{Z}$ while preserving the differentiability of the remainder. 

The change of variables $v=\tf{t}{(t+1)}$ recasts the integral as
\beq
\mc{T}(a,b) \; = \;  \mf{x}^{a + b + 1} \Int{0}{ \tfrac{\de}{\de+ \mf{x}} } v^{a} \cdot (1-v)^{-a-b-2} \cdot \dd v \;. 
\enq
It remains to expand the model integral 
\beq
\wt{\mc{T}}\big(x,y; z \big)\; = \; \Int{0}{z} t^{x-1} \cdot (1-t)^{y-1} \cdot \dd t   \quad, \; \Re(x)>0\, , 
\enq
around $z=1$. Let $p\in \mathbb{N}$ be such that  $\Re(y)+p>0$. Then, using the expansion for $|t-1|<1$
\beq
t^{x-1}\,=\,[1+(t-1)]^{x-1} = \sul{ n \geq 0 }{ } C_{n}(x)(1-t)^n \quad \e{with} \quad  C_0(x)=1 \, ,
\label{definition des coefs Cn de x}
\enq
one has that 
\bem
\wt{\mc{T}}\big(x,y; z \big)\; = \; \Int{0}{z} \Big( t^{x-1} - \sul{n=0}{p-1} C_n(x) (1-t)^n  \Big) \cdot (1-t)^{y-1} \cdot \dd t
\, + \, \sul{n=0}{p-1} \f{1-(1-z)^{y+n}}{y+n} C_n(x) \\
 = \; -\sul{n \geq   0 }{ } C_n(x) \f{(1-z)^{y+n}}{y+n}  \, + \, \wt{\mc{T}}_0
\end{multline}
where 
\beq
 \wt{\mc{T}}_0 \; = \; \sul{n=0}{p-1} \f{ C_n(x) }{ y + n } \, + \, \Int{0}{1} \Big(t^{x-1} - \sul{n=0}{p-1} C_n(x) (1-t)^n \Big) (1-t)^{y-1}  \cdot \dd t  
 \; = \; \f{  \Ga(x) \Ga(y) }{ \Ga(x+y) } 
\enq
has been computed by meromorphic continuation in $y$. The above expansion ensures that there exists a function $h$ that is smooth in $\mf{x}$ belonging to a neighbourhood of $0$, and in  $a>-1$ and  $b \not \in \mathbb{Z}$,  such that 
\beq
\mc{T}(a,b) \; = \;  -  \mf{x}^{a + b + 1}  \f{ \sin [\pi b] }{ \pi } \Ga(a+1)\Ga(b+1)\Ga(-a - b - 1)   \, + \, \f{ (\de+ \mf{x})^{ a+ b  + 1}   }{ a + b +1 } \; + \;  \mf{x}\cdot h( \mf{x} ) \;. 
\label{ecriture DA T alpha beta}
\enq
Owing to the choice of the integer $n$ in \eqref{ecriture choix entier n}, all integrals $\mc{T}(a,b)$ appearing in \eqref{ecriture dvpmt derivee nieme integrale J 1D} diverge in the $\mf{x}\tend 0$ limit. 
Thence, upon using the relation between $\wt{f}_0, \wt{b}_{0}$ and their un-tilded counterparts, one gets 
\bem
\Dp{ \mf{x} }^{n} \Big\{\,  \mc{J}[\msc{F}, A, B](  \mf{x} ) \, \Big\} \;= \;  (-1)^{n+1}   \mf{x}^{ a_0 +  b_0  + 1-n}  f_0 \f{ \sin [\pi b_0 ] }{ \pi } \Ga( a_0+1)\Ga( b_0 +1)\Ga(n-a_0 - b_0 - 1) \\ 
\; + \; \e{O}\bigg( \,   \mf{x}^{ a_0 +  b_0  + 1 + \a -n}   \bigg)  \;. 
\end{multline}
Then, $n$-fold integration in respect to $\mf{x}$ entails the claim.

\begin{lemme}
\label{Lemme integrale type beta reguliere}

 Let $\mf{z}_{\pm}(\la)$ be two real-holomorphic functions in a neighbourhood of $\intff{a}{b}$, $  a \, < \, b $, such that
\begin{itemize}

\item[$\bullet$]  $ \mf{z}_{\pm} >0$  on $\intoo{a}{b}$;
\item[$\bullet$]  $a$, resp. $b$, is, either such that both $\mf{z}_{\pm}(a)>0$, resp. $\mf{z}_{\pm}(b)>0$, or such that it is 
a simple zero of  $\mf{z}_{\eps_{\ell}}$, resp. $\mf{z}_{\eps_{r}}$, ($\eps_{\ell/r}\in \{\pm 1\}$) but not a zero of the other function. 
\end{itemize}

Let $\, \wh{\mf{z}}_{\pm}(\la)=\mf{z}_{\pm}(\la)+\mf{x}$ and let
\begin{itemize}

\item $a(\mf{x})=a$, resp. $b(\mf{x})=b$, in the case when both $\mf{z}_{\pm}(a)>0$, resp. $\mf{z}_{\pm}(b)>0$;
\item   $a(\mf{x})$, resp. $b(\mf{x})$, be the zero of $\, \wh{\mf{z}}_{\eps_{\ell}}$, resp.  $\wh{\mf{z}}_{\eps_{r}}$, such that $a(\mf{x})=a+\e{O}(\mf{x})$, resp. $b(\mf{x}) \, = \,  b + \e{O}(\mf{x})$, if
$\mf{z}_{\eps_{\ell}}(a)=0$, resp.  $\mf{z}_{\eps_{r}}(b)=0$. 
\end{itemize}

\noindent  Let  $\De_{\ups}\geq 0$ be smooth  on $\intff{ a(\mf{x}) }{ b(\mf{x}) }$ uniformly in $\mf{x}$ small enough. Let 
$\msc{G}$ be in the smooth class of $\intff{ a(\mf{x}) }{ b(\mf{x}) }$ with functions $\De_{\pm}$ and constant $\tau$.

 Then, the integral 
\beq
\mc{J}(\mf{x})\, = \, \Int{ a(\mf{x}) }{ b(\mf{x}) }   \msc{G}\Big(\la, \wh{\mf{z}}_{+}(\la) ,  \wh{\mf{z}}_{-}(\la) \Big) \cdot  \pl{\ups= \pm }{} \Big\{ \big[ \, \wh{\mf{z}}_{\ups}(\la) \big]^{ \De_{\ups}(\la)-1 } \Big\} \cdot \dd \la 
\enq
is a smooth function of $\mf{x}$ small enough. In particular, it admits a  Taylor series expansion around $\mf{x}=0$. 
\end{lemme}

\Proof 

To start with, consider the simpler situation when $\De_{\pm}>0$ on $\intff{ a(\mf{x}) }{ b(\mf{x}) }$.

First consider the case when $a$ and $b$ are both a zero of one of the functions $\mf{z}_{\pm}$. 
Then, let $\eps_{\ell/r} \in \{ \pm 1\}$ be such that $\mf{z}_{\eps_{\ell}}(a)=0$, $\mf{z}_{\eps_{r}}(b)=0$. In such a case, for any $\eta>0$ and small enough, the hypotheses of the lemma ensure that
there exists a constant $c>0$ such that
\beq
\left\{ \ba{ccc} \mf{z}_{-\eps_{\ell}}(\la)  \, > \,  c  &   \e{on} & \intff{a}{a+\eta} \\
\mf{z}_{-\eps_{r}}(\la) \, > \, c \      & \e{on} & \intff{b-\eta}{b}  \ea \right. 
\qquad \e{and} \qquad \mf{z}_{\pm}(\la) \, > \, c \;\; \e{on} \;  \intff{a+\eta}{b-\eta} \;. 
\label{ecriture conditions positivite}
\enq
Since $a$, resp. $b$, is a simple zero of $\mf{z}_{\eps_{\ell}}(\la)$, resp.  $\mf{z}_{\eps_{r}}(\la)$,  the function is a local biholomorphism in the neighbourhood of that point. 
Hence, the zeroes $a(\mf{x})$ and $b(\mf{x})$ are analytic in $\mf{x}$ small enough  and one has the factorisation 
\beq
\wh{\mf{z}}_{\eps_{\ell}}(\la) \, =\, (\la-a(\mf{x})) \cdot  h_{ \ell }(\la , \mf{x}) \qquad \e{and} \qquad
\wh{\mf{z}}_{\eps_{r}}(\la) \, =\, (b(\mf{x})-\la) \cdot  h_{ r }(\la , \mf{x}) 
\enq
with $  h_{ r/\ell }(\la , \mf{x}) >0$ and analytic in $\la$ and $\mf{x}$ by the Weierstrass preparation theorem \ref{Theorem Weierstrass preparation theorem}. 
Finally, the inverse  $ \big( \, \wh{\mf{z}}_{\eps_{ \ell/r}}\big)^{-1}$ takes the explicit form
$ \big( \, \wh{\mf{z}}_{\eps_{ \ell/r}}\big)^{-1}(t) \, = \, \mf{z}_{\eps_{ \ell/r}}^{-1}(t-\mf{x})  $

Thence, picking some $\eta>0$ small enough, one can decompose the original integral as  
\beq
 \mc{J}(\mf{x})\, = \, \mc{J}_{\ell}(\mf{x}) \, + \, \mc{J}_{c}(\mf{x})  \, + \, \mc{J}_{r}(\mf{x}) 
\quad \e{with} \quad 
\mc{J}_{c}(\mf{x}) \, = \hspace{-3mm} \Int{ \mf{z}_{\eps_{ \ell}}^{-1}(\eta-\mf{x})  }{ \mf{z}_{\eps_{ r }}^{-1}(\eta-\mf{x}) } \hspace{-3mm}  G_{c}(\la,\mf{x}) \cdot \dd \la 
\enq
and 
\beq
 \mc{J}_{\ell}(\mf{x}) \, =  \hspace{-3mm} \Int{a(\mf{x}) }{ \mf{z}_{\eps_{ \ell}}^{-1}(\eta-\mf{x})  } \hspace{-3mm}  G_{\ell}(\la,\mf{x}) \cdot
\big[ \, \wh{\mf{z}}_{\eps_{\ell}}(\la) \big]^{ \De_{ \eps_{\ell} }(\la)-1 }
 \cdot  \mf{z}_{\eps_{\ell}}^{\prime}(\la)\cdot \dd \la 
\; \quad , \quad   \; 
 \mc{J}_{r}(\mf{x}) \, =  \hspace{-3mm} \Int{  \mf{z}_{\eps_{ r }}^{-1}(\eta-\mf{x}) }{b(\mf{x}) }  \hspace{-3mm}  G_{r}(\la,\mf{x}) 
 \cdot \big[ \, \wh{\mf{z}}_{\eps_{r}}(\la) \big]^{ \De_{ \eps_{r} }(\la)-1 }
 \cdot  \mf{z}_{\eps_{r}}^{\prime}(\la)\cdot \dd \la  \;. 
\enq
Above, I have set 
\beq
G_{c}(\la,\mf{x}) \; = \;  \msc{G}\Big(\la, \wh{\mf{z}}_{+}(\la) ,  \wh{\mf{z}}_{-}(\la) \Big) \cdot \pl{\ups= \pm }{} \Big\{ \big[ \, \wh{\mf{z}}_{\ups}(\la) \big]^{ \De_{\ups}(\la)-1 } \Big\} 
\enq
and
\beq
G_{\ell/r}(\la,\mf{x}) \; = \;  \msc{G}\Big(\la, \wh{\mf{z}}_{+}(\la) ,  \wh{\mf{z}}_{-}(\la) \Big) \cdot   
 \big[ \, \wh{\mf{z}}_{-\eps_{\ell/r}}(\la) \big]^{ \De_{-\eps_{\ell/r}}(\la)-1 } \cdot \f{ 1  }{ \mf{z}_{\eps_{\ell/r}}^{\prime}(\la)  } 
 \;. 
\enq
The lower bounds \eqref{ecriture conditions positivite}, the smoothness of $G_{c}(\la;\mf{x})$  and the fact that the integration runs through a compact,
all together ensure that $\mc{J}_{c}(\mf{x})$ is smooth in $\mf{x}$. Furthermore, a change of variables recasts $ \mc{J}_{a}(\mf{x})$, with $a\in \{\ell, r\}$ as 
\beq
 \mc{J}_{a}(\mf{x}) \, =   \vsg_{a} \Int{ 0  }{ \eta   }    G_{a}\Big(   \mf{z}_{\eps_{ a }}^{-1}(s-\mf{x})   , \mf{x} \Big) \cdot
  s ^{ \De_{ \eps_{a} }\circ \mf{z}_{\eps_{ a }}^{-1}(s-\mf{x}) -1 }
  \dd s
\enq
with $\vsg_{\ell}=1$ and $\vsg_{r}=-1$. The same arguments as for $\mc{J}_{c}(\mf{x})$ then allow one to conclude. 

\vspace{2mm}
The remaining cases of possible values of $\mf{z}_{\pm}(a)$ 
and $\mf{z}_{\pm}(b)$ can be treated quite similarly. 

 \vspace{2mm}

It remains to discuss the situation when one allows for $\De_{\ups}$ to vanish. The latter case remains unchanged relatively to $\mc{J}_{c}(\mf{x})$. 
As for $\mc{J}_{a}(\mf{x})$, $a \in \big\{ \ell, r  \big\}$, by the properties of a smooth class function, one may recast
\beq
  G_{a}\Big(   \mf{z}_{\eps_{ a }}^{-1}(s-\mf{x})   , \mf{x} \Big) \; = \; \De_{\eps_a}\circ   \mf{z}_{\eps_{ a }}^{-1}(s-\mf{x})  \cdot   G_{a}^{(1)}\big(  s-\mf{x} \big)
\; + \; G_{a}^{(2)}\big(  s-\mf{x} \big) \cdot s^{1-\tau}
\enq
with $G_a^{(1)}$, $G_a^{(2)}$ smooth. Thus, 
\bem
 \mc{J}_{a}(\mf{x}) \, =   \vsg_{a} \Int{ 0  }{ \eta   }    G_{a}^{(2)}\big( s-\mf{x}   \big) \cdot  s ^{ \De_{ \eps_{a} }\circ \mf{z}_{\eps_{ a }}^{-1}(s-\mf{x}) - \tau } \cdot \dd s  
\, - \,  \vsg_{a} \Int{ 0  }{ \eta   }  \Dp{s} \Big\{  G_{a}^{(2)}\big( s-\mf{x}   \big) \cdot  v^{ \De_{ \eps_{a} }\circ \mf{z}_{\eps_{ a }}^{-1}(s-\mf{x}) } \Big\}_{\mid v = s } \cdot \dd s \\
\, + \,  \vsg_a G_{a}^{(2)}\big( \eta-\mf{x}   \big) \cdot  \eta ^{ \De_{ \eps_{a} }\circ \mf{z}_{\eps_{ a }}^{-1}(\eta-\mf{x})   } 
\, - \, \vsg_a G_{a}^{(2)}\big(  -\mf{x}   \big) \cdot  { s^{ \De_{ \eps_{a} }\circ \mf{z}_{\eps_{ a }}^{-1}( s-\mf{x})   } }_{\mid s=0} \;. 
\end{multline}
Note that the last term issuing from the integration by parts is present only if $\De_{ \eps_{a} }\circ \mf{z}_{\eps_{ a }}^{-1}(  -\mf{x})=0$ and in that case, 
the contribution is also smooth in $\mf{x}$. Smoothness of all the other terms is clear. \qed




\section{Asymptotics of multi-dimensional $\be$-like integral}
\label{Appendix DA integrales multidimensionnelles}

 \subsection{General assumptions}
\label{SousSection hypothese gles sur consituants integrale modele} 
 
It is convenient to introduce a few notations and objects that will be used throughout this section. 
One assumes to be given:
\begin{itemize}
\item a strictly positive real $\op{v} >0$;
\item  a collection of compact intervals $\msc{I}_{r}$, $r=1,\dots, \ell$ ;
\item smooth functions $\mf{u}_r$ on $\msc{I}_r$ such that $\mf{u}^{\prime}_r$ is strictly monotonous on $\msc{I}_r$, and such that 
\beq
\mf{u}^{\prime}_r(k) \not= \pm \op{v}  \qquad \e{for} \qquad k \in \e{Int}\big(\msc{I}_{r} \big) \;. 
\label{ecriture propriete de vitesse defferente de vF}
\enq
\end{itemize}
Taken the physical interpretation that is discussed in the core of the paper,
\begin{itemize}
\item $k \mapsto \mf{u}_r(k)$ corresponds to the momentum-energy dispersion curve associated with a single particle excitation of "type" $r$;
\item  $\mf{u}^{\prime}_{r}$ corresponds to the velocity of this excitation; 
\item $\msc{I}_r$ is the domain, in  momentum space, where the  dispersion curve $k \mapsto \mf{u}_r(k)$  is strictly convex or concave, \textit{viz}. where $\mf{u}^{\prime\prime}_{r}$ has constant sign in its interior. 
\end{itemize}

Given $n_r \in \mathbb{N}^{*}$, $r=1,\dots, \ell$, define the compact subset $\msc{I}_{\e{tot}}$ of  $\R^{ \ov{\bs{n}}_{\ell} }$ with $ \ov{\bs{n}}_{\ell} \, = \, \sum_{r=1}^{\ell}n_r$,  as
\beq
\msc{I}_{\e{tot}}\; = \; \pl{r=1}{\ell} \msc{I}_r^{n_r} \;. 
\label{definition de intervalle multidimensionnels I}
\enq
It is assumed that the intervals $\msc{I}_r$ partition as 
\beq
\msc{I}_{r}\; = \; \msc{I}_{r}^{(\e{in})}\sqcup \msc{I}_{r}^{(\e{out})}
\label{ecriture partition in out intervalle r}
\enq
with $\msc{I}_{1}\; = \; \msc{I}_{1}^{(\e{in})}$, \textit{i.e.} $\msc{I}_{1}^{(\e{out})} = \emptyset$. The partition is such that 
\beq
\mf{u}_{r}^{\prime}\Big( \e{Int}\big(\msc{I}_{r}^{(\e{out})} \big) \Big) \, \cap  \, \mf{u}_{1}^{\prime}\Big( \e{Int}\big(\msc{I}_{1}^{(\e{in})} \big) \Big) \, = \, \emptyset
\qquad \e{and} \qquad 
\mf{u}_{r}^{\prime}\Big( \e{Int}\big(\msc{I}_{r}^{(\e{in})} \big) \Big) \, =  \, \mf{u}_{1}^{\prime}\Big( \e{Int}\big(\msc{I}_{1}^{(\e{in})} \big) \Big) \;. 
\enq
The hypothesis of strict monotonicity of $\mf{u}^{\prime}_{r}$ ensures that all the sets $\mf{u}_{r}^{\prime}\Big( \e{Int}\big(\msc{I}_{r}^{(\e{in})} \big) \Big)$ are in one-to-one correspondence. More precisely, there exist homeomorphisms 
\beq
t_r \, : \, \msc{I}_{1}^{(\e{in})} \tend \msc{I}_{r}^{(\e{in})} \qquad \e{such} \; \e{that} \qquad 
\mf{u}_1^{\prime}(k) \; = \; \mf{u}_r^{\prime}\big( t_r(k) \big) \;.
\enq
The hypotheses on $\mf{u}_{r}$ ensure that $t_r$ is a smooth diffeomorphism from $\e{Int}\big(\msc{I}_{1}^{(\e{in})} \big)$ onto $\e{Int}\big(\msc{I}_{r}^{(\e{in})} \big)$. 
It will appear useful, sometimes, to denote $t_1(k)=k$. The partitioning \eqref{ecriture partition in out intervalle r} splits the momentum range of type $r$
excitations into an interval $ \msc{I}_{r}^{(\e{in})} $ associated with momenta of type $r$ excitations having a velocity that is also shared by type "1" excitations, 
and an interval $ \msc{I}_{r}^{(\e{out})} $ whose associated velocities never coincide with those of type "1" excitations.  

\vspace{2mm}

Given a choice of signs $\zeta_r \in \{\pm 1\}$, one defines the associated  macroscopic "momentum" and "energy" 
\beq
\mc{P}(k) \; = \; \, \sul{r=1}{\ell }n_r \, \zeta_r  \, t_r(k)   \qquad \e{and} \qquad \mc{E}(k) \; = \; \sul{r=1}{\ell } n_r  \, \zeta_r \,  \mf{u}_r\big( t_r(k) \big)  \;\; ,  \qquad k\in  \msc{I}_{1} \; , 
\label{definition P et E de max multi particule excitation}
\enq
of an agglomeration of equal velocity particles of different types.  It is assumed in the following that $k\mapsto \mc{P}(k)$ is strictly monotonous on $\e{Int}(\msc{I}_1)$, \textit{i.e.} that  
\beq
k \; \mapsto \; \mc{P}^{\prime}(k) \, =\, \sul{r=1}{\ell} \zeta_r \, n_r  \, t_r^{\prime}(k)
\label{ecriture hypothese non vanishing impusion macro}
\enq
does not vanish on $\e{Int}\big(\msc{I}_{1}^{(\e{in})} \big)$.

 Finally, it is convenient to represent vectors in block form relatively to the Cartesian product structure of $\msc{I}_{\e{tot}}$, \textit{c.f.} \eqref{definition de intervalle multidimensionnels I}, 
\beq
\bs{p} \; = \; \big( \bs{p}^{(1)} , \dots ,  \bs{p}^{(\ell)} \, \big)\qquad \e{with} \qquad \bs{p}^{(r)}\, = \, \big( p_1^{(r)}, \cdots, p_{n_r}^{(r)} \big) \in \R^{n_r} \;. 
\label{definition vecteur p}
\enq
Also, given a vector $\bs{p}$ as above, it will be useful to introduce a special notation for a related vector where some of the components of $\bs{p}$ have been dropped:
\beq
\bs{p}^{(r)}_{[a]} \, = \, \big( p_1^{(r)} , \dots, p_{a-1}^{(r)},p_{a+1}^{(r)},\dots , p_{n_r}^{(r)}  \,  \big) \quad \e{and} \quad 
\bs{p}_{[r,a]} \, = \,  \Big( \bs{p}^{(1)} , \dots , \bs{p}^{(r)}_{[a]},   \dots , \bs{p}^{(\ell)} \big) \;. 
\label{definition vecteur p avec composantes omises}
\enq
In the following, $k_0\in \e{Int}(\msc{J}_1)$ will single out a point in $\e{Int}(\msc{J}_1)$. Analogously to the above way of writing vectors, one denotes
\beq
\bs{t}(k_0)\, = \, \big( \bs{t}_1(k_0), \dots, \bs{t}_{\ell}(k_0) \big) \in \R^{\ov{\bs{n}}_{\ell}} \qquad \e{with} \qquad \bs{t}_r(k_0) \, = \, \big( t_r(k_0), \dots, t_r(k_0) \big) \in \R^{n_{r}} 
\label{ecriture definition vecteur tk0}
\enq
as well as 
\beq
\bs{t}_{[\ell,n_{\ell}]}(k_0)\, = \, \big( \bs{t}_1(k_0), \dots, \bs{t}_{\ell,[n_{\ell}]}(k_0) \big) \in \R^{\ov{\bs{n}}_{\ell}-1} \qquad 
\e{with} \qquad  \bs{t}_{\ell,[n_{\ell}]}(k_0)    \, = \, \big( t_{\ell}(k_0), \dots, t_{\ell}(k_0) \big) \in \R^{n_{\ell}-1} 
\enq
and where all the other $\bs{t}_r(k_0)$'s are as given  in \eqref{ecriture definition vecteur tk0}. 

Finally, the set of all possible labels $(r,a)$ arising in the coordinates of $\bs{p}$ given in \eqref{definition vecteur p} is denoted as:
\beq
\mc{M} \; = \; \Big\{ (r,a) \; : \; r\in \intn{1}{\ell} \; \e{and} \; a\in \intn{1}{n_r}  \Big\} \;. 
\enq
Sometimes, the notation 
\beq
\mc{M}_{[\ell, n_{\ell}]} \; = \; \mc{M} \setminus \{(\ell, n_{\ell})\}
\label{definition ensemble M elle n ell enleve}
\enq
will be used.

 \vspace{3mm}

 It is easily seen that properties $Hi)-Hii)$  of a function $\msc{G}$ on $K \times \R^+ \times \R^+$
that is in the  smooth class of $K$ and associated with functions $d_{\pm}$ and a constant $\tau$, \textit{c.f.} Definition \ref{definition smooth class on K}, 
 entail that, for any $(\bs{s},\ell_u,\ell_v)$ as above and for fixed $\eps>0$, it holds  
\begin{itemize}

 \item[H1)] $\big( \bs{x}, u , v  \big) \mapsto \pl{ a=1 }{ n }   \Dp{ x_a  }^{s_a }    \, \cdot \,  \Dp{u}^{\ell_u} \, \msc{G}^{(s)}\big( \bs{x}, u , v  \big)$ is bounded on $  K  \times \intff{ \eps }{ \eps^{-1} } \times \intff{0}{\eps^{-1}}$;
 
 \item[H2)] $\big( \bs{x}, u , v  \big) \mapsto \pl{ a=1 }{ n }\Dp{ x_a  }^{s_a  } \, \cdot \,  \Dp{v}^{\ell_v} \, \msc{G}^{(s)}\big( \bs{x}, u , v  \big)$ is bounded on $ K  \times \intff{0}{\eps^{-1}}\times \intff{ \eps }{ \eps^{-1} }$;
 
 \item[H3)] $\big( \bs{x}, u , v  \big) \mapsto\pl{ a=1}{ n }\Dp{ x_a  }^{s_a } \, \cdot \, \msc{G}^{(s)}\big( \bs{x}, u , v  \big)$ is bounded on $ K  \times \intff{0}{\eps^{-1}}^2$.

\end{itemize}
Note that depending on the values of $s\in\intn{1}{4}$, the $u$ or $v$ variables may or may not be effectively present in the above equations, \textit{viz}. one should understand
in the formulae above $\msc{G}^{(1)}(\bs{x}, u, v)  =  \msc{G}^{(1)}(\bs{x})$ \textit{etc}. 
 
Furthermore, when $n\geq 2$, all the functions appearing in $H1)-H3)$ vanish upon replacing $K \hookrightarrow \Dp{}K$.

 \subsection{The structural theorem in the multidimensional setting}

\begin{theorem}
\label{Theorem principal caractere lisse et non lisse des integrals multi particules}

Let $\msc{I}_{\e{tot}}$ be as defined in \eqref{definition de intervalle multidimensionnels I} and 
$ \De_{\pm} $ be smooth positive functions on $ \msc{I}_{\e{tot}} $ admitting smooth square roots on $ \msc{I}_{\e{tot}} $.  
Let $\msc{G}$ be in the smooth class of $ \msc{I}_{\e{tot}} $ assoiated with the functions $\De_{\pm}$ and a constant $\tau \in \intoo{0}{1}$, 
according to Definition \ref{definition smooth class on K}. 

Finally, let  
\beq
\mf{z}_{\ups}(\bs{p})\;=\; \mc{E}_0 \, -\,  \sul{ (r,a) \in \mc{M} }{  } \zeta_r \mf{u}_r\big( p_a^{(r)} \big)  \, +  \; \ups  \op{v} \, \bigg\{ \mc{P}_0 -  \sul{ (r,a) \in \mc{M} }{  } \zeta_r   p_a^{(r)}  \bigg\} \; , 
\quad \ups \in \{ \pm \} , 
\label{definition fonctions zpm modele}
\enq
with $\zeta_r \in \{\pm 1\}$ as given in \eqref{definition P et E de max multi particule excitation} and where $(\mc{P}_0,\mc{E}_0) \in \R^2$.

\noindent Let $\mc{I}\big( \mf{x} \big) $ be given by the multiple integral 
\beq
\mc{I}\big( \mf{x} \big) \, = \, \Int{ \msc{I}_{\e{tot}} }{}  \dd \bs{p}  \; \msc{G}_{\e{tot}}(\bs{p})   
\enq
where  
\beq
\msc{G}_{\e{tot}}(\bs{p})  \, = \, \msc{G}\Big( \bs{p},  \mf{z}_{+}(\bs{p})+ \mf{x} , \mf{z}_{-}(\bs{p})+ \mf{x}  \Big)  \cdot \pl{\ups=\pm }{} \bigg\{ \Xi\Big( \,  \mf{z}_{\ups}(\bs{p})+ \mf{x}  \Big)
\cdot \Big[ \,   \mf{z}_{\ups }(\bs{p})+ \mf{x}  \Big]^{ \De_{\ups}(\bs{p}) -1 }    \bigg\} \cdot V(\bs{p}) \;, 
\label{definition Gtot dans appendix}
\enq
and 
\beq
 V(\bs{p}) \; = \;   \pl{r=1}{\ell} \pl{a<b}{n_r} \big( p_a^{(r)}-p_b^{(r)} \big)^2  \; . 
\label{defintion V Vdm product squared} 
\enq

  The type of $\mf{x}\tend 0$ asymptotic expansion of $\mc{I}(\mf{x})$ depends on the value of $(\mc{P}_0,\mc{E}_0)$.

\vspace{2mm} 

 {\bf a)} \texttt{The regular case.}

\vspace{2mm} 
 \noindent If the two conditions given below  hold 
\beq
\big( \mc{P}_0,\mc{E}_0 \big) \, \not\in \, \Big\{  \big( \mc{P}(k),\mc{E}(k) \big) \; : \; k \in \msc{I}_1   \Big\}  
%
%
\label{ecriture hypothese forme energie impulsion reguliere}
\enq
and 
\beq
\underset{ \substack{ \a \in \Dp{}\msc{I}_{1}  \\ \ups=\pm}  }{\e{min}} \big| \mc{E}_0\,-\, \mc{E}(\a) +  \ups \op{v}\, (\mc{P}_0\,-\, \mc{P}(\a))\big| \; > \; 0
\label{Hypothese non vanishing of Z ups on boundary of I1}
\enq
then $\msc{G}_{\e{tot}} \in L^1\big( \msc{I}_{\e{tot}} \big)$ and $\mc{I}\big( \mf{x} \big) $ is smooth in  $\mf{x}$, for $|\mf{x}|$ small enough.

\vspace{2mm} 
  {\bf b)}  \texttt{The singular case.} 

\vspace{2mm}  \noindent Let $k_0 \in \e{Int}(\msc{I}_1)$ and recalling $\bs{t}(k_0)$ as defined in \eqref{ecriture definition vecteur tk0}, let 
\beq
\De_{\ups}^{(0)} \, = \, \De_{\ups}\big( \bs{t}(k_0) \big) \qquad and \qquad 
\vth \; = \; \f{1}{2} \sul{r=1}{\ell}n_r^2 \; - \; \f{3}{2}   \, + \, \De_{+}^{(0)} \, + \, \De_{-}^{(0)} \;.
\label{definition de cal theta 0}
\enq

If 
\beq
\big( \mc{P}_0,\mc{E}_0 \big) \, = \,   \big( \mc{P}(k_0),\mc{E}(k_0) \big)    \;, 
\quad \vth \not\in \mathbb{N} \;, \quad and \quad \De_{ \pm }^{(0)} >0
\label{ecriture forme de impulsion energie singuliere}
\enq
then $\msc{G}_{\e{tot}} \in L^1\big( \msc{I}_{\e{tot}} \big)$ and    $\mc{I}(\mf{x})$ admits the $\mf{x}\tend 0^+$ asymptotic expansion:
\bem
\mc{I}\big( \mf{x} \big)   \; = \; 
  \f{ \De_{ +}^{(0)} \, \De_{ - }^{(0)}\,  \mc{G}^{(1)}\big( \bs{t}(k_0) \big)  \cdot  \big(2 \op{v}  \big)^{\De_{ +}^{(0)} + \De_{-}^{(0)} - 1  } }
{ \sqrt{ | \mc{P}^{\prime}(k_0) | } \cdot  \pl{\ups=\pm }{} \big| \op{v} - \ups \mf{u}_1^{\prime}(k_0)  \big|^{ \De_{ \ups }^{(0)} }  }   
   \cdot \Ga\big( \De_{ +}^{(0)} \big)   \Ga\big( \De_{ -}^{(0)} \big) \Ga\big( - \vth   \big)  \cdot 
   \pl{r=1}{\ell} \Bigg\{   \f{ G(2+n_r) \cdot \big( 2\pi\big)^{\frac{ n_r - \de_{r,1} }{2} }  }{   \big|\mf{u}_{r}^{\prime\prime}(t_r(k_0))\big|^{ \frac{1}{2} ( n_r^2 - \de_{r,1} )  }  } \Bigg\}                \\ 
\times  
|\mf{x}|^{ \vth   }    \cdot \Bigg\{ \Xi(\mf{x}) \f{ \sin \big[ \pi  \nu_{+}\big] }{\pi}   \, + \,  \Xi(-\mf{x})\f{ \sin \big[ \pi  \nu_{-}\big] }{\pi}   \Bigg\}  
\, + \,     \mf{r}(\mf{x})  \, + \, \e{O} \Big( |\mf{x}|^{ \vth  +  1 -\tau  }   \Big)\;.
\end{multline}
Above $\mf{r}(\mf{x})$ is smooth in $\mf{x}$, for $|\mf{x}|$ small enough.
Finally, 
\beq
\nu_{ \pm } \; = \;  \f{1}{2}\sul{  \substack{ r=1 \, : \, \\ \veps_r= \mp 1} }{ \ell }  n_r^2 \; - \; \f{ 1 \mp \mf{s} }{ 4 }    
\; +  \hspace{-4mm}  \sul{ \substack{ \ups=\pm  \, : \, \\   \pm [ \op{v} - \ups \mf{u}_1^{\prime}(k_0) ]>0 } }{}  \hspace{-4mm} \De_{ \ups }^{(0)}
\enq
where $\mf{s}=-\e{sgn}\Big(  \frac{ \mc{P}^{\prime}(k_0) }{  \mf{u}_{1}^{\prime\prime}(k_0) }  \Big) $ and $\veps_r=-\zeta_r \e{sgn}\Big(\mf{u}^{\prime\prime}_{r}(t_r(k_0)) \Big) $\;.

\end{theorem}

It is to be expected that the conditions $\De_{ \pm }^{(0)} >0$ are only technical and can be relaxed down to $\De_{ \pm }^{(0)} \geq 0$, upon some improvement 
of the method of analysis. 

\Proof

\vspace{1mm}

\subsubsection*{$\bullet$ $L^{1}(\msc{J}_{\e{tot}})$ character}

The integrand is smooth with the exception of the points where $ \mf{z}_{\ups }(\bs{p})+ \mf{x}=0$. Thus, to conclude on its $L^{1}(\msc{J}_{\e{tot}})$ integrability
it is only necessary to focus on its local behaviour in the vicinity of these points. The local behaviour of the integrand around these points, after an appropriate 
change of variables that rectifies this behaviour, is thoroughly investigated in the core of the proof. It is the integrations over such vicinity 
that generate the non-smooth behaviour in $\mf{x}$. These integrals reduce to the "local" integrals described in \eqref{ecriture integrale I sg parallele cas u lower v 1ere reduction} 
and \eqref{ecriture integrale I sg parallele cas u bigger v 1ere reduction} whose study can be reduced to reasoning on one-dimensional integrals 
by means of appropriate changes of variables. On the level of these representations, it is easy to see that the local $L^{1}$ character, in virtue of 
$\msc{G}$ being in the smooth class of $\De_{\pm}$, reduces to $\De_{\pm}\geq 0$.

\subsubsection*{$\bullet$ A preliminary decomposition into totally collinear and non-collinear parts}

The first step of the analysis consists in decomposing the integral into those  parts which may, under certain conditions on $(\mc{P}_0,\mc{E}_0)$, generate a non-smooth behaviour in $\mf{x}$
and those parts which will always, independently of the value of $(\mc{P}_0,\mc{E}_0)$, produce a smooth behaviour. 
This is achieved by decomposing the integration domain into portions where one can directly apply Lemma \ref{Lemme integrale multidimensionnelle auxiliaire reguliere}, hence guaranteeing
smoothness in $\mf{x}$ of their contribution, 
and those portions which require further study.

\vspace{2mm}

Given  $\eta>0$ small enough, one has the below decomposition of $\msc{I}_{\e{tot}}$ 
\beq
\msc{I}_{\e{tot}}\, = \,  \mc{D}^{(\perp)}_{\eta}  \sqcup \mc{D}^{(\sslash)}_{\eta} 
\enq
where
\beq
\mc{D}^{(\perp)}_{\eta} \, = \, \Big\{ \, \bs{p}\in  \msc{I}_{\e{tot}} \; : \;  \exists \, (r,a) \not= (1,1) \;\;\e{such}\; \e{that} \;\; \big| \mf{u}_1^{\prime}\big( p_1^{(1)} \big) \, - \, \mf{u}_r^{\prime}\big( p_a^{(r)} \big)  \big| \geq \eta \,  \Big\}
\label{definition domaine particules ext}
\enq
contains vectors $\bs{p}$ where at least one variable is associated with a different velocity than the one carried by the first component $p_1^{(1)}$ of $\bs{p}$ and
\beq
\mc{D}^{(\sslash)} _{\eta} \, = \, \Big\{ \, \bs{p}\in  \msc{I}_{\e{tot}} \; : \;  \forall (r,a) \in \mc{M}  \;\; \big| \mf{u}_1^{\prime}\big( p_1^{(1)} \big) \, - \, \mf{u}_r^{\prime}\big( p_a^{(r)} \big)  \big| < \eta \,  \Big\}
\label{definition domaine particules out}
\enq
contains vectors all of whose components have almost equal velocities. Let $\vp^{(\sslash)}$ be smooth and such that 
\beq
0\leq \vp^{(\sslash)} \leq 1  \quad, \qquad \vp^{(\sslash)}\, = \, 1 \quad \e{on} \quad \ov{\mc{D}^{(\sslash)}_{\eta/2}} \quad \e{and} \quad \vp^{(\sslash)}\, = \, 0 \;\; \e{on} \; \; \mc{D}^{(\perp)}_{\eta}  \;. 
\enq
Then set $ \vp^{(\perp)} \, = \,   1 - \vp^{(\sslash)}$. This entails that $ \vp^{(\perp)} \, \not= \, 0 $ only on  $\mc{D}^{(\perp)}_{\tf{\eta}{2}}$
so that one has a partition of unity on $\msc{I}_{\e{tot}}$ : $ \vp^{(\sslash)} + \vp^{(\perp)} \, = \,   1 $ which induces the decomposition
of $\mc{I}\big( \mf{x} \big)  \, = \,  \mc{I}^{(\perp)}\big( \mf{x} \big) \, + \, \mc{I}^{(\sslash)}\big( \mf{x} \big) $ with 
\beq
 \mc{I}^{(\perp)}\big( \mf{x} \big) \; =  \Int{ \mc{D}^{(\perp)}_{ \tf{\eta}{2} } }{}  \hspace{-1mm} \dd \bs{p}  \;   \msc{G}_{\e{tot}}^{(\perp)} (\bs{p})  
\qquad \e{and} \qquad
\mc{I}^{(\sslash)}\big( \mf{x} \big) \; =   \Int{ \mc{D}^{(\sslash)}_{  \eta } }{}  \hspace{-1mm} \dd \bs{p}  \;     \msc{G}_{\e{tot}}^{(\sslash)} (\bs{p})  \;. 
\enq
Above and in the following, we agree upon 
\beq
 \msc{G}_{\e{tot}}^{(\perp)}(\bs{p})   \, = \, \vp^{(\perp)}(\bs{p})  \cdot   \msc{G}_{\e{tot}} (\bs{p})   \qquad \e{and} \qquad 
 \msc{G}_{\e{tot}}^{(\sslash)}(\bs{p})   \, = \, \vp^{(\sslash)}(\bs{p})   \cdot    \msc{G}_{\e{tot}} (\bs{p}) \;. 
\enq

\subsubsection*{$\bullet$ The integral $\mc{I}^{(\perp)}$}

I establish below that $\mc{I}^{(\perp)}\big( \mf{x} \big)$ solely generates a smooth behaviour in $\mf{x}$ small enough. 
 
 Given $ \bs{h} \in \mc{D}^{(\perp)}_{ \tf{\eta}{2} }$, by definition, there exists $(r,a) \not= (1,1)$  such that $\big| \mf{u}_1^{\prime}\big( h_1^{(1)} \big) \, - \, \mf{u}_r^{\prime}\big( h_a^{(r)} \big)  \big| \geq \tf{\eta}{2}$.
Then, the map 
\beq
f_{[r,a]}\big( \bs{p}\big)\, = \, \Big(\bs{p}_{1}^{(1)},\dots ,\bs{p}_{[a]}^{(r)}, \dots, \bs{p}^{(\ell)}, \mf{z}_+\big( \bs{p} \big), \mf{z}_-\big( \bs{p} \big) \Big) 
\enq
satisfies
\beq
\det\Big[ D_{\bs{p}} f_{[r,a]} \Big] \; = \;  2 \op{v}  \, \zeta_1 \, \zeta_r \cdot (-1)^{m_{r,a} }   \cdot \big[ \, \mf{u}_r^{\prime}\big( p_a^{(r)} \big) \, - \, \mf{u}_1^{\prime}\big( p_1^{(1)} \big) \,  \big]   
\quad \e{with} \quad 
m_{r,a}=a+\sul{b=1}{r-1}n_b\;. 
\enq
Hence, for all $\bs{h}\in \mc{D}^{(\perp)}_{ \tf{\eta}{2} }$, 
\beq
\Big| \det\Big[ D_{\bs{h}} f_{[r,a]} \Big] \Big| \, \geq \,    \op{v}  \eta  \; .
\enq
One is thus in position to apply Lemma \ref{Lemme integrale multidimensionnelle auxiliaire reguliere}
so as to conclude that $\mf{x} \mapsto \mc{I}^{(\perp)}\big( \mf{x} \big)$ is smooth in $|\mf{x}|$ small enough.

\noindent As a consequence, it only remains to focus on the $\mf{x}\tend 0$ behaviour of  $\mc{I}^{(\sslash)}\big( \mf{x} \big)$.

\subsubsection*{$\bullet$ Behaviour of $\mc{I}^{(\sslash)}$ in the regular case}

This corresponds to case {\bf a)} appearing in the statement of the theorem. Since $\msc{I}_1$ is compact and $k\mapsto (\mc{P}(k),\mc{E}(k))$ is continuous, 
where $\mc{P}(k),\mc{E}(k)$ are as defined in \eqref{definition P et E de max multi particule excitation}, 
hypotheses \eqref{ecriture hypothese forme energie impulsion reguliere} and \eqref{Hypothese non vanishing of Z ups on boundary of I1} entail that there exists $\varrho>0$ such that  
\beq
\underset{k \in \msc{I}_1  }{\e{inf}}  \Big\{ \; 
\e{d}\Big(   \big( \mc{P}_0,\mc{E}_0 \big) \, , \,  \big( \mc{P}(k),\mc{E}(k) \big) \Big) \; \Big\} \, > \, \varrho   \qquad \e{and} \qquad
\underset{ \substack{ \a \in \Dp{}\msc{I}_{1}  \\ \ups=\pm}  }{\e{min}} \big| \mc{E}_0\,-\, \mc{E}(\a) +  \ups \op{v} (\mc{P}_0\,-\, \mc{P}(\a))\big|  \, > \, \varrho \;. 
\label{ecriture borne inf sur distance a la courbe impulsion energie}
\enq
It is useful to  recast $\mf{z}_{\ups}(\bs{p})$ as:
\beq
\mf{z}_{\ups}(\bs{p}) \; = \; \mc{Z}_{\ups}\big(p_1^{(1)} \big)    \, + \, \de \mf{z}_{\ups}\big( \bs{p} \big)
\enq
where 
\beq
 \mc{Z}_{\ups}(k)
 =    \mc{E}_0\, -\, \mc{E}(k) + \, \ups \op{v} \big[ \mc{P}_0 \, - \, \mc{P}(k)\big] 
\enq
and 
\beq
\de \mf{z}_{\ups}\big( \bs{p} \big) \, = \, 
-\sul{ (r,a) \in \mc{M} }{  } \zeta_r\mf{w}_{\ups}^{(r)}\Big( p_a^{(r)} ; t_r(p_1^{(1)}) \Big) \;. 
\enq
Here, I have introduced 
\beq
 \mf{w}_{\ups}^{(r)}(k ; p) \, = \, \mf{u}_r(k)\, - \, \mf{u}_r(p)\, + \, \ups \op{v}  \big( k - p\big) \;. 
\label{definition fonction w frak}
\enq
One has that $\mc{Z}_{\ups}$ is smooth on $\e{Int}\big( \msc{I}_1 \big)$ and 
\beq
\mc{Z}^{\prime}_{\ups}(k)\,=\,  -  \bigg\{ \sul{r=1}{\ell} n_r \, \zeta_r \,  t^{\prime}_r(k)  \bigg\} \cdot \big(\mf{u}_1^{\prime}(k)+\ups \op{v}  \big) \;. 
\enq
 Thus owing to hypothesis \eqref{ecriture propriete de vitesse defferente de vF} and \eqref{ecriture hypothese non vanishing impusion macro}, 
$\mc{Z}_{\ups}^{\prime}$ does not vanish on $\e{Int}\big( \msc{I}_1 \big)$, so that $\mc{Z}_{\ups}$ is strictly monotonous on $\msc{I}_1$. 
This entails that $\mc{Z}_{\ups}$ has at most one zero on $\msc{I}_1$.

\vspace{2mm}

\noindent There are several cases to discuss depending on whether $\mc{Z}_{\ups}$ has a zero or not on $\msc{I}_1$.

\vspace{2mm}

{\bf i)} $\big(\mc{P}_0,\mc{E}_{0} \big)$ is such that both $\mc{Z}_{\pm}$ do not vanish on $\msc{I}_1$. 

\vspace{2mm}

\noindent In such a case, there exists $C_0$, such that $|\mc{Z}_{\pm}(k)| \geq 2 C_0 $, for any $k \in \msc{I}_1$. 
 I now establish that this property entails the non-vanishing  of $\mf{z}_{\ups}$ on $\mc{D}^{(\sslash)}_{\eta}$. For that purpose, observe that since $\mf{u}^{\prime}_r$ is strictly monotonous on $\msc{I}_r$ and continuous, 
 it is continuously invertible on its image. 
This allows one to recast 
\beq
\de \mf{z}_{\ups}\big( \bs{p} \big) \, = \, 
-\sul{ (r,a) \in \mc{M} }{  } \zeta_r \,  \wt{\mf{w}}_{\ups}^{(r)}\Big(  \mf{u}_r^{\prime}\big( p_a^{(r)} \big) ; \underbrace{ \mf{u}_r^{\prime}\big( t_r(p_1^{(1)}) \big) }_{= \mf{u}_1^{\prime}\big( p_1^{(1)} \big) } \Big)
\enq
where 
\beq
\wt{\mf{w}}_{\ups}^{(r)}\big( k ;  p \big) \; = \;  \mf{u}_r\circ \big( \mf{u}_r^{\prime} \big)^{-1}(k)\, - \,\mf{u}_r\circ \big( \mf{u}_r^{\prime} \big)^{-1} (p)
\, + \, \ups  \,  \op{v}_F \, \Big( \,  \big( \mf{u}_r^{\prime} \big)^{-1}(k)   -  \big( \mf{u}_r^{\prime} \big)^{-1}(p)  \Big) \;. 
\enq
Since  $\mf{u}_r^{\prime} \big( \msc{I}_{r} \big) \times \mf{u}_1^{\prime} \big( \msc{I}_{1}\big) $ is compact, $\wt{\mf{w}}_{\ups}^{(r)}$
is uniformly continuous on this set. This entails that  there exists $s_{\eta}$, with  $s_{\eta} \tend  0^+$ when $\eta \tend 0^+$ such that,
\beq
\e{uniformly} \; \e{in} \quad  \bs{p} \in \mc{D}^{(\sslash)}_{\eta} \qquad  \e{it} \,\e{holds} \qquad  \de \mf{z}_{\ups}\big( \bs{p} \big) \, = \,  \e{O}\big( s_{\eta} \big) \; . 
\label{ecriture estimation sur delta z ups}
\enq
Then, by taking $\eta$ small enough in \eqref{definition domaine particules ext}-\eqref{definition domaine particules out}, 
one gets that for any $\bs{p}\in \mc{D}^{(\sslash)}_{\eta}$
\beq
\big| \mf{z}_{\ups}(\bs{p}) \big| \, > \, \Big| \,  | \mc{Z}^{\prime}_{\ups}( p_1^{(1)} ) |  -  | \de \mf{z}_{\ups}\big( \bs{p} \big) | \, \Big|  \, > \, C_0 \;. 
\enq
This lower bound is enough so as to conclude, by derivation under the integral theorems, that $\mc{I}^{(\sslash)}(\mf{x})$ is smooth, provided that $|\mf{x}|$ is small enough. 

\vspace{2mm}

{\bf ii)}  $\big(\mc{P}_0,\mc{E}_{0} \big)$ is such that least one of the two functions $\mc{Z}_{\pm}$ vanishes on $\msc{I}_1$. 

\vspace{2mm}

First of all, by \eqref{ecriture borne inf sur distance a la courbe impulsion energie}, $\mc{Z}_{\ups}$ cannot vanish on $\Dp{}\msc{I}_1$, and
hence, by continuity, on an open neighbourhood thereof. Thus, if a zero exists, it is at a finite distance from the boundary of $\msc{I}_1$.
Furthermore, $\mc{Z}_{\pm}$ cannot share a common zero on $\e{Int}\big(\msc{I}_1 \big)$. 
Indeed, if that were the case,  then one would have $\mc{Z}_{\pm}(k)=0$ for some $k \in \e{Int}\big( \msc{I}_1 \big)$.
This would then entail that 
\beq
\left\{ \ba{ccc} 0\, = \, \mc{Z}_{+}(k)\,-\, \mc{Z}_{-}(k)& = & 2 \op{v} \,   \big[\mc{P}_0-\mc{P}(k) \big]    \vspace{2mm} \\ 
0\, = \, \mc{Z}_{+}(k)\,+\, \mc{Z}_{-}(k)& = & 2\big[\mc{E}_0-\mc{E}(k) \big]     \ea \right. 
\;. 
\label{ecriture difference impulsions et energie nulle pour zero commun Z pm} 
\enq
However, such a vanishing contradicts \eqref{ecriture borne inf sur distance a la courbe impulsion energie}.

  Denote by  $k_{\ups}\in \e{Int}\big(\msc{I}_1 \big)$ the zeroes of $\mc{Z}_{\ups}$, if these exists. 
Let $\mc{N}_{\ups}$ be an open neighbourhood of $k_{\ups}$ in $\e{Int}(\msc{I}_1)$ such that 
\beq
\ov{\mc{N}}_{\ups} \subset \e{Int}(\msc{I}_1) \qquad \e{and} \qquad 
\ov{\mc{N}}_{+}\cap \ov{\mc{N}}_{-} \,  = \, \emptyset  \;,
\enq
where the last condition only applies if both zeros exist and can be made possible since $k_+\not= k_-$ as argued earlier. Then set
\beq
\mc{K}_{\ups} \; = \;   \Big\{ \, \bs{p}\in  \msc{I}_{\e{tot}} \; : \;    p_{1}^{(1)}\in \mc{N}_{\ups} \quad \e{and} \quad  \forall (r,a)  \in \mc{M} \quad 
\big| \mf{u}_1^{\prime}\big( p_1^{(1)} \big) \, - \, \mf{u}_r^{\prime}\big( p_a^{(r)} \big)  \big| < \eta \,  \Big\} \;. 
\enq
By construction, $k_{\ups} \not \in \e{pr}_{[1,1]}\Big( \mc{D}^{(\sslash)}_{\eta} \setminus \mc{K}_{\ups} \Big)$, where $\e{pr}_{[1,1]}$ is the projection on the first coordinate. 
Thus, $\mc{Z}_{\ups}$ does not vanish on $\e{pr}_{[1,1]}\Big( \mc{D}^{(\sslash)}_{\eta} \setminus \mc{K}_{\ups} \Big)$.

Recall  that, uniformly on $\mc{D}^{(\sslash)}_{\eta}$, one has $\de \mf{z}_{\ups}\big( \bs{p} \big) = \e{O}\big( s_{\eta} \big)$, with $s_{\eta}\tend 0$ as $\eta\tend 0^{+}$.
Reducing $\eta$ if necessary, one concludes, as before, that there exists a constant $C>0$ such that  
\beq
\big| \mf{z}_{\ups}(\bs{p}) \big| > C \qquad \e{for}\, \e{any} \quad  \bs{p}  \in \mc{D}^{(\sslash)}_{\eta} \setminus \mc{K}_{\ups} \;. 
\label{ecriture condition positivite z pm frak}
\enq

It remains to deal with the behaviour of $\mf{z}_{\ups}$ inside of $\mc{K}_{\ups}$. The map 
\beq
f^{(\ups)}_{ [1,1] } \; : \;    \bs{p}   \, \mapsto \,  \big( \bs{p}_1^{(1)},\bs{p}^{(2)}, \dots,  \bs{p}^{(\ell)}, \mf{z}_{\ups}(\bs{p}) \big) 
\label{ecriture chngement vars f ups 11}
\enq
satisfies 
\beq
\det\Big[ D_{\bs{p}} f^{(\ups)}_{ [1,1] } \Big] \, = \, \pl{r=1}{\ell} (-1)^{ n_r } \, \cdot \,  \zeta_{1} \Big( \ups \op{v}  \, + \,  \mf{u}_1^{\prime}\big( p_1^{(1)} \big)   \Big)  \;. 
\enq
Since $\mf{u}_1^{\prime}(k) \not=  \pm \op{v} $ on $\e{Int}\big( \msc{I}_1 \big)$, it follows that $\det\Big[ D_{\bs{p}} f^{(\ups)}_{ [1,1] } \Big] \not=0$
on $\ov{\mc{K}}_{\ups}$. Upon reducing $\eta$ if necessary, by compactness of $\ov{\mc{K}}_{\ups}$ and smoothness of $f^{(\ups)}_{ [1,1] }$ on an open neighbourhood of $\ov{\mc{K}}_{\ups}$, 
there exists: 
\begin{itemize}
\item points $\bs{p}_{k} \in \ov{\mc{K}}_{\ups}$ , $k=1, \dots, m_{\ups}$;

 \item open neighbourhoods  $U_{\ups;k}$ of $\bs{p}_{k}$ forming a finite open cover of $\ov{\mc{K}}_{\ups}$ such that 
\beq
\ov{\mc{K}}_{\ups} \subset \cup_{k=1}^{m_{\ups}} U_{\ups; k} \subset \msc{I}_{\e{tot}}  \; ; 
\enq
\item  open sets  $V_{\ups;k}$ and constants $\de_{\ups;k}>0$ satisfying $\mf{z}_{\ups}(\bs{p}_k)\pm \de_{\ups;k} \not=0$;
\end{itemize}
such that 
\beq
f^{(\ups)}_{ [1,1] } \; : \; U_{\ups; k} \; \tend  \; V_{\ups; k}\times \intoo{ \mf{z}_{\ups}(\bs{p}_k) -\de_{\ups;k} }{ \de_{\ups;k} + \mf{z}_{\ups}(\bs{p}_k) }
\enq
is a diffeomorphism onto and that its inverse $\big( f^{(\ups)}_{ [1,1] }  \big)^{-1}$ extends smoothly to a neighbourhood of 
\beq
V_{\ups;k}\times \intoo{ \mf{z}_{\ups}(\bs{p}_k) - \de_{\ups;k}}{ \de_{\ups;k} + \mf{z}_{\ups}(\bs{p}_k) } \; .
\enq
Note that, if both zeroes exists, the  neighbourhoods $U_{\ups}=\bigcup_{k=1}^{m_{\ups}} U_{\ups;k}$ can and are chosen such that $U_{+}\cap U_{-}=\emptyset$. 
Denote by $\big\{ \vp_{\ups;k} \}_{ k=1 }^{ m_{\ups} }$ the partition of unity associated with the open cover $U_{\ups, k}$. 

\vspace{2mm}

Below, I only discuss the case when both $\mc{Z}_{+}$ and $\mc{Z}_{-}$ have a zero. All other cases are treated analogously. 

\vspace{2mm}

By using that the integrand vanishes outside of $\mc{D}_{\eta}^{(\sslash)}$, one decomposes the integral as 
\beq
\mc{I}^{(\sslash)}(\mf{x}) \, = \, \mc{I}^{(\sslash)}_{\infty}(\mf{x})\, +\,  \mc{I}^{(\sslash)}_{+}(\mf{x})\, + \, \mc{I}^{(\sslash)}_{-}(\mf{x}) \;. 
\enq
There 
\beq
\mc{I}^{(\sslash)}_{\infty}(\mf{x}) \, = \hspace{-3mm} \Int{    \mc{D}^{(\sslash)}_{\eta}   \setminus\big\{ U_{+}\cup U_{-} \big\}    }{} 
\hspace{-4mm} \msc{G}_{ \e{tot }}^{(\sslash)}( \bs{p} ) \cdot  \dd \bs{p}
\enq
and
\beq
\mc{I}_{ \ups }^{(\sslash)}(\mf{x}) \, = \, \sul{k=1}{m_{\ups}} \Int{  V_{\ups;k}   }{} \;  \dd\bs{ v } \hspace{-3mm} \Int{ \mf{z}_{\ups}(\bs{p}_k) -\de_{\ups;k}  +\mf{x} }{  \mf{z}_{\ups}(\bs{p}_k) +\de_{\ups;k}  +\mf{x}  } \hspace{-6mm} \dd u 
\;\; \wt{ \mc{G} }_{ \ups ; k}( \bs{v},u-\mf{x} ) \cdot \Xi\big(u \big) \cdot 
\big[ u \big]^{ \wt{\De}_{\ups}^{(\ups)}( \bs{v},u -\mf{x} )-1 } \; . 
%
%
%
%
%
\label{ecriture integrale I ups sslash}
\enq
Above, I have introduced
\bem
\wt{ \mc{G} }_{ \ups ;k }( \bs{v},u  ) \, = \, \vp_{\ups;k}\circ\big( f^{(\ups)}_{ [1,1]} \big)^{-1}( \bs{v},u  )  
 \cdot  \msc{G}_{\ups}\Big(\big( f^{(\ups)}_{ [1,1]} \big)^{-1}( \bs{v},u  ) , u +\mf{x}, \wt{\mf{z}}_{-\ups}( \bs{v},u  )+\mf{x} \Big)    \cdot \big| \det \Big[    D_{  ( \bs{v},u  ) }   \big( f^{(\ups)}_{ [1,1]} \big)^{-1}  \Big]\big| \\ 
\times \Xi\Big( \mf{x} +\wt{\mf{z}}_{-\ups}( \bs{v},u  )\Big)
\cdot \Big[ \mf{x} + \wt{\mf{z}}_{-\ups}( \bs{v},u  ) \Big]^{ \wt{\De}_{-\ups}^{(\ups)}( \bs{v},u  )-1 } 
\label{expression explicite fct tile ups k G}
\end{multline}
and used the shorthand notation 
\beq
\msc{G}_{+}\Big( \bs{p} , u, v \Big) \; = \; \msc{G}_{\e{tot}}^{(\sslash)}\Big( \bs{p} , u, v \Big)  \qquad , \qquad 
\msc{G}_{-}\Big( \bs{p} , u, v \Big) \; = \; \msc{G}_{\e{tot}}^{(\sslash)}\Big( \bs{p} , v, u \Big) 
\enq
as well as 
\beq
 \wt{\De}_{\pm}^{(\ups)}( \bs{v},u  ) \, = \,  \De_{\pm}\circ \big( f^{(\ups)}_{ [1,1] } \big)^{-1}( \bs{v},u  )  \qquad \e{and} \qquad
\wt{\mf{z}}_{-\ups}( \bs{v},u  )   \, = \,  \mf{z}_{-\ups} \circ \big( f^{(\ups)}_{ [1,1] } \big)^{-1}( \bs{v},u  )\; .
\enq
One can now conclude, individually for each integral.
\begin {itemize}
\item The bound \eqref{ecriture condition positivite z pm frak} along with the smoothness of the integrand allows one to conclude 
that $\mc{I}^{ (\sslash) }_{ \infty } (\mf{x})$ are smooth in $\mf{x}$ belonging to some open neighbourhood of $0$.
\item Regarding to $\mc{I}^{ (\sslash) }_{\ups}(\mf{x})$, one should focus on the contribution of each summand $k$. There are three cases to consider.
If $ \mf{z}_{\ups}(\bs{p}_k) + \de_{\ups;k}  +\mf{x}<0$, the associated integral simply vanishes for $\mf{x}$ small enough
and there is nothing more to do. If $ \mf{z}_{\ups}(\bs{p}_k) -\de_{\ups;k}  >0$, then properties $H1)-H3)$ of a smooth class function as given in Definition \ref{definition smooth class on K}, 
the fact that the Jacobian determinant in \eqref{expression explicite fct tile ups k G} never vanishes and has thus a constant sign, the smoothness of the other
building blocks and the lower bound \eqref{ecriture condition positivite z pm frak} relatively to  $\mf{z}_{-\ups}$ allow one to apply derivation under the integral theorems 
so as to conclude that the corresponding integral generates a smooth behaviour in $\mf{x}$ for $|\mf{x}|$ small enough.
Finally, if $ \mf{z}_{\ups}(\bs{p}_k) + \de_{\ups;k}   >0$  and $ \mf{z}_{\ups}(\bs{p}_k) -\de_{\ups;k}   <0$, then for $\mf{x}$ small enough, the corresponding contribution reduces to 
\beq
\mc{I}_{\ups;k} \, = \, \Int{  V_{\ups;k}   }{} \;  \dd\bs{ v } \hspace{-3mm} \Int{ 0 }{  \mf{z}_{\ups}(\bs{p}_k) +\de_{\ups;k}  +\mf{x}  } \hspace{-6mm} \dd u 
\;\; \wt{ \mc{G} }_{ \ups ; k}( \bs{v}, u - \mf{x} ) \cdot 
\big[ u \big]^{ \wt{\De}_{\ups}^{(\ups)}( \bs{v},u  - \mf{x} ) -1 } \;. 
\enq
By virtue of the decomposition for smooth class functions on $\msc{J}_{\e{tot}}$ associated with $\De_{\pm}$ and the parameter $\tau$, one has the decomposition 
\beq
\wt{ \mc{G} }_{ \ups ;k }( \bs{v},u -\mf{x} ) \, = \,\wt{ \mc{G} }_{ \ups ;k }^{(1)}( \bs{v},u-\mf{x}  ) \cdot [u]^{1-\tau} \, + \, \wt{\De}_{\pm}^{(\ups)}( \bs{v},u -\mf{x} ) \cdot \wt{ \mc{G} }_{ \ups ;k }^{(2)}( \bs{v},u -\mf{x} )
\enq
with $\wt{ \mc{G} }_{ \ups ;k }^{(a)}$ being smooth and bounded in all of their arguments. This allows one for the rewriting 
\bem
\mc{I}_{\ups;k} \, = \, \Int{  V_{\ups;k}   }{} \;  \dd\bs{ v } \hspace{-3mm} \Int{ 0 }{  \mf{z}_{\ups}(\bs{p}_k) +\de_{\ups;k}  +\mf{x}  } \hspace{-6mm} \dd u 
\Bigg\{  \wt{ \mc{G} }_{ \ups ; k}^{(1)}( \bs{v}, u - \mf{x} ) \cdot \big[ u \big]^{ \wt{\De}_{\ups}^{(\ups)}( \bs{v},u  - \mf{x} ) -\tau } 
\, - \, \Dp{u} \Big[ \wt{ \mc{G} }_{ \ups ; k}^{(2)}( \bs{v}, u - \mf{x} ) \cdot \big[ s \big]^{ \wt{\De}_{\ups}^{(\ups)}( \bs{v},u  - \mf{x} )  }  \Big]_{\mid s=u} \Bigg\} \\
+ \Int{  V_{\ups;k}   }{} \hspace{-1mm}  \dd\bs{ v }    \Bigg\{  \wt{ \mc{G} }_{ \ups ; k}^{(2)}( \bs{v}, \mf{z}_{\ups}(\bs{p}_k) +\de_{\ups;k} ) 
\cdot \big[  \mf{z}_{\ups}(\bs{p}_k) +\de_{\ups;k} + \mf{x}  \big]^{ \wt{\De}_{\ups}^{(\ups)}( \bs{v}, \mf{z}_{\ups}(\bs{p}_k) +\de_{\ups;k} )   } 
\, - \, \wt{ \mc{G} }_{ \ups ; k}^{(2)}( \bs{v},   - \mf{x} ) \cdot   \Big[  \big( s \big)^{ \wt{\De}_{\ups}^{(\ups)}( \bs{v},  - \mf{x} )  }  \Big]_{\mid s=0} \Bigg\} \;. 
\label{ecriture integrale I ups k}
\end{multline}
Here, one should note that the terms corresponding to taking the $ s \tend 0$ limit only appear if the exponent $\wt{\De}_{\ups}^{(\ups)}$ is vanishing on a set of positive measure.   
Due to the mentioned properties of the integrand, one may apply derivation under the integral theorems in the above representation so as to infer that the above integral is smooth in $\mf{x}$ small enough. 

\end{itemize}

\subsubsection*{$\bullet$ Behaviour of $\mc{I}^{(\sslash)}$ in the singular case}

This corresponds to case $\bf{b)}$ appearing in the statement of the theorem and is more tricky to deal with. Extracting the $\mf{x}\tend 0$ asymptotics demands several transformations
on the integral $\mc{I}^{ (\sslash)  }(\mf{x})$. I first start with a preliminary decomposition.

Assume that $(\mc{P}_0,\mc{E}_0)$ takes the form \eqref{ecriture forme de impulsion energie singuliere} for some $k_0 \in \e{Int}(\msc{I}_1)$. 
Then, one can recast $\mf{z}_{\ups}(\bs{p})$ as:
\beq
\mf{z}_{\ups}(\bs{p}) \; = \; - \sul{ (r,a) \in \mc{M} }{ } \zeta_{r} \mf{w}_{\ups}^{(r)} \Big( p_a^{(r)} ; t_r(k_0) \Big)  
\enq
where $\mf{w}_{\ups}^{(r)}$  is as in \eqref{definition fonction w frak}. Then,  owing to the proximity of the velocities of the integration variables 
\beq
\mf{u}_1^{\prime}( p_2^{(1)} ),\dots, \mf{u}_1^{\prime}( p_{n_1}^{(1)} )  \quad \e{to} \quad  \mf{u}_1^{\prime}( p_1^{(1)} ) 
\quad \e{and} \quad  \mf{u}_r^{\prime}( p_1^{(r)} ),\dots, \mf{u}_r^{\prime}( p_{n_r}^{(r)} )  \quad \e{to} \quad  \mf{u}_r^{\prime}\big( t_r(p_1^{(1)}) \big) \; , \quad \e{for}  \; r \in \intn{2}{\ell} \;, 
\enq
it is convenient to decompose further $\mf{z}_{\ups}(\bs{p}) $ as
\beq
\mf{z}_{\ups}(\bs{p}) \; = \; Z_{\ups}\big( p_1^{(1)} , k_0 \big) \, + \, \de \mf{z}_{\ups}\big( \bs{p} \big)
\qquad \e{with} \qquad 
Z_{\ups}(k_1,k_0) \, = \,  \mc{E}(k_0)-\mc{E}(k_1) + \ups \op{v}\,  \big[  \mc{P}(k_0) - \mc{P}(k_1)   \big] 
\enq
and
\beq
\de \mf{z}_{\ups}\big( \bs{p} \big) \; = \; - \sul{ (r,a) \in \mc{M} }{} \hspace{-2mm}   \zeta_r \, \mf{w}_{\ups}^{(r)}\Big( p_a^{(r)} ;t_r(  p_1^{ (1) } ) \Big)  \;. 
\enq
The previous estimates ensure that $\de \mf{z}_{\ups}\big( \bs{p} \big) =\e{O}\big( s_{\eta} \big)$ uniformly on $\mc{D}^{(\sslash)}_{\eta}$, \textit{c.f.} \eqref{ecriture estimation sur delta z ups}. 
In other words, $Z_{\ups}$ grasps the dominant part of $\mf{z}_{\ups}(\bs{p}) $. 
By the same arguments as earlier on, one gets that 
\beq
\ups \Dp{k}  Z_{\ups}\big( k , k_0 \big) \; = \; -\ups \cdot \mc{P}^{\prime}(k) \cdot \Big(  \mf{u}_1^{\prime}( k )\, + \,  \ups\op{v}     \Big) \; \not= \; 0
\enq
so that $k \, \mapsto \, Z_{\ups}\big( k , k_0 \big)$ is strictly monotonous on $\msc{I}_1$. One can then rely on this property so as to split, by means of an appropriate  
partition of unity, the integral into one over a domain corresponding to a neighbourhood of the point $\bs{t}(k_0)=\big(\bs{t}_1(k_0), \dots, \bs{t}_{\ell}(k_0)  \big)$ 
with $ \bs{t}_r(k_0) \, = \, (t_r(k_0),\dots, t_r(k_0) ) \in \R^{n_r}$ which will generate a non-smooth behaviour in $\mf{x}$
and an integral over its complement in $\mc{D}^{(\sslash)}_{\eta}$ which will only generate a smooth behaviour. However, the steps for achieving such a decomposition depend on the magnitude of $|\mf{u}^{\prime}_1|$
respectively to $ \op{v} $: one should distinguish between the two  possible situations which can arise due to hypothesis \eqref{ecriture propriete de vitesse defferente de vF}: 
\beq
|\mf{u}^{\prime}_1(k)|<\op{v} \quad \e{on}  \quad \e{Int}(\msc{I}_1) \qquad \e{or} \qquad  |\mf{u}^{\prime}_1(k)|>\op{v}  \quad \e{on} \quad  \e{Int}(\msc{I}_1) \; .
\enq

\subsubsection*{$\bullet$ $|\mf{u}^{\prime}_1(k)|<\op{v}$ on $\e{Int}(\msc{I}_1)$}

Since, $\de \mf{z}_{\ups}(\bs{p})=\e{O}\big( s_{\eta} \big)$ and since $k \mapsto Z_{\ups}(k,k_0)$ is strictly monotonous, the magnitude and sign on $ \mf{z}_{\ups}(\bs{p})$ will depend on whether one is close to a zero of $Z_{\ups}$
or not.

Let 
\beq
\sg \, = \, \e{sgn}\Big( \mc{P}^{\prime}(k) \,  \big( \mf{u}^{\prime}_{1}(k)+  \op{v}  \big) \Big) \, .
\label{definition signe Pprime u plus vF}
\enq
Hypothesis \eqref{ecriture hypothese non vanishing impusion macro} ensures that $\sg$  is constant on $\e{Int}(\msc{I}_1)$. Taking $\eta$ small enough, 
the fact that $Z_{\ups}$ is strictly monotonous and that $\de \mf{z}_{\ups}(\bs{p})=\e{O}\big( s_{\eta} \big)$ both ensure that
there exists $\rho_{\eta}>0$ 
such that $\rho_{\eta}\tend 0^+$ when $\eta\tend 0^+$, and   $\ga_{\eta}$ strictly increasing in $\eta$, $\ga_{\eta}\underset{\eta \tend 0^+}{\tend} 0$,  so that

\beq
\ba{ccccc} \ups \sg \mf{z}_{\ups}\big(\bs{p}\big) &<& -\rho_{\eta} & \e{if} & p_1^{(1)} > k_0 +\ga_{\eta}   \vspace{2mm} \\
\ups \sg \mf{z}_{\ups}\big(\bs{p}\big) &>&  \rho_{\eta}  & \e{if} & p_1^{(1)} < k_0  -  \ga_{\eta}  \ea 
\qquad \e{provided}\; \e{that} \qquad  \bs{p}  \, \in \,  \mc{D}^{(\sslash)}_{\eta}  \;. 
\enq

 The above ensures that, for $\eta$ small enough and $|\mf{x}|<\rho_{\eta}$, $\mf{x} \, + \, \mf{z}_{\pm}\big(\bs{p}\big)$ will have opposite signs if $|p_1^{(1)} - k_0 | \geq \ga_{\eta}$.
The presence of the Heaviside step function in the integrand then allows one to reduce the integration domain in $\mc{I}^{(\sslash)}(\mf{x})$ leading to 
\beq
\mc{I}^{(\sslash)}\big( \mf{x} \big) \, = \, \Int{ \mc{D}^{(\e{sg})}_{\eta, \ga_{\eta} } }{} \dd \bs{p}\; \msc{G}_{\e{tot}}^{(\sslash)}(\bs{p})     \;. 
\enq
Above, I have introduced
\beq
\mc{D}^{(\e{sg})}_{\eta, \ga} \; = \; \Big\{ \bs{p} \in \msc{I}_{\e{tot}} \; : \; |p_1^{(1)}-k_0|<\ga  \;\; \e{and} \;\;
\forall (r,a) \in \mc{M}  \;\; , \;\; \big| \mf{u}_1^{\prime}\big( p_1^{(1)} \big) \, - \, \mf{u}_r^{\prime}\big( p_a^{(r)} \big)  \big| < \eta \,  \Big\} \;. 
\label{definition de D sg eta ga}
\enq
Finally, let $\vp^{(\e{sg})}$ be smooth on $\pl{r=1}{\ell} \R^{n_r}$ and such that 
\beq
0 \leq \vp^{(\e{sg})} \leq 1 \quad, \quad \vp^{(\e{sg})}=1 \quad \e{on} \quad \mc{D}^{(\e{sg})}_{  \eta, \ga_{  \eta } } \quad \e{and} \quad 
 \vp^{(\e{sg})}=0 \quad \e{on} \quad \mc{D}^{(\e{out})}_{ 2\eta, \ga_{2\eta} } \;. 
\label{definition de vp sg indicator function}
\enq
where 
\beq
\mc{D}^{(\e{out})}_{\eta, \ga } \; = \; \Big\{ \bs{p} \in \msc{I}_{\e{tot}} \; : \; |p_1^{(1)}-k_0|>\ga  \;\; 
\e{and} \;\; \forall \; (r,a) \in \mc{M}  \;\; , \;\; \big| \mf{u}_1^{\prime}\big( p_1^{(1)} \big) \, - \, \mf{u}_r^{\prime}\big( p_a^{(r)} \big)  \big| < \eta \,  \Big\}  \;. 
\label{definition de D out eta ga}
\enq
Since, by construction, the integrand vanishes on $\mc{D}^{(\sslash)}_{2\eta}\setminus \mc{D}^{(\sslash)}_{\eta}$, one may recast $\mc{I}^{(\sslash)}\big( \mf{x} \big) $ in the form 
\beq
\mc{I}^{(\sslash)}\big( \mf{x} \big) \; \equiv  \; \mc{I}^{(\sslash)}_{\e{sg}}\big( \mf{x} \big) \; = \hspace{-1mm}   \Int{ \mc{D}^{(\e{sg})}_{ 2\eta, \ga_{2\eta} } }{} \hspace{-2mm} \dd \bs{p}\; \msc{G}_{\e{sg}}(\bs{p})  
\qquad  \e{with} \qquad  
 \msc{G}_{\e{sg}}(\bs{p})   \, = \,  \vp^{(\e{sg})}(\bs{p})\cdot \msc{G}_{\e{tot}}^{(\sslash)}(\bs{p})   \;. 
\label{definition G sg caligraphique}
\enq

\subsubsection*{$\bullet$ $|\mf{u}^{\prime}_1(k)|>\op{v}$ on $\e{Int}(\msc{I}_1)$}

Define  $\sg$ as in \eqref{definition signe Pprime u plus vF}. Then $\sg$ is constant on $\e{Int}(\msc{I}_1)$ by hypotheses \eqref{ecriture propriete de vitesse defferente de vF} and \eqref{ecriture hypothese non vanishing impusion macro}. 
Taking $\eta$ small enough, there exists $\rho_{\eta}>0$ and $\ga_{\eta}>0$, a strictly decreasing function of $\eta$,
such that $\ga_{\eta}\underset{\eta \tend 0^+}{\tend} 0$ so that

\beq
\ba{ccccc}   \sg \mf{z}_{\ups}\big(\bs{p}\big) &<& -\rho_{\eta} & \e{if} & p_1^{(1)} > k_0 +\ga_{\eta}  \vspace{2mm} \\
  \sg \mf{z}_{\ups}\big(\bs{p}\big) &>&  \rho_{\eta}  & \e{if} & p_1^{(1)} <  k_0  -  \ga_{\eta}  \ea 
\qquad \e{provided}\; \e{that} \qquad  \bs{p}  \, \in \,  \mc{D}^{(\sslash)}_{\eta}  \;. 
\label{ecriture condition bornage z dans le cas vitesse superieure a Fermi}
\enq
Taken this into account, it appears convenient to introduce $\vp^{(\e{sg})}$ as in \eqref{definition de vp sg indicator function}.

Then, one gets the decomposition of the integral as  $\mc{I}^{(\sslash)}\big( \mf{x} \big) \, = \, \mc{I}_{\e{sg}}^{(\sslash)}\big( \mf{x} \big)  + \mc{I}_{\e{out}}^{(\sslash)}\big( \mf{x} \big) $
where 
\beq
\mc{I}_{\e{sg}}^{(\sslash)}\big( \mf{x} \big) \, = \, \Int{ \mc{D}^{(\e{sg})}_{2\eta, \ga_{2\eta}} }{} \hspace{-3mm} \dd \bs{p}\; \msc{G}_{\e{sg}}(\bs{p})    
\qquad \e{and} \qquad 
\mc{I}_{\e{out}}^{(\sslash)}\big( \mf{x} \big) \, = \, \Int{ \mc{D}^{(\e{out})}_{   \eta, \ga_{ \eta} } }{} \hspace{-3mm} \dd \bs{p}\; \big(1-\vp^{(\e{sg})}(\bs{p}) \big)  \msc{G}_{\e{tot}}(\bs{p})    \;. 
\enq
There, $ \msc{G}_{\e{sg}}(\bs{p})$ is as appearing in \eqref{definition G sg caligraphique} and $\mc{D}^{(\e{out/sg})}_{\eta, \ga }$
have been defined in \eqref{definition de D sg eta ga} and \eqref{definition de D out eta ga}.

Due to the bound \eqref{ecriture condition bornage z dans le cas vitesse superieure a Fermi}, one has that 
\beq
|\mf{z}_{\ups}\big(\bs{p}\big)|> \rho_{  \eta } \qquad \e{on} \qquad   \mc{D}^{(\e{out})}_{ \eta, \ga_{  \eta } } \;. 
\enq
This lower bound allows one to apply derivation under the integral theorems so as to infer that $\mc{I}_{\e{out}}^{(\sslash)}\big( \mf{x} \big)$ is smooth in $\mf{x}$ belonging to a
sufficiently small neighbourhood of $0$.

\subsubsection*{$\bullet$ Simplified form of  $\mc{I}_{\e{sg}}^{(\sslash)}\big( \mf{x} \big)$}

The fact that the integration domain in $\mc{I}_{\e{sg}}^{(\sslash)}\big( \mf{x} \big)$ has been reduced  $\mc{D}^{(\e{sg})}_{2\eta, \ga_{2\eta} }$ allows one to implement a change of variables which recasts 
the integral in a simplified form. Doing so is an important step towards the analysis of its $\mf{x}\tend 0 $ behaviour.

Observe that given any $\bs{p}\in \mc{D}^{(\e{sg})}_{2\eta, \ga_{2\eta} }$, for any $(r,a)\in \mc{M}$, it holds 
\beq
\big| \mf{u}^{\prime}_{r}(p_a^{(r)}) \, - \, \mf{u}^{\prime}_{r}( t_r(k_0) )  \big|
\, \leq \, \big| \mf{u}^{\prime}_{1}( p_1^{(1)} ) \, - \, \mf{u}^{\prime}_{1}(k_0 )  \big|  \, + \,  \big| \mf{u}^{\prime}_{r}(p_a^{(r)}) \, - \, \mf{u}^{\prime}_{1}( p_1^{(1)} )  \big|
\, < \, 2\eta \, + \, \ga_{2\eta} \cdot \norm{ \mf{u}_1^{\prime} }_{ L^{\infty}(\msc{J}_1) } \;. 
\label{ecriture bornes sur distance entre pts dans image via u prime r}
\enq
Since $k_0 \in \e{Int}(\msc{J}_1)$,  $t_r(k_0) \in \e{Int}(\msc{J}_r)$ and by \eqref{ecriture bornes sur distance entre pts dans image via u prime r}, upon reducing $\eta>0$ if need be, 
it follows that there exists $\veps>0$ such that, for any $\bs{p} \in \mc{D}^{(\e{sg})}_{2\eta, \ga_{2\eta} }$,
\beq
\e{d}\Big(  \mf{u}^{\prime}_r\big( p_a^{(r)} \big) , \Dp{} \mf{u}^{\prime}_r\big( \msc{I}_r \big) \Big) \, > \, \veps \; , 
\enq
for the canonical distance $\e{d}$ between points and subsets of $\R$.  This ensures that there exists a compact $K_r \subset \e{Int}(\msc{J}_r)$ containing an open neighbourhood of $t_r(k_0)$,
such that both $p_a^{(r)}, t_r(p_1^{(1)}) \in K_r$
uniformly in $\bs{p}\in \mc{D}^{(\e{sg})}_{2\eta, \ga_{2\eta} }$. Since $\mf{u}_{r}^{\prime}$ is a smooth diffeomorphism on an open neighbourhood of $K_r$, there exist constants $c_r,C_r>0$ such that 
\beq
c_r \cdot \big| \mf{u}_r^{\prime}\big( x\big) \, - \, \mf{u}_r^{\prime}\big( y \big)  \big|  \; \leq \,  | x  - y  | \; \leq \; C_{r}   \cdot \big| \mf{u}_r^{\prime}\big(x \big) \, - \, \mf{u}_r^{\prime}\big( y \big)  \big| 
\qquad \e{for} \; \e{any} \qquad x,y \in K_r\;. 
\enq
Recall that $k_0 \in \e{Int}(\msc{J}_1)$ so that $t_r(k_0) \in \e{Int}(\msc{J}_r)$. Thus, upon taking $\eta$ small enough, the strict monotonicity of $\mf{u}^{\prime}_r$  on $\msc{J}_r$
and the above bounds ensure that, for any $\bs{p}\in \mc{D}^{(\e{sg})}_{2\eta, \ga_{2\eta} }$, 
\beq
\big| t_r(k_0) \, - \, p_a^{(r)}  \big| \, \leq \, C_r \Big\{  \big|  \mf{u}^{\prime}_{r}(p_a^{(r)}) \, - \, \mf{u}^{\prime}_{1}( p_1^{(1)} )  \big|  \, + \,  \big| \mf{u}^{\prime}_{1}(k_0) \, - \, \mf{u}^{\prime}_{1}( p_1^{(1)} )  \big|  \Big\}
\, \leq \,2 C_r \eta \, + \, \f{ C_r }{ c_1 } \ga_{2\eta} \, \equiv \, \eta_{r} \;. 
\label{definition etar}
\enq
Therefore, upon denoting $B_{\eps}(x_0)=\{ x \in \R \, : \, |x-x_0| < \eps \}$ the open ball in $\R$ of radius $\eps$ centred at $x_0$, one gets 
\beq
\mc{D}^{(\e{sg})}_{2\eta, \ga_{2\eta}} \subset B_{\ga_{2\eta}}(k_0) \times \pl{r=1}{\ell} \Big( B_{\eta_r}\big( t_r(k_0) \big) \Big)^{n_r-\de_{r,1} } \;, 
\label{ecriture inclusion D sg dans boule multi-rayon O eta}
\enq
with $\eta_r$ as given in \eqref{definition etar}.

Define auxiliary functions  on $ \pl{r=1}{\ell} \R^{n_r}$
\beq
\tilde{z}_{\ups}(\bs{x}) \; = \;- \sul{ (r,a)\in \mc{M} }{} \zeta_{r} \bigg\{ \mf{h}_r\big( x_a^{(r)} \big) +\ups \op{v} x_a^{(r)}  \bigg\} \qquad  \e{and}  \qquad 
 \left\{ \ba{ccc}  \mf{h}_r\big( x \big)  & = &   \mf{u}_{r}^{\prime\prime}\big( t_r(k_0) \big) \f{  x^2  }{ 2   } \, + \,   \mf{u}_{1}^{\prime}(k_0)  \,  x  \vspace{1mm}  \\ 
	  t_r^{(0)}(x)  & =& \f{ \mf{u}_{1}^{\prime\prime}\big( t_1(k_0) \big) }{ \mf{u}_{r}^{\prime\prime}\big( t_r(k_0) \big) } \, x    \ea \right.  \;, 
\enq
along with the domain 
\beq
\mc{D}^{(\e{eff})}_{ \eta  } \, = \, \Bigg\{   \bs{x} \in \pl{r=1}{\ell} \R^{n_{r}} \; : \; |x_1^{(1)}| \, \leq \, C \eta  \; , \; 
\forall (r,a )\in \mc{M} \; : \; \big| t_r^{(0)}(x_1^{(1)})-x_a^{(r)}  \big| \, \leq  \,  \f{ \eta  }{  \big| \mf{u}_{r}^{\prime\prime}\big( t_r(k_0) \big) \big|  }  \Bigg\}   \;. 
\enq
By the above discussion, $\mc{D}^{(\e{eff})}_{ \eta  }$ is an open neighbourhood of the origin. 

Then, by virtue of Proposition \ref{Proposition trivialisation locale fct zups}, 
there exists $\mf{x}_0>0$ and $\eta^{\prime}>0$ such that there exists:
\begin{itemize}
 
 \item smooth functions $\mf{f}_{\ups}$ on $\intoo{ -\mf{x}_0 }{ \mf{x}_0 } \, \times \,  \mc{D}^{(\e{eff})}_{\eta^{\prime}}$  satisfying $\mf{f}_{\ups}(\mf{x}; \bs{x})\, = \, 1 + \e{O}\Big( \norm{\bs{x}}+|\mf{x}|\Big)$,
 
\item a smooth diffeomorphism $\Psi_{\mf{x}}: \mc{D}^{(\e{eff})}_{\eta^{\prime}} \;  \tend  \;  \Psi_{\mf{x}}\Big( \mc{D}^{(\e{eff})}_{\eta^{\prime}} \Big)$ satisfying $D_{\bs{0}} \Psi_{\mf{x}} \, = \, \e{id} + \mf{x} \op{N}_{\Psi}$ with 
$\norm{ \op{N}_{\Psi} } \leq C$, for some $\mf{x}$-independent $C>0$, 
 
\end{itemize}
such that 
\beq
\mf{x}+\mf{z}_{\ups} \circ \Psi_{\mf{x}}(\bs{x})  \; = \; \mf{f}_{\ups}(\mf{x};\bs{x}) \cdot \Big(\mf{x} + \tilde{z}_{\ups}(\bs{x}) \Big) \;,
\enq
and, for $|\mf{x}|<\mf{x}_0$,  $\Psi_{\mf{x}}\Big( \mc{D}^{(\e{eff})}_{\eta^{\prime}} \Big) \subset \msc{J}_{\e{tot}}$ contains a $\mf{x}$-independent open neighbourhood of $\bs{t}(k_0) \in \msc{J}_{\e{tot}}$ with $\bs{t}(k_0)$
as given in \eqref{ecriture definition vecteur tk0}. Finally, 
\beq
(\mf{x},\bs{x}) \; \mapsto \; \Psi_{\mf{x}}(\bs{x}) 
\enq
is smooth on $\intoo{ -\mf{x}_0 }{ \mf{x}_0 } \times \mc{D}^{(\e{eff})}_{\eta^{\prime}}$.

Furthermore, by virtue of Proposition \ref{Proposition factorisation jolie en zeroes de tilde z ups}, there exists an invertible linear map $\op{M}$ on $\pl{r=1}{\ell} \R^{n_r}$
and integers  $m_{\pm} \in \mathbb{N}$ satisfying $m_+ + m_- + 1 \, = \, \sum_{r=1}^{\ell}n_r$ such that 
\beq
\tilde{z}_{\ups}\big(   \op{M} \bs{x}  \big) + \mf{x} \, = \,     P_{\ups}(\bs{x})   \qquad \e{with}  \qquad  \bs{x}=\big(y, \bs{z}^{(+)},  \bs{z}^{(-)} \big) \in \R\times  \R^{m_+}\! \times \R^{m_-} 
\label{ecriture effet transformation matrice M}
\enq
 and 
\beq
P_{\ups}(\bs{x}) \; = \; -  \f{  \mf{u}^{\prime\prime}_1(k_0)   }{2   \mc{P}^{\prime}(k_0)  } y^2  \,  - \, \big( \mf{u}^{\prime}_1(k_0) + \ups \op{v} \big) y  \, + \,  \vp_{\mf{x} }(\bs{z})    
\enq
 in which 
\beq
 \vp_{\mf{x} }(\bs{z}) \; = \; \mf{x}  \, + \, \sul{s=1}{m_+} \big( z_s^{(+)} \big)^2 \, - \,  \sul{s=1}{m_-} \big( z_s^{(-)} \big)^2  \quad \e{with} \quad \bs{z}=\big( \bs{z}^{(+)}, \bs{z}^{(-)} \big) \;. 
\label{definition varphi de z et kappa}
\enq

 \vspace{3mm}

To proceed further,  it is convenient to introduce the auxiliary function 
 $F$ which is defined around the origin through its series expansion 
\beq
F(x) \, = \, 1 \, + \,  \f{1}{2 \op{v} }  \sul{k \geq 1 }{} d_k x^k \f{ \big[ \op{v}+\mf{u}_{1}^{\prime}(k_0) \big]^{2k+1} \, - \, \big[  \mf{u}_{1}^{\prime}(k_0) - \op{v}\big]^{2k+1}   }
{ \big[ \op{v}^2 - \big(\mf{u}_{1}^{\prime}(k_0) \big)^2  \big]^{2k}  } \;. 
\label{definition fonction G pour difference des racines de P ups}
\enq
The sequence $d_k$ appearing above can be read-off from its generating series 
\beq
 2 \f{1-\sqrt{1-x}}{x} \, = \, 1+\sul{k\geq 1}{} d_k x^k \;. 
\enq
Then, for $\de_0>0$ and small enough, one sets 
\beq
\mu(\bs{z}) \; = \; \de_0 \,  \wt{F}(  \vp_{\mf{x} }(\bs{z})  ) \qquad \e{with} \qquad 
\wt{F}(x) \, = \,  2   \op{v} \cdot \f{   F\Big(    \tfrac{ - 2 \mf{u}^{\prime\prime}_1(k_0)   }{   \mc{P}^{\prime}(k_0)  }  x \Big)    }{  \big|  \big(\mf{u}_{1}^{\prime}(k_0) \big)^2  \, - \, \op{v}^2 \big|  } \;. 
\label{definition fonction mu de z}
\enq
 Observe that for any
\beq
\bs{z}\, = \, \big( \bs{z}^{(+)}, \bs{z}^{(-)} \big)   \in \big[-\de_{+}^{\tf{1}{2}}  ;  \de_{+}^{\tf{1}{2}} \big]^{m_+} \times  \big[-\de_{-}^{\tf{1}{2}}  ;  \de_{-}^{\tf{1}{2}} \big]^{m_-}
\enq
it holds $ \vp_{\mf{x} }(\bs{z}) \, = \, \e{O}\big( \de_+ + \de_- + |\mf{x}| \big)$. Thus,  provided that $\de_{+}, \de_{-}>0$ and $|\mf{x}|$ are all taken small enough,
$\mu(\bs{z})$ is well-defined for such $\bs{z}$'s and, owing to $F(0)=1$, it holds 
\beq
   \f{     \op{v} \de_0 }{  \big|  \big(\mf{u}_{1}^{\prime}(k_0) \big)^2  \, - \, \op{v}^2 \big|   } \, \leq  \, \mu(\bs{z}) \,  \leq \,    \f{ 4  \op{v} \de_0 }{  \big|  \big(\mf{u}_{1}^{\prime}(k_0) \big)^2  \, - \, \op{v}^2 \big|   }   \;. 
\label{ecriture bornes sur mu de z}
\enq

Enough has now been introduced so as to allow me to define the domain
\beq
  \mc{D} \; = \; \bigg\{  (y, \bs{z} ) \; : \;   \bs{z} \, = \, \big( \bs{z}^{(+)},\bs{z}^{(-)} \big) \in  \big[-\de_{+}^{\tf{1}{2}}  ;  \de_{+}^{\tf{1}{2}} \big]^{m_+} \times  \big[-\de_{-}^{\tf{1}{2}}  ;  \de_{-}^{\tf{1}{2}} \big]^{m_-} 
\;\; \e{and} \; \; y \in \intff{ - \mu(\bs{z}) }{ \mu(\bs{z}) }   \bigg\}  \;. 
\label{definition domaine D pour reduction}
\enq

Then, the inclusion \eqref{ecriture inclusion D sg dans boule multi-rayon O eta} ensures that, given $\de_0>0, \de_{\pm}>0$ small enough, and upon diminishing, if necessary, the parameter $\eta>0$, it holds 
\beq
\mc{D}^{(\e{sg})}_{2 \eta, \ga_{2 \eta}} \; \subset \; \Psi_{\mf{x}}\big( \op{M} \cdot \mc{D} \big) \; \subset \; \Psi_{\mf{x}}\Big( \mc{D}_{\eta^{\prime}}^{(\e{eff})} \Big) \;, 
\enq
 where $\op{M}$ is as in \eqref{ecriture effet transformation matrice M}, this uniformly in $|\mf{x}|<\mf{x}_0$. 

Thus, for such a choice of parameters at play, upon using that the integrand vanishes away from $\mc{D}^{(\e{sg})}_{2 \eta, \ga_{2 \eta}}$ one 
one gets that 
\beq
\mc{I}_{\e{sg}}^{(\sslash)}(\mf{x}) \; = \;  
\hspace{-2mm} \Int{ \Psi_{\op{M}}\big( \mc{D} \big) }{} \hspace{-3mm} \dd \bs{p}\; \msc{G}_{\e{sg}}(\bs{p})   \qquad \e{with} \qquad \Psi_{\op{M}} = \Psi_{\mf{x}} \circ \op{M}\;. 
\enq

The change of variables 
\beq
\bs{p}=\Psi_{\op{M}}(\bs{x}) \qquad \e{with} \qquad  \bs{x} = \big( y, \bs{z}^{(+)},  \bs{z}^{(-)} \big) \in \R \times  \R^{m_+} \times \R^{m_-}
\enq
recasts the integral in the form 
\beq
\mc{I}_{\e{sg}}^{(\sslash)}(\mf{x}) \; = \;   \Int{ \mc{D}  }{}   \dd \bs{x}\; \msc{F}(\bs{x}) \pl{\ups=\pm}{} \bigg\{  \Xi\Big( P_{\ups}(\bs{x}) \Big) \cdot \big[ P_{\ups}(\bs{x}) \big]^{ \mf{d}_{\ups}(\bs{x}) -1 }  \bigg\}
\label{reduction finale de la partie singuliere vers integrale modele}
\enq
where  $\mf{d}_{\ups}(\bs{x}) \, = \, \De_{\ups}\circ\Psi_{\op{M}}(\bs{x})$  and 
\bem
\msc{F}(\bs{x}) \, = \, V\circ \Psi_{\op{M}}(\bs{x}) \cdot \msc{G}\Big( \Psi_{\op{M}}(\bs{x}) \, , \, \mf{f}_{+}(\mf{x};\op{M}\cdot \bs{x}) P_{+}(\bs{x})   \, , \, \mf{f}_{-}(\mf{x};\op{M}\cdot \bs{x}) P_{-}(\bs{x})   \Big) \\
\times  \Big(\vp^{(\sslash)} \vp^{(\e{sg})} \Big)\Big( \Psi_{\op{M}}(\bs{x}) \Big) \cdot \big| \det D_{\bs{x}} \Psi_{\op{M}} \big| 
\cdot \pl{ \ups = \pm }{}  \Big\{  \mf{f}_{\ups}(\mf{x};\op{M}\cdot \bs{x}) \Big\}^{ \mf{d}_{\ups}(\bs{x}) -1 }  \;. 
\end{multline}
Also, I remind that $V$ has been defined in \eqref{defintion V Vdm product squared}.

\subsubsection*{$\bullet$ Properties of the polynomials $P_{\ups}(\bs{x})$}

One may put the integral $\mc{I}_{\e{sg}}^{(\sslash)}\big( \mf{x} \big)$ into a canonical form by focalising more on the structure of the polynomial $P_{\ups}(\bs{x})$. It is easy to see that, uniformly in 
$\bs{x} \in \mc{D}$ with $\de_{\pm}$ small enough and $\mc{D}$ as defined through \eqref{definition domaine D pour reduction}, it admits the factorisation 
\beq
P_{\ups}( \bs{x}) \; = \; -   \f{  \mf{u}^{\prime\prime}_1(k_0)   }{2   \mc{P}^{\prime}(k_0)  } \cdot \Big( y-y_+^{(\ups)}  \Big) \cdot \Big( y-y_-^{(\ups)}  \Big) \;. 
\enq
Upon setting $\sg_{\ups}=\e{sgn}\big( \mf{u}^{\prime}_1(k_0) + \ups \op{v} \big)$, one has that 
\beq
y_{- \sg_{\ups}}^{(\ups)}  \; = \; \f{ \vp_{ \mf{x} }(\bs{z}) }{ \mf{u}^{\prime}_1(k_0) + \ups \op{v}   } 
\, U\Bigg(  -  \f{  2  \mf{u}^{\prime\prime}_1(k_0)  \, \vp_{\mf{x}}(\bs{z})  }{  \mc{P}^{\prime}(k_0)  \, \big[  \mf{u}^{\prime}_1(k_0) + \ups \op{v}  \big]^2 } \Bigg) \;, 
\quad U(x) \, = \, 2 \f{  1 \, - \, \sqrt{ 1 - x } }{ x } \;, 
\label{ecriture racine y ups moins sg ups}
\enq
where  I remind that $\bs{z}=\big( \bs{z}^{(+)}, \bs{z}^{(-)} \big)$  and $ \vp_{\mf{x} }(\bs{z}) $ is as introduced in \eqref{definition varphi de z et kappa}. Also, it holds
\beqa
y_{ \sg_{\ups}}^{(\ups)}  & = &  \f{-  \mc{P}^{\prime}(k_0)  }{   \mf{u}^{\prime\prime}_1(k_0)   } \Big( \mf{u}^{\prime}_1(k_0) + \ups \op{v} \Big)
\cdot  \Bigg( 1 \, + \, \sqrt{ 1+ \tfrac{  2  \mf{u}^{\prime\prime}_1(k_0)  \vp_{\mf{x}}(\bs{z})  }{  \mc{P}^{\prime}(k_0)  [  \mf{u}^{\prime}_1(k_0) + \ups \op{v}  ]^2 }  }  \Bigg)  \label{ecriture racine y ups sg ups}\\
 & = &  \f{- 2 \mc{P}^{\prime}(k_0)  }{   \mf{u}^{\prime\prime}_1(k_0)   } \Big( \mf{u}^{\prime}_1(k_0) + \ups \op{v} \Big) \cdot  \Big( 1 \, + \, \e{O}\big( \de_+ + \de_-  +|\mf{x}|\big) \Big) \;. 
\nonumber
\eeqa
Note that the expressions \eqref{ecriture racine y ups moins sg ups} and \eqref{ecriture racine y ups sg ups} entail that $y_{\pm}^{(\ups)}$ are functions of the 
variable $\bs{z}$ only through the combination $ \vp_{\mf{x} }(\bs{z})$. In the following, unless it will be necessary, this $\bs{z}$-dependence  of $y_{\pm}^{(\ups)}$
will be kept implicit.

Let 
\beq
\mf{s}= \e{sgn}\Big( -   \tfrac{  \mf{u}^{\prime\prime}_1(k_0)   }{  \mc{P}^{\prime}(k_0)  }    \Big) \;. 
\label{definition sign mf s}
\enq
One has that $ \sg_{\ups} \, \mf{s}  \, y_{\sg_{\ups}}^{(\ups)} \,  >  \,  \sg_{\ups}  \, \mf{s}  \, y_{-\sg_{\ups}}^{(\ups)} $ so that 
\beq
\ba{c|c|c|c|c}
				&  \sg_{\ups} \mf{s} >0     						&  \sg_{\ups} \mf{s}<0  \\ \hline   
\mf{s} P_{\ups}(\bs{x}) < 0  &   \Big] y_{-\sg_{\ups}}^{(\ups)} ; y_{\sg_{\ups}}^{(\ups)} \Big[^{}   &     \Big] y_{\sg_{\ups}}^{(\ups)} ; y_{-\sg_{\ups}}^{(\ups)} \Big[    \\ \hline  
\mf{s} P_{\ups}(\bs{x}) > 0  &    \R \setminus \Big[ y_{-\sg_{\ups}}^{(\ups)} ; y_{\sg_{\ups}}^{(\ups)} \Big]   &      \R \setminus \Big[ y_{\sg_{\ups}}^{(\ups)} ; y_{-\sg_{\ups}}^{(\ups)} \Big]    \\ \hline 
\ea
\label{ecriture tableau signe de P ups}
\enq
gives the domains, in the $y$ variable and for fixed $\bs{z}$, of positivity and negativity of the polynomials  $P_{\ups}(\bs{x})$.

To proceed further,  one should distinguish between the two cases where $|\mf{u}_1^{\prime}(k_0)|<\op{v}$ and $|\mf{u}_1^{\prime}(k_0)|>\op{v}$, since their treatment slightly differs.

\subsubsection*{$\bullet$ Joint positivity interval in the $|\mf{u}_1^{\prime}(k_0)|<\op{v}$ regime}

 \hspace{4mm}  $\circledast$ If $\mf{s}>0$, then by \eqref{ecriture tableau signe de P ups}, one will have 
\beqa
P_+(\bs{x}) > 0  \qquad \e{on} \qquad \big]-\infty;  y_{-}^{(+)} \,  \big[ \cup \big] \, y_{+}^{(+)}; +\infty \big[  & \quad \e{with}  \quad & 
\left\{ \ba{c} y_{+}^{(+)} >0  \\  y_{-}^{(+)} \, = \, \e{O}\Big(\de_+ + \de_- +|\mf{x}| \Big) \ea \right.    \\
P_-(\bs{x}) > 0  \qquad \e{on} \qquad \big]-\infty;  y_{-}^{(-)} \,  \big[ \cup \big] \, y_{+}^{(-)}; +\infty \big[  & \e{with} &  \left\{ \ba{c} y_{-}^{(-)} <0  \\  y_{+}^{(-)} \, = \, \e{O}\Big(\de_+ + \de_- +|\mf{x}| \Big) \ea \right.  \;. 
\eeqa
Thus, both polynomials $P_{\pm}(\bs{x})$ will be positive on the union of intervals
\beq
\big]-\infty;  y_{-}^{(-)} \,  \big[ \cup   \big] \,  y_{+}^{(-)} \, ;  \, y_{-}^{(+)} \,  \big[ \cup \big] \, y_{+}^{(+)}; +\infty \big[ 
\enq
where the central interval is present only if the subsidiary consition $ y_{-}^{(+)}  \, - \,   y_{+}^{(-)} \, > \, 0 $ holds. In fact, this is the sole interval that will be included in the 
$y$ integration domain present in $\mc{D}$ \eqref{definition domaine D pour reduction},  \textit{viz}. $\intff{ -\mu(\bs{z}) }{\mu(\bs{z})}$, where $\mu(\bs{z})$ 
given in \eqref{definition fonction mu de z} is fixed upon choosing the $\bs{z}$ variables and is small enough as in  \eqref{ecriture bornes sur mu de z}.

Then, using the local positivity on this interval of the various building blocks present in the factorisation of the polynomials $P_{\ups}$, one gets that, for $\bs{x} \in \mc{D}$,  
\bem
\pl{\ups=\pm}{} \bigg\{  \Xi\Big( P_{\ups}(\bs{x}) \Big) \cdot \big[ P_{\ups}(\bs{x}) \big]^{ \mf{d}_{\ups}(\bs{x}) -1 }  \bigg\} \; = \; \Xi\Big( \, y_{-}^{(+)}  \, - \,   y_{+}^{(-)} \Big)
\cdot \bs{1}_{  \big[ \,  y_{+}^{(-)} \, ;  \, y_{-}^{(+)} \,  \big]  }(y)   \\ 
\times \pl{\ups = \pm }{} \bigg\{    \tfrac{ - \ups \mf{u}^{\prime\prime}_1(k_0)   }{2   \mc{P}^{\prime}(k_0)  }  \cdot  \big(\,  y_{\ups}^{(\ups)}- y  \big)  \bigg\}^{ \mf{d}_{\ups}(\bs{x}) -1 } 
\cdot \pl{\ups = \pm }{} \bigg\{ \ups \big( \, y_{-\ups}^{(\ups)}- y  \big)  \bigg\}^{ \mf{d}_{\ups}(\bs{x}) -1 } \;. 
\label{factorisation partie petite et grande factorisation polynome cas mod u lower v}
\end{multline}

\par $\circledast$ If $\mf{s}<0$, then by \eqref{ecriture tableau signe de P ups}, one will have 
\beqa
P_+(\bs{x}) > 0  \qquad \e{on} \qquad \big] y_{+}^{(+)} ;  y_{-}^{(+)} \,  \big[    & \quad \e{with}  \quad &  \left\{ \ba{c} y_{+}^{(+)} <0  \\  y_{-}^{(+)} \, = \, \e{O}\Big(\de_+ + \de_- +|\mf{x}| \Big) \ea \right. \;,  \\
P_-(\bs{x}) > 0  \qquad \e{on} \qquad \big] y_{+}^{(-)} ;  y_{-}^{(-)} \,  \big[   & \e{with} &  \left\{ \ba{c} y_{-}^{(-)} >0  \\  y_{+}^{(-)} \, = \, \e{O}\Big(\de_+ + \de_- +|\mf{x}| \Big) \ea \right.  \; .
\eeqa
Thus, both polynomials will be simultaneously positive if any only if $ y_{-}^{(+)}  \, - \,   y_{+}^{(-)} \, > \, 0 $ and then the interval of joint positivity is 
\beq
   \big] \,  y_{+}^{(-)} \, ;  \, y_{-}^{(+)} \,  \big[  \;. 
\enq
Upon using the local positivity, on this interval, of the various building blocks present in the factorisation of the polynomials $P_{\ups}$, one gets 
that the factorisation \eqref{factorisation partie petite et grande factorisation polynome cas mod u lower v} also holds  in the present case.

\subsubsection*{$\bullet$ Joint positivity interval in the $|\mf{u}_1^{\prime}(k_0)|>\op{v}$ regime}

In this regime, one has that 
\beq
\sg_{\ups} \, = \, \e{sgn}\big( \mf{u}^{\prime}_{1}(k_0)+\ups \op{v} \big) = \vsg \;, 
\label{definition signe varsigma}
\enq
\textit{i.e.}  $\sg_{\ups}$ does not depend on $\ups\in \{\pm 1\}$. 

\par $\circledast$ If $\mf{s}>0$, then by \eqref{ecriture tableau signe de P ups}, one will have 
\beq
P_{\pm}(\bs{x}) > 0  \qquad \e{on} \qquad \big]-\infty;  y_{-}^{(\pm)} \,  \big[ \cup \big] \, y_{+}^{(\pm)}; +\infty \big[   
\enq
where, for some $c>0$ 
\beq
  \ba{cccc} y_{+}^{(\pm)} >c> 0  &, \quad   y_{-}^{(\pm)} \, = \, \e{O}\Big(\de_+ + \de_- +|\mf{x}| \Big)    & \e{if} & \vsg = + \vspace{2mm} \;  \\ 
  y_{-}^{(\pm)} <-c<0  &,  \quad  y_{+}^{(\pm)} \, = \, \e{O}\Big(\de_+ + \de_- +|\mf{x}| \Big)    & \e{if} &  \vsg = -  \ea \;. 
\enq

Thus, the polynomials $P_{\pm}(\bs{x})$ will be simultaneously positive on the union of intervals
\beq
\Big]-\infty;  \e{min}\Big(y_{-}^{(+)}, y_{-}^{(-)} \Big)  \,  \Big[  \cup \Big] \,  \e{max}\Big(y_{+}^{(+)}, y_{+}^{(-)} \Big)  ; +\infty \Big[  \;.
\enq
Indeed, this is a consequence of the fact that the roots $y^{(\pm)}_{\vsg}$ have both "large" absolute value -in respect to $\mu(\bs{z})$ and this uniformly in $\bs{z}$ by virtue of \eqref{ecriture bornes sur mu de z}-, 
whereas the roots $y^{(\pm)}_{-\vsg}$ are both close to the origin. 
Thus, taken that the function $\mu(\bs{z})$ \eqref{definition fonction mu de z} delimiting the $y$ integration domain in $\mc{D}$ \eqref{definition domaine D pour reduction} is small enough \eqref{ecriture bornes sur mu de z}, 
for any fixed $\bs{z}$, the interval $\intff{ - \mu(\bs{z}) }{ \mu(\bs{z}) }$ defining the $y$-integration in \eqref{reduction finale de la partie singuliere vers integrale modele} will reduce to 
\beq
J_{\vsg}(\bs{z}) \; = \; \vsg \Big] - \vsg \mu(\bs{z}) \, ; \, \vsg \e{min}\big(\vsg \, y_{-\vsg}^{(+)} , \vsg\, y_{-\vsg}^{(-)} \big) \Big[
\label{definition intervalle J vsg}
\enq
in which the prefactor $\vsg$ indicates the orientation of the interval. 

Then, using the local positivity, on this interval, of the various building blocks present in the factorisation of the polynomials $P_{\ups}$, one gets that 
\beq
\pl{\ups=\pm}{} \bigg\{  \Xi\Big( P_{\ups}(\bs{x}) \Big) \cdot \big[ P_{\ups}(\bs{x}) \big]^{ \mf{d}_{\ups}(\bs{x}) -1 }  \bigg\} \; = \; 
 \bs{1}_{  J_{\vsg}(\bs{z})  }(y)    
\cdot \pl{\ups = \pm }{} \bigg\{    \tfrac{ - \vsg \mf{u}^{\prime\prime}_1(k_0)   }{2   \mc{P}^{\prime}(k_0)  }     \,  \big(\,  y_{\vsg}^{(\ups)}- y  \big)  \bigg\}^{ \mf{d}_{\ups}(\bs{x}) -1 } 
\cdot \pl{\ups = \pm }{} \bigg\{ \vsg \big( \, y_{-\vsg}^{(\ups)}- y  \big)  \bigg\}^{ \mf{d}_{\ups}(\bs{x}) -1 } \;. 
\label{factorisation partie petite et grande factorisation polynome cas mod u greater v}
\enq

\par $\circledast$ If $\mf{s}<0$, then by \eqref{ecriture tableau signe de P ups}, one will have 
\beqa
P_{\ups}(\bs{x}) > 0  \qquad \e{on} \qquad \big] y_{+}^{(\ups)} ;  y_{-}^{(\ups)} \,  \big[    
\eeqa
where, for some $c>0$,  
\beq
  \ba{cccc} y_{+}^{(\ups)} < -c < 0  &, \quad   y_{-}^{(\ups)} \, = \, \e{O}\Big(\de_+ + \de_- +|\mf{x}| \Big)    & \e{if} & \vsg  = + \vspace{2mm} \\ 
  y_{-}^{(\ups)} > c >0  &,  \quad  y_{+}^{(\ups)} \, = \, \e{O}\Big(\de_+ + \de_- +|\mf{x}| \Big)    & \e{if} &   \vsg  = -  \ea \;. 
\enq
Thus, both polynomials will be simultaneously positive only on the interval
\beq
   \Big] \,  \e{max}\big(y_{+}^{(+)}, y_{+}^{(-)} \big)  \, ;  \, \e{min}\big(y_{-}^{(+)}, y_{-}^{(-)} \big)\,  \Big[  \;. 
\enq
Since $\mu(\bs{z})$ is taken small enough, \textit{c.f.} \eqref{ecriture bornes sur mu de z} and in particular such that $ 0 < \mu(\bs{z}) < |y_{\vsg}^{(\ups)}| $, $\ups=\pm$, one will have that 
the presence of Heaviside functions of $P_{\ups}(\bs{x})$ will, effectively, result in a reduction of the $y$-integration domain $\intff{ - \mu(\bs{z}) }{ \mu(\bs{z}) }$ in
$\mc{D}$ \eqref{definition domaine D pour reduction} to the interval $J_{\vsg}(\bs{z})$ already introduced in \eqref{definition intervalle J vsg}.  

Furthermore, using the local positivity on this interval of the various building blocks present in the factorisation of the polynomials $P_{\ups}$,  
 the factorisation \eqref{factorisation partie petite et grande factorisation polynome cas mod u greater v} also holds in the present case.

\subsubsection*{$\bullet$ Canonical form of $\mc{I}_{\e{sg}}^{(\sslash)}(\mf{x}) $  in the $|\mf{u}_1^{\prime}(k_0)|<\op{v}$ regime}

The factorisation \eqref{factorisation partie petite et grande factorisation polynome cas mod u lower v} entails that, irrespectively of the value of $\mf{s}$ introduced in \eqref{definition sign mf s}, 
$\mc{I}_{\e{sg}}^{(\sslash)}(\mf{x}) $ as given in  \eqref{reduction finale de la partie singuliere vers integrale modele} now takes the form 
\beq
\mc{I}_{\e{sg}}^{(\sslash)}(\mf{x}) \; = \;  \pl{ \ups = \pm }{} \bigg\{  \Int{ -\sqrt{\de_{\ups}}  }{ \sqrt{\de_{\ups}} } \hspace{-2mm} \dd^{m_{\ups}}z^{(\ups)}  \bigg\}  
\Int{  y_{+}^{(-)}  }{ y_{-}^{(+)} } \dd y  \cdot \Xi\Big( y_{-}^{(+)}  \, - \,   y_{+}^{(-)} \Big) \cdot 
\, \msc{F}^{(1)}(\bs{x})  \cdot \pl{\ups = \pm }{} \bigg\{ \ups \big(\,  y_{-\ups}^{(\ups)}- y  \big)  \bigg\}^{ \mf{d}_{\ups}(\bs{x}) -1 } \;, 
\label{ecriture integrale I sg parallele cas u lower v 1ere reduction}
\enq
where  $\bs{x}$ is parameterised in terms of the integration variables as in \eqref{ecriture effet transformation matrice M} and, for short, 
\beq
\msc{F}^{(1)}(\bs{x}) \, = \, \msc{F} (\bs{x}) \cdot \pl{\ups = \pm }{} \bigg\{    \tfrac{ - \ups  \mf{u}^{\prime\prime}_1(k_0)   }{2   \mc{P}^{\prime}(k_0)  }  \big(\,  y_{\ups}^{(\ups)}- y  \big)  \bigg\}^{ \mf{d}_{\ups}(\bs{x}) -1 }  \;. 
\enq
Note that the integration domain for the $\bs{z}^{(\pm)}$ variables is symmetric. Hence, only the totally even part of the integrand in respect to these variables does
contribute to the value of $\mc{I}_{\e{sg}}^{(\sslash)}(\mf{x})$. Hence, one has 
\beq
\mc{I}_{\e{sg}}^{(\sslash)}(\mf{x}) \; = \;  \pl{ \ups = \pm }{} \bigg\{  \Int{ -\sqrt{\de_{\ups}}  }{ \sqrt{\de_{\ups}} } \hspace{-2mm} \dd^{m_{\ups}}z^{(\ups)}  \bigg\}  
\Int{  y_{+}^{(-)}  }{ y_{-}^{(+)} } \dd y  \cdot \Xi\Big( y_{-}^{(+)}  \, - \,   y_{+}^{(-)} \Big) \cdot 
\, \bigg[ \msc{F}^{(1)}(\bs{x})  \cdot \pl{\ups = \pm }{} \bigg\{ \ups \big(\,  y_{-\ups}^{(\ups)}- y  \big)  \bigg\}^{ \mf{d}_{\ups}(\bs{x}) -1 }  \bigg]_{(\bs{z}^{(+)},\bs{z}^{(-)})-\e{even} }\;, 
%
\enq
where $ \big[ \cdot \big]_{(\bs{z}^{(+)},\bs{z}^{(-)})-\e{even} }$ stands for the totally even part of a function in respect to the mentioned variables
\beq
\Big[  g(\bs{z}, \bs{v}) \Big]_{ \bs{z}-\e{even} } \; = \; \f{1}{ 2^{d} } \sul{ \substack{ \eps_a=\pm \\  a=1,\dots, d }  }{ }  g(\bs{z}^{(\eps)}, \bs{v})
\qquad \e{with} \qquad 
\bs{z}^{(\eps)}\; = \; \big(\eps_1 z_1, \dots, \eps_d z_d \big) \;. 
\enq
At this stage, one observes that 
\beq
 y_{-}^{(+)}  \, - \,   y_{+}^{(-)} \, = \, \f{ 2 \op{v} \vp_{\mf{x}}(\bs{z}) }{ \op{v}^2 - \big(\mf{u}_{1}^{\prime}(k_0) \big)^2  } F\Big(    \tfrac{ - 2 \mf{u}^{\prime\prime}_1(k_0)   }{    \mc{P}^{\prime}(k_0)  } \vp_{\mf{x}}(\bs{z})  \Big) 
\label{ecriture forme explicite difference y+ moins y-}
\enq
where $F$ is as defined in \eqref{definition fonction G pour difference des racines de P ups}.  Note that the series defining $F$ is convergent since $\vp_{\mf{x}}(\bs{z}) = \e{O}\Big( \de_+ + \de_- + |\mf{x}| \Big)$ 
and $\de_{\pm}, |\mf{x}|$ are all taken small enough. 
Furthermore, since $F(0)=1$, the estimate on $\vp_{\mf{x}}(\bs{z})$ ensures that  the $F$-dependent term in \eqref{ecriture forme explicite difference y+ moins y-} will be bounded from below by a strictly positive constant,
this throughout the whole integration domain $\mc{D}$.

Further, setting, 
\beq
\bs{t}=\Big( t_0, \bs{z}  \Big)  \quad \e{with} \quad  \left\{ \ba{c} 
\bs{z} \; = \; \Big( \bs{z}^{(+)}, \bs{z}^{(-)} \Big) \in \Big[-\sqrt{\de_+} ; \sqrt{\de_+} \Big]^{m_+}\!\times \Big[-\sqrt{\de_-} ; \sqrt{\de_-} \Big]^{m_-} \vspace{2mm} \\ 
\quad t_0 \, = \, y_{+}^{(-)} \, + \;  t \cdot \big[\,  y_{-}^{(+)}  \, - \,   y_{+}^{(-)} \big] \ea \right. 
\label{definition variable integration t}
\enq
entails that 
\beq
P_{\ups}(\bs{t}) \;= \; \vp_{\mf{x}}\big( \bs{z} \big) \cdot F_{\ups} \Big( \vp_{\mf{x}}\big( \bs{z} \big), t_0  \Big) \cdot
\left\{ \ba{cc} (1-t) & \ups= + \\
		t	 & \ups= -  \ea \right. 
\enq
with 
\beq
F_{\ups} \Big( \vp_{\mf{x}}\big( \bs{z} \big) , t_0 \Big) \; = \;  \f{  - \ups \mf{u}^{\prime\prime}_1(k_0) \, \op{v}   }{   \mc{P}^{\prime}(k_0) \big[ \op{v}^2 - \big(\mf{u}_{1}^{\prime}(k_0) \big)^2 \big]   } 
F\Big(    \tfrac{ - 2 \mf{u}^{\prime\prime}_1(k_0)   }{   \mc{P}^{\prime}(k_0)  } \vp_{\mf{x}}(\bs{z})  \Big) 
\cdot  \Big( y_{\ups}^{(\ups)}- y_{-}^{(+)} \, - \;  t \cdot \big[\,  y_{-}^{(+)}  \, - \,   y_{+}^{(-)} \big]  \Big) \;,  
\label{definition G ups pour prefacteur f ups}
\enq
and $t_0$ as in \eqref{definition variable integration t}.  Note that $F_{\ups}$ is indeed a sole function of $\vp_{\mf{x}}\big( \bs{z} \big)$ and $t_0$ since the roots $y_{\pm}^{(\ups)}$ only depend on  $\vp_{\mf{x}}\big( \bs{z} \big)$, \textit{c.f.} 
\eqref{ecriture racine y ups moins sg ups} and \eqref{ecriture racine y ups sg ups}.

Thus, the change of variables 
\beq
y \, = \, y_{_-}^{(+)} \, + \;  t \cdot \big[\,  y_{-}^{(+)}  \, - \,   y_{+}^{(-)} \big] 
\enq
recasts $\mc{I}_{\e{sg}}^{(\sslash)}(\mf{x})$ in the form
\beq
\mc{I}_{\e{sg}}^{(\sslash)}(\mf{x}) \; = \;  \pl{ \ups = \pm }{} \bigg\{  \Int{ -\sqrt{\de_{\ups}}  }{ \sqrt{\de_{\ups}} } \hspace{-2mm} \dd^{m_{\ups}}z^{(\ups)}  \bigg\}  
\Int{  0  }{ 1 } \dd t   \,  \bigg[ (1-t)^{  \mf{d}_{+}(\bs{t}) -1 }\, t^{  \mf{d}_{-}(\bs{t}) -1 }
\, \msc{F}^{(2)}(\bs{t})  \cdot  \Xi\Big( \vp_{\mf{x}}(\bs{z}) \Big) \cdot  \Big[ \vp_{\mf{x}}(\bs{z})  \Big]^{ \mf{d}_{+}(\bs{t})+\mf{d}_{-}(\bs{t}) -1 } \bigg]_{\bs{z}-\e{even}}
\label{ecriture integrale I sg parallele cas u lower v 2eme reduction}
\enq
where $\bs{t}$ is as in \eqref{definition variable integration t} and
\beq
 \msc{F}^{(2)}(\bs{t}) \, = \, \msc{F} (\bs{t}) \cdot 
   \f{ 2 \op{v} }{ \op{v}^2 - \big(\mf{u}_{1}^{\prime}(k_0) \big)^2  } F\Big(    \tfrac{  - 2 \mf{u}^{\prime\prime}_1(k_0)   }{    \mc{P}^{\prime}(k_0)  } \vp_{\mf{x}}(\bs{z})  \Big)   
   \cdot \pl{\ups = \pm }{} \Big[ F_{\ups} \Big( \vp_{\mf{x}}\big( \bs{z} \big), t_0  \Big) \Big]^{ \mf{d}_{\ups}(\bs{t}) -1 }  \;. 
\enq
%
%
%

 
 At this stage, one decomposes the integral into domains where a square root change of variables is well defined:
\beq
\Big[- \sqrt{\de_{\ups}} \, ;\,   \sqrt{\de_{\ups}} \,  \Big]^{ m_{\ups} }  \, = \, 
\bigsqcup_{ \bs{\eps}^{(\ups)} \in \{ \pm 1 \}^{ m_{\ups} } }  \Bigg\{ \pl{a=1}{m_{\ups}} \Big\{ \eps_{a}^{(\ups)} \Big[ 0 ;  \eps_a^{(\ups)} \sqrt{\de_{\ups}}  \Big]  \Big\}  \Bigg\}
\qquad \e{with} \qquad \bs{\eps}^{(\ups)} \; = \; \big( \eps^{ (\ups) }_{ 1 }, \dots,  \eps^{ (\ups) }_{ m_{\ups} } \big)
\enq
in which the sign prefactor in front of each interval indicates its orientation. Then, in each of the sets building up the partition, one sets 
\beq
z_a^{(\ups)} \, = \, \eps_{a}^{(\ups)} \cdot \big[ w_a^{(\ups)} \big]^{\f{1}{2}} \qquad a=1,\dots, m_{\ups} \;. 
\enq
This yields 
\beq
\mc{I}_{\e{sg}}^{(\sslash)}(\mf{x}) \; = \;   \sul{  \substack{ \bs{\eps}^{(\ups)}  \in \{ \pm 1 \}^{ m_{\ups} }  \\  \ups = \pm }  }{}  \mc{J}_{\bs{\eps}}\Big[\msc{F}^{(3)}, \mf{d}_{+}-1, \mf{d}_{-}-1  \Big]
\label{ecriture I sg slash decompose sur les divers eps pm}
\enq
where the building block integral is defined as
\bem
\mc{J}_{\bs{\eps}}\Big[\mc{F} , A , B    \Big] \; = \;   \pl{\ups= \pm }{}\Bigg\{ \Int{ 0 }{ \de_{\ups} }  \f{ \dd^{ m_{\ups} } w^{(\ups)} }{ \sqrt{ w_a^{(\ups)} } } \Bigg\}  \Int{0}{1}\dd t \; 
\bigg[ \mc{F}\Big( \bs{u}^{(\bs{\eps})} , (1-t)\vp_{\mf{x}}\big(\bs{u}^{(\bs{\eps})}_{\bs{w}} \big) ,   t \vp_{\mf{x}}\big(\bs{u}^{(\bs{\eps})}_{\bs{w}} \big)   \Big)  \\
\times   (1-t)^{ A(\bs{u}^{(\bs{\eps})}) } \cdot t^{ B(\bs{u}^{(\bs{\eps})}) }      \cdot 
\Xi\big[ \vp_{\mf{x}}\big(\bs{u}^{(\bs{\eps})}_{\bs{w}} \big)  \big]  \cdot
\big[ \vp_{\mf{x}}\big(\bs{u}^{(\bs{\eps})}_{\bs{w}} \big)  \big]^{ A(\bs{u}^{(\bs{\eps})})+B(\bs{u}^{(\bs{\eps})})+1 } \bigg]_{\bs{u}^{(\bs{\eps})}_{\bs{w}}-\e{even} }\;. 
\label{ecriture integrale modele pour eps decomposition de l'integrale I sg u lower v}
\end{multline}
 Above, it is undercurrent 
\beq
\bs{u}^{(\bs{\eps})}\, = \, \bigg(u_0^{(\bs{\eps})} ,
\bs{u}^{(\bs{\eps})}_{\bs{w}}\Big)
\quad \e{with} \quad 
\bs{u}^{(\bs{\eps})}_{\bs{w}}\, = \, \Big( \bs{u}^{(\bs{\eps};+)}_{\bs{w}}, \bs{u}^{(\bs{\eps};-)}_{\bs{w}}  \Big) \;, \quad 
u_0^{(\bs{\eps})} \, = \, y_{-}^{(+)}\Big(\vp_{\mf{x}}(\bs{u}_{\bs{w}}^{(\bs{\eps})}) \Big)  \, + \;  t \cdot \Big[\,  y_{-}^{(+)}\Big(\vp_{\mf{x}}(\bs{u}_{\bs{w}}^{(\bs{\eps})}) \Big) 
\, - \,   y_{+}^{(-)}\Big(\vp_{\mf{x}}(\bs{u}_{\bs{w}}^{(\bs{\eps})}) \Big) \Big]
\enq
and, finally,
\beq
\bs{u}^{(\bs{\eps};\ups)}_{\bs{w}} \, = \, \Big( \eps^{ (\ups) }_{ 1 }  \big[ w_1^{(\ups)} \big]^{\f{1}{2}}  , \dots,  \eps^{ (\ups) }_{ m_{\ups} }   \big[ w_{ m_{\ups} }^{(\ups)} \big]^{\f{1}{2}}   \Big) \;. 
\enq
Here, I have made explicit the fact that the functions $y_{\mp}^{(\pm)}$ only  depend on the $\bs{u}^{(\bs{\eps})}_{\bs{w}}$ integration variables through the function $\vp_{\mf{x}}(\bs{u}_{\bs{w}}^{(\bs{\eps})})$. 
In fact, after this change of variables, it holds 
\beq
\vp_{\mf{x}}(\bs{u}_{\bs{w}}^{(\bs{\eps})}) \; = \; \mf{x} \; + \; \sul{a=1}{m_+} w_a^{(+)}  \; - \; \sul{a=1}{m_-} w_a^{(-)} \;. 
\enq
 The integrand appearing in \eqref{ecriture I sg slash decompose sur les divers eps pm} takes the form
\bem
 \msc{F}^{(3)}\Big( \bs{u}^{(\bs{\eps})}   , (1-t)\vp_{\mf{x}}\big(\bs{u}^{(\bs{\eps})}_{\bs{w}} \big) ,   t \vp_{\mf{x}}\big(\bs{u}^{(\bs{\eps})}_{\bs{w}} \big)   \Big) \, = \,  \,  
 \msc{G}\Big( \Psi_{\op{M}}(\bs{u}^{(\bs{\eps})}) \, , \, (1-t) \cdot \wt{\mf{f}_{+}}( \bs{u}^{(\bs{\eps})}) \, \vp_{\mf{x}}\big( \bs{u}^{(\bs{\eps})}_{\bs{w}} \big)  
\, , \, t \cdot \wt{\mf{f}_{-}}( \bs{u}^{(\bs{\eps})})\, \vp_{\mf{x}}\big( \bs{u}^{(\bs{\eps})}_{\bs{w}} \big) \Big) \hspace{2cm} \\
\times    \f{ 2^{2-\ov{\bs{n}}_{\ell}} \op{v} }{ \op{v}^2 - \big(\mf{u}_{1}^{\prime}(k_0) \big)^2  } F\Big(    \tfrac{  - 2 \mf{u}^{\prime\prime}_1(k_0)   }{    \mc{P}^{\prime}(k_0)  } \vp_{\mf{x}}(\bs{u}^{(\bs{\eps})}_{\bs{w}} )  \Big)    \cdot 
\Big(V \vp^{(\sslash)} \vp^{(\e{sg})} \Big)\Big( \Psi_{\op{M}}(\bs{u}^{(\bs{\eps})}) \Big) \\
\times \big| \det D_{\bs{u}^{(\bs{\eps})}} \Psi_{\op{M}} \big| 
\cdot \pl{ \ups = \pm }{}  \Big\{  \wt{\mf{f}}_{\ups}\big( \bs{u}^{(\bs{\eps})}  \big)  \Big\}^{ \mf{d}_{\ups}(\bs{u}^{(\bs{\eps})}) -1 }  \;, 
\end{multline}
where 
\beq
  \wt{\mf{f}}_{\ups}\big( \bs{u}^{(\bs{\eps})}   \big) \, = \,  
  \mf{f}_{\ups}\big( \mf{x}; \op{M}\cdot \bs{u}^{(\bs{\eps})} \big) \cdot  F_{\ups} \Big( \vp_{\mf{x}}\big( \bs{u}^{(\bs{\eps})}_{\bs{w}} \big) , u_0^{(\bs{\eps})}    \Big) \;, 
\enq
and I remind that  $\ov{\bs{n}}_{\ell}\, = \, \sul{r=1}{\ell} n_r$. 
 
 Here, one observes that
\beq
\mf{d}_{\ups}(\bs{t}) = \De_{\ups}(\bs{t}(k_0)) + \e{O}\Big( \sqrt{\de_+} + \sqrt{\de_-} + |\mf{x}|  \Big) \;. 
\enq
By hypothesis one has $\De_{\ups}(\bs{t}(k_0)) >0$ so that reducing $\de_{\pm}$ and $|\mf{x}|$ if need be, one gets that $\mf{d}_{\pm}>0$
throughout the integration domain.

 The expansion \eqref{ecriture I sg slash decompose sur les divers eps pm} of $\mc{I}_{\e{sg}}^{(\sslash)}(\mf{x}) $
decomposes this integral into a sum of elementary integrals \eqref{ecriture integrale modele pour eps decomposition de l'integrale I sg u lower v} whose
$\mf{x}\tend 0$ asymptotic behaviour is analysed in Lemma \ref{Lemme integrale beta multi-dim auxiliaire locale cas u less than v}. Also,
the $L^1$-nature of the integrand is part of the conclusions of that lemma. Moreover, upon invoking Lemma \ref{Lemme VdM local expansion} so as to access to the small $\norm{ \bs{u}^{(\bs{\eps})} }$ expansion of $ \msc{F}^{(3)}$,
 Lemma \ref{Lemme integrale beta multi-dim auxiliaire locale cas u less than v}, specialised to the function $ \msc{F}^{(3)}$, ensures that there exists a smooth function $ \mf{x} \tend \mf{ R }_{ \bs{\eps} }( \mf{x} ) $
around $\mf{x}=0$ such that 
\beq
 \mc{J}_{\bs{\eps}}\Big[\msc{F}^{(3)}, \mf{d}_{+}-1, \mf{d}_{-}-1  \Big] \; = \; \De^{(0)}_{+}\,\De^{(0)}_{-}\,\msc{G}^{(1)}\big( \bs{t}(k_0) \big) \mc{J}_{\bs{\eps}}\Big[\msc{F}_{\e{eff}}, \De^{(0)}_{+}-1, \De^{(0)}_{-}-1  \Big] 
\; + \; \e{O}\Big( |\mf{x}|^{\vth  +   1-\tau   }  \Big) \; + \;    \mf{R}_{ \bs{\eps} }(\mf{x})
\enq
where $\De^{(0)}_{\ups}= \De_{\ups}\big( \bs{t}(k_0) \big)$, $\vth$ is as defined in \eqref{definition de cal theta 0}, $\msc{G}^{(1)}$ corresponds to the first term of the expansion of 
$\msc{G}$ as given in \eqref{ecriture decomposition smooth class K}, and 
\bem
\msc{F}_{\e{eff}} \big( \bs{u}^{(\bs{\eps})} , x, y  \big)  \; = \; \exp\bigg\{ -\Big(  \op{M}\,  \bs{u}^{(\bs{\eps})} ,  \op{D} \cdot \op{M} \,  \bs{u}^{(\bs{\eps})} \Big)  \bigg\} \cdot 
\Big(V \vp_{\e{eff}}^{(\sslash)} \vp_{\e{eff}}^{(\e{sg})} \Big)\Big( \op{M} \, \bs{u}^{(\bs{\eps})}   \Big)  \cdot |\det[\op{M}]| \\
\times \f{ 2^{2-\ov{\bs{n}}_{\ell}} \op{v} }{ \op{v}^2 - \big(\mf{u}_{1}^{\prime}(k_0) \big)^2  } F\Big(    \tfrac{  - 2 \mf{u}^{\prime\prime}_1(k_0)   }{    \mc{P}^{\prime}(k_0)  } \vp_{\mf{x}}(\bs{z})  \Big)
\cdot \pl{ \ups = \pm }{}  \Big\{ F_{\ups}\big( \bs{u}^{(\bs{\eps})} ,  u_0^{(\bs{\eps})}  \big)  \Big\}^{ \De^{(0)}_{\ups}  - 1 } \;. 
\end{multline}
Above, $\op{M}$ is as introduced in \eqref{ecriture effet transformation matrice M} while the diagonal matrix $\op{D}$ defining the Gaussian weight reads 
\beq
\op{D} \, = \, \left( \ba{ccc}   |\mf{u}^{\prime\prime}_{1}(t_1(k_0) ) | \op{I}_{n_1}  & 	0 		  		&			 0 \\ 
					  0						&  \ddots &     0  					\\
						0					&  0      &  |\mf{u}^{\prime\prime}_{\ell}(t_{\ell}(k_0) ) | \op{I}_{n_{\ell}}   \ea \right)
\label{definition matrice D pour poid gaussien integrale modele}
\enq
In particular, $\msc{F}_{\e{eff}}$ is $x,y$ independent. Furthermore, $\vp_{\e{eff}}^{(\sslash)}$ and  $\vp_{\e{eff}}^{(\e{sg})}$ are as appearing in \eqref{definition fcts approx unite sg et sslash integrale modele}
of Proposition \ref{Proposition reduction vers voisinage singularite integrale modele}. Namely, they are smooth on $\pl{r=1}{\ell}\R^{n_r}$ and such that
\beqa
0 \leq \vp_{\e{eff}}^{(\e{sg})} \leq 1 		& \quad   \vp_{\e{eff}}^{(\e{sg})}=1  \quad \e{on} \quad \ov{ \mc{D}^{(\e{eff})}_{\eta^{\prime}} }   & \quad  \vp_{\e{eff}}^{(\e{sg})}=0  \quad \e{on} \quad 
\pl{r=1}{\ell}\R^{n_r} \setminus  \mc{D}^{(\e{eff})}_{2\eta^{\prime}}  \nonumber \\
0 \leq \vp_{\e{eff}}^{(\sslash)} \leq 1 		&  \quad   \vp_{\e{eff}}^{(\sslash)}=1  \quad \e{on} \quad \ov{ \mc{D}^{(\sslash;\e{eff})}_{\frac{1}{2} \eta^{\prime} } }   & \quad  \vp_{\e{eff}}^{(\sslash)}=0  \quad \e{on} \quad 
\pl{r=1}{\ell}\R^{n_r} \setminus  \mc{D}^{(\sslash;\e{eff})}_{ \eta^{\prime}  } \;. 
\label{definition des fcts vp eff slash et sg}
\eeqa
The domain $\mc{D}^{(\e{eff})}_{\eta^{\prime}}$, resp. $ \mc{D}^{(\sslash;\e{eff})}_{ \eta^{\prime}  } $, is as defined in \eqref{definition du domaine D eff}, resp. \eqref{definition domaine De eff sslash}. 

Thus, by performing backwards, on the level of $\mc{J}_{\bs{\eps}}\Big[\msc{F}_{\e{eff}}, \De^{(0)}_{+}-1, \De^{(0)}_{-}-1  \Big]$, 
all the transformations that were carried out starting from \eqref{reduction finale de la partie singuliere vers integrale modele}, one gets that 
there exists a smooth function $\mf{x} \tend \mf{R}(\mf{x})$
around $\mf{x}=0$ such that 
\beq
\mc{I}_{\e{sg}}^{(\sslash)}(\mf{x}) \; = \;  \De^{(0)}_{+}\,\De^{(0)}_{-}\,\msc{G}^{(1)}\big( \bs{t}(k_0) \big)  \cdot \mc{J}_{\e{sg};\e{eff}}^{(\sslash)}(\mf{x})
\; + \; \e{O}\Big( |\mf{x}|^{\vth   +  1-\tau  }  \Big) \; + \;    \mf{R} (\mf{x})
\label{ecriture reduction de I sg slash vers J sg slash}
\enq
in which 
\beq
 \mc{J}_{\e{sg};\e{eff}}^{(\sslash)}(\mf{x}) \; = \; \Int{ \mc{D}^{(\e{eff})}_{2\eta^{\prime}}  }{} \hspace{-2mm} \dd \bs{x} \;  \ex { - \, (  \bs{x} ,  \op{D} \bs{x} ) }
\cdot \Big(V \vp_{\e{eff}}^{(\sslash)} \vp_{\e{eff}}^{(\e{sg})} \Big)\big(\bs{x}  \big) \cdot 
\pl{\ups= \pm }{} \bigg\{ \Xi\big[\mf{x} + \wt{z}_{\ups}(\bs{x}) \big] \cdot \big[\mf{x} + \wt{z}_{\ups}(\bs{x}) \big]^{ \De^{(0)}_{\ups}-1} \bigg\} \;. 
\enq

 Now, by virtue of Proposition \ref{Proposition reduction vers voisinage singularite integrale modele} there exists a smooth function $\wt{\mf{R}} (\mf{x})$ such that 
\beq
\mc{J}_{\e{sg};\e{eff}}^{(\sslash)}(\mf{x}) \; = \;\wt{\mf{R}} (\mf{x}) \, + \, 
 \Int{ \pl{r=1}{\ell} \R^{n_r}  }{} \hspace{-2mm} \dd \bs{x} \;  
\f{ \ex { - \, (  \bs{x} ,   \bs{x} ) } \cdot  V  \big(\bs{x}  \big) }{   \pl{r=1}{\ell} \big| \mf{u}^{\prime\prime}_r(t_r(k_0)) \big|^{\frac{1}{2} n_r^2 }  }\cdot 
\pl{\ups= \pm }{} \bigg\{ \Xi\big[\mf{x} + z_{\ups}(\bs{x}) \big] \cdot \big[\mf{x} + z_{\ups}(\bs{x}) \big]^{ \De^{(0)}_{\ups}-1} \bigg\} 
\label{reecriture integrale sg locale modele et son extension a R ov n ell}
\enq
in which $z_{\ups}$ is as defined in \eqref{definition z ups etape 1} where the below identification of parameters  has been made 
\beq
\veps_{r} \, = \,  - \zeta_r \e{sgn}\Big(  \mf{u}^{\prime\prime}_r(t_r(k_0))  \Big) \quad ,  \quad 
\xi_{r}  \, = \,    \big|  \mf{u}^{\prime\prime}_r(t_r(k_0))  \big|^{-\frac{1}{2}}  \quad , \quad 
 \op{u} \, = \, \mf{u}^{\prime}_1(k_0)   \;. 
\enq
By tracking the previous transformations backwards, one gets that there exists a smooth function  $ \check{\mf{R}}$ in the vicinity of the origin
\bem
\mc{I}(\mf{x}) \, = \, \Int{ \pl{r=1}{\ell} \R^{n_r}  }{} \hspace{-2mm} \dd \bs{x} \;  
\f{ \ex { - \, (  \bs{x} ,   \bs{x} ) } \cdot  V  \big(\bs{x}  \big) }{   \pl{r=1}{\ell} \big| \mf{u}^{\prime\prime}_r(t_r(k_0)) \big|^{\frac{1}{2} n_r^2 }  }\cdot 
\pl{\ups= \pm }{} \bigg\{ \Xi\big[\mf{x} + z_{\ups}(\bs{x}) \big] \cdot \big[\mf{x} + z_{\ups}(\bs{x}) \big]^{ \De^{(0)}_{\ups}-1} \bigg\}  \\
\; + \, \check{\mf{R}} (\mf{x}) \; + \; \e{O}\Big( |\mf{x}|^{\vth   +   1-\tau   }  \Big) \; .
\end{multline}
Then, it remains to apply Proposition \ref{Proposition asymptotique integrale modele} so as to get the form of the $\mf{x}\tend 0$ asymptotic expansion of 
the multiple integral appearing above, what yields the claimed form of the $\mf{x}\tend 0$ asymptotics of $\mc{I}(\mf{x})$ in the regime $ | \mf{u}^{\prime}_1(k_0)  | \, < \, \op{v}$.

\subsubsection*{$\bullet$ Canonical form of $\mc{I}_{\e{sg}}^{(\sslash)}(\mf{x}) $  in the $|\mf{u}_1^{\prime}(k_0)| > \op{v}$ regime}

The factorised form of the products of polynomials $P_{\ups}(\bs{x})$ given in \eqref{factorisation partie petite et grande factorisation polynome cas mod u greater v} 
entails that $\mc{I}_{\e{sg}}^{(\sslash)}(\mf{x}) $ takes the form 
\beq
\mc{I}_{\e{sg}}^{(\sslash)}(\mf{x}) \; = \;  \pl{ \ups = \pm }{} \bigg\{  \Int{ -\sqrt{\de_{\ups}}  }{ \sqrt{\de_{\ups}} } \hspace{-2mm} \dd^{m_{\ups}}z^{(\ups)}  \bigg\}  
\Int{  J_{\vsg}(\bs{z}) }{ } \dd y   
\, \msc{F}^{(1)}(\bs{x})  \cdot \pl{\ups = \pm }{} \bigg\{ \vsg \big(\,  y_{-\vsg}^{(\ups)}- y  \big)  \bigg\}^{ \mf{d}_{\ups}(\bs{x}) -1 }
\label{ecriture integrale I sg parallele cas u bigger v 1ere reduction}
\enq
where $\vsg$ has been introduced,  $J_{\vsg}(\bs{z})$ in \eqref{definition intervalle J vsg}, the argument $\bs{x}$ is expressed in terms of the integration variables $y, \bs{z}^{(\pm)}$ 
as in \eqref{ecriture effet transformation matrice M}, and, for short, I agree upon
\beq
\msc{F}^{(1)}(\bs{x}) \, = \, \msc{F} (\bs{x}) \cdot \pl{\ups = \pm }{} \bigg\{    \tfrac{ - \vsg  \mf{u}^{\prime\prime}_1(k_0)   }{2   \mc{P}^{\prime}(k_0)  }  \big(\,  y_{\vsg}^{(\ups)}- y  \big)  \bigg\}^{ \mf{d}_{\ups}(\bs{x}) -1 }  \;. 
\enq
As earlier on, the symmetry of the $(\bs{z}^{(+)}, \bs{z}^{(-)})$ integration domain entails that 
\beq
\mc{I}_{\e{sg}}^{(\sslash)}(\mf{x}) \; = \;  \pl{ \ups = \pm }{} \bigg\{  \Int{ -\sqrt{\de_{\ups}}  }{ \sqrt{\de_{\ups}} } \hspace{-2mm} \dd^{m_{\ups}}z^{(\ups)}  \bigg\}  
\Int{  J_{\vsg}(\bs{z}) }{ } \dd y   
\, \bigg[ \msc{F}^{(1)}(\bs{x})  \cdot \pl{\ups = \pm }{} \bigg\{ \vsg \big(\,  y_{-\vsg}^{(\ups)}- y  \big)  \bigg\}^{ \mf{d}_{\ups}(\bs{x}) -1 } \bigg]_{ \bs{z}-\e{even} }
\enq
where the $\bs{z}$-even part of a function is as defined in \eqref{definition even part of a function}.

At this stage, one observes that 
\beq
 \vsg y_{-\vsg}^{(+)}  \, - \,   \vsg y_{-\vsg}^{(-)} \, = \, - \vsg \vp_{\mf{x}}(\bs{z}) \cdot  \wt{F}\big( \vp_{\mf{x}}(\bs{z})  \big) \quad \e{with} \quad 
  \wt{F}(x) \, = \,   2   \,  \op{v} \f{   F\Big(    \tfrac{ - 2 \mf{u}^{\prime\prime}_1(k_0)   }{   \mc{P}^{\prime}(k_0)  }  x \Big)    }{  \big(\mf{u}_{1}^{\prime}(k_0) \big)^2  \, - \, \op{v}^2  } 
\label{definition tilde G}
\enq
and where $F$ is as defined in \eqref{definition fonction G pour difference des racines de P ups}. Just as earlier on, one has that, uniformly on $\mc{D}$, 
it holds $\wt{F}\big( \vp_{\mf{x}}(\bs{z})  \big) >0$ so that
\beq
\mf{p} \, = \, \e{sgn}\Big(  \vsg y_{-\vsg}^{(+)}  \, - \,   \vsg y_{-\vsg}^{(-)}  \Big) \; = \; - \vsg \e{sgn}\big( \vp_{\mf{x}}(\bs{z})\big) \;, 
\enq
meaning that 
\beq
a_{\mf{p}} \, = \,   \vsg y_{-\vsg}^{(\mf{p})}  \, - \,   \vsg y_{-\vsg}^{(-\mf{p})} \; = \;  \big| \vp_{\mf{x}}(\bs{z}) \big| \cdot \wt{F}\big( \vp_{\mf{x}}(\bs{z})  \big)  \;  \geq  \; 0 \;. 
\enq
%
%
%
%
%
%
%
%
%
%
%
%
%
%
The change of variables 
\beq
y \, = \, b_{\mf{p}}\big( \vp_{\mf{x}}(\bs{z})  \big)  \, - \, \vsg t \,  \wt{F}\big( \vp_{\mf{x}}(\bs{z})  \big)  			\qquad \e{with} \qquad    		
b_{\mf{p}}\big( \vp_{\mf{x}}(\bs{z})  \big) \, = \,   \vsg \e{min}\Big(  \vsg y_{-\vsg}^{(+)}  \, , \,   \vsg y_{-\vsg}^{(-)}  \Big) \, = \,  y_{-\vsg}^{(-\mf{p})} \;, 
\enq
in the $y$-integration recasts $\mc{I}_{\e{sg}}^{(\sslash)}(\mf{x})$ in the form
\beq
\mc{I}_{\e{sg}}^{(\sslash)}(\mf{x})  \; = \; \sul{ \mf{y} = \pm }{   }\mc{I}_{\e{sg};\mf{y}}^{(\sslash)}(\mf{x}) 
\enq
where the two building blocks read
\beq
\mc{I}_{\e{sg};\mf{y}}^{(\sslash)}(\mf{x}) \; = \;  \pl{ \ups = \pm }{} \bigg\{  \Int{ -\sqrt{\de_{\ups}}  }{ \sqrt{\de_{\ups}} } \hspace{-2mm} \dd^{m_{\ups}}z^{(\ups)}  \bigg\}  
\Int{  0  }{ \de_0 } \dd t   \,  \bigg[  t^{  \mf{d}_{-\mf{p}}(\bs{t}) -1 }  \Big[ t +  |\vp_{\mf{x}}\big(\bs{z}\big)|  \Big]^{  \mf{d}_{\mf{p}}(\bs{t}) -1 }  \cdot  \Xi\big[ \mf{y}   \vp_{\mf{x}}\big(\bs{z}\big) \big]
\, \msc{F}^{(2)}(\bs{t})   \bigg]_{\bs{z}-\e{even}}\;. 
\enq
There, 
\beq
\bs{t}=\Big(t_0 , \bs{z}  \Big) \; , \qquad 
\bs{z} \; = \; \Big( \bs{z}^{(+)}, \bs{z}^{(-)} \Big) \in \R^{m_+}\!\times \R^{m_-} \;, \quad 
 t_0 \, = \, b_{ \mf{p}}\big( \vp_{\mf{x}}(\bs{z})  \big)  \, - \, \vsg t\,  \wt{F}\big( \vp_{\mf{x}}(\bs{z})  \big) \;. 
\enq
Also, 
\beq
 \msc{F}^{(2)}(\bs{t}) \; = \;  \msc{F}(\bs{t}) \cdot \wt{F}\big( \vp_{\mf{x}}(\bs{z})  \big) \pl{\ups = \pm }{}  \Big[ \wt{F}_{\ups}\big( \vp_{\mf{x}}(\bs{z}) ,t_0  \big)  \Big]^{ \mf{d}_{\ups}(\bs{t}) -1  }
\enq
and
\beq
\wt{F}_{\ups}\big( \vp_{\mf{x}}(\bs{z}) ,t_0  \big)   \; = \; \tfrac{ - \vsg  \mf{u}^{\prime\prime}_1(k_0)   }{2   \mc{P}^{\prime}(k_0)  } \wt{F}\big( \vp_{\mf{x}}(\bs{z})  \big) 
\cdot \Big(  \,  y_{\vsg}^{(\ups)} \, + \, t \vsg \, \wt{F}\big( \vp_{\mf{x}}(\bs{z})  \big)\, - \, b_{\mf{p}}  \Big) \;. 
\enq

\vspace{4mm}

Observe that it holds
\beq
 \mf{y}   \vp_{\mf{x}}\big(\bs{z}\big) \, = \,  \mf{y}\mf{x} + \sul{\ups=\pm}{} \ups \sul{s=1}{m_{\mf{y} \ups} } \big( z_{s}^{(\mf{y}\ups)} \big)^{2}  \;. 
\enq
Thus, denoting
\beq
 \qquad \wh{m}_{\ups} \, = \,  m_{\mf{y} \ups}\;,    \quad \e{and} \quad
\wh{\de}_{\ups} \, = \; \de_{\mf{y} \ups} \;, 
\enq
the change of variables $\bs{z}^{(\ups)} \hookrightarrow \bs{z}^{(\ups \mf{y})}$ 
leads to 
\beq
\mc{I}_{\e{sg};\mf{y}}^{(\sslash)}(\mf{x}) \; = \;  \pl{ \ups = \pm }{} \bigg\{  \Int{ -\sqrt{ \, \wh{\de}_{\ups} }  }{ \sqrt{ \, \wh{\de}_{\ups}} } \hspace{-2mm} \dd^{ \wh{m}_{\ups}}z^{(\ups)}  \bigg\}  
\Int{  0  }{ \de_0 } \dd t   \, \bigg[ t^{  \mf{d}_{-\mf{p}}( \, \wh{\bs{t}} \, ) -1 } \,  \Big[ t +  \wh{\vp_{\mf{x}}}\big(\bs{z}\big)   \Big]^{  \mf{d}_{\mf{p}}( \, \wh{\bs{t}} \, ) -1 } \cdot 
\Xi\big[ \,  \wh{\vp_{\mf{x}}}\big(\bs{z}\big) \, \big]
\, \msc{F}^{(2)}\big( \, \wh{\bs{t}} \, \big)  \bigg]_{\bs{z}-\e{even}} \;. 
\enq
There 
\beq
\wh{\bs{t}}=\Big(t_0 , \bs{z}  \Big) \; , \qquad 
\bs{z} \; = \; \Big( \bs{z}^{(\mf{y})}, \bs{z}^{(-\mf{y})} \Big) \in \R^{ \wh{m}_+ }\!\times \R^{ \wh{m}_- } \;, \quad 
 t_0 \, = \, b_{ \mf{p} }\big( \wh{\vp_{\mf{x}}}(\bs{z})  \big)  \, - \, \vsg t\,  \wt{F}\big( \wh{\vp_{\mf{x}}}(\bs{z})  \big) \;, 
\enq
and 
\beq
\wh{\vp_{\mf{x}}}(\bs{z}) \, = \, \mf{y} \mf{x} \, + \, \sul{s=1}{ \wh{m}_{ + } } \big( z_{s}^{(+)} \big)^{2} \, - \, \sul{s=1}{ \wh{m}_{ - } } \big( z_{s}^{(-)} \big)^{2}  \;. 
\enq

 At this stage, one decomposes the integral into domains where a square root change of variables is well defined:
\beq
\Big[ - \big[ \; \wh{\de}_{\ups}  \big]^{\f{1}{2}} \, ; \,   \big[ \; \wh{\de}_{\ups}  \big]^{\f{1}{2}} \;  \Big]^{ \wh{m}_{\ups} }  \, = \, 
\bigsqcup_{ \bs{\eps}^{(\ups)} \in \{ \pm 1 \}^{ \wh{m}_{\ups} } }  \Bigg\{ \pl{a=1}{ \wh{m}_{\ups}} \bigg\{ \eps_{a}^{(\ups)} \Big[ 0 ;  \eps_a^{(\ups)} \cdot \big[ \; \wh{\de}_{\ups}  \big]^{\f{1}{2}} \;  \Big]  \bigg\} \Bigg\}
\qquad \e{with} \qquad \bs{\eps}^{(\ups)} \; = \; \Big( \eps^{ (\ups) }_{ 1 }, \dots,  \eps^{ (\ups) }_{ \, \wh{m}_{\ups} } \Big)
\enq
in which the sign pre-factor indicates the orientation of the interval. Then, in each of the sets building up the partition, one sets 
\beq
z_a^{(\ups)} \, = \, \eps_{a}^{(\ups)} \cdot \big[ w_a^{(\ups)} \big]^{\f{1}{2}} \qquad a=1,\dots, \wh{m}_{\ups} \;. 
\enq
This yields 
\beq
\mc{I}_{\e{sg};\mf{y}}^{(\sslash)}(\mf{x}) \; = \; \sul{ \substack{ \bs{\eps}^{(\ups)}  \in \{ \pm 1 \}^{ \wh{m}_{\ups} }  \\ \ups= \pm }  }{} 
\chi_{\bs{\eps};\mf{y}}\Big[\msc{F}^{(3)}, \mf{d}_{-\mf{p}}-1, \mf{d}_{ \mf{p} }-1  \Big]
\label{decomposition I slash mf y sur integrales elementaires}
\enq
where the building block integral is defined as
\bem
 \chi_{\bs{\eps};\mf{y}}\Big[\mc{F} , A , B    \Big] \; = \;   \pl{\ups= \pm }{}\Bigg\{ \Int{ 0 }{ \wh{\de}_{\ups} }  \f{ \dd^{ \wh{m}_{\ups} } w^{(\ups)} }{  \sqrt{ w_a^{(\ups)} }  } \Bigg\}  \Int{ 0 }{ \de_0 } \dd t \; 
\bigg[ \mc{F}\Big( \bs{u}^{(\bs{\eps})} , t   ,   t + \wh{\vp_{\mf{x}}}\big(\bs{u}^{(\bs{\eps})}_{\bs{w}} \big)   \Big)  \\
\times   \big[t + \wh{\vp_{\mf{x}}}\big(\bs{u}^{(\bs{\eps})}_{\bs{w}}\big) \big]^{ A(\bs{u}^{(\bs{\eps})}) } \cdot t^{ B(\bs{u}^{(\bs{\eps})}) }     \cdot 
\Xi\Big[ \wh{\vp_{\mf{x}}}\big(\bs{u}^{(\bs{\eps})}_{\bs{w}} \big)  \Big]    \bigg]_{\bs{u}^{(\bs{\eps})}_{\bs{w}}-\e{even}}\;. 
\label{ecriture integrale modele pour eps decomposition de l'integrale I sg u greater v}
\end{multline}
 Above, it is undercurrent 
\beq
\bs{u}^{(\bs{\eps})}\, = \, \Big( u^{(\bs{\eps})}_0 , \bs{u}^{(\bs{\eps})}_{\bs{w}}\Big)
\quad \e{with} \quad 
\bs{u}^{(\bs{\eps})}_{\bs{w}}\, = \, \Big( \bs{u}^{(\bs{\eps};\mf{y})}_{\bs{w}}, \bs{u}^{(\bs{\eps};-\mf{y})}_{\bs{w}}  \Big) \;, \quad 
 u^{(\bs{\eps})}_0 \, = \, b_{\mf{p}}\Big( \wh{\vp_{\mf{x}}}( \bs{u}^{(\bs{\eps})}_{\bs{w}} )  \Big)  \, - \, \vsg t\,  \wt{F}\Big( \wh{\vp_{\mf{x}}}( \bs{u}^{(\bs{\eps})}_{\bs{w}} )  \Big) \;, 
\enq
and, finally,
\beq
\bs{u}^{(\bs{\eps};\ups)}_{\bs{w}} \, = \, \Big( \eps^{ (\ups) }_{ 1 }  \big[ w_1^{(\ups)} \big]^{\f{1}{2}}  , \dots,  \eps^{ (\ups) }_{ \wh{m}_{\ups} }   \big[ w_{ \wh{m}_{\ups} }^{(\ups)} \big]^{\f{1}{2}}   \Big) \;. 
\enq
Also, after the change of variables, one has that  
\beq
\wh{\vp_{\mf{x}}}(\bs{u}^{(\bs{\eps})}_{\bs{w}}) \; = \; \mf{y} \mf{x} \; + \; \sul{a=1}{ \wh{m}_+} w_a^{(+)}  \; - \; \sul{a=1}{ \wh{m}_-} w_a^{(-)} \;. 
\enq
 The integrand appearing in \eqref{decomposition I slash mf y sur integrales elementaires} takes the form
\bem
 \msc{F}^{(3)}\Big( \bs{u}^{(\bs{\eps})} , t   ,   t + \wh{\vp_{\mf{x}}}\big(\bs{u}^{(\bs{\eps})}_{\bs{w}} \big)   \Big)  \, = \, 2^{1-\ov{\bs{n}}_{\ell}}  \cdot 
  \wt{F}\Big(   \wh{\vp_{\mf{x}}}\big(\bs{u}^{(\bs{\eps})}_{\bs{w}} \big)  \Big)    \cdot 
\Big(V \vp^{(\sslash)} \vp^{(\e{sg})} \Big)\Big( \Psi_{\op{M}}(\bs{u}^{(\bs{\eps})}) \Big) \cdot \big| \det D_{\bs{u}^{(\bs{\eps})}} \Psi_{\op{M}} \big|  \vspace{8mm}  \\
\times  \pl{ \ups = \pm }{}  \Big\{ \,  \wt{\mf{f}}_{\ups}\big( \bs{u}^{(\bs{\eps})}, u^{(\bs{\eps})}_0 \big)  \Big\}^{ \mf{d}_{\ups}(\bs{u}^{(\bs{\eps})}) -1 }  
 \; \cdot  \; \left\{ \ba{ccc}  \msc{G}\Big( \Psi_{\op{M}}(\bs{u}^{(\bs{\eps})}) \, , \,   \wt{\mf{f}_{+}}( \bs{u}^{(\bs{\eps})}, u^{(\bs{\eps})}_0) \cdot \big[ t \, +  \,  \wh{\vp_{\mf{x}}}\big(\bs{u}^{(\bs{\eps})}_{\bs{w}} \big)  \big]
										  \, , \, t  \cdot \wt{\mf{f}_{-}}( \bs{u}^{(\bs{\eps})} ,u^{(\bs{\eps})}_0 )  \Big)  & \e{if} & \mf{p} = +  \vspace{3mm} \\ 
 \msc{G}\Big( \Psi_{\op{M}}(\bs{u}^{(\bs{\eps})}) \, , \,  t \cdot  \wt{\mf{f}_{+}}( \bs{u}^{(\bs{\eps})}, u^{(\bs{\eps})}_0 ) 
\, , \,  \wt{\mf{f}_{-}}( \bs{u}^{(\bs{\eps})}, u^{(\bs{\eps})}_0) \cdot \big[ t \, +  \,  \wh{\vp_{\mf{x}}}\big(\bs{u}^{(\bs{\eps})}_{\bs{w}} \big)  \big]  \Big)  & \e{if} & \mf{p} = -    \ea \right.  \;, 
\end{multline}
where 
\beq
  \wt{\mf{f}}_{\ups}\big( \bs{u}^{(\bs{\eps})} , u^{(\bs{\eps})}_0 \big) \, = \, 
  \mf{f}_{\ups}\big( \op{M}\cdot \bs{u}^{(\bs{\eps})} \big) \cdot  \wt{F}_{\ups} \Big( \vp_{\mf{x}}\big( \bs{u}^{(\bs{\eps})}_{\bs{w}} \big) , u^{(\bs{\eps})}_0 \Big) \;, 
\enq
and I remind that  $\ov{\bs{n}}_{\ell}\, = \, \sul{r=1}{\ell} n_r$.

As in the previous case,  one gets that $\mf{d}_{\pm}>0$
throughout the integration domain.  The expansion \eqref{decomposition I slash mf y sur integrales elementaires} of $\mc{I}_{\e{sg}}^{(\sslash)}(\mf{x}) $
decomposes this integral into a sum of elementary integrals \eqref{ecriture integrale modele pour eps decomposition de l'integrale I sg u greater v} whose
$\mf{x}\tend 0$ asymptotic behaviour is analysed in Lemma \ref{Lemme integrale beta multi-dim auxiliaire locale cas u greater than v}. 
The conclusions of this lemma, specialised to the function $ \msc{F}^{(3)}$, ensure that the integrand is in $L^1$ and that there exists a smooth function $ \mf{x} \tend \mf{ R }_{ \bs{\eps} }( \mf{x} ) $
around $\mf{x}=0$ such that 
\beq
 \chi_{\bs{\eps};\mf{y}}\Big[\msc{F}^{(3)}, \mf{d}_{ -\mf{p} }-1, \mf{d}_{\mf{p}}-1  \Big] \; = \; \De^{(0)}_{+}\, \De^{(0)}_{-} \, \msc{G}^{(1)}\big( \bs{t}(k_0) \big)\cdot
 \chi_{\bs{\eps};\mf{y}}\Big[\msc{F}_{\e{eff}}, \De^{(0)}_{-\mf{p}}-1, \De^{(0)}_{\mf{p}}-1  \Big] 
\; + \; \e{O}\Big( |\mf{x}|^{\vth  +   1-\tau   }  \Big) \; + \;    \mf{R}_{ \bs{\eps} }(\mf{x})
\enq
where $\De^{(0)}_{\ups}= \De_{\ups}\big( \bs{t}(k_0) \big)$, $\vth$ is as defined in \eqref{definition de cal theta 0} and 
\bem
\msc{F}_{\e{eff}} \big( \bs{u}^{(\bs{\eps})} , x, y  \big)  \; = \; \exp\bigg\{ -\Big(  \op{M}\,  \bs{u}^{(\bs{\eps})} ,  \op{D} \op{M} \,  \bs{u}^{(\bs{\eps})} \Big)  \bigg\} \cdot 
\Big(V \vp_{\e{eff}}^{(\sslash)} \vp_{\e{eff}}^{(\e{sg})} \Big)\Big( \op{M} \, \bs{u}^{(\bs{\eps})}   \Big)   \\
\times   2^{1-\ov{\bs{n}}_{\ell}}  \wt{F}\Big(     \vp_{\mf{x}}(\bs{z})  \Big)
\cdot \pl{ \ups = \pm }{}  \Big\{  \wt{F}_{\ups}\big( \bs{u}^{(\bs{\eps})},u_0^{(\bs{\eps})}  \big)  \Big\}^{ \de_{\ups}  - 1 } \;. 
\end{multline}
Above, $\op{M}$ is as introduced in \eqref{ecriture effet transformation matrice M} and $\op{D}$  as in \eqref{definition matrice D pour poid gaussien integrale modele}. 
Finally, $\vp_{\e{eff}}^{(\sslash)}$ and  $\vp_{\e{eff}}^{(\e{sg})}$ are as appearing in \eqref{definition fcts approx unite sg et sslash integrale modele}, \textit{c.f.} 
also \eqref{definition des fcts vp eff slash et sg}. 
By performing backwards, on the level of $ \chi_{\bs{\eps};\mf{y}}\Big[\msc{F}_{\e{eff}}, \De^{(0)}_{ -\mf{p} }-1, \De^{(0)}_{\mf{p}}-1  \Big]$, 
all the transformations that were carried out starting from \eqref{reduction finale de la partie singuliere vers integrale modele}, one 
arrives to the conclusions stated in \eqref{ecriture reduction de I sg slash vers J sg slash}. From there, one concludes as in the 
regime $ | \mf{u}^{\prime}_1(k_0)  | \, < \, \op{v}  $. \qed

\section{Auxiliary results}
\label{Appendix Section Auxiliary results}

\subsection{A regularity lemma}

Given $\bs{x}\in \R^n$ with $n\geq 2$ and integers $1\leq a<b \leq n$ denote 
\beq
\bs{x}_{a,b} \, = \, \big( x_1,\dots, x_{a-1},x_{a+1},\dots,   x_{b-1},  x_{b+1},\dots, x_n \big) \;. 
\label{definition vecteur simple ss 2 composantes}
\enq

\begin{lemme}
\label{Lemme integrale multidimensionnelle auxiliaire reguliere}

Let $K$ be a compact subset of $\R^n$, $n\geq 2$,  and let $\mf{z}_{\pm}, \psi_{\pm}$ be smooth functions on $K$.
Let $0<\tau <1$ and let $\msc{G}$ be in the smooth class of $K$ with functions $\De_{\pm} \, = \, \big[ \psi_{\pm}\big]^2$
and constant $\tau$, \textit{c.f.} Definition \ref{Definition fct lisse sur ferme}

Assume that for any $\bs{x}\in K$ there exists $a<b$, $a,b \in \intn{1}{n}$, such that the differential $D_{\bs{x}}f_{a,b}$ of the map 
\beq
f_{a,b} \, : \, \bs{x} \mapsto \Big( \bs{x}_{a,b},  \mf{z}_+\big(\bs{x}\big),  \mf{z}_-\big(\bs{x}\big) \Big) \;, 
\enq
with $\bs{x}_{a,b}$ as in \eqref{definition vecteur simple ss 2 composantes}, is invertible. Then, the integral  
\beq
\mc{J}(\mf{x}) \, = \, \Int{ K }{} \dd^n x  \;  \msc{G}\Big( \bs{x}, \wh{\mf{z}}_{+}(\bs{x}) , \wh{\mf{z}}_{-}(\bs{x}) \Big) \cdot
\pl{\ups=\pm }{} \bigg\{ \Xi\big[ \,  \wh{\mf{z}}_{\ups}(\bs{x})   \big] \cdot \big[\,   \wh{\mf{z}}_{\ups}(\bs{x})  \big]^{ \De_{\ups}(\bs{x}) -1} \bigg\} 
\quad with \quad
\wh{\mf{z}}_{\ups}(\bs{x})=\mf{z}_{\ups}(\bs{x})+\mf{x} \,  , 
\enq
is a smooth function of $\mf{x}$, provided that $|\mf{x}|$ is small enough. 
\end{lemme}

\proof 

By virtue of the Whintey extension theorem, $\mf{z}_{\pm}$ and $\psi_{\pm}$ admit smooth extensions to an open neighbourhood $U_K$ of $K$. Thus, so does $\De_{\ups}=\psi_{\ups}^2$
and one obviously has that $\De_{\pm}(U_K) \subset \intff{0}{+\infty}$. One may also extend $\msc{G}$ to $U_K\times \R^+\times \R^+$ smoothly by setting 
\beq
\msc{G}_{\mid \{ U_K \setminus K \} \times \R^+\times \R^+ }  \, = \, 0
\enq
where the smoothness of this extension is ensured by the smooth vanising, to all orders in the derivatives, of $\msc{G}$ on $\Dp{} K  \times \R^+\times \R^+$.

It follows from the hypothesis of the lemma that, for any $\bs{x}\in K$, there exists integers  $a_{\bs{x}}<b_{\bs{x}}$, an open, relatively compact, neighbourhood $U_{\bs{x}}$ of $\bs{x}$, 
an open, relatively compact, neighbourhood $V_{\bs{x}}$ of  $\bs{x}_{a_{\bs{x}},b_{\bs{x}}}$ in $\R^{n-2}$ and $\eta_{\bs{x}} >0$
such that 
\beq
f_{a_{\bs{x}},b_{\bs{x}}} \; : \; U_{\bs{x}}  \; \tend  \; f_{a_{\bs{x}},b_{\bs{x}}} \big( U_{\bs{x}} \big) \, = \, 
V_{ \bs{x} } \, \times \, \intoo{  \mf{z}_{+}(\bs{x}) - \eta_{\bs{x}}  }{  \mf{z}_{+}(\bs{x}) + \eta_{\bs{x}} } \,  \times \,  \intoo{  \mf{z}_{-}(\bs{x}) - \eta_{\bs{x}} }{ \mf{z}_{-}(\bs{x}) +\eta_{\bs{x}} } 
\enq
is a smooth diffeomorphism onto its image. Furthermore, reducing $\eta_{\bs{x}}$ if necessary, one may always assume that  $\mf{z}_{\ups}(\bs{x}) \pm \eta_{\bs{x}} \not=0$ for both values of $\ups \in \{\pm\}$. 
Finally, the sets can always be chosen such that $f_{a_{\bs{x}},b_{\bs{x}}}^{-1}$ has a smooth extension to an open neighbourhood of $\ov{ f_{a_{\bs{x}},b_{\bs{x}}} \big( U_{\bs{x}} \big) }$ and such that 
$U_{\bs{x}} \subset U_K$.

Then, $\cup_{\bs{x} \in K} U_{\bs{x}} \subset U_K$ is an open cover of $K$ and hence there exists a finite sub-cover $\bigcup_{k=1}^{\ell} U_{\bs{x}_k} \subset U_K$
with associated diffeomorphisms $f_{a_{\bs{x}_k},b_{\bs{x}_k}}$ mapping $U_{\bs{x}_k}$ onto $f_{a_{{\bs{x}_k}},b_{{\bs{x}_k}}} \big( U_{\bs{x}_k} \big)$. 
Let $\{\vp_k\}_{k=1}^{\ell}$ be the partition of unity subordinate to the cover $\bigcup_{k=1}^{\ell} U_{\bs{x}_k} $. Then, using that the integrand extends by zero outside of 
$K$, one has
\beq
\mc{J}(\mf{x}) \, = \, \Int{ U_K }{} \dd^n x  \;  \msc{G}\Big( \bs{x}, \wh{\mf{z}}_{+}(\bs{x}) , \wh{\mf{z}}_{-}(\bs{x}) \Big) \cdot
\pl{\ups=\pm }{} \bigg\{ \Xi\big[ \,  \wh{\mf{z}}_{\ups}(\bs{x})   \big] \cdot \big[\,   \wh{\mf{z}}_{\ups}(\bs{x})  \big]^{ \De_{\ups}(\bs{x}) -1} \bigg\} 
\enq
what allows one to decompose the integral as 
$\mc{J}(\mf{x}) \, = \, \sum_{k=1}^{\ell} \mc{J}_k(\mf{x})$ where 
\beq
\mc{J}_k(\mf{x}) \; = \; \Int{ V_{\bs{x}_k} }{} \dd^{n-2} w \cdot \pl{\ups=\pm}{} \bigg\{  \Int{  \mf{x} + \mf{z}_{\ups}(\bs{x}_k) - \eta_{\bs{x}_k} }{ \mf{x}+\mf{z}_{\ups}(\bs{x}_k) + \eta_{\bs{x}_k}  } \dd \mf{z}_{\ups} \bigg\}
\cdot \wt{\mc{G}}_k(\bs{u}_{\mf{x}})  \cdot \pl{\ups=\pm }{} \bigg\{ \Xi\big(  \mf{z}_{\ups} \big) \cdot \big[  \mf{z}_{\ups}  \big]^{ \wt{\De}_{\ups}(\bs{u}_{\mf{x}}) -1} \bigg\} 
\enq
with $\bs{u}_{\mf{x}} \, = \, ( \bs{w},  \mf{z}_+ - \mf{x} , \mf{z}_- - \mf{x})$ and 
\beq
\wt{\mc{G}}_k( \bs{u}_{\mf{x}} ) \, = \,   \vp_{k}\Big( f_{ a_{\bs{x}_k},b_{\bs{x}_k}}^{-1}( \bs{u}_{\mf{x}} ) \Big)  \cdot    \big| \e{det}\big[ D_{\bs{u}_{\mf{x}}} f^{-1}_{ _{a_{\bs{x}_k},b_{\bs{x}_k}}  }\big] \big|  
\cdot \msc{G}\big(f_{ a_{\bs{x}_k},b_{\bs{x}_k}}^{-1}( \bs{u}_{\mf{x}} ) , \mf{z}_+, \mf{z}_- \big)   \;.
%
%
\label{definition fct tilde G cal k}
\enq
Finally,  $\wt{\De}_{\ups}(\bs{u} )  \, = \,  \De_{\ups}\circ f_{ a_{\bs{x}_k},b_{\bs{x}_k}}^{-1}(\bs{u})$. 
This representation is obtained, by restricting the integration domain to $U_{\bs{x}_k}$ due to the presence of $\vp_{k}$ followed by making the change of variables 
$f_{a_{\bs{x}_k},b_{\bs{x}_k}}^{-1}\big( \, \wt{\bs{u}} \, \big) \, = \, \bs{x}$.  Finally, one shifts the last two integration variables by $ -\mf{x}$. 
Note that $\wt{\mc{G}}_k$ is smooth since the determinant never vanishes and has thus constant sign.

There are four cases to distinguish depending on whether $
0\in \intoo{ \mf{x}  +  \mf{z}_{\ups}(\bs{x}_k) - \eta_{\bs{x}_k} }{  \mf{x} + \mf{z}_{\ups}(\bs{x}_k) + \eta_{\bs{x}_k}  }$ or not.

\begin{itemize}
 
 \item[$i)$] If  $\mf{z}_{\ups}(\bs{x}_k) + \eta_{\bs{x}_k}<0$ for at least one $\ups \in \{ \pm\}$, then $\mc{J}_k(\mf{x})$ vanishes for $|\mf{x}|$ small enough. 
 
 \item[$ii)$] If $\mf{z}_{\ups}(\bs{x}_k) - \eta_{\bs{x}_k}>0$ for $\ups = \pm$, then, for $|\mf{x}|$ small enough,  the integral reduces to
\beq
\mc{J}_k(\mf{x}) \; = \; \Int{ V_{\bs{x}_k} }{} \dd^{n-2} w \cdot \pl{\ups=\pm}{} \bigg\{  \Int{  \mf{x} + \mf{z}_{\ups}(\bs{x}_k) - \eta_{\bs{x}_k} }{ \mf{x}+\mf{z}_{\ups}(\bs{x}_k) + \eta_{\bs{x}_k}  }\hspace{-5mm}  \dd \mf{z}_{\ups} \bigg\}
\cdot \wt{\mc{G}}_k(\bs{u}_{\mf{x}})  \cdot \pl{\ups=\pm }{} \Big\{   \big[  \mf{z}_{\ups}  \big]^{ \wt{\De}_{\ups}(\bs{u}_{\mf{x}}) -1} \Big\} \;. 
\enq
By construction, the endpoints of integration in the $\mf{z}_{\pm}$ variables are uniformly away from $0$, for $|\mf{x}|$ small enough.
The hypotheses of the lemma ensure that  $\wt{\mc{G}}_k(\bs{u}_{\mf{x}})$ is smooth in $\mf{x}$ small enough
and in $(\bs{w},\mf{z}_+,\mf{z}_-)$ provided that $\bs{u}_{\mf{x}}\in  f_{a_{\bs{x}_k},b_{\bs{x}_k}} \big( U_{\bs{x}_k} \big)$. 
Taken that the only singularities of the integrand, which are at $\mf{z}_{\pm}=0$, are uniformly away from the integration domain  and taken that 
the integral runs through a compact set, derivation under the integral -and in respect to endpoints of integration- theorems entail 
 that $\mc{J}_k(\mf{x})$ is a smooth function of $\mf{x}$, for $|\mf{x}|$ small enough. 

\item[$iii)$] If $\mf{z}_{\ups}(\bs{x}_k) - \eta_{\bs{x}_k} < 0$ and  $\mf{z}_{\ups}(\bs{x}_k) + \eta_{\bs{x}_k}>0$ for both values of $\ups$, then the integral splits as
$\mc{J}_k(\mf{x}) \; = \; \sul{b=1}{4} \mc{J}_k^{(b)}(\mf{x})  $ where 
\beq
\mc{J}_k^{(b)}(\mf{x}) \; = \; \Int{ V_{\bs{x}_k} }{} \dd^{n-2} w \cdot \pl{\ups=\pm}{} \bigg\{  \Int{  0  }{ \mf{x}+\mf{z}_{\ups}(\bs{x}_k) + \eta_{\bs{x}_k}  } \hspace{-5mm}  \dd \mf{z}_{\ups} \bigg\}
\cdot \wt{\mc{G}}_k^{(b)}(\bs{u}_{\mf{x}})  \cdot \pl{\ups=\pm }{} \Big\{  \big[  \mf{z}_{\ups}  \big]^{ \wt{\De}_{\ups}(\bs{u}_{\mf{x}}) -1 } \Big\} 
\cdot \left\{\ba{cc}   \wt{\De}_{+}(\bs{u}_{\mf{x}}) \,  \wt{\De}_{-}(\bs{u}_{\mf{x}})  &  b=1  \vspace{2mm} \\ 
			\wt{\De}_{-}(\bs{u}_{\mf{x}}) \cdot \big[  \mf{z}_{+}  \big]^{1-\tau} & b=2   \vspace{2mm} \\
			\wt{\De}_{+}(\bs{u}_{\mf{x}}) \cdot  \big[  \mf{z}_{-}  \big]^{1-\tau} & b=3   \vspace{2mm}  \\
			  \pl{\ups=\pm }{} \Big\{  \big[  \mf{z}_{\ups}  \big]^{ 1-\tau } \Big\}  & b=4    \ea \right. \;. 
\nonumber
\enq
The function $\wt{\mc{G}}_k^{(b)}$ are obtained from $\wt{\mc{G}}_k$ defined in \eqref{definition fct tilde G cal k} upon substituting  $\msc{G} \hookrightarrow \msc{G}^{(b)}$
with $ \msc{G}^{(b)}$ arising in the expansion \eqref{ecriture decomposition smooth class K}. 

Taken that $\msc{G}^{(4)}$ fulfils property $H3)$ stated below of \eqref{definition ensemble M elle n ell enleve} and that $ \mf{z} \mapsto   \mf{z}^{ \de-\tau } $ is integrable on $\intff{0}{\eps}$ for any $\de\geq 0$,
one readily  concludes that $\mc{J}_k^{(4)}(\mf{x}) $ produces a smooth function of $\mf{x}$. The analysis of the remaining integrals demands one more step which I detail for 
$\mc{J}_k^{(2)}(\mf{x})$, the other cases being tractable in a similar way. In the case of $\mc{J}_k^{(2)}(\mf{x})$, an integration by parts yields:
\bem
\mc{J}_k^{(2)}(\mf{x}) \; = \; - \Int{ V_{\bs{x}_k} }{} \dd^{n-2} w \cdot \pl{\ups=\pm}{} \bigg\{  \Int{  0  }{ \mf{x}+\mf{z}_{\ups}(\bs{x}_k) + \eta_{\bs{x}_k}  } \hspace{-5mm}  \dd \mf{z}_{\ups} \bigg\}
\cdot  \Dp{\mf{z}_{-}}\bigg\{  \wt{\mc{G}}_k^{(2)}(\bs{u}_{\mf{x}}) \cdot \big[  \mf{z}_{+}  \big]^{ \wt{\De}_{+}(\bs{u}_{\mf{x}}) - \tau }  \cdot \big[  s  \big]^{ \wt{\De}_{-}(\bs{u}_{\mf{x}})   } \bigg\}_{\mid s = \mf{z}_-} \\
\hspace{-1cm} + \Int{ V_{\bs{x}_k} }{} \dd^{n-2} w   \hspace{-6mm}  \Int{  0  }{ \mf{x}+\mf{z}_{+}(\bs{x}_k) + \eta_{\bs{x}_k}  } \hspace{-6mm}  \dd \mf{z}_{+} 
\Bigg\{   \bigg(  \wt{\mc{G}}_k^{(2)}(\bs{u}_{\mf{x}}) \cdot \big[  \mf{z}_{+}  \big]^{ \wt{\De}_{+}(\bs{u}_{\mf{x}}) - \tau }  \cdot \big[ \mf{z}_-  \big]^{ \wt{\De}_{-}(\bs{u}_{\mf{x}})   } \bigg)_{\mid \mf{z}_-= \mf{x}+\mf{z}_{-}(\bs{x}_k) + \eta_{\bs{x}_k} }  
 \hspace{-8mm} - \qquad    \bigg(  \wt{\mc{G}}_k^{(2)}(\bs{u}_{\mf{x}}) \cdot \big[  \mf{z}_{+}  \big]^{ \wt{\De}_{+}(\bs{u}_{\mf{x}}) - \tau }  \cdot \big[  \mf{z}_-  \big]^{ \wt{\De}_{-}(\bs{u}_{\mf{x}})   } \bigg)_{\mid \mf{z}_-=0}  \Bigg\} \;. 
\nonumber
\end{multline}
Note that the last term occurring in this expression is only present if $\wt{\De}_{-}(\bs{u}_{\mf{x}})_{\mid \mf{z}_-=0} =0 $ on a set of non-zero measure. 
Property $H3)$ fulfilled by $\msc{G}^{(2)}$, the fact that $\msc{G}^{(2)}$ does not depend on the $v$ variables as given in \eqref{ecriture decomposition smooth class K}
and the integrability of the $\mf{z}_{\ups}$-related part of the integrand all together ensure that 
the resulting integrals produce smooth contributions in $\mf{x}$ for $|\mf{x}|$ small enough.

\item[$iv)$] The situation is quite similar when only one of the $\mf{z}_{\ups}$ changes sign, \textit{viz}. 
 $\mf{z}_{+}(\bs{x}_k) - \eta_{\bs{x}_k} < 0$ and  $\mf{z}_{+}(\bs{x}_k) + \eta_{\bs{x}_k}>0$ but $\mf{z}_{-}(\bs{x}_k) - \eta_{\bs{x}_k}>0$ or the analogous situation when $+\leftrightarrow -$. 
In such a case,  one should decompose the integral similarly to $iii)$ and then invoke 
 one of the two properties $H1)$ or $H2)$, \textit{c.f.} below of \eqref{definition ensemble M elle n ell enleve},  depending on which of among the two integration domains passes though zero, 
 relative to the boundedness of the function $\msc{G}$ and its partial derivatives so as to conclude on the smoothness of the associated integral.

\end{itemize}

Thus the claim. \qed

\subsection{Local rectification of $\mf{z}_{\ups}$}

\begin{prop}
\label{Proposition trivialisation locale fct zups}
 
Let the assumptions and notation given in Subsection \ref{SousSection hypothese gles sur consituants integrale modele} hold. 
Given $k_0 \in \e{Int}{\msc{J}}_1$ and given any $ \zeta_r \in \{\pm 1\}$,   let 
\beq
\mf{z}_{\ups}(\bs{p})\;=\;   \, - \,  \sul{ (r,a) \in \mc{M} }{  } \zeta_r  \mf{w}_{\ups}^{(r)}\big(  p_a^{(r)} ; t_r(k_0) \big) \quad , \qquad \ups \in \{ \pm \} \;, 
\enq
where 
\beq
 \mf{w}_{\ups}^{(r)}(k ; p) \, = \, \mf{u}_r(k)\, - \, \mf{u}_r(p)\, + \, \ups \op{v}  \big( k - p\big) \qquad and \qquad \op{v} \in \R^{+*}\;. 
\enq
Further, let
\beq
\tilde{z}_{\ups}(\bs{x}) \; = \;- \sul{ (r,a)\in \mc{M} }{} \zeta_{r} \bigg\{ \mf{h}_r\big( x_a^{(r)} \big) +\ups \op{v} x_a^{(r)}  \bigg\} \qquad  where  \qquad 
 \mf{h}_r\big( x \big) \; = \;  - \zeta_r \veps_r \f{ x^2  }{ 2 \xi_r^2 } \, + \,  \op{u}\,  x \;, 
\label{definition fonction z tilde ups}
\enq
$\veps_{r}\in \{ \pm \}$, $\xi_r \in \R^*$ and $\op{u}$ are such that 
\beq
-  \f{ \zeta_r \veps_r  }{  \xi_r^2 } \; = \; \mf{u}_{r}^{\prime\prime}\big( t_r(k_0) \big) \quad and \quad  \op{u}=\mf{u}_{1}^{\prime}(k_0) \;. 
\label{definition parametres zetar vepsr}
\enq
Finally, let 
\beq
t_r^{(0)}(x)  \, = \,  \f{ \zeta_1 \veps_1 }{ \zeta_r \veps_r } \Big( \f{ \xi_r }{ \xi_1 } \Big)^2 x 
\enq
and for $C>0$, $\eta>0$, consider the domain 
\beq
\mc{D}^{(\e{eff})}_{ \eta  } \, = \, \bigg\{   \bs{x} \in \pl{r=1}{\ell} \R^{n_{r}} \; : \; |x_1^{(1)}| \, \leq \, C \eta  \; , \; 
\forall (r,a )\in \mc{M} \; : \; \big| t_r^{(0)}(x_1^{(1)})-x_a^{(r)}  \big| \, \leq  \, \xi_{r}^2 \eta   \bigg\}   \;. 
\label{ecriture domaine D eff eta prime the rectification des z ups}
\enq
Then, there exists $\mf{x}_0>0$, $\eta^{\prime}>0$ and 
\begin{itemize}
 
 \item smooth functions $\mf{f}_{\ups}$ on $ \intoo{ - \mf{x}_0 }{ \mf{x}_0 }\times \mc{D}^{(\e{eff})}_{\eta^{\prime}}$  satisfying $\mf{f}_{\ups}(\mf{x}; \bs{x})\, = \, 1 + \e{O}\Big( \norm{\bs{x}}+|\mf{x}|\Big)$,
 
\item a smooth diffeomorphism $\Psi_{\mf{x}}:\mc{D}^{(\e{eff})}_{\eta^{\prime}} \tend  \Psi\Big( \mc{D}^{(\e{eff})}_{\eta^{\prime}} \Big)$ satisfying $D_{\bs{0}} \Psi \, = \, \e{id} + \mf{x} \op{N}_{\Psi}$ with 
$\norm{ \op{N}_{\Psi} } \leq C$, for some $\mf{x}$-independent $C>0$, 
 
\end{itemize}
such that
\beq
\mf{x}+\mf{z}_{\ups} \circ \Psi_{\mf{x}}(\bs{x})  \; = \; \mf{f}_{\ups}(\mf{x};\bs{x}) \cdot \Big(\mf{x} + \tilde{z}_{\ups}(\bs{x}) \Big) \;, 
\label{ecriture fondamentale de rectification}
\enq
and $\Psi_{\mf{x}}\Big( \mc{D}^{(\e{eff})}_{\eta^{\prime}} \Big) \subset \msc{J}_{\e{tot}}$ contains a $\mf{x}$-independent open neighbourhood of $\bs{t}(k_0)$. 
Furthermore, the map 
\beq
(\mf{x},\bs{x}) \mapsto \Psi_{\mf{x}}(\bs{x})
\enq
is smooth on $\intoo{ - \mf{x}_0 }{ \mf{x}_0 }\times  \mc{D}^{(\e{eff})}_{\eta^{\prime}}$. 

\end{prop}

\Proof

\subsubsection*{$\bullet$ Canonical form of  $\mf{z}_{\ups}$}

Let  $I_{\ell}\subset \e{Int}(\msc{J}_{\ell})$ be  a segment such that $t_{\ell}(k_0) \in  \e{Int}(I_{\ell})$. Since $\Dp{k}\mf{w}_{\ups}^{(\ell)}\big( k; t_{\ell}(k_0) \big) \, = \, \mf{u}_{\ell}(k)+\ups \op{v}\not=0$ 
on $I_{\ell}$, $k\mapsto \mf{w}_{\ups}^{(\ell)}\big( k; t_{\ell}(k_0) \big)$  is strictly monotone on $I_{\ell}$ and thus 
admits $t_{\ell}(k_0)$ as its unique zero on $I_{\ell}$. Furthermore, this also implies that there exists 
\beq
c>0 \quad \e{such}\, \e{that} \quad  \intff{-c}{c} \subset \mf{w}_{\ups}^{(\ell)}\big( I_{\ell} ; t(k_0) \big) \;. 
\label{introduction parametre c}
\enq
Given $\eps>0$, set 
\beq
\mc{B}_{\eps}\Big( \bs{t}_{[\ell, n_{\ell}]}(k_0) \Big) \; = \; \bigg\{ \bs{p}_{[\ell, n_{\ell}]} \in \prod_{r=1}^{\ell} \msc{J}_r^{n_r-\de_{r,\ell}} \; : \; 
\; \;   \big|p_a^{(r)} - t_r(k_0) \big| < \eps \;  , \; \;  \forall (a,r) \in \mc{M}_{[\ell,n_{\ell}]}  \bigg\}  \;,  
\enq
where $\mc{M}_{[\ell,n_{\ell}]}=\mc{M}\setminus \{ (\ell,n_{\ell}) \}$ has been introduced in \eqref{definition ensemble M elle n ell enleve} while  $\bs{p}_{[\ell, n_{\ell}]}$ 
is as defined in \eqref{definition vecteur p avec composantes omises}.

Given the function $\mf{z}_{\ups}$ of $\sul{r=1}{\ell}n_{r}$ variables, it is of use to agree to denote its analogue on $\mc{B}_{\eps}\Big( \bs{t}_{[\ell, n_{\ell}]}(k_0) \Big) $, \textit{viz}. when the last variable is deleted, as :
\beq
\mf{z}_{\ups}^{([\ell, n_{\ell}])}( \bs{p}_{[\ell, n_{\ell}]} ) \; \equiv \;   \, - \,  \sul{ (r,a) \in \mc{M}_{[\ell,n_{\ell}]} }{  } \zeta_r  \mf{w}_{\ups}^{(r)}\big(  p_a^{(r)} ; t_r(k_0) \big) \quad , \qquad \ups \in \{ \pm \} \;.
\enq

Let $\bs{t}_{[\ell, n_{\ell}]}(k_0)$ be as defined in \eqref {ecriture definition vecteur tk0} and let $V_{[\ell,n_{\ell}]}$ be any open neighbourhood of $\bs{t}_{[\ell, n_{\ell}]}(k_0)$ such that $V_{[\ell,n_{\ell}]} \subset \mc{B}_{\eps}\Big( \bs{t}_{[\ell, n_{\ell}]}(k_0) \Big)$.
Then, 
\beq
\mf{x}+\mf{z}_{\ups}^{([\ell, n_{\ell}])}( \bs{p}_{[\ell, n_{\ell}]} )  \; = \; \e{O}\Big( |\mf{x}|+\eps \Big)   \qquad \e{uniformly} \; \e{on} \;    V_{[\ell,n_{\ell}]}  \;. 
\label{ecriture eqn  pour zero dans fct Heaviside} 
\enq
Thus, provided that $|\mf{x}|$ and $\eps$ are taken small enough, one has that, for any 
\beq
\bs{p}_{[\ell, n_{\ell}]} \in V_{[\ell,n_{\ell}]}, \quad \e{it} \; \e{holds} \quad 
\mf{x} \, + \, \mf{z}_{\ups}\big(\bs{p}_{[\ell, n_{\ell}]} \big) \in \intff{- \tf{c}{2} }{ \tf{c}{2} } \;, 
\enq
with $c>0$ as appearing in \eqref{introduction parametre c}. Then, the monotonicity of 
$k \mapsto \mf{w}_{\ups}^{(\ell)}\big(k; t_{\ell}(k_0) \big) $ on $I_{\ell}$ ensures that there exists a unique $\mc{V}_{\ups}\big( \bs{p}_{[\ell, n_{\ell}]}  \big) \in I_{\ell}$
such that 
\beq
\mf{x}+\mf{z}_{\ups}\big( \bs{p}_{\ups} \big) \, = \, 0 \quad \e{with} \quad  \bs{p}_{\ups}\; = \; \Big( \bs{p}_{[\ell, n_{\ell}]} , \mc{V}_{\ups}\big( \bs{p}_{[\ell, n_{\ell}]} \big) \Big)
\qquad \e{for}\; \e{any} \quad   \bs{p}_{[\ell, n_{\ell}]} \in V_{[\ell,n_{\ell}]} \;. 
\label{definition vecteur p ups}
\enq
Here, for simplicity, the $\mf{x}$ dependence of $\mc{V}_{\ups}$ has been kept implicit. 
The function $\mc{V}_{\ups}$ takes the explicit form 
\beq
\mc{V}_{\ups}\big(  \bs{p}_{[\ell, n_{\ell}]} \big) \; = \; \Big( \mf{w}_{\ups}^{(\ell)}\Big)^{-1}\bigg(  \f{ \mf{x} \, +\, \mf{z}_{\ups}^{([\ell, n_{\ell}])}\big( \bs{p}_{[\ell, n_{\ell}]} \big) }{ \zeta_{\ell} }; t_{\ell}(k_0) \bigg) \;. 
\label{ecriture forme explicite pour zero Vups}
\enq
By construction, the function   $\mc{V}_{\ups}$ is smooth on $V_{[\ell,n_{\ell}]}$ as a composition of smooth functions.

By the Malgrange preparation Theorem \ref{Theorem Malgrange preparation theorem} applied to the function 
\beq
\big( \mf{x}, \bs{p} \big)  \; \mapsto  \; \mf{x}+\mf{z}_{\ups}\big( \bs{p} \big)
\enq
of $\ov{\bs{n}}_{\ell}+1$ variables, $\ov{\bs{n}}_{\ell}=\sul{r=1}{\ell} n_r$, at the point $(0,\bs{t}(k_0))$,  one concludes that there exist 
\begin{itemize}
\item $\mf{x}_0>0$;

\item an open neighbourhood $V_{[\ell,n_{\ell}]}^{\prime} \subset V_{[\ell,n_{\ell}]}$ of  $\bs{t}_{[\ell, n_{\ell}]}(k_0)$,
\item an open neighbourhood  $I^{\prime}_{\ell}$ of $t_{\ell}(k_0)$,  
\item a smooth, non-vanishing, function $h_{\ups}$ on $ \intoo{-\mf{x}_0}{ \mf{x}_0 } \times V_{[\ell,n_{\ell}]}^{\prime} \times I^{\prime}_{\ell}$ , 
\end{itemize}
such that, for $\bs{p}\in V_{[\ell,n_{\ell}]}^{\prime} \times I^{\prime}_{\ell}$  and $|\mf{x}| < \mf{x}_0$, it holds 
\beq
\mf{x}   +   \mf{z}_{\ups}\big(\bs{p} \big) \, = \, \vsg_{\ups;\ell} \cdot  \Big[ p_{n_{\ell}}^{(\ell)} - \mc{V}_{\ups}\big(  \bs{p}_{[\ell, n_{\ell}]}   \big) \Big] \cdot  h_{\ups}(\mf{x};\bs{p}) 
\qquad \e{with} \quad \vsg_{\ups;\ell} \, =\, -\zeta_{\ell} \, \e{sgn}\Big( \mf{u}_{1}^{\prime}(k_0) + \ups \op{v} \Big) \;. 
\label{ecriture factorisation x plus z pm}
\enq
Furthermore, given $\bs{p}_{\ups}$ as in \eqref{definition vecteur p ups},  partial differentiation of the relation \eqref{ecriture factorisation x plus z pm} in respect to $p_{n_{\ell}}^{(\ell)}$ allows one to conclude that  
\beq
 h_{\ups}\big( \mf{x}; \bs{p}_{\ups} \big) \; = \;  \Big| \ups  \op{v}  +  \mf{u}_{\ell}^{\prime}\Big( \mc{V}_{\ups}\big( \bs{p}_{[\ell, n_{\ell}]} \big) \Big) \Big| \, > \, 0\;. 
\label{ecriture simplification locale de hups}
\enq
Let $  \bs{v}  \, = \, \bs{p}_{[\ell, n_{\ell}]}  \, - \,\bs{t}_{[\ell, n_{\ell}]}(k_0) $. The explicit expression for $\mc{V}_{\ups}$ given in \eqref{ecriture forme explicite pour zero Vups}
warrants that one has the small $\norm{   \bs{v} }$ expansion
\beq
\mc{V}_{\ups}\big( \bs{p}_{[\ell, n_{\ell}]}  \big) \; = \; \mc{V}_{\ups;0} \, + \, L \cdot  \mc{V}_{\ups;1}\,+ \, \mc{V}_{\ups;2}\big(L,Q\big) \, + \, \e{O}\big( \norm{\bs{v}}^3 \big)  \;.  
\label{ecriture developpement nu ups}
\enq
There, I have set
\beq
L \, = \, \f{1}{\zeta_{\ell} } \, \big( \bs{y}, \bs{v} \big) \qquad \e{with} \quad 		 
\bs{y}^{\op{t}} \, = \, \Big( \big( \bs{y}^{(1)} \big)^{\op{t}} , \dots,  \big( \bs{y}^{(\ell)} \big)^{\op{t}}\Big) \qquad \e{and} \qquad  \big( \bs{y}^{(r)} \big)^{\op{t}}\, = \,   \zeta_{r}  \,   \big( 1,\dots, 1 \big)  \in \R^{ n_r  -  \de_{r,\ell} } \;,
\label{definition L et Q et vecteur y}
\enq
while 
\beq
 Q \, = \, \bigg(\bs{v}, \left(\ba{ccc}   \ddots & 0 &  0  \\ 
 0   &   \tfrac{ \zeta_{r} }{ \zeta_{\ell} } \mf{u}^{\prime\prime}_{r}(t_r(k_0)) \op{I}_{n_r-\de_{\ell, r} } & 0 \\ 0 & 0 & \ddots    \ea \right) \bs{v} \bigg)   \;. 
\enq
Above,  $\op{I}_n$ denotes the identity matrix on $\R^n$,  $(\cdot, \cdot)$  denotes the canonical scalar product on $\pl{r=1}{\ell}\R^{ n_{r}-\de_{r,\ell}}$ and $^{\op{t}}$ denotes the transposition. 
The first three coefficients of the expansion \eqref{ecriture developpement nu ups} are given by 
\beq
\mc{V}_{\ups;0} \, = \, \Big(\mf{w}_{\ups}^{(\ell)} \Big)^{-1}\big(   \mf{x} ; t_{\ell}(k_0) \big)  \, = \,  t_{\ell}(k_0) \, + \, \f{ \tf{ \mf{x} }{ \zeta_{\ell}} }{ \ups \op{v}  + \mf{u}_1^{\prime}\big( k_0 \big)  } \, + \, \e{O}\big( \mf{x}^2 \big) \;, 
\enq
\beq
\mc{V}_{\ups;1} \, = \, - \f{ \ups \op{v}  + \mf{u}_1^{\prime}\big( k_0 \big)  }{  \ups \op{v}  + \mf{u}_{\ell}^{\prime}\big(\mc{V}_{\ups}^{(0)}\big)   } \, = \,  
 -1 + \f{ \mf{x} \cdot  \mf{u}_{\ell}^{\prime\prime}\big( t_{\ell}(k_0) \big) }{ \zeta_{\ell} \big[ \ups \op{v}  + \mf{u}_1^{\prime}\big( k_0 \big) \big]^2   } \, + \, \e{O}\big( \mf{x}^2 \big) 
\enq
and 
\bem
\mc{V}_{\ups;2}\big(L,Q\big) \, = \, - \f{     Q  }{ 2 \Big( \ups \op{v} + \mf{u}_{\ell}^{\prime}\big( \mc{V}_{\ups}^{(0)} \big) \,  \Big)  }
 -    \mf{u}_{\ell}^{\prime\prime}\big(\mc{V}_{\ups}^{(0)} \big)\cdot \f{  L^2 \, \big[ \ups \op{v}  + \mf{u}_1^{\prime}\big( k_0 \big) \big]^2 }{  2 \big[ \ups \op{v} + \mf{u}_{\ell}^{\prime}\big( \mc{V}_{\ups}^{(0)} \big) \big]^3   } \\
\, = \, -  \f{  Q  +  L^2  \cdot   \mf{u}_{\ell}^{\prime\prime}\big( t_{\ell}(k_0) \big)    }{ 2 \Big( \ups \op{v}  + \mf{u}_1^{\prime}\big( k_0 \big) \Big)  }  \; + \; \e{O}\Big(\mf{x} \cdot \big[ |Q| +L^2 \big] \Big)\; . 
\end{multline}

These expansions ensure that 
\beq
\mc{V}_{-;0}-\mc{V}_{+;0} \; = \; - \f{ 2 \mf{x} \, \op{v} \cdot \zeta_{\ell}^{-1} }{ \op{v}^2-\big( \mf{u}_1^{\prime}(k_0) \big)^2 } \; + \; \e{O}\big(\mf{x}^2\big) \; , \qquad
\mc{V}_{-;1}-\mc{V}_{+;1}  \; = \;     4  \op{v}   \zeta_{\ell}^{-1}  \mf{x} \cdot \f{   \mf{u}_{\ell}^{\prime\prime}\big( t_{\ell}(k_0) \big) \cdot  \mf{u}_1^{\prime} (k_0) }
{ \big[ \op{v}^2 - \big( \mf{u}_1^{\prime} (k_0) \big)^2 \big]^2 } \; + \; \e{O}\big(\mf{x}^2\big)
\enq
and 
\beq
\mc{V}_{-;2}\big(L,Q\big)  \, - \, \mc{V}_{+;2}\big(L,Q\big) \; = \;   \f{ \op{v}      }{  \op{v} ^2 - \big( \mf{u}_1^{\prime} (k_0) \big)^2   }\cdot \Big( Q + L^2 \cdot  \mf{u}_{\ell}^{\prime\prime}\big( t_{\ell}(k_0) \big) \Big) 
\; + \; \e{O}\Big(\mf{x} \cdot \big[ |Q| + L^2 \big] \Big)\; . 
\enq
Thus, provided that $|\mf{x}|$ is small enough, there exist smooth functions $\mc{U}_1, \mc{U}_2$ on $V_{[\ell,n_{\ell}]}^{\prime}$ such that 
\beq
\mc{V}_{-}\big(  \bs{p}_{[\ell, n_{\ell}]}  \big)\, - \, \mc{V}_{+}\big(  \bs{p}_{[\ell, n_{\ell}]}  \big) \; = \; \mf{x} \, \mc{U}_1\big( \bs{p}_{[\ell, n_{\ell}]} \big) \, + \, \mc{U}_2\big( \bs{p}_{[\ell, n_{\ell}]}   \big)  \;. 
\enq
 The functions $\mc{U}_{a}$, which may depend on $\mf{x}$,  are such that 
\beq
\mc{U}_1\big( \bs{p}_{[\ell, n_{\ell}]}  \big) \; = \; \f{ -2\op{v} \zeta_{\ell}  }{ \op{v} ^2 \, - \, \big( \mf{u}_1^{\prime}(k_0) \big)^2  } \; + \;\e{O}\big( |\mf{x}| + \norm{ \bs{v} } \big) \; , \quad
\mc{U}_2\big(\bs{p}_{[\ell, n_{\ell}]} \big) \; = \; \f{ - \zeta_{\ell} \, \op{v}  \, \big(\bs{v},\op{M}  \bs{v}  \big)  }{  \op{v}^2 -  \big( \mf{u}_1^{\prime}(k_0) \big)^2   }    \; + \;\e{O}\big(  \norm{\bs{v}}^3 \big) \;, 
\label{ecriture des dvpmnts de U1 et U2 en t de k0}
\enq
where I remind that $\bs{v} =  \bs{p}_{[\ell, n_{\ell}]} -  \bs{t}_{[\ell, n_{\ell}]}(k_0) $,  $(\cdot , \cdot )$ is the canonical
scalar product on $\pl{r=1}{\ell}\R^{ n_{r}-\de_{r,\ell}}$  and the matrix $\op{M}$ takes the form
\beq
\op{M} \; = \; - \zeta_{\ell} \mf{u}^{\prime\prime}_{\ell}(t_{\ell}(k_0)) \;  \bs{y}\cdot \bs{y}^{\op{t}}  \, + \,  \left(\ba{ccc}   \ddots & 0 &  0  \\ 
 0   &   - \zeta_{r} \mf{u}^{\prime\prime}_{r}(t_r(k_0)) \op{I}_{n_r-\de_{\ell, r} } & 0 \\ 0 & 0 & \ddots    \ea \right)
\label{definition matrice M}
\enq
with $\bs{y}$ as given by \eqref{definition L et Q et vecteur y}.  Since $\op{M}$ is symmetric, it is diagonalisable and has real eigenvalues. 
For further utility, one still needs to establish that these are non-vanishing. For that purpose, it is enough to show that $\op{M}$ has a non-zero determinant. 

\noindent Upon factorising the diagonal part, one gets that
\beq
\det\big[ \op{M}  \big] \; = \; \pl{r=1}{\ell}\Big\{ - \zeta_{r} \mf{u}^{\prime\prime}_{r}(t_r(k_0))  \Big\}^{ n_{r}-\de_{r,\ell}} \cdot \det\big[ \e{id} + \bs{w}\cdot \bs{y}^{\op{t}} \big]
\enq
with $\bs{y}$ as in \eqref{definition L et Q et vecteur y}, 
\beq
 \bs{w}^{\op{t}} \, = \, \Big( \big( \bs{w}^{(1)} \big)^{\op{t}} , \dots,  \big( \bs{w}^{(\ell)} \big)^{\op{t}}\Big)   
\qquad \e{and} \qquad \big( \bs{w}^{(r)} \big)^{\op{t}}\, = \,   \zeta_{\ell}  \cdot \f{ \mf{u}^{\prime\prime}_{\ell}(t_{\ell}(k_0))  }{ \mf{u}^{\prime\prime}_{r}(t_r(k_0))   }   \cdot    \big( 1,\dots, 1 \big)  \in \R^{ n_r  -  \de_{r,\ell} }    \;. 
\enq
The determinant can be computed explicitly and, upon using the relation $\tf{ \mf{u}^{\prime\prime}_{\ell}(t_{\ell}(k_0))  }{ \mf{u}^{\prime\prime}_{r}(t_r(k_0))   }  = \tf{ t_{r}^{\prime}(k_0) }{ t_{\ell}^{\prime}(k_0) }$,
which follows from a differentiation of $ \mf{u}^{\prime}_{\ell}(t_{\ell}(k))  = \mf{u}^{\prime}_{r}(t_r(k)) $ at $k=k_0$, one eventually obtains that 
\beq
\det\big[ \op{M}  \big] \; = \; - \pl{r=1}{\ell}\Big\{ - \zeta_{r} \mf{u}^{\prime\prime}_{r}(t_r(k_0))  \Big\}^{ n_{r} } \cdot \f{ \mc{P}^{\prime}(k_0) }{ \mf{u}_1^{\prime \prime}(k_0) } \, \not = 0 
\enq
since, by hypothesis \eqref{ecriture hypothese non vanishing impusion macro}, $\mc{P}^{\prime}(k_0)\not= 0$. 

The above ensures that
\beq
\f{ \mc{U}_2\big( \bs{t}_{[\ell, n_{\ell}]}(k_0) \big) }{ \mc{U}_1\big( \bs{t}_{[\ell, n_{\ell}]}(k_0) \big) } \, = \, 0 \; \; , \quad
D_{\bs{t}_{[\ell, n_{\ell}]}(k_0)}\bigg( \f{ \mc{U}_2  }{ \mc{U}_1 } \bigg) \, = \, 0 \quad \e{and} \quad 
D_{\bs{t}_{[\ell, n_{\ell}]}(k_0)}^{2} \bigg( \f{ \mc{U}_2  }{ \mc{U}_1 } \bigg)  (\bs{s},\bs{s}^{\prime})\, = \, \big[ \mf{c}(\mf{x}) \big]^2 \cdot \big(\bs{s}, \op{M} \bs{s}^{\prime} \big)
\enq
for $\bs{s},\bs{s}^{\prime} \in  \pl{r=1}{\ell}\R^{ n_{r}-\de_{r,\ell}}$ and where $\mf{c}(\mf{x})=1+\e{O}(\mf{x})$ is smooth in the neighbourhood of $\mf{x}=0$.

Thus, by virtue of the Morse lemma, Theorem \ref{Theorem Morse Lemma}, followed by a dilatation of variables by $\mf{c}(\mf{x})$, one infers that there exists 
\begin{itemize}
\item an open neighbourhood $V^{\prime \prime }_{[\ell, n_{\ell}]}\subset V^{\prime}_{[\ell, n_{\ell}]}$ of $\bs{t}_{[\ell, n_{\ell}]}(k_0)$,
\item an open neighbourhood $W_{\bs{\phi}}$ of $\bs{0}$ in $\pl{r=1}{\ell}\R^{ n_{r}-\de_{r,\ell}}$,
\item a smooth diffeormorphism $\bs{\phi}_{\mf{x}}: \, W_{\bs{\phi}} \; \tend \;  V^{\prime\prime}_{[\ell, n_{\ell}]} $, with $\bs{\phi}_{\mf{x}}(\bs{0})=\bs{t}_{ [\ell, n_{\ell}] }(k_0)$
 
 \end{itemize}
such that 
\beq
\f{ \mc{U}_2\big(  \bs{\phi}_{\mf{x}}(\bs{v})   \big) }{ \mc{U}_1\big( \bs{\phi}_{\mf{x}}(\bs{v})  \big) }  \; = \; \big(\bs{v}, \op{M} \bs{v} \big)\;. 
\label{ecriture rectification de Morse U2 sur U1}
\enq
In particular, one readily infers from \eqref{ecriture rectification de Morse U2 sur U1} that $  \Big(D_{\bs{0}}\bs{\phi}_{\mf{x}} \Big)^{\op{t}}\cdot \op{M} \cdot D_{\bs{0}}\bs{\phi}_{\mf{x}} \, = \,  2 \, \op{M}$. 

It is clear that the size of all the domains appearing above may be taken to be $\mf{x}$-independent, at least provided that $|\mf{x}|$ is small enough, say $|\mf{x}|<\mf{x}_0$, and that then 
\beq
(\mf{x}, \bs{v} ) \; \mapsto \; \bs{\phi}_{\mf{x}}(\bs{v}) \;, 
\enq
is smooth on $\intoo{ - \mf{x}_0 }{ \mf{x}_0 } \times W_{\bs{\phi}}$. Clearly, upon adjusting the parameters, one may take $\mf{x}_0$ as introduced earlier on.

\subsubsection*{$\bullet$ Canonical form of  $\tilde{z}_{\ups}$} 

The very same reasoning applied to the function $\tilde{z}_{\ups}(\bs{x})$, as defined in \eqref {definition fonction z tilde ups}, ensures that there exist
\begin{itemize}

 \item  an open neighbourhood $V^{(0)}_{[\ell,n_{\ell}]}$ of $ \bs{0}   \in  \pl{r=1}{\ell}\R^{ n_{r}-\de_{r,\ell}}$,

 \item a segment $I_{\ell}^{(0)}$ containing an open neighbourhood of $0\in \R$ ,

 \item a smooth, non-vanishing,  function  $h_{\ups}^{(0)}$ on $ \intoo{ - \mf{x}_0  }{  \mf{x}_0  } \times V^{(0)}_{[\ell,n_{\ell}]} \times I_{\ell}^{(0)}$,
\item a smooth function  $\mc{V}_{\ups}^{(0)}$ on $\intoo{ - \mf{x}_0  }{  \mf{x}_0  } \times V^{(0)}_{[\ell,n_{\ell}]}$,
\end{itemize}
such that 
\beq
\mf{x}   +  \tilde{z}_{\ups}(\bs{x}) \, = \, \vsg_{\ups;\ell} \cdot  \Big[ x_{n_{\ell}}^{(\ell)} - \mc{V}_{\ups}^{(0)}\big(  \bs{x}_{[\ell, n_{\ell}]}   \big) \Big] \cdot  h_{\ups}^{(0)}(\mf{x}; \bs{x}) 
\qquad \e{with} \quad \vsg_{\ups;\ell} \, =\, -\zeta_{\ell} \, \e{sgn}\big( \mf{u}_{1}^{\prime}(k_0) + \ups \op{v} \big) \;. 
\label{ecriture factorisation x plus z pm modele effectif}
\enq
Here, again, I kept the $\mf{x}$-dependence of $ \mc{V}_{\ups}^{(0)}$ implicit. 

Note that here, $\vsg_{\ups;\ell}$ is exactly as defined in \eqref{ecriture factorisation x plus z pm} owing to the very choice of the parameters 
$\veps_a$, $\xi_a$, $\op{u}$, defining the effective function $\tilde{z}_{\ups}(\bs{x})$. The function $h_{\ups}^{(0)}$ enjoys the identity
\beq
 h_{\ups}^{(0)}\big( \mf{x}; \bs{x}_{\ups} \big) \; = \;  \Big| \ups  \op{v}  +  \mf{h}_{\ell}^{\prime}\Big( \mc{V}_{\ups}^{(0)}\big( \bs{x}_{[\ell, n_{\ell}]} \big) \Big) \Big| \, > \, 0
\quad \e{with} \quad 
 \bs{x}_{\ups} \, = \, \Big( \bs{x}_{[\ell, n_{\ell}]} ,   \mc{V}_{\ups}^{(0)}\big( \bs{x}_{[\ell, n_{\ell}]} \big) \Big) \;. 
\label{ecriture simplification locale de hups effectif}
\enq
Furthermore, there exist two smooth functions on $V^{(0)}_{[\ell,n_{\ell}]}$ such that 
\beq
\mc{V}_{-}^{(0)}\big(  \bs{x}_{[\ell, n_{\ell}]}  \big)\, - \, \mc{V}_{+}^{(0)}\big(  \bs{x}_{[\ell, n_{\ell}]}  \big) \; = \; \mf{x} \, \mc{U}_1^{(0)}\big( \bs{x}_{[\ell, n_{\ell}]} \big) \, + \, \mc{U}_2^{(0)}\big( \bs{x}_{[\ell, n_{\ell}]}   \big)    
\enq
and satisfying 
\beqa
\mc{U}_1^{(0)}\big( \bs{x}_{[\ell, n_{\ell}]}  \big) & = & \f{ -2\op{v} \zeta_{\ell}  }{ \op{v} ^2 \, - \, \big( \mf{u}_1^{\prime}(k_0) \big)^2  } \; + \;\e{O}\big( |\mf{x}|+\norm{ \bs{x}_{[\ell, n_{\ell}]} } \big) \label{ecriture des dvpmnts de U1 0 et U20 en 0} \\
\mc{U}_2^{(0)}\big(\bs{x}_{[\ell, n_{\ell}]} \big) & = & \f{ - \zeta_{\ell} \, \op{v}  }{  \op{v}^2 -  \big( \mf{u}_1^{\prime}(k_0) \big)^2   }  \,  \big( \bs{x}_{[\ell, n_{\ell}]} ,\op{M} \, \bs{x}_{[\ell, n_{\ell}]}  \big)    
			  \; + \;\e{O}\big(  \norm{\bs{x}_{[\ell, n_{\ell}]}}^3 \big) \;. 
\eeqa
Following the above reasoning, and re-adjusting the domains $W_{\bs{\phi}}, V^{\prime\prime}_{[\ell, n_{\ell}]}$ appearing above if necessary, one eventually concludes that there exists 
 a smooth diffeormorphism $\bs{\phi}^{(0)}_{\mf{x}}: \, W_{\bs{\phi}} \; \tend \;  V^{(0)}_{[\ell, n_{\ell}]} $ such that 
\beq
\f{ \mc{U}_2^{(0)}\big(  \bs{\phi}^{(0)}_{\mf{x}}(\bs{v})   \big) }{ \mc{U}_1^{(0)}\big( \bs{\phi}^{(0)}_{\mf{x}}(\bs{v})  \big) }  \; = \; \big(\bs{v}, \op{M} \bs{v} \big) \quad \e{and} \quad 
		      \left\{ \ba{ccc}  D_{\bs{0}}\bs{\phi}^{(0)}_{\mf{x}} & = & D_{\bs{0}}\bs{\phi}_{\mf{x}}  \\ 
							\bs{\phi}^{(0)}_{\mf{x}}\big( \bs{0} \big) & = & \bs{0}    \ea \right. \;. 
\enq
Likewise to the previous situation, $(\mf{x}, \bs{v}) \mapsto \bs{\phi}^{(0)}_{\mf{x}}(\bs{v})$ is smooth on  $\intoo{ - \mf{x}_0 }{ \mf{x}_0 } \times W_{\bs{\phi}}$. 

I stress that  the open neighbourhood  $W_{\bs{\phi}}$ appearing above coincides exactly with the domain of the diffeomorphism $\bs{\phi}_{\mf{x}}$ introduces earlier on. 
Also, I should comment relatively to the possibility of choosing $\bs{\phi}^{(0)}_{\mf{x}}$ such that $D_{\bs{0}}\bs{\phi}^{(0)}_{\mf{x}} \, = \, D_{\bs{0}}\bs{\phi}_{\mf{x}}$. 
Just as for the case of $\bs{\phi}_{\mf{x}}$, one deduces that any Morse function $\bs{\phi}^{(0)}_{\mf{x}}$ rectifying $\tf{ \mc{U}_2^{(0)} }{  \mc{U}_1^{(0)} }$ has to satisfy
$ \Big(D_{\bs{0}}\bs{\phi}^{(0)}_{\mf{x}} \Big)^{\op{t}} \cdot \op{M}  \cdot D_{\bs{0}}\bs{\phi}^{(0)}_{\mf{x}} \, = \,  2 \op{M}$. 
\textit{A priori} this equation has a space of solutions that is isomorphic to $SO(p,q)$ where 
$(p,q)$ is the signature of $\op{M}$. However, upon looking at the proof of the Morse Lemma, one constructs 
a Morse function from a given choice of a solution to this equation. Thus, when constructing $\bs{\phi}^{(0)}_{\mf{x}}$, the latter can always be chosen
so that $D_{\bs{0}}\bs{\phi}^{(0)}_{\mf{x}} \, = \, D_{ \bs{0} }\bs{\phi}_{\mf{x}}$.

\subsubsection*{$\bullet$ The \textit{per-se} rectification}

With all the ingredients being introduced, one may define the smooth diffeomorphism 
\beq
\Phi_{\mf{x}} \; : \; \left\{ \ba{ccc}   V^{(0)}_{[\ell,n_{\ell}]} & \tend &   V^{\prime\prime}_{[\ell,n_{\ell}]}   \vspace{3mm} \\ 
				\bs{x}_{[\ell,n_{\ell}]}  & \mapsto & \bs{\phi}_{\mf{x}} \circ \Big( \bs{\phi}^{(0)}_{\mf{x}} \Big)^{-1} \big( \bs{x}_{[\ell,n_{\ell}]} \big) \ea  \right. 
\quad \e{which} \; \e{satisfies} \quad 
\f{\mc{U}_2}{\mc{U}_1}\circ \Phi_{\mf{x}} \, = \, \f{ \mc{U}_2^{(0)} }{ \mc{U}_1^{(0)}  } \;. 
\enq
One is now in position to introduce the smooth map
\beq
\Psi_{\mf{x}} \; : \; \left\{ \ba{ccc}   V^{(0)}_{[\ell,n_{\ell}]} \times I_{\ell}^{(0)} & \tend & \Psi_{\mf{x}}\Big(  V^{(0)}_{[\ell,n_{\ell}]} \times I_{\ell}^{(0)} \Big)     \\ 
				\bs{x}   & \mapsto & \Psi_{\mf{x}} (\bs{x}) \, = \, \bigg( \Phi_{\mf{x}}(\bs{x}_{[\ell,n_{\ell}]} ), 
			   \mc{V}_{-}\circ\Phi_{\mf{x}} (\bs{x}_{[\ell,n_{\ell}]} )  \, + \,  \Big( x_{n_{\ell}}^{(\ell)}  -   \mc{V}_{-}^{(0)}(\bs{x}_{[\ell,n_{\ell}]} )  \Big) 
			   \cdot \f{\mc{U}_1\circ \Phi_{\mf{x}} (\bs{x}_{[\ell,n_{\ell}]} ) }{\mc{U}_1^{(0)}(\bs{x}_{[\ell,n_{\ell}]} )  } 
				      \bigg) \ea  \right. \;. 
\enq
I first establish that $\Psi_{\mf{x}}$ is a diffeomorphism.  

Indeed, from its very construction, one has that $\Phi_{\mf{x}}(\bs{0})=\bs{t}_{[\ell,n_{\ell}]}(k_0)$ and that $D_{ \bs{0} } \Phi_{\mf{x}} \, = \, \op{I}_{ \ov{\bs{n}}_{\ell}-1}$, with  $\ov{\bs{n}}_{\ell}=\sul{r=1}{\ell}n_r$. 
Thence, the expansions \eqref{ecriture des dvpmnts de U1 et U2 en t de k0} and \eqref{ecriture des dvpmnts de U1 0 et U20 en 0} ensure that  
\beq
\left. \f{\mc{U}_1\circ \Phi_{\mf{x}} (\bs{x}_{[\ell,n_{\ell}]} ) }{\mc{U}_1^{(0)}(\bs{x}_{[\ell,n_{\ell}]} )  } \right|_{ \bs{x}_{[\ell,n_{\ell}]} = \bs{0} } \; = \; 
1+\e{O}\big( \mf{x} \big) \;. 
\label{ecriture valeur en zero ration U1 sur U1 effectif}
\enq
Furthermore, since $\mf{z}_{\ups}^{([\ell,n_{\ell}])} \big( \bs{t}_{[\ell,n_{\ell}]}(k_0)  \big)=0$, one has 
\beq
 \mc{V}_{\ups} \Big( \bs{t}_{[\ell,n_{\ell}]}(k_0)  \Big) \; = \; \big( \mf{w}^{(\ell)}_{\ups} \big)^{-1}\Big( \f{ \mf{x} }{ \zeta_{\ell}} ; t_{\ell}(k_0)  \Big) \, = \, t_{\ell}(k_0)+\e{O}(\mf{x})
\label{ecriture dvpmt local zero V ups}
\enq
and, similarly, $\mc{V}_{\ups}^{(0)} ( \bs{0}  ) \; = \; \e{O}(\mf{x})$. All of the above put together entails that 
\beq
{\Psi_{\mf{x}}(\bs{x})}_{\mid_{\bs{x}=0}} \; = \; \Big(  \bs{t}_{[\ell,n_{\ell}]}(k_0) , t_{\ell}(k_0) + \e{O}(\mf{x}) \Big) \, = \, \bs{t}(k_0) \, + \, \big( \underbrace{ \bs{0} }_{ \in \R^{ \ov{\bs{n}}_{\ell}-1 } } , \e{O}(\mf{x}) \big) \;. 
\label{ecriture valeur Psi en origine}
\enq
Furthermore, denote by $\big[\Psi_{\mf{x}}(\bs{x})\big]_{n_{\ell}}^{(\ell)} $ the ultimate scalar entry of $\Psi_{\mf{x}}(\bs{x})$. Then, the expansion \eqref{ecriture developpement nu ups}
and an analogous one for $\mc{V}_{-}^{(0)}(\bs{x}_{[\ell,n_{\ell}]})$, yields for $\bs{u}\in \R^{\ov{\bs{n}}_{\ell}-1}$ and $s\in \R$
\beq
D_{\bs{0}} \Big( \big[\Psi_{\mf{x}}(\bs{x})\big]_{n_{\ell}}^{(\ell)}  \Big) \cdot \big(\bs{u},s \big) \, = \, \zeta_{\ell}^{-1}\, \mc{V}_{-;1} \,   \big(D_{\bs{0}}\Phi_{\mf{x}} \cdot \bs{u},\bs{y} \big)
+\Big(s-\zeta_{\ell}^{-1}\, \mc{V}_{-;1}^{(0)} \,  \big(  \bs{u},\bs{y} \big) \Big) \, \f{\mc{U}_1\circ \Phi_{\mf{x}} (\bs{0} ) }{\mc{U}_1^{(0)}(\bs{0} )  } \; + \; 
\underbrace{ \mc{V}_{-;0}^{(0)} }_{=\e{O}( \mf{x} ) } \,
D_{\bs{0}} \bigg( \f{\mc{U}_1\circ \Phi_{\mf{x}}   }{\mc{U}_1^{(0)}  }   \bigg) \cdot \bs{u} \;. 
\enq
Since $\mc{V}_{-;0}^{(0)}  = \e{O}( \mf{x} ) $ and $\mc{V}_{-;1} -\mc{V}_{-;1}^{(0)} = \e{O}(\mf{x}^2)$, it holds that there exists a linear form $\mc{L}$ on $\R^{ \ov{\bs{n}}_{\ell} }$
such that 
\beq
D_{\bs{0}} \Big( \big[\Psi_{\mf{x}}(\bs{x})\big]_{n_{\ell}}^{(\ell)}  \Big) \cdot \big(\bs{u},s \big) \, = \, s \, + \,  \mf{x} \, \mc{L}\cdot \big(\bs{u},s \big) \;. 
\enq
Thence, there exists an endomorphism $\op{N}_{\Psi}$ on $\R^{\ov{\bs{n}}_{\ell} }$, with $ \norm{ \op{N}_{\Psi} } \leq C $ for some $\mf{x}$-independent constant, 
such that $D_{\bs{0}}  \Psi_{\mf{x}} \, = \, \e{id} + \mf{x} \op{N}_{\Psi} $ \;. 
Thus, $\Psi_{\mf{x}}$ is invertible in some open neighbourhood of $\bs{0}$ which, upon reducing  $V^{(0)}_{[\ell,n_{\ell}]}$ and $I_{\ell}^{(0)}$ if necessary, 
may be taken to be $V^{(0)}_{[\ell,n_{\ell}]} \times I_{\ell}^{(0)}$. Note that the estimates on the differential $D_{\bs{0}}  \Psi_{\mf{x}} \, = \, \e{id} + \mf{x} \op{N}_{\Psi} $ 
and \eqref{ecriture valeur Psi en origine} ensures that $ \bs{t}(k_0)\in \Psi_{\mf{x}}\Big(  V^{(0)}_{[\ell,n_{\ell}]} \times I_{\ell}^{(0)} \Big)$ 
and that the latter set contains a $\mf{x}$-independent open neighbourhood of $ \bs{t}(k_0)$ in $\R^{ \ov{\bs{n}}_{\ell} }$.

All is now in place so as to establish the rectification relation. Observe that there exists a direct relation between the zeroes  $\mc{V}_{\pm} $ and  $\mc{V}_{\pm}^{(0)}$:
\bem
 \mc{V}_{-}\circ\Phi_{\mf{x}} (\bs{x}_{[\ell,n_{\ell}]} )  \, - \,  \mc{V}_{+}\circ\Phi_{\mf{x}} (\bs{x}_{[\ell,n_{\ell}]} )
 \, = \, \mc{U}_1\circ \Phi_{\mf{x}} (\bs{x}_{[\ell,n_{\ell}]} ) \Bigg[ \mf{x} + \f{\mc{U}_2\circ \Phi_{\mf{x}} (\bs{x}_{[\ell,n_{\ell}]} )}{\mc{U}_1\circ \Phi_{\mf{x}} (\bs{x}_{[\ell,n_{\ell}]} )} \Bigg] \\
 \, = \, \mc{U}_1\circ \Phi_{\mf{x}} (\bs{x}_{[\ell,n_{\ell}]} ) \cdot \bigg[ \mf{x} + \Big( \big(\bs{\phi}^{(0)}_{\mf{x}}\big)^{-1}(\bs{x}_{[\ell,n_{\ell}]} ), \op{M}  \, \big(\bs{\phi}^{(0)}_{\mf{x}}\big)^{-1}(\bs{x}_{[\ell,n_{\ell}]} ) \Big) \bigg]   \\
\, = \,  \mc{U}_1\circ \Phi_{\mf{x}} (\bs{x}_{[\ell,n_{\ell}]} ) \cdot \Bigg[ \mf{x} + \f{ \mc{U}_2^{(0)}\big( \bs{x}_{[\ell,n_{\ell}]} \big) }{ \mc{U}_1^{(0)} \big( \bs{x}_{[\ell,n_{\ell}]} \big) }  \Bigg] 
\; = \; \f{ \mc{U}_1\circ \Phi_{\mf{x}} (\bs{x}_{[\ell,n_{\ell}]} )  }{ \mc{U}_1^{(0)} \big( \bs{x}_{[\ell,n_{\ell}]} \big) }  \cdot 
\Big[ \mc{V}_{-}^{(0)}  (\bs{x}_{[\ell,n_{\ell}]} )  \, - \,  \mc{V}_{+}^{(0)} (\bs{x}_{[\ell,n_{\ell}]} ) \Big] \;. 
\end{multline}
This identity entails that, for any $\ups \in \{\pm 1\}$,  
\beq
\big[\Psi_{\mf{x}}(\bs{x})\big]_{n_{\ell}}^{(\ell)}  -  \mc{V}_{\ups}\circ\Phi_{\mf{x}} (\bs{x}_{[\ell,n_{\ell}]} )  \; = \; 
\Big[ x_{n_{\ell}}^{(\ell)} - \mc{V}_{\ups}^{(0)} (\bs{x}_{[\ell,n_{\ell}]} )  \Big] \cdot \f{\mc{U}_1\circ \Phi_{\mf{x}} (\bs{x}_{[\ell,n_{\ell}]} ) }{\mc{U}_1^{(0)}(\bs{x}_{[\ell,n_{\ell}]} )  } \;. 
\label{ecriture trivialisation coordonnel ell nelle de Psi}
\enq
Thus, starting from the factorisation \eqref{ecriture factorisation x plus z pm} and applying the equality \eqref{ecriture trivialisation coordonnel ell nelle de Psi} followed by 
an application of the factorisation \eqref{ecriture factorisation x plus z pm modele effectif} backwards, one gets that 
\beq
\mf{x} + \mf{z}_{\ups}\circ\Psi_{\mf{x}}(\bs{x}) \, = \, \mf{f}_{\ups}(\mf{x};\bs{x}) \cdot \Big( \mf{x} + \tilde{z}_{\ups}(\bs{x}) \Big) 
\qquad \e{with} \qquad 
\mf{f}_{\ups}(\mf{x}; \bs{x}) \, = \, \f{\mc{U}_1\circ \Phi_{\mf{x}} (\bs{x}_{[\ell,n_{\ell}]} ) }{\mc{U}_1^{(0)}(\bs{x}_{[\ell,n_{\ell}]} )  }  \cdot  
\f{ h_{\ups}\Big(\mf{x}; \Psi (\bs{x} ) \Big) }{ h_{\ups}^{(0)}(\mf{x};\bs{x} )  } \;. 
\enq
Clearly, $\mf{f}_{\ups}$  is smooth. I now  establish that $\mf{f}_{\ups}$ has the claimed form of the expansion.  

Putting \eqref{ecriture dvpmt local zero V ups} and \eqref{ecriture valeur Psi en origine} together, one infers that, for any $\ups\in \{ \pm \}$,
\beq
\Psi_{\mf{x}}(\bs{0}) \, = \,\Big( \bs{t}_{[\ell, n_{\ell}]}(k_0),  \mc{V}_{\ups}\big(\bs{t}_{[\ell, n_{\ell}]}(k_0) \big)   \Big) + \e{O}\big( \mf{x} \big)  \;. 
\enq
This, along with $\mc{V}^{(0)}_{\ups}(0)=\e{O}(\mf{x})$ and the smoothness of $h_{\ups}$ and $h_{\ups}^{(0)}$, 
allows one to use the expressions \eqref{ecriture simplification locale de hups} and \eqref{ecriture simplification locale de hups effectif} so as to 
deduce that 
\bem
\left. \f{ h_{\ups}\Big(\mf{x};\Psi_{\mf{x}} (\bs{x}) \Big)  }{ h_{\ups}^{(0)}(\mf{x}; \bs{x})  }  \right|_{\bs{x}=0}  \; = \; 
\f{ h_{\ups}(\mf{x};\bs{p}_{\ups})\mid_{\bs{p}_{[\ell,n_{\ell}]}= \bs{t}_{[\ell,n_{\ell}]}(k_0) }+ \e{O}(\mf{x})  }
{ h_{\ups}^{(0)}(\mf{x};\bs{x}_{\ups})\mid_{\bs{x}_{[\ell,n_{\ell}]}= \bs{0} } +  \e{O}(\mf{x} )   }  \, = \, 
\f{ \big| \ups \op{v}+\mf{u}^{\prime}_{\ell}\big( t_{\ell}(k_0)\big)  + \e{O}(\mf{x})   \big| + \e{O}(\mf{x} )      }
{   \big| \ups \op{v}+\mf{h}^{\prime}_{\ell}\big( 0 \big) + \e{O}(\mf{x})  \big| + \e{O}(\mf{x})     }  \\
\, = \,  \f{   \ups \op{v}+\mf{u}^{\prime}_{1}\big( k_0  \big)   + \e{O}(\mf{x})      }
{     \ups \op{v}+ \op{u} + \e{O}(\mf{x})     } 
\, = \, 1  + \e{O}(\mf{x}) \;, 
\end{multline}
where $\bs{p}_{\ups}$, resp. $\bs{x}_{\ups}$, are as defined through \eqref{definition vecteur p ups}, resp. \eqref{ecriture simplification locale de hups effectif}. 
Furthermore, I used that $\mf{u}^{\prime}_1(k_0)=\op{u}$. 
Thence, a similar result for the ratio of $\mc{U}_1$'s established in \eqref{ecriture valeur en zero ration U1 sur U1 effectif}, and smoothness in $\bs{x}$ all together, 
entail that  one has $ \mf{f}_{\ups}(\mf{x};\bs{x}) = 1+ \e{O}\big( \norm{\bs{x}}+|\mf{x}| \big) $. 

Recall the definition \eqref{ecriture domaine D eff eta prime the rectification des z ups} of the domain $\mc{D}^{(\e{eff})}_{\eta^{\prime}}$. 
To complete the proof, it remains to establish that 
\beq
\mc{D}^{(\e{eff})}_{\eta^{\prime}} \subset V^{(0)}_{[\ell,n_{\ell}]} \times I_{\ell}^{(0)} \qquad  \e{provided}\; \e{that} \qquad  0<\eta^{\prime}<\eta_0
\enq
for some $\eta_0>0$ small enough. The map 
\beq
G \; : \;  \R^{ \ov{\bs{n}}_{\ell}} \tend \R^{\ov{\bs{n}}_{\ell}}  \quad \e{such}\, \e{that} \quad  
\big[G(\bs{x})\big]_{a}^{(r)} = t_{r}^{(0)}\big( x_1^{(1)} \big) - x_a^{(r)} (1-\de_{a,1}\de_{r,1})  \;, 
\enq
is obviously continuous, and thus,  upon agreeing to denote $B_{\eps}(0)= \big\{x \in \R \; : \; |x| < \eps \big\}$ the open ball around $0$ in $\R$ of radius $\eps$, one gets that 
\beq
\mc{D}^{(\e{eff})}_{\eta^{\prime}} \; = \; G^{-1} \bigg(  B_{C \eta^{\prime}}(0) \times \pl{r=1}{\ell}   \Big( B_{ \xi_r^2 \eta^{\prime} }(0) \Big)^{ n_r- \de_{r,1} }   \bigg)
\enq
is open as a pre-image of an open set by a continuous function. Since $\bs{0} \in \mc{D}^{(\e{eff})}_{\eta^{\prime}} $, it is an open neighbourhood of that point in $\R^{ \ov{\bs{n}}_{\ell} }$. Since its diameter
shrinks to $0$ as $\eta^{\prime}\tend 0$, and since $V^{(0)}_{[\ell,n_{\ell}]} \times I_{\ell}^{(0)}$ is also an open neighbourhood of $\bs{0}$ in $\R^{ \ov{\bs{n}}_{\ell} }$, the claim follows. \qed

\subsection{Factorisation of the maps $\tilde{z}_{\ups}$}

\begin{prop}
\label{Proposition factorisation jolie en zeroes de tilde z ups}

Let $\tilde{z}_{\ups}$ correspond to  the below multivariate polynomial on $\pl{r=1}{\ell}\R^{n_r}$:
\beq
\tilde{z}_{\ups}(\bs{x}) \; = \;- \sul{ (r,a)\in \mc{M} }{} \zeta_{r} \bigg\{ \mf{h}_r\big( x_a^{(r)} \big) +\ups \op{v} x_a^{(r)}  \bigg\} \qquad  where  \qquad 
 \mf{h}_r\big( x \big) \; = \;  - \zeta_r \veps_r \f{ x^2  }{ 2 \xi_r^2 } \, + \,  \op{u}\,  x \;, 
\enq
$\veps_{r}\in \{ \pm \}$, $\xi_r \in \R^*$ and $(\op{u}, \op{v}) \in \R \times \R^+$.  
Then, there exists a linear map $\op{M} $ on  $\pl{r=1}{\ell}\R^{n_r} $  such that:
\begin{itemize}
 \item $\op{M}$ is invertible;
 \item there exist integers $m_{\pm} \in \mathbb{N}$ satisfying $m_++m_-+1=\sul{r=1}{\ell}n_r$ such that   
it holds 
\beq
\tilde{z}_{\ups}\Big(  \op{M} (y,\bs{z})  \Big)  \, = \,    \wt{P}_{\ups}(y,\bs{z})   \quad with  \quad  \bs{z}=\big( \bs{z}^{(+)},  \bs{z}^{(-)} \big) \in \R^{m_+}\times \R^{m_-} 
\enq
 and 
\beq
\wt{P}_{\ups}(y,\bs{z}) \; = \;  \f{   y^2  }{2   \mc{P}_{\e{eff}}  }  -\big( \op{u}+\ups \op{v} \big) y    \, + \,    \sul{s=1}{m_+} \big( z_s^{(+)} \big)^2  \, - \; \sul{s=1}{m_-} \big( z_s^{(-)} \big)^2 \;, 
\quad with \quad \mc{P}_{\e{eff}}=\sul{r=1}{\ell} \veps_r n_r \xi_r^2  \;. 
\enq
\end{itemize}

 \end{prop}
 
 \Proof 

 One may recast the polynomial $\tilde{z}_{\ups}$ in the form  
\beq
\tilde{z}_{\ups}(\bs{x}) \; = \;  - \big( \op{u}+\ups \op{v} \big) \, \ov{\bs{x}}_{\zeta} \, + \,   \sul{ (r,a)\in \mc{M} }{} \f{\veps_{r}}{2 \xi_r^2} \big( x_a^{(r)} \big)^2 \qquad \e{with} \qquad 
\ov{\bs{x}}_{\zeta} \; = \;  \sul{ (r,a)\in \mc{M} }{}  \zeta_{r}  x_a^{(r)} \;. 
\enq
 Then, let $\wt{\op{M}} \, = \, \wt{\op{D}}  \, + \, \bs{g}\cdot \bs{e}^{\op{t}}$ with $\bs{e}^{\op{t}} = \big(1, \dots, 1 \big) \in  \R^{ \ov{\bs{n}}_{\ell} }$, 
$\bs{g}^{\op{t}} = \big(  \big(\bs{g}^{(1)}\big)^{\op{t}} , \dots,  \big( \bs{g}^{(\ell)} \big)^{\op{t}} \big) $  and where 
\beq
\Big( \bs{g}^{(r)} \Big)^{\op{t}} =  \f{ \zeta_1 \zeta_r \veps_r }{ \mc{P}_{\e{eff}} } \cdot \xi_r^2  \cdot \big(1, \dots, 1 \big) \in  \R^{ n_r }
\qquad \e{and} \qquad \wt{\op{D}} \, = \, \left( \ba{ccccc}      0 & \cdots      &  \cdots  &   \cdots  &   0  \\ 
                                                        \vdots & -I_{n_1-1} & 0 & \cdots  \\  
							    \vdots   &   0         & -\zeta_1 \zeta_2 I_{n_2} & 0 & \vdots   \\  
							 \vdots	&  		&  0 			& \ddots & 0    \\      
							  0	&   \cdots	&  0		&  0  &    -\zeta_1 \zeta_{\ell} I_{n_{\ell}}    \\       \ea \right) \;. 
\enq
It is straightforward to see that $|\det[\wt{\op{M}}]|=| \mc{P}_{\e{eff}} |^{-1} \not=0$. 

Then, a straightforward calculation shows that, given $\bs{y} \in \pl{r=1}{\ell} \R^{n_{r} }$ 
\beq
\tilde{z}_{\ups}\big( \, \wt{\op{M}}  \bs{y} \, \big) \, = \,  + \f{  \big( y_1^{(1)} \big)^2  }{2   \mc{P}_{\e{eff}}  } 
- \zeta_1 \big( \op{u} + \ups \op{v} \big) y_1^{(1)}   + \mc{Q} \big( \bs{y}_{[1,1] }  \big)
\enq
in which I employed the convention introduced in \eqref{definition vecteur p avec composantes omises}, while the quadratic form $\mc{Q}$ reads
\beq
 \mc{Q}\big( \bs{y}_{[1,1] }  \big)  \, = \,    \sul{ \substack{ (r,a) \in \\ \mc{M}_{[1,1]} }  }{}  \f{ \zeta_r }{ 2 \xi_r^2 }  \big( y_a^{(r)} \big)^2  
 \, - \; \f{1}{2 \mc{P}_{\e{eff}} }  \bigg( \sul{ \substack{ (r,a) \in \\ \mc{M}_{[1,1]} }  }{} y_a^{(r)} \bigg)^2  \;. 
\enq
Here $\mc{M}_{[1,1]}$ is as defined in \eqref{definition ensemble M elle n ell enleve}. 
Representing the quadratic form as $  \mc{Q}\big( \bs{y}_{[1,1] }  \big)   \, = \,  \Big(  \bs{y}_{[1,1] } , \op{M}_{\mc{Q}}   \bs{y}_{[1,1] }  \Big) $, one gets that 
the matrix $\op{M}_{\mc{Q}}$ is a rank one perturbation of a diagonal matrix:
\beq
 \op{M}_{\mc{Q}}\; = \; \op{D}_{\mc{Q}}   \,  - \, \f{ 1 }{ 2 \mc{P}_{ \e{eff} } } \bs{e} \cdot \bs{e}^{\op{t}}  \quad \e{ with} \quad  \bs{e}^{\op{t}} = \big(1, \dots, 1 \big) \in  \R^{ \ov{\bs{n}}_{\ell} -1 } 
\enq
and where I denoted
\beq
 \op{D}_{\mc{Q}}\, = \, \left( \ba{ccccc}       \\ 
                                                           \tfrac{1}{2} \veps_1 \cdot \xi_1^{-2} \cdot  I_{n_1-1} & 0 & \cdots &0 \\  
							       0         &   \tfrac{1}{2} \veps_2 \cdot \xi_2^{-2}  I_{n_2} & 0 & \vdots   \\  
							   		&  0 			& \ddots & 0    \\      
							  0	&   \cdots	&  \cdots		&        \tfrac{1}{2} \veps_{\ell} \cdot \xi_{\ell}^{-2} I_{n_{\ell}}    \\       \ea \right) \;. 
\enq
The determinant of $\mc{M}_{\mc{Q}}$ can thus be computed in a closed form 
\beq
\det\big[  \op{M}_{\mc{Q}}  \big] \, = \, \f{ \veps_1 \, \xi_1^2 }{ \mc{P}_{\e{eff}} } \pl{r=1  }{\ell} \bigg\{ \f{  \veps_r }{ 2 \xi_r^2} \bigg\}^{n_r- \de_{r,1} } \; \not= \;  0 \;. 
\enq
$ \op{M}_{\mc{Q}}$ being invertible and symmetric, there exists an orthogonal linear map $\op{N}$ such that 
\beq
\op{M}_{\mc{Q}} \; = \; \op{N} \left( \ba{cc}  \op{I}_{m_+} & 0  \\  0 & - \op{I}_{m_-}\ea \right) \op{N}^{\op{t}}
\enq
in which $(m_{+},m_{-})$ is the signature of $\op{M}_{\mc{Q}} $. Thus the map 
\beq
\op{M}\, = \, \wt{\op{M}} \cdot \left( \ba{cc}   \zeta_1   &    0     \\   0   &    \op{N}    \ea \right)
\enq
does the job. \qed

\subsection{Local expansion of a Vandermonde determinant}

Recall the notations for norms and partial order on vectors of integers \eqref{notation norme et ordre partiel sur entiers vecteurs} and the one for exponents $\bs{x}^{\bs{\a}}$ \eqref{notation exposant polynomial vectoriel}
with $\bs{x}\in  \mathbb{R}^{\ov{\bs{n}}_{\ell} } \, = \, \pl{r=1}{\ell} \R^{ n_{r}} $  and $\bs{\a}\in  \mathbb{N}^{\ov{\bs{n}}_{\ell} }$.  
$\ov{\bs{n}}_{\ell}$ is as defined in \eqref{definition bs ne ell}.

\begin{lemme}
\label{Lemme VdM local expansion}
 
 Let $\Psi: U \tend \Psi(U)$ be a smooth diffeomorphism on a open neighbourhood $U$ of $\bs{0}\in \mathbb{R}^{\ov{\bs{n}}_{\ell} }$ such that 
\begin{itemize}
 \item $\Psi(\bs{0})=\bs{v} \in \mathbb{R}^{\ov{\bs{n}}_{\ell} }$  with $\bs{v}=\big(\bs{v}^{(1)},\dots, \bs{v}^{(\ell)} \big)$, each entry $\bs{v}^{(r)}\, = \, \big(v^{(r)},\dots, v^{(r)} \big) \in \R^{n_r}$  having equal coordinates;
\item $D_{\bs{x}}\Psi = \e{id}+\mf{x} \op{N}_{\Psi}$, with $\op{N}_{\Psi} \in \mc{L}\big(  \mathbb{R}^{\ov{\bs{n}}_{\ell} } \big)$ such that $\norm{ \op{N}_{\Psi} } \leq C$, for a $|\mf{x}|$-independent constant $C$. 
\end{itemize}
Let 
\beq
V(\bs{x})=\pl{r=1}{\ell} \pl{a<b}{n_r} \big(x_a^{(r)}-x_b^{(r)} \big)^2
\enq
be a product of Vandermonde determinants relative to each of the $r$-coordinates. 
Then, there exists a smooth map $Q:U \tend \R$ such that $V\big( \Psi( \bs{x} ) \big) \, = \, V(\bs{x}) \, + \, Q(\bs{x})$. The map $Q$ has the expansion around $\bs{x}=\bs{0}$ of the form 
\beq
Q(\bs{x}) \, = \, \sul{  \substack{ \bs{\a}, |\bs{\a}|\geq m \\  \bs{\a}_0 \geq  \bs{\a} }  }{} \mf{x} C_{\bs{\a}} \bs{x}^{ \bs{\a}  }
\, + \,  \sul{  \substack{ \bs{\a} , |\bs{\a}|\geq m+1 \\  \bs{\a}_1 \geq  \bs{\a} }  }{}  D_{ \bs{\a} } \bs{x}^{ \bs{\a} }
\, + \,  \e{O}\Big( |\mf{x}| \bs{x}^{\bs{\a}_{0}} + \bs{x}^{\bs{\a}_1} \Big) \qquad with  \qquad m=\sul{r=1}{\ell} n_r(n_r-1) \;. 
\enq
In the above expansion, the even integer coordinate vectors $\bs{\a}_0, \bs{\a}_{1} \in \big( 2 \mathbb{N}\big)^{\ov{\bs{n}}_{\ell} }$ can be taken arbitrary provided that $|\bs{\a}_a|\geq m+a$ , and 
$C_{\bs{\a}}, D_{\bs{\a}} \in \R$ are coefficients that are bounded uniformly in $\mf{x}$.

\end{lemme}

\Proof

The hypotheses on $\Psi$ entail that $\Psi(\bs{x}) \, =\, \bs{v} \, + \,  \bs{x} \, + \, \mf{x} \op{N}_{\Psi} \cdot \bs{x} \, + \, \de\Psi(\bs{x})$, with $\de \Psi(\bs{x}) = \e{O}\big(  \norm{\bs{x}}^2 \big)$. 
Then, one can write 
\beq
\Big( \Psi(\bs{x}) \Big)^{(r)}_{a} \; = \; x_a^{(r)}+y_a^{(r)} \qquad \e{with} \quad  y_a^{(r)} \; = \; \e{O}\Big( \mf{x} \norm{ \bs{x} } + \norm{ \bs{x} }^2  \Big)
\enq
and smooth in $\bs{x}$. 
Then, one has
\beq
V\big( \Psi( \bs{x} ) \big) \, = \, \pl{r=1}{\ell} \det^2_{ n_r } \Big[  \big(x_a^{(r)}+y_a^{(r)} \big)^{b-1} \Big] \;. 
\enq
Upon expanding the power-law, one gets that 
\beq
  \big(x_a^{(r)}+y_a^{(r)} \big)^{b-1} \, = \,   \big(x_a^{(r)} \big)^{b-1} \, + \, \sul{k=1}{b-1} C_{b-1}^{k} \, \big(x_a^{(r)} \big)^{b-1-k}  \, \big(y_a^{(r)} \big)^{k}
\, = \, \big(x_a^{(r)} \big)^{b-1} \, + \, P_{a,b}^{(r)}(\bs{x}) \;, 
\enq
where $ C_{b-1}^{k}$ are binomial coefficients. The smoothness of $y_a^{(r)}$ and the estimates in $\mf{x}$ ensure that $P_{a,b}^{(r)}(\bs{x})$ takes the form:
\beq
P_{a,b}^{(r)}(\bs{x}) \; = \;  \sul{  \substack{ \bs{\a}, |\bs{\a}| \geq b-1 \\  \bs{\a} \leq  \bs{\a}_0 }  }{} \mf{x} \, C_{\bs{\a},a,b}^{(r)} \,  \bs{x}^{ \bs{\a}  }
\, + \,  \sul{  \substack{ \bs{\a} , |\bs{\a}|\geq b \\  \bs{\a} \leq  \bs{\a}_1 }  }{}  D_{\bs{\a},a,b}^{(r)} \, \bs{x}^{ \bs{\a} }
\, + \,  \e{O}\Big( |\mf{x}| \, \bs{x}^{ \bs{\a}_{0} } + \bs{x}^{ \bs{\a}_1 } \Big)
\enq
for some coefficients $C_{\bs{\a},a,b}^{(r)}, D_{\bs{\a},a,b}^{(r)}$ and $\bs{\a}_0,\bs{\a}_1 \in (2\mathbb{N})^{ \ov{\bs{n}}_{\ell} }$. Developing the determinant in respect to the sum appearing in each column yields
\beq
\det_{ n_r } \Big[  \big(x_a^{(r)}+y_a^{(r)} \big)^{b-1} \Big]  \; = \; \det_{ n_r } \Big[  \big(x_a^{(r)} \big)^{b-1} \Big]  \; + \; R^{(r)}(\bs{x}) 
\label{ecriture dvpment binomial du determinant}
\enq
with
\beq
R^{(r)}(\bs{x}) \; = \; \sul{k=1}{n_r} \sul{  \substack{ \intn{1}{n_r} = L\sqcup \ov{L}  \\ |\ov{L} |=k }  }{} \det_{n_r}\big[ \op{M}_{ L,\ov{L} }\big] \quad  , 
\qquad \Big( \op{M}_{ L,\ov{L} } \Big)_{ab} \; = \; \left\{ \ba{cc}  \big(x_a^{(r)} \big)^{b-1}  & \e{if} \, a \in L   \vspace{2mm}  \\
								      P_{a,b}^{(r)}(\bs{x})    &    \e{if} \, a \in \ov{L}   \ea \right.  \;. 
\enq
Above, the sum runs through all partitions $L\sqcup \ov{L}$ of $\intn{1}{n_r}$  such that  $\ov{L}$  has fixed cardinality $k$. 
Upon using the expansion
\beq
\det_{n_r}\big[ \op{M}_{ L,\ov{L} }\big] \, = \, \sul{ \sg \in \mf{S}_{N} }{} (-1)^{\sg} \pl{a \in L}{}  \big(x_a^{(r)} \big)^{ \sg(a)-1}  \cdot  \pl{a \in \ov{L} }{} P_{a,\sg(a) }^{(r)}(\bs{x})
\enq
a direct exponent counting argument entails that there exist constants $ C_{\bs{\a},L, \ov{L} }^{(r)}, D_{\bs{\a},L, \ov{L} }^{(r)} \in \R$ such that 
\beq
\det_{n_r}\big[ \op{M}_{ L,\ov{L} }\big] \, = \, \sul{  \substack{ \bs{\a}, |\bs{\a}|\geq \tfrac{n_r(n_r-1)}{2} \\  \bs{\a} \leq  \bs{\a}_0 }  }{} \mf{x} \, C_{\bs{\a},L, \ov{L} }^{(r)}  \, \bs{x}^{ \bs{\a}  }
\quad  + \,  \sul{  \substack{ \bs{\a} , |\bs{\a}|\geq \tfrac{n_r(n_r-1)}{2} +1 \\  \bs{\a} \leq  \bs{\a}_1 }  }{}  D_{\bs{\a},L, \ov{L} }^{(r)} \,  \bs{x}^{ \bs{\a} } 
\, + \,  \e{O}\Big( |\mf{x}| \,  \bs{x}^{\bs{\a}_{0}} + \bs{x}^{\bs{\a}_1} \Big) \;. 
\enq
All of this being established, it remains to take the square of the expression in \eqref{ecriture dvpment binomial du determinant} and then the product over $r$ so as to get the claim. \qed



\subsection{Asymptotic behaviour of a local integral}

\subsubsection{The integral associated with the $|\mf{u}_1^{\prime}(k_0)|< \op{v}$ regime}

Given $\de_{\pm}>0$ and $m_{\pm} \in \mathbb{N}^*$ define
\beq
I_{\de_{\ups},m_{\ups}} \; = \; \big[0 ;  \sqrt{\de}_+ \, \big]^{m_+}\times \big[0 ;  \sqrt{\de}_- \, \big]^{m_-} \;. 
\label{definition intervalle I de pm m pm}
\enq

\begin{lemme}
\label{Lemme integrale beta multi-dim auxiliaire locale cas u less than v}

Let $\de_{\pm}, \eta>0$ be fixed and small enough, $m_{\pm} \in \mathbb{N}$. Let $a,b$ be two smooth functions on $\intff{ - 2 m_+ \de_+ - 2 m_- \de_- }{ 2 m_+ \de_+ + 2 m_- \de_- } $ 
depending, possibly, on an auxiliary parameter $\mf{x}$ and such that 
\beq
a(0)     \, = \,   b(0) \, = \, 0  \qquad \e{and}  \qquad |a(s)| \, + \, |b(s)|   \, \leq  \,  \eta   
\label{proprietes fcts a et b}
\enq
uniformly  in  $s \in \intff{ - 2 m_+ \de_+ - 2 m_- \de_- }{ 2 m_+ \de_+ + 2 m_- \de_- }$. Let $ A, B > -1 $ be smooth function on  $I_{\de_{\ups},m_{\ups}} \times \intff{-\eta}{ \eta}$ and 
let $\mc{G}$ be a smooth  function on  $I_{\de_{\ups},m_{\ups}}  \times \intff{-\eta}{ \eta} \times \R^+ \times \R^+$ 
such that 
\beq
\mc{G}\big( \bs{u}, x, y \big) \; = \; G(\bs{u})   +\e{O}\big( |x|^{1-\tau}  \, + \,   |y|^{1-\tau}   \big)   \qquad with \qquad 
0< \tau <1 \;, 
\label{ecriture propriete fct G etape 1}
\enq
$G$ being a smooth function on $I_{\de_{\ups},m_{\ups}}  \times \intff{-\eta}{ \eta}$ and the remainder being differentiable in the sense of Definition \ref{Defintion reste differentiable}. 
Further, assume that $\mc{G}(\bs{u},x,y)=0$ whenever 
\beq
\bs{u}=\big( \bs{u}^{(+)}, \bs{u}^{(-)}, s \big) \qquad  with  \qquad \sul{a=1}{m_{+}} \big( u^{(+)}_a \big)^{2} \; > \; \de_{+} \quad  or  \quad 
 \sul{a=1}{m_{-}} \big( u^{(-)}_a \big)^{2} \; > \; \de_{-} \;. 
\label{ecriture propriete fct G etape 2}
\enq
Let  $\mc{W}$ be smooth on $I_{\de_{\ups},m_{\ups}}  \times \intff{-\eta}{ \eta} \times \R^+ $
and admit the expansion around the origin
\beq
\mc{W}\big( \bs{u}, \kappa \big) \; = \; \sul{ \substack{ \bs{\a}, | \bs{\a} |\geq m_0 \\ \bs{\a}=(\a_0,\bs{\be}) }  }{} c_{\bs{\a}} \cdot \kappa^{\a_0} \, \bs{u}^{\bs{\be} }
\qquad  with  \qquad m_0 \in 2 \mathbb{N} \;. 
\label{ecriture propriete fct W}
\enq
Consider the integral 
\bem
\mc{J}[\mc{G}_{\e{tot}},A,B]( \mf{x} ) \;= \; \pl{\ups= \pm }{}\Bigg\{ \Int{ 0 }{ \de_{\ups} }  \f{ \dd^{ m_{\ups} } w^{(\ups)}  }{  \pl{a=1}{m_{\ups}} \sqrt{ w_a^{(\ups)} }  }  \Bigg\}  \Int{0}{1}\dd t \; 
\bigg[ \mc{G}_{\e{tot}}\Big( \bs{u} , ( 1- t )\vp_{\mf{x}}\big(\bs{u}_{\bs{w}}\big), t  \vp_{\mf{x}}\big(\bs{u}_{\bs{w}}\big) \Big)  \\
\times   ( 1- t )^{ A(\bs{u}) } \cdot t^{ B(\bs{u}) }\cdot \Xi\big[ \vp_{\mf{x}}\big(\bs{u}_{\bs{w}}\big)  \big]  \cdot
\big[ \vp_{\mf{x}}\big(\bs{u}_{\bs{w}}\big) \big]^{ A(\bs{u})+B(\bs{u})+1 } \bigg]_{ \bs{u}_{\bs{w}}-\e{even} }
\end{multline}
where the even part of a function is as defined in \eqref{definition even part of a function} and vectors $\bs{u}$, $\bs{u}_{\bs{w}}$ appearing under the integral sign are parameterised in terms of $\bs{w}^{(\pm)}$, $t$ as 
\beq
\bs{u} \, = \, \Big(  \bs{u}_{\bs{w}} ,  a\circ\vp_{\mf{x}}(\bs{u}_{\bs{w}}) + t b\circ\vp_{\mf{x}}(\bs{u}_{\bs{w}}) \Big) \quad  with \quad  \bs{u}_{\bs{w}}  \, = \,  \Big( \bs{u}_{\bs{w}}^{(+)}, \bs{u}_{\bs{w}}^{(-)} \Big) 
\quad and \quad \bs{u}_{\bs{w}}^{(\ups)} \, = \,  \Big( \sqrt{ w_1^{(\ups)} },\dots,\sqrt{ w_{m_{\ups}}^{(\ups)} } \Big) \;, 
\label{definition variable u full et u index w}
\enq
while 
\beq
\vp_{\mf{x}}\big(\bs{u}_{\bs{w}}\big) \; = \;  \mf{x} \, + \, \sul{a=1}{m_+} w_a^{(+)} \, - \, \sul{a=1}{m_-} w_a^{(-)} \;. 
\label{definition fct varphi de u index w}
\enq
Finally, the main building block of the integrand reads 
\beq
\mc{G}_{\e{tot}}\big( \bs{u} ,x, y  \big) \, = \,  \mc{W}\big( \bs{u}, \vp_{\mf{x}}\big(\bs{u}_{\bs{w}}\big) \big)\cdot \mc{G} \big( \bs{u} , x,y  \big)  \;. 
\enq

 Then, the integrand belongs to $L^1(\intff{0}{\de_+}^{m_+}\times \intff{0}{\de_-}^{m_-}   \times \intff{ 0 }{ 1 })$ and  for any smooth function $F$ on $I_{\de_{\ups},m_{\ups}}  \times \intff{-\eta}{ \eta}$ satisfying $F(\bs{0})=1$, 
there exists a smooth function   $\mc{R}$ around $0$ such that one has the $  \mf{x} \tend 0$ behaviour 
\beq
\mc{J}[\mc{G}_{\e{tot}},A,B]( \mf{x} ) \;= \;  G(\bs{0}) \, \mc{J}\big[ \mc{W}_0 \, F ,A(\bs{0}),B(\bs{0}) \big]( \mf{x} )  \; + \; \mc{R}(\mf{x}) \; + \; 
\e{O}\big( \, |\mf{x}|^{ \varrho   }  \,  \big)
\enq
where
\beq
\varrho \, = \, \frac{ 1 }{2}( m_+ + m_0 + m_- ) + 1+ (A+B)(\bs{0}) +   1-\tau      \;, 
\enq
and 
\beq
\mc{W}_0\big( \bs{u}, \kappa \big) \; = \; \sul{ \substack{ \bs{\a}, |\bs{\a}|= m_0 \\ \bs{\a}=(\a_0,\bs{\be}) }  }{} c_{\bs{\a}} \cdot \kappa^{\a_0} \, \bs{u}^{\bs{\be} }
\qquad  with  \qquad m_0 \in 2 \mathbb{N} \;. 
\enq

\end{lemme}

\Proof 

One first implements the   change of the $\bs{w}$-integration variables $\bs{w}^{(\ups)} \hookrightarrow \bs{s}^{(\ups)}$  with  
\beq
 s_k^{(\ups)}\,=\, \sul{ a = k }{ m_{\ups} } w_a^{(\ups)} \qquad i.e. \qquad  w_k^{(\ups)}=s_k^{(\ups)} - s_{k+1}^{(\ups)} \; \;  \e{for} \;   \;
k=1,\dots, m_{\ups}-1 \;\;  \e{and} \;\; w_{m_{\ups}}^{(\ups)}=s_{m_{\ups}}^{(\ups)}   
\enq
whose Jacobian equals to $1$. The inequalities $0 \leq w_k^{(\ups)} \leq \de_{\ups} $ defining the integration domain in the original variables can be recast in terms of an equivalent set of encased 
inequalities defining the integration domain in the $\bs{s}^{(\ups)}$ variables:
\beq
\left\{ \ba{c}    0\leq s_1^{(\ups)} \leq m_{\ups} \de_{\ups}  \vspace{2mm}  \\ 
\mf{s}_{k-1}^{(\ups;-)}  \, \leq \,  s_k^{(\ups)} \, \leq \, \mf{s}_{k-1}^{(\ups;+)}  \quad \e{for} \;  k=2,\dots, m_{\ups}  \ea \right. 
\qquad \e{with} \qquad  \left\{ \ba{ccl}   \mf{s}_{k-1}^{(\ups;-)} &=& \e{max}\big\{ 0 , s_{k-1}^{(\ups)}-\de_{\ups} \big\} \vspace{1mm} \\ 
			      \mf{s}_{k-1}^{(\ups;+)}  &=& \e{min}\big\{ s_{k-1}^{(\ups)}, ( m_{\ups} +1-k) \de_{\ups} \big\} \ea  \right. 
\;.  
\label{ecriture domaine integration variables s ups}
\enq
This recasts the original integral in the form 
\bem
\mc{J}[\mc{G}_{\e{tot}},A,B](  \mf{x} ) \;= \;  \pl{\ups= \pm }{}  \Bigg\{ \Int{ 0 }{  m_{\ups} \de_{\ups} }\dd s^{(\ups)}_1 \pl{k=2}{m_{\ups}}  \Int{  \mf{s}_{k-1}^{(\ups;-)} }{  \mf{s}_{k-1}^{(\ups;+)}  }  \dd s_{k}^{(\ups)} \Bigg\}
\Int{0}{1}\dd t \; \pl{\ups=\pm}{} \pl{a=1}{m_{\ups}} \Bigg\{ \f{ 1  }{  \sqrt{ s_a^{(\ups)} - s_{a+1}^{(\ups)} }  }  \Bigg\}  \\
\times  \bigg[ ( 1- t )^{ A(\bs{x}) } \cdot t^{ B(\bs{x}) }  \mc{G}_{\e{tot}}\Big( \bs{x} , ( 1- t ) \vp_{\mf{x}}\big(\bs{x}_{\bs{s}}\big), t   \vp_{\mf{x}}\big(\bs{x}_{\bs{s}}\big)  \Big) 
\cdot \Xi\big[ \vp_{\mf{x}}\big(\bs{x}_{\bs{s}}\big) \big]  \cdot
\big[ \vp_{\mf{x}}\big(\bs{x}_{\bs{s}}\big) \big]^{ A(\bs{x})+B(\bs{x})+1 } \bigg]_{ \bs{x}_s-\e{even} }
\end{multline}
where it is understood that $ s_{m_{\ups}+1}^{(\ups)} \equiv 0$, while $\bs{x}\, =\, \big( \bs{x}_{\bs{s}}, a\circ\vp_{\mf{x}}(\bs{x}_{\bs{s}})+t b\circ\vp_{\mf{x}}(\bs{x}_{\bs{s}}) \big)$,  in which  
\beq
\bs{x}_{\bs{s}}  \, = \,  \Big( \bs{x}_{\bs{s}}^{(+)}, \bs{x}_{\bs{s}}^{(-)} \Big) \qquad \e{where} \qquad 
\bs{x}_{\bs{s}}^{(\ups)} \, = \, \bigg( \sqrt{s_1^{(\ups)}-s_2^{(\ups)}}, \dots, \sqrt{s_{m_{\ups}-1}^{(\ups)}-s_{m_{\ups}}^{(\ups)} }, \sqrt{ s_{m_{\ups}}^{(\ups)} } \bigg) \; .
\enq
Finally, one has $\vp_{\mf{x}}\big(\bs{x}_{\bs{s}}\big) \; = \; \mf{x} \, + \, s_1^{(+)}\, - \, s_1^{(-)}$.

Recall that the integrand vanishes when, either, $s_1^{(+)}\geq \de_{+}$ or $s_1^{(-)}\geq \de_{-}$. Thus, one may reduce the $\big( s_1^{(+)} ,   s_1^{(-)} \big)$ integration from 
$\intff{0}{m_+ \de_{+} } \times \intff{0}{m_- \de_{-} }$ to 
the rectangle $\intff{0}{\de_{+}}\times \intff{0}{\de_{-}}$. However,  as soon as it holds $0 \leq s_1^{(\ups)} \leq \de_{\ups} $, one can readily check that the endpoints of integration 
$  \mf{s}_{k-1}^{(\ups;-)},   \mf{s}_{k-1}^{(\ups;+)}$ in \eqref{ecriture domaine integration variables s ups} for the variable $s_{k}^{(\ups)} $ reduce to  
\beq
 \mf{s}_{k-1}^{(\ups;-)}=0 \quad \e{and} \quad  \mf{s}_{k-1}^{(\ups;+)}=s_{k-1}^{(\ups)} \quad \e{for} \quad  k=2,\dots,m_{\ups} \; .
\enq
Upon this reduction, one may implement another change of variables $\bs{s}^{(\ups)}  \; \hookrightarrow \;  \bs{u}^{(\ups)}$ with
\beq
 s_{a}^{(\ups)}=u_1^{(\ups)}\dots u_a^{(\ups)} \qquad \e{and} \qquad 
\det\Big[  D_{\bs{u}^{(\ups)}}\bs{s}^{(\ups)} \Big] \, = \, \pl{ a=2 }{ m_{\ups} }\big\{ u_1^{(\ups)} \cdots u_{a-1}^{(\ups)} \big\} = \pl{a=1}{ m_{\ups} } \big\{ u^{(\ups)}_a \big\}^{m_{\ups}-a} \;. 
\enq
Since
\beq
\pl{ a = 1 }{ m_{\ups} } \sqrt{ s_a^{(\ups)}-s_{a+1}^{(\ups)} } \; = \; \pl{a=2}{m_{\ups}}\sqrt{1-u_a^{(\ups)}} \cdot \pl{a=1}{m_{\ups}} \big\{ u_a^{(\ups)} \big\}^{ \tfrac{m_{\ups}+1-a }{2} }  
\enq
the integral takes the form 
\bem
\mc{J}[\mc{G}_{\e{tot}},A,B](  \mf{x} ) \;= \;  \pl{\ups= \pm }{} \Bigg\{ \Int{ 0 }{   \de_{\ups} }\dd u^{(\ups)}_1  \big[ u_1^{(\ups)} \big]^{ \tfrac{m_{\ups}-2 }{2} }  
\pl{k=2}{m_{\ups}}  \Int{ 0 }{  1  }  \dd u_{k}^{(\ups)}  \f{ \big( u_k^{(\ups)} \big)^{ \tfrac{m_{\ups}-1-k }{2} }  }{ \sqrt{ 1-u_{k}^{(\ups)} } }  \Bigg\} \\
\times \Int{0}{1}\dd t \;  \bigg[  \mc{G}_{\e{tot}}\Big( \bs{y} , ( 1- t )\vp_{\mf{x}}\big(\bs{y}_{\bs{u}}\big),  t   \vp_{\mf{x}}\big(\bs{y}_{\bs{u}}\big) \Big)\cdot
  ( 1- t )^{ A(\bs{y}) } \cdot t^{ B(\bs{y}) }   \cdot \Xi\big[ \vp_{\mf{x}}\big(\bs{y}_{\bs{u}}\big) \big]  \cdot
\big[ \vp_{\mf{x}}\big(\bs{y}_{\bs{u}}\big) \big]^{ A(\bs{y})+B(\bs{y})+1 } \bigg]_{\bs{y}_{\bs{u}}-\e{even} }\;. 
\end{multline}

There, one should identify  $\bs{y}=\big( \bs{y}_{\bs{u}}, a\circ\vp_{\mf{x}}(\bs{y}_{\bs{u}}) + t b\circ\vp_{\mf{x}}(\bs{y}_{\bs{u}}) \big)$ where  $\bs{y}_{\bs{u}}  \, = \,  \big( \bs{y}_{\bs{u}}^{(+)}, \bs{y}_{\bs{u}}^{(-)} \big) $ and
\beq
\bs{y}_{\bs{u}}^{(\ups)} \, = \, \bigg( \sqrt{u_1^{(\ups)} \Big(1-u_2^{(\ups)} \Big) }, \dots, \sqrt{u_1^{(\ups)}\cdots u_{m_{\ups}-1}^{(\ups)} \Big(1-u_{m_{\ups}}^{(\ups)} \Big) } ,
\sqrt{u_1^{(\ups)}\cdots u_{m_{\ups} }^{(\ups)}  } \bigg) \; .
\enq
Finally, $ \vp_{\mf{x}}\big(\bs{y}_{\bs{u}}\big) \; = \; \mf{x} \, + \, u_1^{(+)}\, - \, u_1^{(-)}$. 

At this stage, one may perform explicitly the reduction of the integration domain due to the presence of the Heaviside function. One has $ \vp_{\mf{x}}\big(\bs{y}_{\bs{u}}\big) \geq 0 $ on
\beq
 \bigg\{ \Big( u_{1}^{(+)}, u_{1}^{(-)} \Big) \in \intff{ 0 }{ \de_{+} }\times \intff{ 0 }{ \de_{-} }\; , \; -\e{min}(0,\mf{x}) \leq u_{1}^{(+)} \leq \de_+ \;\;  
 \e{and} \;\; 0 \leq u_{1}^{(-)} \leq \e{min}\big(u_{1}^{(+)} + \mf{x}, \de_{-} \big) \bigg\} \;. 
\enq
Since the integrand vanishes when $u_{1}^{(-)}\geq \de_-$, one may just as well, for fixed $u_1^{(+)}$, extend the $u_{1}^{(-)}$ integration to the segment $ \intff{0}{ u_{1}^{(+)} + \mf{x}}$. 

This form of the integration domain takes explicitly into account the constraints implied by the presence of the Heaviside function. This reduced form of the 
integration domain leads naturally to  the last change of variables 
\beq
u_a^{(+)} =z_{a}^{(+)}\;\;, \;\; a=1,\dots, m_{+} \; , \quad u_1^{(-)}=z_{1}^{(-)} \cdot \Big( z_{1}^{(+)} + \mf{x}\Big) \quad \e{and} \quad  u_a^{(-)} =z_{a}^{(-)}\;\;, \;\; a=2,\dots, m_{-} \;. 
\enq
The integration variables are then gathered in $\bs{r}=\big( \bs{r}_{\bs{z}}, a\circ\vp_{\mf{x}}(\bs{r}_{\bs{z}}) + t b\circ\vp_{\mf{x}}(\bs{r}_{\bs{z}}) \big)$,  with 
\beq
\bs{r}_{\bs{z}}  \, = \,  \Big( \bs{r}_{\bs{z}}^{(+)}, \big[ z_{1}^{(+)} + \mf{x}\big]^{\f{1}{2}}  \cdot  \bs{r}_{\bs{z}}^{(-)} \Big) 
\label{definition variable r index z}
\enq
and 
\beq
\bs{r}_{\bs{z}}^{(\ups)} \, = \, \bigg( \sqrt{ z_1^{(\ups)} \Big( 1 - z_2^{(\ups)} \Big) }, \dots, \sqrt{z_1^{(\ups)}\cdots z_{m_{\ups}-1}^{(\ups)} \Big( 1 - z_{m_{\ups}}^{(\ups)} \Big) } ,
\sqrt{z_1^{(\ups)}\cdots z_{m_{\ups} }^{(\ups)}  } \bigg) \; .
\label{definition variable r full}
\enq
All of this recasts the integral in the form 
\bem
\mc{J}[\mc{G}_{\e{tot}},A,B](  \mf{x} ) \;= \hspace{-4mm}   \Int{ -\e{min}(0,\mf{x}) }{   \de_{+} } \hspace{-2mm} \dd z^{(+)}_1      \Int{ 0 }{   1 } \dd z^{(-)}_1    
\pl{\ups= \pm }{} \Bigg\{ \pl{k=2}{m_{\ups}}  \Int{ 0 }{  1  }  \dd z_{k}^{(\ups)}  \f{ \big( z_k^{(\ups)} \big)^{ \tfrac{m_{\ups}-1-k }{2} }  }{ \sqrt{ 1-z_{k}^{(\ups)} } }  \Bigg\}\cdot
 \big[z_1^{(+)}+\mf{x} \big]^{ \frac{m_-}{2}   } \cdot  \pl{\ups = \pm }{} \big[ z_1^{(\ups)} \big]^{ \tfrac{m_{\ups} }{2}-1 }  \\
\times   \Int{0}{1}\dd t \;  \bigg[  ( 1- t )^{ A(\bs{r}) } \cdot t^{ B(\bs{r}) }     \cdot 
  \Big[ (z_1^{(+)}+\mf{x})(1-z_1^{(-)}) \Big]^{   A(\bs{r})+B(\bs{r}) +1   }
  \cdot \mc{G}_{\e{tot}}\Big( \bs{r} , \,  ( 1- t ) \vp_{\mf{x}}\big(\bs{r}_{\bs{z}}\big) , \,  t \vp_{\mf{x}}\big(\bs{r}_{\bs{z}}\big)\Big) \bigg]_{ \bs{r}_{\bs{z}}-\e{even}  } \;. 
\label{ecriture representation integrale quasi 1D pour integrale modele u prime lower than v}
\end{multline}
in which $\vp_{\mf{x}}\big(\bs{r}_{\bs{z}}\big) \; = \; \big( \mf{x} \, + \, z_1^{(+)} \big) \cdot \big( 1\, - \, z_1^{(-)} \big)$. 
The $L^1$ nature of the integrand is manifest on the level of \eqref{ecriture representation integrale quasi 1D pour integrale modele u prime lower than v}.

By composing the various expansions at $\bs{0}$, it is easy to see that 
\bem
\mc{G}_{\e{tot}}\Big( \bs{r} , (1-t) \vp_{\mf{x}}\big(\bs{r}_{\bs{z}}\big) , t \vp_{\mf{x}}\big(\bs{r}_{\bs{z}}\big) \Big)  \, = \, \sul{ \substack{ k, \ell \\ 2k+2\ell=m_0}  }{} C_{k,\ell}\Big(\bs{z}^{(+)}_2,\bs{z}^{(-)},t \Big)
\cdot \big( z_1^{(+)} \big)^{ \ell }\cdot \big( z_1^{(+)}+\mf{x} \big)^{ k + 1 + A(\bs{0}) + B(\bs{0}) }  \\
\; + \;  \e{O}\bigg( \sul{ \substack{ k, \ell  \\ 2k+2\ell  = m_0 + 1 }  }{} D_{k,\ell,p}\Big(\bs{z}^{(+)}_2,\bs{z}^{(-)},t \Big)
\cdot \big( z_1^{(+)} \big)^{ \ell } \cdot \big( z_1^{(+)}+\mf{x} \big)^{ k + 1 + A(\bs{0}) + B(\bs{0})} \Big\{  1 \, + \, \big| \ln \big( z_1^{(+)}+\mf{x} \big) \, \big| \Big\}  \bigg) \\ 
\; + \; 
\e{O}\bigg( \sul{ \substack{ k, \ell  \\ 2k+2\ell  = m_0   }  }{} \wt{D}_{k,\ell,p}\Big(\bs{z}^{(+)}_2,\bs{z}^{(-)},t \Big)
\cdot  \big( z_1^{(+)} \big)^{ \ell } \cdot \big( z_1^{(+)}+\mf{x} \big)^{ k + 2-\tau  + A(\bs{0}) + B(\bs{0})}   \bigg) \;. 
\end{multline}
There, $\bs{z}^{(+)}_2 \, = \, \big( z_{2}^{(+)}, \dots, z_{m_+}^{(+)} \big)$ and the functions $C_{k,\ell,p}\Big(\bs{z}^{(+)}_2,\bs{z}^{(-)},t \Big)$, 
$ D_{k,\ell,p}\Big(\bs{z}^{(+)}_2,\bs{z}^{(-)},t \Big)$, $\wt{D}_{k,\ell,p}\Big(\bs{z}^{(+)}_2,\bs{z}^{(-)},t \Big)$ are all continuous on $\intff{0}{1}^{m_++m_-}$. 
Finally, the remainders are differentiable. 

An application of Lemma \ref{Lemme integrale beta auxiliaire} relatively to the $z_1^{(+)}$ integration then leads to the claim. \qed



\subsubsection{The integral associated with the $|\mf{u}_1^{\prime}(k_0)| > \op{v}$ regime}

\begin{lemme}
\label{Lemme integrale beta multi-dim auxiliaire locale cas u greater than v}

Assume  the notations and hypotheses outlined in Lemma \ref{Lemme integrale beta multi-dim auxiliaire locale cas u less than v}, with the exception that $b$ does not have to vanish at $0$. 
Pick $\de_0$ small enough and such that $\de_0>2 \de_+$ and assume further that 
\beq
\mc{G}(\bs{u},x,y)=0 \quad if \quad  s>\eta^{\prime} \qquad where  \qquad  \bs{u}=\big( \bs{u}^{(+)}, \bs{u}^{(-)}, s\big) \;, 
\label{ecriture condition de support compact par rapport a variable t de G cal}
\enq
such that 
\beq
a\circ\vp_{\mf{x}}(\bs{v})  \;+\; t \,  b\circ\vp_{\mf{x}}(\bs{v}) > 2 \eta^{\prime} \qquad uniformly \; in \quad t> \f{\de_0 }{ 2 }\; ,  \quad \bs{v} \in \intff{ 0 }{ \de_+ }^{ m_{+} } \times \intff{ 0 }{ \de_- }^{ m_{-} } \;. 
\enq
Consider the integral 
\bem
\chi[\mc{G}_{\e{tot}},A,B]( \mf{x} ) \;= \; \pl{\ups= \pm }{}\Bigg\{ \Int{ 0 }{ \de_{\ups} }  \f{ \dd^{ m_{\ups} } w^{(\ups)} }{  \pl{ a = 1 }{ m_{\ups} } \sqrt{ w_a^{(\ups)} }  }  \Bigg\}  \Int{0}{\de_0}\dd t \; 
 \\
\times 
\bigg [\mc{G}_{\e{tot}}\Big( \bs{u} , t , t \, + \,  \vp_{\mf{x}}\big(\bs{u}_{\bs{w}}\big) \Big)  t^{ A(\bs{u}) } \cdot \big[ t \, + \,  \vp_{\mf{x}}\big(\bs{u}_{\bs{w}}\big) \big]^{ B(\bs{u}) }  
\cdot \Xi\big[ \vp_{\mf{x}}\big(\bs{u}_{\bs{w}}\big)  \big] \bigg]_{ \bs{u}_{\bs{w}}-\e{even} } 
\end{multline}
where the vectors $\bs{u}$, $\bs{u}_{\bs{w}}$ appearing under the integral sign are as defined in 
\eqref{definition variable u full et u index w}, the even part of a function is as defined in \eqref{definition even part of a function}, 
while $\vp_{\mf{x}}\big(\bs{u}_{\bs{w}}\big)$ 
has been defined in \eqref{definition fct varphi de u index w}.

 Then, the integrand belongs to $L^1(\intff{0}{\de_+}^{m_+}\times \intff{0}{\de_-}^{m_-}   \times \intff{ 0 }{ \de_0 })$  and, for any smooth function $F$ on $I_{\de_{\ups},m_{\ups}}  \times \intff{-\eta}{ \eta}$ satisfying $F(\bs{0})=1$, 
there exists a smooth function   $\mc{R}$ around $0$ such that one has the $  \mf{x} \tend 0$ behaviour 
\beq
\mc{J}[\mc{G}_{\e{tot}},A,B]( \mf{x} ) \;= \;  G(\bs{0}) \, \mc{J}\big[ \mc{W}_0 \, F ,A(\bs{0}),B(\bs{0}) \big]( \mf{x} )  \; + \; \mc{R}(\mf{x}) \; + \; 
\e{O}\big( \, |\mf{x}|^{ \varrho   }  \,  \big)
\enq
where
\beq
\varrho \, = \, \frac{ 1 }{2}( m_+ + m_0 + m_- ) + 1+ (A+B)(\bs{0}) +  1-\tau    \;, 
\enq
and 
\beq
\mc{W}_0\big( \bs{u}, \kappa \big) \; = \; \sul{ \substack{ \bs{\a}, |\bs{\a}|= m_0 \\ \bs{\a}=(\a_0,\bs{\be}) }  }{} c_{\bs{\a}} \cdot \kappa^{\a_0} \, \bs{u}^{\bs{\be} }
\qquad  with  \qquad m_0 \in \mathbb{N} \;. 
\enq

\end{lemme}

\Proof

The very same chain of transformation that was implemented in the proof of Lemma \ref{Lemme integrale beta multi-dim auxiliaire locale cas u less than v}
and the vanishing condition \eqref{ecriture condition de support compact par rapport a variable t de G cal}
allows one to recast the original integral as 
\bem
\mc{J}[\mc{G}_{\e{tot}},A,B](  \mf{x} ) \;=  
\Int{ -\e{min}(0,\mf{x}) }{   \de_{+} } \hspace{-2mm} \dd z^{(+)}_1      \Int{ 0 }{  \de_- } \dd z^{(-)}_1    \Int{ 0 }{ \de_0 }\dd t \;   
  \pl{\ups= \pm }{} \Bigg\{ \pl{k=2}{m_{\ups}}  \Int{ 0 }{  \de_{\ups} }  \dd z_{k}^{(\ups)}  \f{ \big( z_k^{(\ups)} \big)^{ \tfrac{m_{\ups}-1-k }{2} }  }{ \sqrt{ 1-z_{k}^{(\ups)} } }  \Bigg\}   \\
 \times \;  \big[z_1^{(+)}+\mf{x} \big]^{ \frac{m_-}{2} }  \pl{\ups=\pm}{} \Big\{ \big[ z_1^{(\ups)} \big]^{ \tfrac{m_{\ups} }{2}-1 }   \Big\} \cdot  \phi(t)
 \cdot  \bigg[ t^{ A(\bs{r}) } \cdot \Big[t\, + \, (z_1^{(+)}+\mf{x})(1-z_1^{(-)} ) \Big]^{ B(\bs{r}) }   \mc{G}_{\e{tot}}\Big( \bs{r} , t , t \, + \, (z_1^{(+)}+\mf{x})(1-z_1^{(-)} ) \Big)  \bigg]_{ \bs{r}_{\bs{z}}-\e{even} }
\label{ecriture simplification vers integrales quasi 1D de integrale modele a u prime bigger than v}
\end{multline}
in which  $\bs{r}$, $\bs{r}_{\bs{z}}$ are as defined in \eqref{definition variable r full}, \eqref{definition variable r index z} and 
$\phi \geq 0$ is smooth with compact support on $\intff{0}{\de_0}$ and such that $\phi_{\mid \intff{ 0 }{ \tf{\de_0}{2} } } = 1$. 
The $L^{1}$ nature of the integrand is already manifest on the level of \eqref{ecriture simplification vers integrales quasi 1D de integrale modele a u prime bigger than v}.

Then a direct application of Lemma \ref{Lemme integrale triple type beta auxiliaire} leads to the claim. \qed



\begin{lemme}
 \label{Lemme integrale triple type beta auxiliaire}
 
Let $\de_0> 2 \de_+ >0$, $m_{\pm} \in \mathbb{N}^*$ and $\mf{x} \in \R^{*}$. Consider the integral
\bem
\mc{L}(\mf{x}) \,  = \, \Int{ -\e{min}(0,\mf{x}) }{ \de_+ } \dd z_+ \Int{ 0 }{ 1 } \dd z_- \Int{0}{\de_0} \dd t \;  \phi(t) \cdot 
\big[ z_+ + \mf{x} \big]^{ \f{m_-}{2} } \cdot \big[ z_+ \big]^{ \f{m_+}{2} -1  } \\
\times \bigg[ t^{ A(\bs{z}) } \cdot \Big[t\, + \, (z_++\mf{x})(1-z_- ) \Big]^{ B(\bs{z}) } 
\mc{F}\Big( \bs{z}_0 ; t,  t +  (z_++\mf{x})(1-z_- )   \Big) \bigg]_{\bs{z}_0-\e{even} }
\end{multline}
 in which 
\beq
\bs{z} \, = \, \Big(\bs{z}_{0}, g_1\big(  (z_++\mf{x})(1-z_- )  \big)  \, + \,  t g_{2}\big(  (z_++\mf{x})(1-z_- )  \big) \Big) 
\quad \bs{z}_0\, = \, \big(  \sqrt{z_+}, \big[ z_-(z_+ + \mf{x}) \big]^{\f{1}{2}} \big) \; , 
\enq
$g_1, g_2$ are smooth, such that $g_1(0)=0$, $g_2(0)=1$ while the even part of a function is as given in \eqref{definition even part of a function}. 
Furthermore, the function $\mc{F}$ is assumed smooth and has the small argument expansion, with a differentiable remainder: 
\beq
\mc{F}\big( x,  y ; u, v   \big)   \; = \;\,  x^{2 s_+}  \,  y^{2 s_-}\Big(  F_0  \, + \, \e{O}\big(  u^{1-\a} + v^{1-\a} + x + y \big) \, \Big)  \quad  for  \,  some  \quad 0 < \tau < 1 \;, 
\enq
and integers $s_{\pm}$. The functions $A,B$ are smooth, $A,B>-1$, while $\phi \geq 0$ is smooth with compact support on $\intff{0}{\de_0}$ and such that $\phi_{\mid \intff{ 0 }{ \tf{\de_0}{2} } } = 1$

Then, for any function of two variables $G$ such that $G(\bs{0})=1$,  there exists a smooth 
function $\mf{r}$ around $\mf{x}=0$ such that 
\bem
\mc{L}(\mf{x}) \,  = \,   F_0   \hspace{-3mm} \Int{ -\e{min}(0,\mf{x}) }{ \de_+ }  \hspace{-3mm} \dd z_+ \Int{ 0 }{ 1 } \dd z_- \Int{0}{\de_0} \dd t \; \phi(t) \cdot 
\big[ z_+ + \mf{x} \big]^{ \f{m_-}{2} } \cdot \big[ z_+ \big]^{ \f{m_+}{2} -1  } 
\bigg[  t^{ A(\bs{0}) } \cdot \Big[t\, + \, (z_++\mf{x})(1-z_- ) \Big]^{ B(\bs{0}) } 
 \cdot G\big( \bs{z}_0 \big)  \bigg]_{\bs{z}_0-\e{even} }\; \\
\; + \;  \mf{r}(\mf{x}) \; + \; \e{O}\Big( \mf{x}^{3 + a_0 + b_0 + \a_+ + \a_- } |\ln \mf{x} | \, + \,   \mf{x}^{3 + a_0 + b_0 + \a_+ + \a_- -\tau} \Big) \;, 
\end{multline}
 with 
\beq
\a_+ \, = \, s_+\, + \, \frac{m_+}{2} -1 \;, \quad \a_-\, = \, s_- \, + \, \frac{ m_- }{2 } \;.  
\enq

\end{lemme}

\Proof

We only discuss the proof in the case of $\mf{x}>0$ and small enough in that the case $\mf{x}<0$ can be dealt with in much the same way. 

For further convenience, set $a_0 =  A(\bs{0})$ and $b_0 =  B(\bs{0})$. Let $n$ be such that 
\beq
-2 \,  < \,  1 + a_0 + b_0 + \a_- + \a_+ - n \, < \, -1 \; . 
\label{ecriture bon choix parametres pour DA L} 
\enq
It is obvious from the form of the integrand that $\mc{L}(\mf{x})$ defines a smooth function of $\mf{x}$ on $\R^+$ and that its derivative can be obtained directly by carrying the derivative under the integral sign. 
Then, it holds 
\bem
\Dp{\mf{x}}^n\bigg[  t^{ A(\bs{z}) } \cdot \Big[t\, + \, (z_++\mf{x})(1-z_- ) \Big]^{ B(\bs{z}) } \cdot \big[ z_+ + \mf{x} \big]^{ \f{m_-}{2} } \cdot \big[ z_+ \big]^{ \f{m_+}{2} -1  }  \cdot 
\mc{F}\Big( \sqrt{z_+}, \big[ z_-(z_+ + \mf{x}) \big]^{\f{1}{2}} ; t +  (z_++\mf{x})(1-z_- )   \Big) \bigg]_{\bs{z}_0-\e{even} }  \\
\; = \; \sul{p=0}{n} C_{n}^{p} \cdot \big(\a_- \big)_{n-p} \big( b_0 \big)_{p}  \, z_+^{\a_+} \cdot z_-^{\a_-} \big(z_+ + \mf{x} \big)^{\a_--n+p}
\Big[t\, + \, (z_++\mf{x})(1-z_- ) \Big]^{ b_0 -p} (1-z_-)^p \cdot  t^{a_0}\\
\times \bigg\{ F_0  +\e{O}\bigg( t^{ 1 - \tau }+\big[t\, + \, (z_++\mf{x})(1-z_- ) \big]^{1-\tau}  \\
 \, + \;     \Big\{ z_+ + z_-(z_++\mf{x}) + t \Big\} \cdot  \Big\{ 1 + |\ln t| + \ln\big[t\, + \, (z_++\mf{x})(1-z_- )\;  \big]   \Big\}    \bigg) \;  \bigg\} 
\end{multline}
where $(x)_p=x (x-1) \cdots (x-p+1)$ refers to the descending Pochhammer symbol.

Thus, upon setting 
\beq
\mc{T}(a,b,c,d,e,f) \; = \; \Int{0}{\de_+} \dd z_+ \Int{0}{1} \dd z_- \Int{0}{\de_0} \dd t \; \phi(t) \cdot t^a \big[t\, + \, (z_++\mf{x})(1-z_- ) \big]^{b} \, z_+^c \, z_-^{d} \, (\mf{x}+z_+)^e (1-z_-)^{f}
\enq
all together, one gets that 
\bem
\Dp{\mf{x}}^n\mc{L}(\mf{x}) \, = \, \sul{p=0}{n} C_{n}^{p} F_0 (b_0)_{p}(\a_-)_{n-p} \mc{T}\Big( \, a_0 , b_0-p , \a_+ , \a_-, \a_--n+p , p \Big) \\
\, + \, \sul{  \ups\in \{ \pm,  0 \} }{}\e{O}\bigg(  \Big[|\Dp{a}\mc{T}| \, + \, |\Dp{b}\mc{T}| \, + \, |\mc{T}| \Big]\Big( \, a_0  + \de_{\ups,0} , b_0-p , \a_+ + \de_{\ups, +} ,\a_- + \de_{\ups, -}, \a_--n+p + \de_{\ups, -} ,p \, \Big) \bigg) \\
\, + \,\sul{\ups= \pm }{} \e{O}\bigg(  \,  \big|\mc{T}\Big( \, a_0 + \a   \de_{\ups, +} , b_0-p + \a \de_{\ups, -} ,\a_+,\a_-,\a_--n+p,p \, \Big) \big| \, \bigg)  \;. 
\label{ecriture derivee neme integraleL}
\end{multline}
At this stage, it remains to focus on $\mc{T}$. There, one implements the change of variables
\beq
t= \f{ \zeta \, \nu }{ 1- \nu}  \qquad \e{with} \qquad \zeta= (z_++\mf{x})(1-z_- ) \quad \e{i.e.} \quad  \nu \, = \, \f{t}{t+\zeta} \; , 
\enq
leading to 
\beq
\mc{T}(a,b,c,d,e,f) \; = \; \Int{0}{\de_+} \dd z_+ \Int{0}{1} \dd z_-  z_+^c \; z_-^{d} \, (1-z_-)^{f+a+b+1}   (z_++\mf{x})^{e+a+b+1}  h\big( (z_++\mf{x})(1-z_- )  \big) 
\enq
where 
\beq
h(\zeta) \;= \; \Int{0}{  \frac{\de_0}{\zeta + \de_0 } } 
\dd \nu \;   \f{ \nu^a  }{ (1-\nu)^{2+a+b} }  \phi\Big(  \f{ \zeta \, \nu }{ 1- \nu}  \Big)  \;. 
\enq
The fact that $\phi$ is smooth with compact support on $\intff{0}{\de_0}$ entails that for $\Re(a)>-1$, and irrespectively of the value of $b$, $h$ is smooth in the neighbourhood of the 
origin and $h(0)=  \tf{\Ga(a+1)\Ga(-1-a-b) }{ \Ga(-b) }$, so that  upon making use of hypergeometric like notations, an application of Lemma \ref{Lemme integrale beta auxiliaire} yields 
\bem
\mc{T}(a,b,c,d,e,f) \; = \;    \Ga\left( \ba{c} -2-a-b-c-e , \,   -1-a-b, \, f+a+b+2  \\  -b , \,  f+a+b+d+3 , \, -1-a-b-e \ea \right)        \\
\times   \Ga( a+1, \,  d+1,  \, 1+c ) \cdot      \mf{x} ^{2+e+a+b+c} \; + \; \mc{R}(\mf{x})
 \; + \; \e{O}\big( \mf{x}^{3+e+a+b+c} \big) \;,  
\end{multline}
in which $\mc{R}(\mf{x})$ is smooth in $\mf{x}$.  Upon inserting this expansion into \eqref{ecriture derivee neme integraleL}, one gets 
\bem
\Dp{\mf{x}}^n\mc{L}(\mf{x}) \, = \, \sul{p=0}{n} C_{n}^{p} F_0 (b_0)_{p}(\a_-)_{n-p} \cdot  \mf{x}^{ 2+ a_0+b_0  + \a_- + \a_+ - n      } \\
 \times  \Ga\left( \ba{c}  a_0+1, \,  \a_+ + 1,  \, \a_- + 1 \, , p-1-b_0-a_0, \, a_0+b_0+2 , \,  n - a_0 - b_0 - \a_- - \a_{+} - 2 \\  
p-b_0 , \,  a_0+b_0  + \a_- + 3 , \, n-1-a_0-b_0 - \a_-    \ea \right)     \\ 
\; + \; \wt{\mc{R}}(\mf{x})  \; + \e{O}\bigg(\mf{x}^{ 3+ a_0+b_0  + \a_- + \a_+ - n      }  |\ln \mf{x} |  \, + \, \mf{x}^{ 3+ a_0 + b_0  + \a_- + \a_+ - n    + \tau  }\bigg) \;. 
\end{multline}
Then, the bounds \eqref{ecriture bon choix parametres pour DA L} followed by a direct integration of the above expansion lead to the claim.

\qed

\section{Asymptotic behaviour of a model integral}
\label{Appendix Section Asymptotics of model integral}

\subsection{Reduction of the model integral into regular and singular parts}

Recall that
\beq
V(\bs{x}) \; = \; \pl{r=1}{\ell} \pl{a<b}{n_r} \Big( x_a^{(r)}-x_b^{(r)}\Big)^2 \;. 
\label{rappel definition V et P alpha}
\enq
%
%
%

\begin{prop}
\label{Proposition reduction vers voisinage singularite integrale modele}

Let $\de_{\ups}>0$, $\sum_{r=1}^{\ell}n_r\geq 2$, $\xi_{r} \in \R^{*}$ and $\veps_r \in \{\pm 1\}$ be such that 
\beq
\sum_{r=1}^{\ell} \veps_{r}\, \xi_r^2 \, n_r  \not=0 \;.
\label{hypothese positivite contrainte sur nr epsr zetar squared}
\enq
Consider the integral 
\beq
\mc{J}(\mf{x}) \, = \,  \Int{ \mc{D}_{\bs{n}} }{}  \ex{-(\bs{x},\bs{x}) } V(\bs{x}) \pl{\ups=\pm }{} \Big\{ \Xi\big(\, \mf{x} + z_{\ups}(\bs{x}) \big)\cdot \big[ \,  \mf{x} + z_{\ups}(\bs{x}) \big]^{\de_{\ups}-1}  \Big\} \cdot \dd \bs{x}
\quad over \quad 
\mc{D}_{\bs{n}}=\pl{r=1}{\ell} \R^{n_r} 
\enq
where the functions $ z_{\ups}(\bs{x})$, $\ups=\pm$, are  quadratic forms
\beq
z_{\ups}(\bs{x}) \; = \; \f{1}{2} \big( \bs{x},\op{E} \bs{x} \big) \, + \, \big( \bs{x}, \bs{e}) \cdot \f{ \op{u}+\ups \op{v}   }{ 2\op{v}  }
\quad with \quad 
\op{E}=\left( \ba{ccc} \veps_1 \e{I}_{n_1} & 0& \ddots     \\
			  0  &  \ddots &  0       \\   
			  \ddots & 0 &   \veps_{\ell} \e{I}_{n_{\ell} }  \ea  \right)\;.
\label{definition z ups etape 1}
\enq
The parameters $(\op{u} , \op{v})\in \R \times \R^+$ are such that $\op{u} \not= \pm \op{v}$ while, one has  
\beq
\bs{e} \, = \, \big( \bs{e}^{(1)},\cdots, \bs{e}^{(\ell)}  \big) \quad with \quad \bs{e}^{(r)}\, = \, 
- 2\op{v} \xi_{r}\cdot \big(1,\cdots, 1 \big) \in \R^{n_r}   \;. 
\label{definition z ups etape 2}
\enq

\vspace{2mm}
\noindent Then,  for $|\mf{x}|\not=0$ and small enough,
\beq
 \bs{x}  \; \mapsto \;    \ex{-(\bs{x},\bs{x}) } V(\bs{x}) \pl{\ups=\pm }{} \Big\{ \Xi\big(\, \mf{x} + z_{\ups}(\bs{x}) \big)\cdot \big[ \,  \mf{x} + z_{\ups}(\bs{x}) \big]^{\de_{\ups}-1}  \Big\}  \, \in \,  L^{1}\big( \mc{D}_{\bs{n}} \big) \;, 
\enq
ensuring that $\mc{J}(\mf{x})$ is well-defined. 

\vspace{2mm} \noindent Furthermore,  there exists a smooth function $\mc{S}(\mf{x})$ in an open neighbourhood of $\mf{x}=0$ such that it holds 
\beq
\mc{J}(\mf{x}) \, = \, \mc{J}_{\e{eff}}(\mf{x}) \, + \,   \mc{S}(\mf{x})
\enq
where
\beq
\mc{J}_{\e{eff}}(\mf{x}) \, = \,     \pl{r=1}{\ell}  |\xi_r|^{-n_r^{2} } \cdot  \Int{ \mc{D}^{(\e{eff})}_{ 2 \eta^{\prime} } }{}  \ex{-(\bs{x},\op{M} \bs{x}) }  \Big( \vp^{(\e{\sslash})}_{\e{eff}}\vp^{(\e{sg})}_{\e{eff}}V \Big)(\bs{x})   
\pl{\ups=\pm }{} \Big\{ \Xi\big(\, \mf{x} + \tilde{z}_{\ups}(\bs{x}) \big)\cdot \big[ \,  \mf{x} + \tilde{z}_{\ups}(\bs{x}) \big]^{\de_{\ups}-1}  \Big\} \cdot \dd \bs{x} \;. 
\label{definition integrale effective J eff modele}
\enq
The integral above runs through the domain
\beq
\mc{D}^{(\e{eff})}_{ \eta^{\prime} } \, = \, \bigg\{   \bs{x} \in \mc{D}_{\bs{n}} \; : \; |x_1^{(1)}| \, \leq \, C \eta^{\prime} \; , \; 
\forall (r,a )\in \mc{M} \; : \; \big| t^{(0)}_r(x_1^{(1)})-x_a^{(r)}  \big| \, \leq  \, \xi_{r}^2 \eta^{\prime} \bigg\}   
\label{definition du domaine D eff}
\enq
where $C>0$ is some constant while $t^{(0)}_r(x) = \tfrac{ \veps_1 \zeta_1}{ \veps_r \zeta_r } \big( \tfrac{\xi_r^2}{ \xi_1^2 } \big) \, x $. Furthermore,  $\op{M}$ stands for the positive definite matrix 
\beq
\op{M}=\left( \ba{ccc} \xi_1^{-2} \e{I}_{n_1} & 0& \ddots     \\
			  0  &  \ddots &  0       \\   
			  \ddots & 0 &   \xi_{\ell}^{-2} \e{I}_{n_{\ell} }  \ea  \right)\;, 
\label{definition matrice positive definie M}
\enq
and the function $\tilde{z}_{\ups}$ takes the explicit form 
\beq
\tilde{z}_{\ups}(\bs{x}) \; = \;- \sul{ (r,a)\in \mc{M} }{} \zeta_{r} \Bigg\{ \mf{h}_r\big( x_a^{(r)} \big) +\ups \op{v} x_a^{(r)}  \Bigg\} \qquad where \qquad 
 \mf{h}_r\big( x \big) \; = \;  - \zeta_r \veps_r \f{ x^2  }{ 2 \xi_r^2 } \, + \,  \op{u} x \;. 
\label{definition fonction z tilde effective}
\enq
 The variables $\zeta_r\in \{ \pm \}$ are arbitrary. 
 
 Finally, one has that $\vp^{(\e{sg})}_{\e{eff}}$, $\vp^{(\sslash)}_{\e{eff}}$ are arbitrary smooth functions on $\mc{D}_{\bs{n}}$, such that 
 $0\leq\vp^{(*)}_{\e{eff}} \leq 1 $   
\beq
\left\{ \ba{ccc}  \vp^{(\e{sg})}_{\e{eff}} \; = \; 1 & on & \ov{\mc{D}^{(\e{eff})}_{ \eta^{\prime} }} \vspace{2mm} \\ 
		  \vp^{(\sslash)}_{\e{eff}} \; = \; 1 & o n&  \ov{ \mc{D}^{(\sslash;\e{eff})}_{ \frac{1}{2}\eta^{\prime} } }  \ea \right.    \qquad and \qquad 
\left\{ \ba{ccc}    \vp^{(\e{sg})}_{\e{eff}} \; = \; 0 &  on & \mc{D}_{\bs{n}} \setminus \mc{D}^{(\e{eff})}_{ 2\eta^{\prime} }  \vspace{2mm} \\
\vp^{(\sslash)}_{\e{eff}} \; = \; 0 &  on &  \mc{D}_{\bs{n}} \setminus\mc{D}^{(\sslash;\e{eff})}_{  \eta^{\prime} }   \ea \right. 
\label{definition fcts approx unite sg et sslash integrale modele}
\enq
 with 
\beq
 \mc{D}^{(\sslash;\e{eff})}_{  \eta^{\prime} } \; = \; \bigg\{   \bs{x} \in \mc{D}_{\bs{n}} \; : \;   \; \forall (r,a )\in \mc{M} \; : \; \big| t^{(0)}_r(x_1^{(1)})-x_a^{(r)}  \big| \, \leq  \, \xi_{r}^2  \eta^{\prime}   \bigg\}  \;. 
\label{definition domaine De eff sslash}
\enq

\end{prop}

\Proof

To start with, I take for granted that 
\beq
 \bs{x} \; \mapsto  \, \ex{-(\bs{x},\bs{x}) } V(\bs{x}) \pl{\ups=\pm }{} \Big\{ \Xi\big(\, \mf{x} + z_{\ups}(\bs{x}) \big)\cdot \big[ \,  \mf{x} + z_{\ups}(\bs{x}) \big]^{\de_{\ups}-1}  \Big\} \; \in \; L^{1}\big( \mc{D}_{\bs{n}} \big) \;. 
\enq
This issue will be dealt with at the end of the proof. 

It is convenient to introduce $\check{\xi}_r \, = \, \zeta_r  \xi_r$ with $\zeta_r\in \{ \pm \}$ and change variables through a rescaling
\beq
 \bs{y}^{(r)}   \; = \; \check{\xi}_r \bs{x}^{(r)}  \;. 
\enq
This yields 
\beq
\mc{J}(\mf{x}) \, = \,  \Int{ \mc{D}_{\bs{n}} }{} \mc{F}_{\e{tot}}(\bs{x})\cdot \dd \bs{x}    
\enq
with 
\beq
 \mc{F}_{\e{tot}}(\bs{x})  \; = \;   \pl{r=1}{\ell}  |\xi_r|^{-n_r^{2} } \cdot   \ex{-(\bs{x},\op{M} \bs{x}) } V(\bs{x})  
 \cdot  \pl{\ups=\pm }{} \Big\{ \Xi\big(\, \mf{x} + \tilde{z}_{\ups}(\bs{x}) \big)\cdot \big[ \,  \mf{x} + \tilde{z}_{\ups}(\bs{x}) \big]^{\de_{\ups}-1}  \Big\} 
\enq
and the positive definite matrix $\op{M}$ is as given in \eqref{definition matrice positive definie M}. Finally, the 
functions $\tilde{z}_{\ups}$ have been introduced in \eqref{definition fonction z tilde effective}.

Observe that the functions $\mf{h}_r$ appearing as building blocks of $\tilde{z}_{\ups}$ are such that $\mf{h}_r^{\prime}$ is strictly monotonous. 
Furthermore, it is readily checked that $t_r^{(0)}$, as given above of \eqref{definition du domaine D eff}, satisfies
$\mf{h}_r^{\prime}\big( t_r^{(0)}(x) \big) \, =  \, \mf{h}_1^{\prime}\big(x \big)$ on $\R$.

One has the decomposition $\mc{D}_{\bs{n}}=\mc{D}^{(\perp ;\e{eff})}_{\eta^{\prime}}\sqcup \mc{D}^{(\sslash ;\e{eff})}_{\eta^{\prime}}$, with 
\beq
\mc{D}^{(\perp ;\e{eff})}_{\eta^{\prime}} \, = \, \Big\{ \bs{x} \in \mc{D}_{\bs{n}} \; : \; \exists (r,a) \in \mc{M}  \;\; , \;\; \big| \mf{h}_1^{\prime}( x_1^{(1)} )  \, - \,  \mf{h}_r^{\prime}( x_a^{(r)} ) \big| > \eta^{\prime}  \Big\}
\enq
and 
\beq
\mc{D}^{(\sslash ;\e{eff})}_{\eta^{\prime}} \, = \, \Big\{ \bs{x} \in \mc{D}_{\bs{n}} \; : \; \forall (r,a) \in \mc{M}  \;\; , \;\; \big| \mf{h}_1^{\prime}( x_1^{(1)} )  \, - \,  \mf{h}_r^{\prime}( x_a^{(r)} ) \big|  \, \leq \,  \eta^{\prime}  \Big\} \;. 
\enq

Let $\vp^{(\sslash)}_{\e{eff}}  $ be smooth and such that $0 \leq \vp^{(\sslash)}_{\e{eff}} \leq 1$
\beq
\vp^{(\sslash)}_{\e{eff}}  \,  = \, 1 \quad \e{on} \quad   \ov{ \mc{D}^{(\sslash;\e{eff})}_{ \frac{ 1 }{ 2 } \eta^{\prime} } } \qquad \e{and} \qquad 
\vp^{(\sslash)}_{\e{eff}}  \,  = \, 0 \quad \e{on} \quad \mc{D}^{(\perp;\e{eff})}_{ \eta^{\prime} } \;. 
\enq
This allows one to split the original integral as  $\mc{J}(\mf{x}) \, = \, \mc{J}^{(\perp)}(\mf{x})\, + \, \mc{J}^{(\sslash)}(\mf{x})$, with 
\beq
\mc{J}^{(\perp)}(\mf{x}) \, = \,  \Int{  \mc{D}^{(\perp ;\e{eff})}_{ \frac{ 1 }{ 2 } \eta^{\prime} }    }{} \mc{F}_{\e{tot}}^{(\perp)}(\bs{x})\cdot \dd \bs{x}    
\qquad \e{and} \qquad 
\mc{J}^{(\sslash)}(\mf{x}) \, = \,  \Int{  \mc{D}^{(\sslash ;\e{eff})}_{  \eta^{\prime} }   }{} \mc{F}_{\e{tot}}^{(\sslash)}(\bs{x})\cdot \dd \bs{x}    \;,  
\enq
where it is understood that 
\beq
\mc{F}_{\e{tot}}^{(\perp)}(\bs{x}) \; = \; \Big(1-\vp^{(\sslash)}_{\e{eff}}(\bs{x})\Big) \cdot \mc{F}_{\e{tot}} (\bs{x}) \qquad \e{and} \qquad 
\mc{F}_{\e{tot}}^{(\sslash)}(\bs{x}) \; = \; \vp^{(\sslash)}_{\e{eff}}(\bs{x}) \cdot \mc{F}_{\e{tot}} (\bs{x}) \;. 
\enq

\subsubsection*{$\bullet$ The integral $\mc{J}^{(\perp)}(\mf{x})$}

Pick $R>0$ large enough and let $\vp^{(\perp)}_{\e{eff}}$ be a smooth function on $\R$ satisfying 
\beq
0 \, \leq \,  \vp^{(\perp)}_{\e{eff}}  \, \leq  \, 1 \; , \quad 
\vp^{(\perp)}_{\e{eff}}(x) = 1 \quad \e{for} \quad  |x| \, \leq \, R \qquad \e{and} \qquad \vp^{(\perp)}_{\e{eff}}(x) = 0 \quad \e{for} \quad |x| \, \geq \,  R+1 \;. 
\enq
Thus, by writing $1=1\, - \, \vp^{(\perp)}_{\e{eff}}\big(x_a^{(r)} \big) \, + \, \vp^{(\perp)}_{\e{eff}}\big(x_a^{(r)} \big)$,  $\vp^{(\perp)}_{\e{eff}}$ allows one 
to build a partition of unity on $\mc{D}_{ \bs{n} }$ which separates, in each variable, the pieces containing $\infty$ in some of the variables and those being bounded
\beq
%
1\, = \hspace{-3mm}
 \sul{ \substack{ \mc{M} = \\  \mc{M}_{\e{in}} \sqcup \mc{M}_{\e{out}} } }{}  \hspace{-3mm} \Phi_{ \mc{M}_{\e{in}} ; \mc{M}_{\e{out}} }(\bs{x})
\qquad \e{with} \qquad 
 \Phi_{ \mc{M}_{\e{in}} ; \mc{M}_{\e{out}} }(\bs{x}) \, = \,  \pl{(a,r) \in  \mc{M}_{\e{in}} }{} \Big\{ \vp^{(\perp)}_{\e{eff}}\big(x_a^{(r)} \big) \Big\} \; \cdot \hspace{-3mm}
  \pl{(a,r) \in  \mc{M}_{\e{out}} }{} \Big\{ 1-\vp^{(\perp)}_{\e{eff}}\big(x_a^{(r)} \big) \Big\}  
\enq
where the sum runs through all partitions of $\mc{M}$ into two disjoint sets $\mc{M}_{\e{in}}$ and $\mc{M}_{\e{out}}$.  
 This partition of unity leads to the decomposition 
\beq
\mc{J}^{(\perp)}(\mf{x}) \; = \;  \sul{ \mc{M} = \mc{M}_{\e{in}} \sqcup \mc{M}_{\e{out}}  }{}  \mc{J}^{(\perp)}_{ \mc{M}_{\e{in}} ; \mc{M}_{\e{out}} }(\mf{x})  
\enq
where, upon making the change of variables 
\beq
y_a^{(r)}\, = \, (x_a^{(r)})^{-1} \quad \e{if} \quad  (r,a) \in  \mc{M}_{\e{out}} \; , \qquad 
y_a^{(r)}\, = \, x_a^{(r)}  \quad \e{if}  \quad (r,a) \in  \mc{M}_{\e{in}}
\enq
and denoting $\bs{x}(\bs{y}; \mc{M}_{\e{out}} )$ the obtained vector, one has 
\beq
 \mc{J}^{(\perp)}_{ \mc{M}_{\e{in}} ; \mc{M}_{\e{out}} }(\mf{x})  \; = \; 
 \Int{ \mc{D}^{(\perp)}_{R; \mc{M}_{\e{out}} } }{} \f{ \mc{F}_{\e{tot}}^{(\perp)}\Big( \bs{x}(\bs{y}; \mc{M}_{\e{out}} ) \Big) }{ \pl{(a,r) \in\mc{M}_{\e{out}} }{} \big( y_a^{(r)} \big)^2   }  \cdot \dd \bs{y} 
\label{ecriture integrale J perp ramenee a un interval compact}
\enq
with
\beq
 \mc{D}^{(\perp)}_{R; \mc{M}_{\e{out}} } \; = \; \Bigg\{   \bs{y}  \; : \; \bs{x}(\bs{y}; \mc{M}_{\e{out}} ) \in \mc{D}^{(\perp)}_{\eta^{\prime}} \quad \e{and} \quad 
 \ba{cc} |y_a^{(r)}| \, \leq \,  R +1   & \forall (r,a) \in  \mc{M}_{\e{in}}  \vspace{2mm} \\
  |y_a^{(r)}| \, \leq  \, R^{-1}   & \forall (r,a) \in  \mc{M}_{\e{out}} \ea  \Bigg\} \;. 
\enq
Note that $\bs{x}(\bs{y}; \mc{M}_{\e{out}} )$ does depend on the given choice of the partition. 

It is easy to see that the integrand in \eqref{ecriture integrale J perp ramenee a un interval compact} is smooth and vanishes on $\Dp{} \mc{D}^{(\perp)}_{R; \mc{M}_{\e{out}} }  $. The smoothness at the origin
follows from the Gaussian decay of $\mc{F}_{\e{tot}}^{(\perp)}$ at infinity. 
Furthermore, for any $\bs{k}\in \mc{D}^{(\perp)}_{R; \mc{M}_{\e{out}} }  $, there exists $(r,a)\not= (1,1)$  such that 
\beq
\bigg| \mf{h}^{\prime}_1\Big(x_1^{(1)}(\bs{k}; \mc{M}_{\e{out}} ) \Big) - \mf{h}^{\prime}_r\Big(x_a^{(r)}(\bs{k}; \mc{M}_{\e{out}} ) \Big)   \bigg| \,  > \,   \eta^{\prime}  \;.
\enq
Then, set 
\beq
f_{[r,a]}(\bs{y}) \, = \, \Big( \bs{y}_1^{(1)},\bs{y}^{(2)},\cdots, \bs{y}^{(r-1)}, \bs{y}_{[a]}^{(r)}, \bs{y}^{(r+1)}, \cdots, \bs{y}^{(\ell)},
\tilde{z}_{+}\big( \bs{x}(\bs{y}; \mc{M}_{\e{out}} )  \big),    \tilde{z}_{-}\big( \bs{x}(\bs{y}; \mc{M}_{\e{out}} )  \big)  \Big)
\enq
so that it holds
\beq
\det\Big[  D_{\bs{k}} f_{[r,a]} \Big] \; = \; (-1)^{m_{r,a}} \zeta_1 \zeta_r 2 \op{v}  \cdot \bigg(  \f{ -1 }{ (k_1^{(1)})^2  } \bigg)^{\bs{1}_{ \mc{M}_{\e{out}}(1,1) }} \cdot 
\bigg(  \f{ -1 }{ (k_a^{(r)})^2  } \bigg)^{\bs{1}_{ \mc{M}_{\e{out}}(a,r) }}
\cdot \Big(  \mf{h}^{\prime}_r   \big( x_a^{(r)}(\bs{k}; \mc{M}_{\e{out}} )  \big)   \, - \, \mf{h}^{\prime}_1 \big( x_1^{(1)}(\bs{k}; \mc{M}_{\e{out}} )  \big)     \Big) \;, 
\enq
where $m_{r,a}=a+\sul{b=1}{r-1}n_b$. The latter ensures the local invertibility of $f_{[r,a]}(\bs{y})$ around $\bs{k}$ and thus the applicability of Lemma \ref{Lemme integrale multidimensionnelle auxiliaire reguliere} to the integral of interest. 
Hence, for any given partition $ \mc{M}_{\e{in}}\sqcup \mc{M}_{\e{out}} $ of $\mc{M}$, $ \mc{J}^{(\perp)}_{ \mc{M}_{\e{in}} ; \mc{M}_{\e{out}} }(\mf{x}) $ is smooth in $\mf{x}$ and thus so is $\mc{J}^{(\perp)}(\mf{x}) $
as a finite sum of smooth functions.

\subsubsection*{$\bullet$ The integral $\mc{J}^{(\sslash)}(\mf{x})$}

For the purpose of further reasoning, it is convenient to define
\beq
\msc{P}(x) \; = \; \f{ x }{ \veps_1 \zeta_1 \xi_1^2 } \sul{r=1}{\ell} n_r \veps_r \xi_r^2 \qquad \e{and} \qquad  
\msc{E}(x) \; = \; \sul{r=1}{\ell} n_r \, \zeta_r \cdot  \mf{h}_r\Big( t_r^{(0)} (x ) \Big)
\label{definition impulsion et energie effectives locales}
\enq
so that, by assumption \eqref{hypothese positivite contrainte sur nr epsr zetar squared}, $\big| \msc{P}^{ \prime }(x) \big|\geq c>0$ for any $x \in \R$ and some $c>0$.

To start with, one observes that $\tilde{z}_{\ups}(\bs{x}) \; = \; \mc{Z}_{\ups}(x_1^{(1)}) \, + \,  \de \tilde{z}_{\ups}(\bs{x}) $ with 
\beq
\mc{Z}_{\ups}(x) \; = \; - \Big\{  \msc{E}(x) + \ups \op{v}  \msc{P}(x) \Big\}  \qquad \e{and} \qquad 
 \de \tilde{z}_{\ups}(\bs{x}) \; = \;  - \sul{ (r,a)\in \mc{M} }{} \zeta_{r} \mf{w}_{\ups}^{(r)}\Big( x_a^{(r)}, t_r^{(0)}(x_1^{(1)}) \Big)  \;, 
\enq
in which $\mf{w}_{\ups}^{(r)}\big( x, y \big) \, = \, \mf{h}_{r}(x) - \mf{h}_{r}(y) \, + \, \ups \op{v} (x-y) $.

Observe that the bound 
\beq
 \big| \mf{h}_1^{\prime}( x_1^{(1)} )  \, - \,  \mf{h}_r^{\prime}( x_a^{(r)} ) \big|  \, \leq \,  \eta^{\prime}  
\qquad \e{is} \; \e{equivalent} \; \e{to} \qquad 
| t_r^{(0)}(x_1^{(1)})-x_a^{(r)} |\, \leq \,  \xi_r^2 \eta^{\prime} \;, 
\enq
 what ensures that $ \mf{w}_{\ups}^{(r)}\Big( x_a^{(r)}, t_r^{(0)}(x_1^{(1)}) \Big)  = \e{O}(\eta^{\prime})$
under such bounds and thus, it holds uniformly in $\bs{x}\in \mc{D}^{(\sslash)}_{\eta^{\prime}}$ that  $ \de \tilde{z}_{\ups}(\bs{x}) \, = \, \e{O}(\eta^{\prime})$.

It is readily seen that 
\beq
\mc{Z}_{\ups}^{\prime}(x) \; = \;- \msc{P}^{\prime}(x) \Big( \mf{h}_{1}^{\prime}(x)+  \ups \op{v}  \Big)
\enq
so that $\mc{Z}_{\ups}^{\prime}$ vanishes at the points  $s_{ \ups}= \zeta_1 \veps_1 \xi^2_1 \big( \op{u}+\ups \op{v} \big)$. 
However, one has that for $\ups, \ups^{\prime} \in \{\pm 1\}$, 
\beq
\mc{Z}_{\ups}(s_{\ups^{\prime}}) \; = \;  - \f{ 1 }{ 2 } \big( \op{u} + \ups^{\prime} \op{v} \big) \big( \op{u}+(2\ups-\ups^{\prime}) \op{v} \big) \cdot \sul{r=1}{\ell} \veps_r n_r \xi_r^2 \; \not= \; 0
\label{ecriture evalutation Z ups sur s ups prime}
\enq
owing to the hypotheses of the proposition.

Thus, it follows from the above that $\mc{Z}_{\ups}$ is strictly monotonous on $\intoo{-\infty}{ s_{\ups} }$ and on $\intoo{ s_{\ups} }{+\infty}$. Furthermore, one has $s_{+}\not= s_{-}$. 
Thus one may introduce the three intervals
\beq
I^{(-)} \, = \,  \intoo{-\infty}{  \min_{\ups= \pm} \{ s_{\ups} \}  }\; \; ,  \qquad 
I^{(c)} \, = \,  \intoo{  \min_{\ups= \pm} \{ s_{\ups} \}  }{  \max_{\ups= \pm} \{ s_{\ups} \}  }  \; \; , \qquad 
I^{(+)} \, = \,  \intoo{  \max_{\ups= \pm} \{ s_{\ups} \}  }{ + \infty }   \;. 
\enq
In each of these intervals $x \mapsto \mc{Z}_{\pm}(x)$ are both strictly monotonous. First assume that $\op{u} \not= \pm 3 \op{v}$. Then, the previous calculations ensure that $\mc{Z}_{\pm} \not=0$
on $\Dp{} I^{(\tau)}$, for any $\tau \in \{c, \pm \}$. 
Let $\tau_0\in  \{c, \pm \}$ be such that $0 \in I^{(\tau_0)}$. Note that $\tau_{0}$ is well defind since $s_{\pm} \not=0$. This entails that $0 \not \in \Dp{} I^{(\tau)}$ for any $\tau \in \{c, \pm \}$. 

Finally, one decomposes 
\beq
\mc{D}^{(\sslash;\e{eff})}_{\eta^{\prime}} \, = \, \bigcup_{ \tau \in \{c, \pm \} } \mc{D}^{(\sslash;\e{eff})}_{_{\eta^{\prime}}; \tau} \qquad \e{with} \qquad 
 \mc{D}^{(\sslash;\e{eff})}_{_{\eta^{\prime}};\tau}\, = \, \Big\{ \bs{x} \in \mc{D}^{(\sslash;\e{eff})}_{\eta^{\prime}} \; : \; x_1^{(1)} \in \ov{I^{(\tau)}} \Big\}
\enq
what induces the decomposition of the original integral $\mc{J}^{(\sslash)}(\mf{x})= \sul{ \tau \in \{c, \pm \} }{} \mc{J}^{(\sslash)}_{\tau} (\mf{x})$
with 
\beq
\mc{J}^{(\sslash)}_{\tau}(\mf{x}) \, = \,  \Int{  \mc{D}^{(\sslash)}_{_{\eta^{\prime}};\tau}  }{} \mc{F}_{\e{tot}}^{(\sslash)}(\bs{x})\cdot \dd \bs{x} \;. 
\enq

\subsubsection*{$\bullet$ The integral $\mc{J}^{(\sslash)}_{\tau_0}(\mf{x})$}

The analysis depends on whether $|\op{u}|>\op{v}$ or $|\op{u}|<\op{v}$. 

\vspace{2mm}

{\bf i)} The  $|\op{u}|<\op{v}$ case. 
\vspace{2mm}

\noindent Let $\sg=\e{sgn}\big( \msc{P}^{\prime}(0) \, (\op{u}+\op{v}) \big)$. Since $x \mapsto \mc{Z}_{\ups}(x)$ is strictly monotonous on $I^{(\tau_0)}$,  $0 \in I^{(\tau_0)}$
it follows that 
\beq
\e{sgn}\Big( \mc{Z}_{\ups}^{\prime}(x) \Big) = \e{sgn}\Big( \mc{Z}_{\ups}^{\prime}(0) \Big) = -\ups \sg   \quad \e{for} \quad x \in I^{(\tau_0)}  \; .
\enq
By using that $\mc{Z}_{\ups}(0)=0$, since $x \mapsto \mc{Z}_{\ups}(x)$ is strictly monotonous and since $\mc{Z}_{\ups}^{\prime}(x)$ vanishes, at most, on $\Dp{}I^{(\tau_0)}$, 
one infers that 
\beq
\ba{ccclccc} \sg \ups \tilde{z}_{\ups}(\bs{x}) & < & - \eta^{\prime}   &  \e{if} &  x_1^{(1)} & > & C \eta^{\prime}   \vspace{3mm}  \\
 \sg \ups \tilde{z}_{\ups}(\bs{x}) & > &  \eta^{\prime}   &  \e{if} &  x_1^{(1)} & < &  - C \eta^{\prime}
\ea  \;, 
\label{region positivite tilde z ups}
\enq
this uniformly in $\bs{x} \in  \mc{D}^{(\sslash)}_{\eta^{\prime};\tau_0}$ and for some constant $C>0$. 

Therefore \eqref{region positivite tilde z ups} entails that, at least for $|\mf{x}|<\tf{ \eta^{\prime} }{2}$,  
$\mf{x}+\tilde{z}_{\ups}(\bs{x})$ have both opposite signs on  $\mc{D}^{(\sslash)}_{\eta^{\prime};\tau_0}$ as soon as $ |x_1^{(1)}| \,  >  \, C \eta^{\prime} $. 
The presence of Heaviside functions in the integrand $\mc{F}_{\e{tot}}(\bs{x})$ entails that one has the reduction of the integration domain so that it holds 
\beq
\mc{J}^{(\sslash)}_{\tau_0}(\mf{x}) \, = \, \Int{  \mc{D}^{(\e{eff})}_{ \eta^{\prime} }  }{}  \mc{F}_{\e{tot}}^{(\sslash)}(\bs{x}) \cdot \dd \bs{x}  
\qquad \e{for} \;\; |\mf{x}|<\tf{ \eta^{\prime} }{2}. 
\enq
The domain $\mc{D}^{(\e{eff})}_{ \eta^{\prime} }$ appearing above is as defined in \eqref{definition du domaine D eff}. 

Finally, let $\vp^{(\e{sg})}_{\e{eff}}$ be smooth on $\pl{r=1}{\ell} \R^{n_r}$ and such that 
\beq
0 \leq \vp^{(\e{sg})}_{\e{eff}} \leq 1 \quad, \quad \vp^{(\e{sg})}_{\e{eff}}=1 \quad \e{on} \quad \ov{\mc{D}^{(\e{eff})}_{ \eta^{\prime} }}  \quad \e{and} \quad 
 \vp^{(\e{sg})}_{\e{eff}}=0 \quad \e{on} \quad \mc{D}_{\bs{n}} \setminus \mc{D}^{(\e{eff})}_{ 2 \eta^{\prime} }   \;. 
\label{definition fct charact lisse ensemble D eff eta prime}
\enq
Since the integrand vansihes anyway outside of $\mc{D}^{(\e{eff})}_{ \eta^{\prime} } $,  it holds
\beq
\mc{J}^{(\sslash)}_{\tau_0}(\mf{x}) \, = \, \Int{  \mc{D}^{(\e{eff})}_{ 2 \eta^{\prime} }  }{} \vp^{(\e{sg})}_{\e{eff}}(\bs{x}) \mc{F}_{\e{tot}}^{(\sslash)}(\bs{x}) \cdot \dd \bs{x}   \; ,
\enq
what corresponds exactly to $\mc{J}^{(\sslash)}_{\e{eff}}(\mf{x})$ as given in \eqref{definition integrale effective J eff modele}.

\vspace{2mm}

{\bf ii)} The  $|\op{u}|>\op{v}$ case. 
\vspace{2mm}

Keeping the   definition for $\sg$ as above, the same reasonings ensure that, now, $\e{sgn}\Big( \mc{Z}_{\ups}^{\prime}(x) \Big)   = - \sg $. 
This then leads to 
\beq
\ba{ccc} \sg   \tilde{z}_{\ups}(\bs{x}) \, < \, - \eta^{\prime}   &  \e{if} &  x_1^{(1)} > C \eta^{\prime}   \vspace{3mm} \\
 \sg   \tilde{z}_{\ups}(\bs{x}) \, > \,  \eta^{\prime}   &  \e{if} &  x_1^{(1)} < - C \eta^{\prime}
\ea  \;, 
\enq
this uniformly in $\bs{x} \in  \mc{D}^{(\sslash)}_{\eta^{\prime};\tau_0}$ and for some constant $C>0$. One then introduces $\vp^{(\e{sg})}_{\e{eff}}$
as in \eqref{definition fct charact lisse ensemble D eff eta prime} and using that 
\beq
\mc{J}^{(\sslash)}_{\tau_0}(\mf{x}) \, = \,  \Int{  \mc{D}^{(\sslash;\e{eff})}_{ 2\eta^{\prime} ;\tau_0}  }{} \mc{F}_{\e{tot}}^{(\sslash)}(\bs{x})\cdot \dd \bs{x} 
\enq
since $ \mc{F}_{\e{tot}}^{(\sslash)}$ vanishes on $ \mc{D}^{(\sslash;\e{eff})}_{2\eta^{\prime};\tau_0}  \setminus  \mc{D}^{(\sslash;\e{eff})}_{\eta^{\prime};\tau_0} $, one may 
decompose the integral as 
\beq
\mc{J}^{(\sslash)}_{\tau_0}(\mf{x}) \, = \, \mc{J}_{\e{eff}}(\mf{x})    \; + \;     \mc{J}_{\tau_0; \e{out}}(\mf{x})  
\qquad \e{with} \qquad 
\mc{J}_{\tau_0; \e{out}}(\mf{x})  \,  =   \hspace{-2mm}     \Int{  \mc{D}^{(\e{out})}_{ \eta^{\prime} ; \tau_0 }  }{} \hspace{-2mm} \big( 1  -  \vp^{(\e{sg})}(\bs{x}) \big) \cdot  \mc{F}_{\e{tot}}^{(\sslash)}(\bs{x}) \cdot \dd \bs{x} 
\enq
in which $\mc{J}_{\e{eff}}(\mf{x})$ is as given in \eqref{definition integrale effective J eff modele}, while 
\beq
 \mc{D}^{(\e{out})}_{ \eta^{\prime}; \tau_0 } \; = \;   \bigg\{   \bs{x} \in \mc{D}^{(\sslash)}_{\eta^{\prime};\tau_0} \; : \; |x_1^{(1)}| \, > \, C \eta^{\prime} \; , \; \forall (r,a )\in \mc{M} \; : \;
 \big| t_r^{(0)}(x_1^{(1)})-x_a^{(r)}  \big| \, < \, 2 \xi_{r}^2 \eta^{\prime} \bigg\} \;. 
\enq

The above bounds ensure that if $|\mf{x}|< \tf{\eta^{\prime}}{2}$, then $\big| \mf{x}+\tilde{z}_{\ups}(\bs{x}) \big| \geq \tf{\eta^{\prime}}{2}$ uniformly on $ \mc{D}^{(\e{out})}_{ \eta^{\prime}; \tau_0 }$. 
Then, derivation under the integral theorems ensure that $\mf{x} \mapsto \mc{J}_{\tau_0; \e{out}}(\mf{x}) $ is a smooth function of $\mf{x}$
in an open neighbourhood of $\mf{x}=0$.

\subsubsection*{$\bullet$ The integral $\mc{J}^{(\sslash)}_{ \tau }(\mf{x})$ with $\tau \not= \tau_0$}

When $\tau \not= \tau_0$, by construction, it holds that $\e{d}(I^{(\tau)},0)>0$, in which $\e{d}(A,x)$ stands for the Euclidian distance from the set $A$ to the point $x$. 
This property, along with the explicit form for $\msc{P}$ and $\msc{E}$ given in  \eqref{definition impulsion et energie effectives locales} both ensure that  it holds
\beq
  (0,0) \not\in    \Big\{  (\msc{P}(x),\msc{E}(x)) \; : \; x \in I^{(\tau)}   \Big\} \;. 
\enq
Furthermore, the previous arguments ensure that 
\beq
\min_{ \substack{ x \in \Dp{} I^{(\tau)}  \\  \ups = \pm  } } \Big| \msc{E}(x) + \ups  \msc{P}(x) \Big| >0 \;. 
\enq
The above properties guarantee that the integral $\mc{J}^{(\sslash)}_{\tau}(\mf{x})$ is a particular example of the general class 
of integrals considered in the section "Behaviour of $\mc{I}^{(\sslash)}$ in the regular case"
of the proof of Theorem \ref{Theorem principal caractere lisse et non lisse des integrals multi particules}.

One should simply make the identification  $\mf{u}_r \hookrightarrow \mf{h}_r$. Then, that very same analysis ensures that 
$\mf{x} \mapsto \mc{J}^{(\sslash)}_{\tau}(\mf{x})$ is smooth in $\mf{x}$.

\vspace{2mm}

It remains to comment on the case when $\op{u} =  3 \ups^{\prime} \op{v}$, for some  $\ups^{\prime} \in \{\pm 1\}$. Then, by  \eqref{ecriture evalutation Z ups sur s ups prime}, 
for any $\ups \in \{ \pm \}$ it holds that $\mc{Z}_{\ups}(s_{\ups})\not=0$. However, one has that  $\mc{Z}_{- \ups^{\prime}}(s_{\ups^{\prime}}) =0$ 
and  $\mc{Z}_{ \ups^{\prime}}(s_{-\ups^{\prime}}) \not=0$.  
One should then split the integration domain as
\beq
\intoo{ -\infty }{ s_{-\ups^{\prime}} } \cup \intoo{  s_{-\ups^{\prime}}  }{   s_{ \ups^{\prime}} -\eps } \cup \intoo{   s_{ \ups^{\prime}} -\eps }{   s_{ \ups^{\prime}} +\eps } \cup 
\intoo{   s_{ \ups^{\prime}} +\eps }{+\infty} 
\enq
where $\eps>0$ is taken small enough and, for simplicity, I assumed that $s_{\ups^{\prime}}>s_{-\ups^{\prime}}$, the other situation being tractable in a similar way. 
The analysis on all the intervals other that $\intoo{   s_{ \ups^{\prime}} -\eps }{   s_{ \ups^{\prime}} +\eps } $ goes along the lies described above, while on 
$\intoo{   s_{ \ups^{\prime}} -\eps }{   s_{ \ups^{\prime}} +\eps } $, one should proceed by implementing a change of variables analogous to \eqref{ecriture chngement vars f ups 11}. 
Reasonings as in \eqref{ecriture integrale I ups sslash}-\eqref{ecriture integrale I ups k} then allow one to conclude on the smoothness of such contributions.

\subsubsection*{$\bullet$ The  $L^1(\mc{D}_{\bs{n}})$ character}

It remains to prove the $L^{1}(\mc{D}_{\bs{n}})$ nature of the integrand. Since 
\beq
V(\bs{x}) \pl{\ups=\pm }{} \Big\{ \Xi\big(\, \mf{x} + z_{\ups}(\bs{x}) \big)\cdot \big[ \,  \mf{x} + z_{\ups}(\bs{x}) \big]^{\de_{\ups}-1}  \Big\} 
\enq
grows algebraically in $\norm{\bs{x}}$ at infinity, the Gaussian prefactor $ \ex{-(\bs{x},\bs{x}) } $ ensures  integrability at $\infty$. 
Furthermore, if both $\de_{\pm}\geq 1$, then the integrand is bounded on compact subsets of $\mc{D}_{ n }$ what entails its
$L^{1}(\mc{D}_{\bs{n}})$ nature. If at least one inequality $0<\de_{\ups} <1$ holds, then integrability issues may arise 
from a neighbourhood of the points where $\mf{x} + z_{\ups}(\bs{x})  = 0$.  Moreover, since the integrand is strictly positive, it is enought to prove 
local integrability. By following the integral reduction steps that are outlined in the last part of the proof of Theorem \ref{Theorem principal caractere lisse et non lisse des integrals multi particules},
one eventually ends up with one-dimensional integrals whose direct inspection shows that the local $L^1$-character boils down to the condition $0<\de_{\ups} <1$. 

\qed




\subsection{Asymptotic behaviour of the model integral}

\begin{prop}
 \label{Proposition asymptotique integrale modele}
 
Let  $V$ be as given in \eqref{rappel definition V et P alpha}. Consider the integral 
\beq
\mc{J}(\mf{x}) \, = \,  \Int{ \mc{D}_{\bs{n}} }{}  \ex{-(\bs{x},\bs{x}) } V(\bs{x}) \pl{\ups=\pm }{} \Big\{ \Xi\big(\, \mf{x} + z_{\ups}(\bs{x}) \big)\cdot \big[ \,  \mf{x} + z_{\ups}(\bs{x}) \big]^{\de_{\ups}-1}  \Big\} \cdot \dd \bs{x}
\qquad over \qquad 
\mc{D}_{\bs{n}}=\pl{r=1}{\ell} \R^{n_r} \;. 
\label{definition integrale modele pour extraire le comportement singulier}
\enq
Here $\de_{\ups}>0$   and it is assumed that $\sum_{r=1}^{\ell}n_r\geq 2$. Further, the functions $ z_{\ups}(\bs{x})  $ are quadratic forms
\beq
z_{\ups}(\bs{x}) \; = \; \f{1}{2} \big( \bs{x},\op{E} \bs{x} \big) \, + \, \big( \bs{x}, \bs{e}) \cdot \f{ \op{u}+\ups \op{v}   }{ 2\op{v}  }
\qquad with \qquad 
\op{E}=\left( \ba{ccc} \veps_1 \e{I}_{n_1} & 0& \ddots     \\
			  0  &  \ddots &  0       \\   
			  \ddots & 0 &   \veps_{\ell} \e{I}_{n_{\ell} }  \ea  \right)\;.
\enq
There $\veps_r \in \{\pm 1\}$, $(\cdot, \cdot)$ is the canonical scalar product on $\mc{D}_{\bs{n}}$,  $(\op{u} , \op{v})\in \R \times \R^+$ are such that $\op{u}\not= \pm \op{v}$ and, given $\xi_{r} \in \R$ one has  
\beq
\bs{e} \, = \, \big( \bs{e}^{(1)},\cdots, \bs{e}^{(\ell)}  \big) \quad with \quad \bs{e}^{(r)}\, = \, - 2\op{v} \xi_{r}\cdot \big(1,\cdots, 1 \big) \in \R^{n_r}   \;. 
\enq
The parameters in play are such that 
\beq
\sum_{r=1}^{\ell} \veps_{r} \xi_r^2 n_r  \not=0 \;.
\label{ecriture hypothese sur les nr veps et zetar}
\enq
Then, there exists a smooth function $\mc{S}$ in a neighbourhood of $0$ such that the $\mf{x}\tend 0$ asymptotic expansion holds:
\bem
\mc{J}(\mf{x}) \, = \, | \mf{x} |^{ \vth  } \Ga(\de_+) \Ga(\de_-) \Ga(-\vth ) \cdot \f{ [2\op{v}]^{\de_++\de_--1}   }{ \pl{\ups= \pm }{} \big| \op{v}-\ups \op{u} \big|^{\de_{\ups}}    }
\cdot \f{ \pl{r=1}{\ell} \Big\{ G(2+n_r) \cdot \big( 2\pi  \big)^{ \tfrac{ n_r-\de_{r,1} }{2} } \Big\} }{  \sqrt{ \big| \sum_{r=1}^{\ell} \veps_r \xi_r^2 n_r \big|   } } \\
\times \bigg\{  \Xi(\mf{x}) \f{ \sin[\pi \nu_+] }{  \pi } \, + \, \Xi(-\mf{x}) \f{ \sin[\pi \nu_-] }{  \pi }  \bigg\}    \Big(  1 + \e{O}(\mf{x}) \Big)  \; + \; \mc{ S }( \mf{x} ) \;. 
\label{ecriture DA kappa zero de integrale modele}
\end{multline}
where 
\beq
\vth \; = \; \f{1}{2}\sul{r=1}{\ell}n_r^2 \; - \; \f{3}{2}   \, + \, \de_+ \, + \, \de_-  \; 
\label{definition de cal theta}
\enq
and given $\eps \in \{ \pm \}$
\beq
\nu_{ \eps } \; = \;  \f{1}{2}\sul{  \substack{ r=1 \, : \, \\ \eps \veps_r= -1} }{ \ell }  n_r^2 \; - \; \f{ 1 +\eps \vsg }{ 4 }    \, + \, \sul{ \substack{ \ups=\pm  \\   \eps \sg_{\ups}>0 } }{} \de_{\ups}
\qquad with \qquad \left\{ \ba{ccc} 
					\vsg & =& \e{sgn}\Big( - \sul{r=1}{\ell} \veps_r n_r \xi_r^2 \Big)   \vspace{2mm} \\
					\sg_{\ups} &= &  1-\ups \f{ \op{u} }{ \op{v} }   \ea \right. \;. 
\label{definition du signe xi}
\enq

\end{prop}

\Proof

Recall the notation \eqref{notation vecteur dans produit cartesien} for the vector $\bs{x}$. 
Recall that, for $\om \in \R^{*}$, one has the integral representation 
\beq
\f{ \Ga(\a) }{ 2\pi } \Int{ \msc{C}_{\om} }{}  \dd t \f{ \ex{-\i \om t} }{ [-\i t]^{\a} }  \, = \, \om^{\a-1} \Xi(\om)
\label{ecriture rep int fonction saut fois puissance}
\enq
where the contour $\msc{C}_{\om}$ passes slightly above $0$ and then goes to infinity in either of the two directions $\Re(t)\tend \pm \infty$
along two rays $\mc{R}_{\om}$ which enjoy  the property that 
$\Im(\om t) \tend -\infty$, linearly in $|t|$,  when $ t \tend \infty$ along $\mc{R}_{\om}$. 

The maps  $\wh{z}_{\ups}=\mf{x}+z_{\ups}$ are such that 
\beq
D_{\bs{x}} \wh{z}_{\ups}\cdot \bs{h} = \big( \bs{x},\op{E}\bs{h}) \, + \, \f{\op{u}+\ups \op{v} }{2\op{v} } \big( \bs{e},\bs{h} \big) 
\enq
so that $D_{\bs{x}} \wh{z}_{\ups}$ is  surjective with the exception of the point $\bs{x}_{\ups}=-\op{E}^{-1}\bs{e}\tfrac{\op{u}+\ups \op{v} }{2\op{v} }$. 
Still, as ensured by the assumption \eqref{ecriture hypothese sur les nr veps et zetar}, one has, for $|\mf{x}|$ small enough,   
\beq
\wh{z}_{\ups}(\bs{x}_{\ups}) \, = \, \mf{x} -\f{1}{2} \Big( \f{\op{u}+\ups \op{v} }{2\op{v} } \Big)^2 \Big( \bs{e}, \op{E}^{-1} \bs{e} \Big) \not=0\;. 
\enq
Therefore, $\wh{z}_{\ups}$ is a submersion in an open neighbourhood of  $\mc{M}_{\ups}=(\,\wh{z}_{\ups})^{-1}(0)$. Hence, $\mc{M}_{\ups}$
is a $\sum_{r=1}^{\ell}n_r-1$ dimensional sub-manifold of $\mc{D}_{\bs{n}}$
and, as such,  has Lebesgue measure zero. It follows that upon setting $\mc{M}=\mc{M}_{+}\cup\mc{M}_-$, one has the representation
\beq
\f{ \mc{J}(\mf{x}) }{ \Ga(\de_+) \, \Ga( \de_-) } \; = \; 
%
%
%
 \Int{ \mc{D}_{\bs{n}} \setminus \mc{M} }{}  \hspace{-3mm} \dd \bs{x}\;  \ex{-(\bs{x},\bs{x}) } V(\bs{x}) \Int{ \msc{C}_{\,\wh{z}_+(\bs{x})} }{} \! \f{ \dd \la }{2 \pi} \, \f{ \ex{-\i  \la \, \wh{z}_+(\bs{x}) } }{  [-\i \la]^{\de_{+}} }
\Int{ \msc{C}_{\,\wh{z}_-(\bs{x})} }{} \!  \f{ \dd \mu }{2 \pi} \, \f{ \ex{-\i  \mu \, \wh{z}_-(\bs{x}) } }{  [-\i \mu]^{\de_{-}} }  \;. 
\enq
This representation follows by, first, replacing $\mc{D}_{\bs{n}}$ by $\mc{D}_{\bs{n}}\setminus\mc{M}$ in $\mc{J}(\mf{x})$ as given by \eqref{definition integrale modele pour extraire le comportement singulier}
and then using the integral representation \eqref{ecriture rep int fonction saut fois puissance} for the products involving the $\wh{z}_{\ups}$ functions. 
The form of the $\la,\mu$ contours ensures exponential decay at infinity of these integrals. Furthermore, it is easy to check that 
\beq
\bigg| \Int{ \msc{C}_{\,\wh{z}_{\pm}(\bs{x})} }{} \f{ \dd \mu }{2 \pi} \f{ \ex{-\i  \mu \, \wh{z}_{\pm}(\bs{x}) } }{  [-\i \mu]^{\de_{\pm}} }  \bigg| \; \leq \; C \, | \wh{z}_{\pm}(\bs{x}) |^{\de_{\pm}-1}
\enq
for some constant $C$. By using  Proposition \ref{Proposition reduction vers voisinage singularite integrale modele}, it is easy to see that 
\beq
\bs{x}  \; \mapsto  \; \ex{-(\bs{x},\bs{x}) } V(\bs{x}) \pl{\ups=\pm }{} \Big\{| \wh{z}_{\pm}(\bs{x}) |^{\de_{\pm}-1} \Big\}
\enq
is in $L^{1}( \mc{D}_{\bs{n}} )$. Thence, one can apply dominated convergence to get 
\beq
\f{ \mc{J}(\mf{x}) }{\Ga(\de_+) \, \Ga( \de_-) } \; = \; \lim_{\tau_t, \tau_s\tend 0^+} \Int{ \mc{D}_{\bs{n}} \setminus \mc{M} }{} \hspace{-3mm} \dd \bs{x}\;  \ex{-(\bs{x},\bs{x}) } V(\bs{x}) \hspace{-4mm}
\Int{ \msc{C}_{\,\wh{z}_+(\bs{x})} \times \msc{C}_{\,\wh{z}_-(\bs{x})} }{} \hspace{-5mm} \f{ \dd \la \,  \dd \mu }{(2 \pi)^2} \f{ \ex{-\i  \la \, \wh{z}_+(\bs{x}) } }{  [-\i \la]^{\de_{+}} }
\cdot  \f{ \ex{-\i  \mu \, \wh{z}_-(\bs{x}) } }{  [-\i \mu]^{\de_{-}} }   \cdot \exp\bigg\{ -\tau_t\Big(\f{\la+\mu}{2}\Big)^2-\tau_s\Big(\f{\la-\mu}{2}\Big)^2 \bigg\}  \;. 
\enq
Since the integrand under the limit has now Gaussian convergence in $\la, \mu \tend \infty$, one can deform the contours $\msc{C}_{\,\wh{z}_{\ups}(\bs{x})} \hookrightarrow \R+\i \a$
for some $\a>0$ and small enough. Since the new $\la, \mu$ contours become $\bs{x}$ independent and one has a rapid convergence of the integrand at infinity, by Fubbini's theorem, one can swap the order of integration and take the 
$\bs{x}$ integration first. Then, using again that $\mc{M}$ has Lebesgue measure $0$,  yields
\beq
\f{ \mc{J}(\mf{x}) }{ \Ga(\de_+) \, \Ga( \de_-) } \; = \;  \lim_{\tau_t, \tau_s\tend 0^+} \Int{ (\R+\i\a)^2  }{}  \hspace{-3mm}  \f{ \dd \la \, \dd \mu }{(2 \pi)^2}  
\Int{ \mc{D}_{\bs{n}}   }{}   \dd \bs{x}\;  \ex{-(\bs{x},\bs{x}) } V(\bs{x})  
\f{ \ex{-\i  \la \, \wh{z}_+(\bs{x}) } }{  [-\i \la]^{\de_{+}} }
\cdot  \f{ \ex{-\i  \mu \, \wh{z}_-(\bs{x}) } }{  [-\i \mu]^{\de_{-}} }   \cdot \exp\bigg\{ -\tau_t\Big(\f{\la+\mu}{2}\Big)^2-\tau_s\Big(\f{\la-\mu}{2}\Big)^2 \bigg\}  \;. 
\enq
Then the change of variables 
\beq
\la\, = \, t\Big( 1 - \tfrac{ \op{u} }{ \op{v} }\Big) \, + \, s   \quad \e{and}  \quad \mu \, = \, t\Big( 1 + \tfrac{ \op{u} }{ \op{v} }\Big) \, - \, s
\enq
recasts the integral in the form 
\beq
 \mc{J}(\mf{x})  \; = \;     \lim_{\tau_t, \tau_s\tend 0^+}
\Int{ \R+\i\a  }{}  \hspace{-2mm}   \dd t   \Int{ \R +\i \a_{\op{u}}  }{}    \hspace{-1mm}     \dd s   \ex{ -\tau_t t^2-\tau_s (s-\f{\op{u}}{\op{v}}t) ^2 }
\chi(t,s) \; \mc{I}_{V}(t,s)
\enq
where $\a_{\op{u}}=\f{ \op{u} }{ \op{v} } \a$, 
\beq
\chi(t,s) \, = \, \pl{\ups=\pm}{} \bigg\{ \f{ \i }{   t \cdot \sg_{\ups}+ \ups s   } \bigg\}^{ \de_{\ups} } \quad \e{and} \quad 
\sg_{\ups}=1-\ups \f{ \op{u} }{ \op{v} } \;. 
\label{definition chi et sigma ups}
\enq
while 
\beq
\mc{I}_{V}(t,s)  \, = \, \f{ \Ga(\de_+) \, \Ga( \de_-) }{  2\pi^2  } \Int{ \mc{D}_{\bs{n}}   }{}   \dd \bs{x}\;   V(\bs{x})  \cdot 
  \ex{-  \i s   ( \bs{x}, \bs{e})  }   \cdot  \ex{-(\bs{x},[\e{id}+\i t \op{E}]\bs{x}) -2 \i t \mf{x} } 
\enq
Above, $\e{id}$ refers to the identity matrix acting on $\mc{D}_{\bs{n}}$. 

\vspace{2mm}
The $\bs{x}$ integral can already be taken explicitly. Indeed, for $t \in \R+\i\a$ with $|\a|<1$, upon dilating the variables and then shifting them, \textit{viz}. leading to the substitution 
\beq
\bs{x}^{(r)} \; = \;  \f{ \bs{y}^{(r)} - \i s \, \bs{e}^{(r)}  }{  2 \sqrt{1+\i \veps_r t } }  \quad \bs{y}^{(r)} \in \R^{n_r} \, , 
\enq
one obtains
\beq
\mc{I}_{V}(t,s)  \, = \, \f{  \ex{ - \be_{t} s^2   }  \Ga(\de_+) \, \Ga( \de_-)  }{  2\pi^2  \pl{r=1}{\ell}    \big[ 1+\i \veps_r t \big]^{ \frac{ 1 }{2}  n_r^2   }    } \cdot \ex{-2\i t \mf{x} }
 \cdot \Int{ \mc{D}_{\bs{n}} }{} \dd \bs{y} \;  V(\bs{y})  \ex{  - ( \bs{y}     , \bs{y}   ) }   \;,  
\enq
where  
\beq
\be_t\;= \; \sul{r=1}{\ell} \f{\op{v}^2  \xi_r^2 n_r }{ 1+\i\veps_r t } \;.
\label{definition constante beta de t et suite mr}
\enq
The remaining integrals can be taken by means of the Gaudin-Mehta formula \eqref{formule integrale Gaudin-Mehta}, leading to 
\beq
\mc{I}_{V}(t,s)  \, = \,  \f{    \Ga(\de_+) \, \Ga( \de_-)  }{  2\pi^2  \cdot  \ex{  \be_{t} s^2   }    } 
\pl{r=1}{\ell}  \Bigg\{  \f{   \big( 2\pi \big)^{ \f{n_r}{2} } G(2+n_r)  }{   \big[ 2(1+\i \veps_r t) \big]^{ \f{1}{2} n_r^2 }    }   \Bigg\} \cdot \ex{-2\i t \mf{x} }\;. 
\enq

Thus, all-in-all, one gets 
\beq
 \mc{J}(\mf{x})  \; = \;     \f{    \Ga(\de_+) \, \Ga( \de_-)  }{  2\pi^2       } 
										\pl{r=1}{\ell}  \Bigg\{  \f{   \big( 2\pi \big)^{ \f{n_r}{2} } G(2+n_r)  }{    2 ^{ \f{1}{2} n_r^2 }    } \Bigg\}     \cdot \mc{K}(\mf{x})   \;, 
\enq
where 
\beq
 \mc{K}(\mf{x})  \, = \,  \lim_{\tau_t, \tau_s\tend 0^+}
\Int{ \R+\i\a  }{}  \hspace{-1mm}   \dd t    \Int{ \R +\i \a_{\op{u}}  }{}    \hspace{-1mm}     \dd s  \;  \ex{ -\tau_t t^2-\tau_s (s-\f{\op{u}}{\op{v}}t) ^2 }
\f{ \chi(t,s)  \,  \ex{-\be_{t}s^2-2 \i t \mf{x} }     }{ \prod_{r=1}^{\ell}  \big[  1+\i \veps_r t  \big]^{ \frac{1}{2}n_r^2 }  }  \;. 
\label{formule intermediaire pour J kappa modele}
\enq

Since the double limit  $ \lim_{\tau_t, \tau_s\tend 0^+}$ exists, it can be computed in any way, in particular, by taking the successive limits 
$ \lim_{\tau_t \tend 0^+} \lim_{\tau_s\tend 0^+}$. For $(t,s)\in (\R+\i \a)\times (\R+\i\a_{\op{u}})$ one has the lower bound
$|t \sg_{\ups}+\ups s| \geq \a  >0$ and thus 
\beq
 \big| \chi(t,s) \big| \, \leq \, \pl{\ups= \pm }{}  \a^{-\de_{\ups}} \;. 
\enq
For such $t,s$, owing to $|\be_{t}|\leq C$ for some $C>0$, one thus has the bound
\beq
 \Big| \ex{ -\tau_t t^2-\tau_s (s-\f{\op{u}}{\op{v}}t) ^2 }
\f{ \chi(t,s)  \, \ex{-\be_{t}s^2-2 \i t \mf{x} } }{ \prod_{r=1}^{\ell}  \big[  1+\i \veps_r t  \big]^{ \frac{1}{2} n_r^2  }  } \Big| \; \leq \; W(t,s)
\label{ecriture borne pour dominantion limite taus vers zero dans s integrale}
\enq
with 
\beq
 W(t,s)
=C   \cdot 
 \ex{-\tau_t (\Re(t))^2 } \ex{-\Re(\be_{t}) (\Re(s))^2 +2 \a_{\op{u}} \Re(s) \Im(\be_{t}) } \;. 
\enq
Observe that $\Re(\be_t)>0$ for finite $t$ and 
\beq
\be_t= -\f{\i}{t}\op{v}^2   \sul{r=1}{\ell} \veps_r \xi_r^2 n_r +  \f{1}{t^2}\sul{r=1}{\ell} \op{v}^2 n_r \xi_r^2 + \e{O}(t^{-3}) \quad \e{when} \quad \Re(t) \tend \pm \infty\, . 
\enq
Thus, for $\a$ small enough,  one has the bound
\beq
\Int{\R+\i\a }{} \hspace{-2mm}  \dd t \Int{ \R + \i\a_{\op{u}} }{} \hspace{-2mm} \dd s \; W(t,s) \; \leq  \; 
 C \Int{\R+\i\a }{} \hspace{-2mm} \dd t \ex{-\tau_t (\Re(t))^2 }   \ex{   \a_{\op{u}}^2  \f{ (\Im(\be_{t}))^2 }{ \Re(\be_{t}) }  } \cdot  \Int{ \R + \i\a_{\op{u}} }{} \hspace{-2mm} \dd s \;   \ex{-\Re(\be_{t}) (\Re(s))^2 }   \;. 
\enq
Then, owing to the asymptotics at large $\Re(t)$ of $\be_t$, observe that for  some $C_0$
\beq
  \Big| \tfrac{ \Im(\be_{t})  }{ \sqrt{ |\Re(\be_{t})| } }  \Big| \, \leq  \, C_0   \, . 
\enq
Upon the rescaling $\Re(s) \hookrightarrow   |\Re(\be_{t})|^{-\tfrac{1}{2} } \Re(s)$, the above bounds entail that, for some constant $C^{\prime}$ 
\beq
\Int{\R+\i\a }{} \hspace{-2mm}  \dd t \Int{ \R + \i\a_{\op{u}} }{} \hspace{-2mm} \dd s \; W(t,s) \; \leq  \;
 C^{\prime} \Int{\R  }{}  \dd t \ex{-\tau_t t^2 }    \cdot \big( 1+|t| \big) \cdot  \Int{ \R   }{}   \dd s \;   \ex{-  s^2 }  
		    \; < \; +\infty\;. 
\enq
Hence, since the bounding function in \eqref{ecriture borne pour dominantion limite taus vers zero dans s integrale} is positive, by Fubbini's theorem, 
the above estimate ensures that it is in $L^1\Big( (\R+\i \a)\times (\R+\i\a_{\op{u}}) \Big)$, so that one can apply dominated convergence so as to take the $\tau_{s}\tend 0^+$ limit in \eqref{formule intermediaire pour J kappa modele}, 
hence yielding 
\beq
 \mc{K}(\mf{x})  \; = \;   \lim_{\tau_t \tend 0^+}
\Int{ \R+\i\a  }{}  \hspace{-2mm}   \dd t  
\f{ \ex{ -\tau_t t^2  } \ex{ -2 \i t \mf{x} }  }{ \prod_{r=1}^{\ell}  \big[  1+\i \veps_r t  \big]^{ \frac{1}{2}n_r^2 }   } \cdot \mc{F}(t)
\qquad \e{where} \qquad 
\mc{F}(t) \, = \, \Int{ \R +\i \a_{\op{u}}  }{}    \hspace{-2mm}    \dd s  \; \chi(t,s)  \ex{-\be_{t}s^2} \;. 
\enq
The  function $\mc{F} (t)$ is analysed in Lemma \ref{Lemme cpmyt asymptotique de F}, which ensures that there exist three functions $\mc{F}^{(a)}(t)$
such that $\mc{F}=\mc{F}^{(1)}+\mc{F}^{(2)}+\mc{F}^{(3)}$ where 
\begin{itemize}

 \item $\mc{F}^{(1)}$ and  $\mc{F}^{(2)}$ are holomorphic on the set 
\beq
\mc{S}_{\th_0;A} \, = \, \Big\{ t \in \Cx : |\Re(t)|\, > \, A \; \;  \e{and} \; \; t=\rho \ex{\i \th} \quad \e{with} \quad  \rho \in \R \quad \e{and} \quad  |\th|<\th_0  \Big\} 
\enq
where $A$ is large enough while $\th_0$ is small enough;

\item when $ \Re(t) \tend + \infty $
\beq
\mc{F}^{(1)}(t) \; = \; \sqrt{\pi} \cdot \chi(t,0) \cdot \Big( \tfrac{1}{\be_t} \Big)^{ \f{1}{2} }  
\cdot \Big( 1+ \e{O}\Big( \f{1}{ t  }\Big) \Big)
\label{ecriture DA partie algebrique fct F varrho un}
\enq
 with a remainder that is uniform and holomorphic on $\mc{S}_{\th_0;A}$;  

\item there exists $C_1,C_2>0$ such that, for $\rho \tend  \pm \infty$, $\th>0$ and $\vsg=-\e{sgn}\big( \sum_{r=1}^{\ell} \veps_r \xi_r^2 n_r \big) $,  
\beq
\Big|  \mc{F}^{(2)}\Big( \rho \ex{-\i\th \vsg \e{sgn}(\rho) } \Big) \Big| \, \leq \, C_1 \ex{ - C_2 |\rho| \th } \quad \e{and} \quad
\big|  \mc{F}^{(3)}( t ) \big| \, \leq \, C_1 \ex{ - C_2 |\Re(t)| } \quad \e{for} \quad t\in \R+\i\a \; . 
\enq

\end{itemize}

Define the contours $\msc{C}^{(1)} \, = \,  \mc{R}_{0}^{(\mf{x})}\cup_{\ups=\pm} \mc{R}_{\ups}^{(\mf{x})} $, where $\mc{R}_{0}^{(\mf{x})}$ is a curve joining $-A$ to $A$ in the upper half-plane but having a sufficiently small imaginary part, 
while $\mc{R}_{\pm}^{(\mf{x})}$ are two rays  
\beq
\mc{R}_{\pm}^{(\mf{x})} \; = \; \bigg\{  z\; = \; \pm A \pm  \rho \ex{ \mp \i \th  \e{sgn}(\mf{x}) } \;, \quad  \rho \in \R^+    \bigg\} 
\enq
going to $\infty$ in the direction $\Re(z)\tend \pm \infty$ with a slight angle $\th>0$ small enough, so that $\Im( t \mf{x}) \tend +\infty$ linearly in $|t|$ along these rays. 
The contour  $\msc{C}^{(2)}$ has a similar structure:  $\msc{C}^{(2)} \, = \,  \mc{R}_{0}^{(\mf{x})}\cup_{\ups=\pm} \mc{R}_{\ups;\th} $ with the 
two rays $ \mc{R}_{\ups;\th}$ given as 
\beq
\mc{R}_{\ups;\th} \; = \; \bigg\{  z\; = \; \ups A + \ups  \rho \ex{ -\ups \vsg \i \th   } \;, \quad  \rho \in \R^+    \bigg\} 
\enq
 for $\th >0$ and small enough. Finally, take $\msc{C}^{(3)}\, = \, \R + \i\a$.

Upon 
\begin{itemize}
\item[i)] inserting the decomposition $\mc{F}  \, = \, \sum_{a=1}^{3}\mc{F}^{(a)}$ into $\mc{K}(\mf{x})$;
\item[ii)] splitting the integrations for each piece ;
\item[iii)] deforming the $t$-integration contour to $\msc{C}^{(a)}$ in the integrals associated with $\mc{F}^{(a)}$;
\end{itemize}
one obtains three integrals whose respective integrands decay, uniformly in $\tau_t$ small enough, exponentially fast to $0$ along $\msc{C}^{(a)}$.  
Thus, one can invoke dominated convergence so as to send $\tau_t\tend 0^+$ and obtain that
\beq
\mc{K}(\mf{x})=\sul{a=1}{3} \mc{K}^{(a)}(\mf{x})  \qquad  \e{with}  \qquad 
 \mc{K}^{(a)}(\mf{x}) \; = \;  \Int{ \msc{C}^{(a)} }{}    \dd t  
\f{ \ex{ -2 \i t \mf{x} } \, \mc{F}^{(a)}(t) }{ \prod_{r=1}^{\ell}  \big[  1+\i \veps_r t  \big]^{ \frac{1}{2} n_r^2 }  }  \;. 
\enq

Since, the integrands in $ \mc{K}^{(2)}(\mf{x})$ and $ \mc{K}^{(3)}(\mf{x})$ are bounded and decay exponentially fast to $0$ at $\infty$, this uniformly in $|\mf{x}|$ small enough, 
one can apply derivation under the integral theorems so as to infer that $ \mc{K}^{(2)}+ \mc{K}^{(3)}$ is smooth in $\mf{x}$ around $0$.

Hence, it remains to focus on the $\mf{x} \tend 0^+ $ behaviour  of $ \mc{K}^{(1)}(\mf{x})$ which can be decomposed as 
\beq
 \mc{K}^{(1)}(\mf{x}) \, = \,  \sul{ c \in \{ \pm , 0 \} }{} \mc{K}^{(1)}_{c}(\mf{x}) \quad \e{with} \quad 
 \mc{K}^{(1)}_{ c}(\mf{x}) \, = \,  \Int{ \mc{R}_{ c }^{(\mf{x})} }{} \hspace{-2mm} \dd t \;   \f{ \ex{ -2 \i t \mf{x} } \, \mc{F}^{(1)}(t) }{ \prod_{r=1}^{\ell}  \big[  1+\i \veps_r t  \big]^{ \frac{1}{2} n_r^2 }  }   \;. 
\enq

Since the integration in $\mc{K}^{(1)}_{0}$ runs through a compact set and since the integrand in bounded, it follows that $\mc{K}^{(1)}_{0}(\mf{x})$ is a smooth function of $\mf{x}$ by derivation under the integral theorems. 
It thus remains to estimate $\mc{K}^{(1)}_{\pm}(\mf{x})$. The properties of $\mc{F}^{(1)}(t)$ ensure that 
\beq
\f{   \mc{F}^{(1)}(t) }{ \prod_{r=1}^{\ell}  \big[  1+\i \veps_r t  \big]^{ \frac{1}{2}n_r^2 }  } \; = \; \vp_{\mf{s}_t}(t) \; + \; \psi_{\mf{s}_t}(t) \qquad \e{with} \qquad \psi_{\mf{s}_t}(t)=\e{O}\Big( \f{ \vp_{\mf{s}_t}(t) }{ t } \Big)
\;\; , \; \mf{s}_t \,=\, \e{sgn}\big[ \Re(t) \big]
\enq
and $\vp_{\mf{s}_t }(t)\, = \,\ga_{\mf{s}_t}  \cdot \big(\mf{s}_t \cdot t)^{ -\vth - 1  }$ with  $\vth$ as defined in \eqref{definition de cal theta}. 
The constant prefactor takes the form 
\beq
\ga_{\mf{s}_t} \; = \;  \f{ \sqrt{\pi}  \cdot \ex{ -\i \vsg \mf{s}_t \f{\pi}{4} }   }
				    { \Big| \op{v}^2 \sum_{r=1}^{\ell} n_r \veps_r \xi_r^2  \Big|^{\frac{1}{2} }    }
\cdot \pl{\ups = \pm }{} \bigg\{   \f{ \ex{ \i \mf{s}_t \mf{s}_{\sg_{\ups}} \f{\pi}{2} } }{ |\sg_{\ups}| }   \bigg\}^{\de_{\ups}}
\cdot \pl{r=1}{\ell} \Big\{ \ex{-\i \veps_r \mf{s}_t n_r^2 \f{\pi}{4} }  \Big\} \;. 
\enq
$\vsg$ has been introduced in \eqref{definition du signe xi}, $\sg_{\ups}$ is given by \eqref{definition chi et sigma ups}, 
 while $\mf{s}_{\sg_{\ups}}=\e{sgn}\big( \sg_{\ups} \big)$. 

This being settled, one splits the integrals as
\beq
\mc{K}^{(1)}_{\pm}(\mf{x}) \; = \; \msc{J}^{(1)}_{\pm}(\mf{x})  \, + \, \de \! \msc{J}^{(1)}_{\pm}(\mf{x}) 
\enq
where 
\beq
\msc{J}^{(1)}_{ \ups}(\mf{x})  \, = \, \Int{ \mc{R}_{ \ups}^{(\mf{x}) } }{}    \ex{ -2 \i t \mf{x} } \,   \vp_{\ups  }(t) \cdot  \dd t   \qquad \e{and} \qquad 
\de \! \msc{J}^{(1)}_{ \ups}(\mf{x}) \, = \, \Int{ \mc{R}_{ \ups}^{(\mf{x}) } }{}     \ex{ -2 \i t \mf{x} }  \,  \psi_{\ups}(t)\cdot  \dd t \;. 
\enq
The rest depends on how large $\vth$ is. Let $n \in \mathbb{N}$ and $0\leq \a <1$ be such that $\vth+2=\a+n$.  
First, focus on $\de \! \msc{J}^{(1)}_{ \ups}(\mf{x})$ and introduce $\psi_{\ups;0} \, = \, \psi_{\ups}$, and, for $p\geq 1$, 
\beq
\psi_{\ups;p}(t)\; = \; \Int{ \mc{R}_{ \ups}^{(\mf{x}) }  }{ t }  \psi_{\ups;p-1}(s) \cdot  \dd s 
\enq
where the integration runs from $\infty$, along $\mc{R}_{ \ups}^{(\mf{x}) }$, up to $t \in \mc{R}_{ \ups}^{(\mf{x}) }$. 
Then, integrating by parts $n$ times, one has
\beq
\de \! \msc{J}^{(1)}_{ \ups}(\mf{x}) \, = \, -\ups \sul{p=0}{n-1} (2\i \mf{x} )^{p} \, \psi_{ \ups ; p +1}( \ups A ) \, \ex{-2\i \ups \mf{x} A t } 
\, + \,  (2\i \mf{x} )^{n}  \Int{ \mc{R}_{ \ups}^{(\mf{x}) } }{}      \ex{ -2 \i t \mf{x} }  \,  \psi_{ \ups ; n }( t ) \cdot  \dd t \;. 
\label{ecriture dvpmt de J 1 ups}
\enq
One has  $ \psi_{ \ups ; n }( t )= \e{O}\big( |t|^{-\a} \big)$. Upon setting $\mf{s}_{\mf{x}}=\e{sgn}(\mf{x})$ and taking $\th>0$ and small enough, one gets 
\bem
\bigg| \Int{ \mc{R}_{ \ups}^{(\mf{x}) } }{}      \ex{ -2 \i t \mf{x} }   \psi_{ \ups ; n }( t ) \cdot  \dd t  \bigg| \; = \; 
\bigg| \Int{ 0 }{ + \infty}      \ex{ -\ups 2 \i \rho \mf{x} \ex{ -\ups \i \mf{s}_{\mf{x}}\th}  }    \psi_{ \ups ; n }\Big( \ups A  + \ups \rho \ex{ -\ups \i \mf{s}_{\mf{x}}\th} \Big) \cdot  \dd  \rho  \bigg| \\
\; \leq \; C  \Int{ 0 }{ + \infty}    \f{  \ex{ - 2  \rho  |\mf{x}| \sin(\th)   } }{ \big[ A + \cos(\th) \rho  \big]^{\a} }     \cdot  \dd \rho  
\; = \;   C \,  |\mf{x}|^{ \a - 1 }    \Int{ 0 }{ + \infty}      u_{\mf{x}}( \rho )     \cdot  \dd \rho  \;, 
\end{multline}
for some constant $C>0$.  In the last integral, I have set 
\beq
u_{\mf{x}}(\rho )  =  \ex{ - 2  \rho    \sin(\th)   } \cdot \big[  |\mf{x}| A + \cos(\th) \rho  \big]^{-\a}  \; \limit{ \mf{x}  }{ 0 } \, u_{0}(\rho ) \in L^1(\R^+)
\enq
point-wise on $\R^{+}\setminus\{ 0 \}$. Since $u_{\mf{x}}(\rho) \leq u_{0}(\rho)$, by dominated convergence, the expansion \eqref{ecriture dvpmt de J 1 ups} ensures that there exist smooth functions $g_{\pm}$ such that 
\beq
\de \! \msc{J}^{(1)}_{ \ups }(\mf{x}) \, = \, g_{\ups}(\mf{x})  \, + \, \e{O}\Big( |\mf{x}|^{ \vth +1}  \Big) \;. 
\enq

The integral $ \msc{J}^{(1)}_{ \ups }(\mf{x})$ can be dealt with analogously by doing $n-1$ integration by parts, where $\vth + 1 = n - 1 + \a$. Namely, one has 
\bem
  \Int{ \mc{R}_{ \ups }^{(\mf{x})} }{}  \f{   \ex{ -2 \i  \mf{x} t  }   }{ (\ups t)^{\vth+1} }   \cdot \dd t 
\; = \;    \Ga\bigg( \ba{c} \vth  \\ \vth+1 \ea \bigg) \f{ 1 }{ A^{\vth } }    \ex{ -2 \i \ups  \mf{x} A  }
\, -\, 2\i \mf{x} \ups \Ga\bigg( \ba{c} \vth  \\ \vth +1 \ea \bigg) \Int{ \mc{R}_{ \ups }^{(\mf{x})} }{}  \f{  \ex{ -2 \i  \mf{x} t  }    }{ (\ups t)^{\vth} }     \cdot \dd t  \\
\; = \;   \ex{ -2 \i \ups  \mf{x} A  } \sul{p=0}{n-2} \f{ \big(-2\i \mf{x} \ups\big)^{p }  }{ A^{\vth-p} }   \Ga\bigg( \ba{c} \vth-p  \\ \vth+1 \ea \bigg)       
\, +\,   \big(-2\i \mf{x} \ups\big)^{n-1 }  \Ga\bigg( \ba{c} \vth -n +2 \\ \vth +1 \ea \bigg) \Int{ \mc{R}_{ \ups }^{(\mf{x})} }{}  \f{  \ex{ -2 \i  \mf{x} t  }    }{ (\ups t)^{\a} }     \cdot \dd t  \;. 
\end{multline}
All this leads to the representation 
\beq
 \msc{J}^{(1)}(\mf{x})= \sul{\ups=\pm}{} \msc{J}^{(1)}_{\ups}(\mf{x}) \, = \,  \msc{J}^{(1)}_{\e{reg}}(\mf{x}) \, + \,  \msc{J}^{(1)}_{\e{sing}}(\mf{x})
\enq
where $\msc{J}^{(1)}_{\e{reg}}$ is smooth and given by 
\beq
\msc{J}^{(1)}_{\e{reg}}(\mf{x}) \; = \;   \sul{p=0}{n-2}  \sul{\ups=\pm}{}    \ga_{ \ups }  \cdot   \Ga\bigg( \ba{c} \vth - p  \\ \vth + 1 \ea \bigg)  \,  \cdot \, 
\ex{ -2 \i \ups  \mf{x} A  } \f{ \big(-2\i \mf{x} \ups\big)^{p }  }{ A^{ \vth - p } }  
\enq
while, upon deforming the contours $\mc{R}_{ \ups }^{(\mf{x})}$ to $\ups A -\i \mf{s}_{\mf{x}} \R^+_{\ups}$,  where $\R^+_{\ups}$ corresponds to $\R^+$ oriented with the sign $\ups$,  
\bem
 \msc{J}^{(1)}_{\e{sing}}(\mf{x}) \, = \,  \Ga\bigg( \ba{c} \vth  -n +2 \\ \vth +1 \ea \bigg) \cdot |2 \mf{x}|^{n-1} \Bigg\{
\ga_{ - }  \cdot  \big( \i \mf{s}_{\mf{x}} \big)^{n  }    \ex{2\i \mf{x} A }  \Int{ 0 }{ + \infty }  \f{  \ex{ -2 |\mf{x}| t  }    }{ (A + \i t \mf{s}_{\mf{x}}  )^{\a} }     \cdot \dd t \\
\hspace{7cm} \, + \,  \ga_{ + }  \cdot  \big(- \i \mf{s}_{\mf{x}} \big)^{n  }    \ex{-2\i \mf{x} A }  \Int{ 0 }{ + \infty }  \f{  \ex{ -2 |\mf{x}| t  }    }{ (A -\i t \mf{s}_{\mf{x}}  )^{\a} }     \cdot \dd t 
\Bigg\} \\
 \, = \, (-1)^n |2 \mf{x}|^{ \vth  }  \cdot \Ga\bigg( \ba{c} \vth -n +2 \\ \vth + 1 \ea \bigg)  \sul{\ups=\pm}{}
\ga_{\ups}  \cdot  \big( \ups\i \mf{s}_{\mf{x}} \big)^{n+\a  }    \ex{-2\i \mf{x}  \ups A }  \Int{ 0 }{ + \infty }  \f{  \ex{ - t  }    }{ (t +\ups \i 2\mf{x} A  )^{\a} }     \cdot \dd t \\
\noindent 
\end{multline}
By dominated convergence, one has that 
\beq
\Int{ 0 }{ + \infty }  \f{  \ex{ - t  }    }{ (t +\ups \i 2\mf{x} A  )^{\a} }     \cdot \dd t \; = \;  \Ga(1-\a) \cdot (1+\e{o}(1) ) 
\enq
when $\mf{x} \tend 0$. With some more efforts, one can even establish that  it is a
$\e{O}(\mf{x})$. Since $\Ga(1-\a)=\Ga(n-1-\vth )$, straightforward calculation lead to 
\beq
 \msc{J}^{(1)}_{\e{sing}} (\mf{x}) \, = \,  2  | 2 \mf{x} |^{ \vth  }\cdot  \f{ \sqrt{\pi} \cdot \Ga( -\vth ) \cdot  \prod_{\ups= \pm }^{}  | \sg_{\ups} |^{-\de_{\ups}}   }
  {   \Big| \op{v}^2 \sum_{r=1}^{\ell} n_r \veps_r \xi_r^2  \Big|^{\frac{1}{2} }      }
\sul{\ups=\pm}{} \bigg\{    \Xi( \ups \mf{x})  \cdot \sin[\pi \nu_{\ups} ] 
%
%
\bigg\}  \cdot   \Big(  1 + \e{O}(\mf{x}) \Big)   \;. 
\enq
Above, it is understood that 
\beq
\nu_{ \eps }  \; = \;  \frac{1}{2}\sul{  \substack{ r=1 \, : \, \\ \eps \veps_r= -1} }{ \ell }  n_r^2 \; - \; \f{ 1 +\eps \vsg }{ 4 }    \; +  \sul{ \substack{ \ups=\pm  \\   \eps \sg_{\ups}>0 } }{} \de_{\ups} \;. 
\enq

Finally, the obtained estimates are readily seen to be differentiable. Thus, upon putting all the results together, the asymptotic expansion given in \eqref{ecriture DA kappa zero de integrale modele} follows.

\subsection{Asymptotics of auxiliary functions}

\begin{lemme}
 \label{Lemme cpmyt asymptotique de F}

Let $\a>0$,  $t \in \R + \i\a $, $(\op{u},\op{v}) \in \R \times \R^+$ be such that $\op{u}\not= \pm \op{v}$ and set $\a_{\op{u}} \, = \, \f{ \op{u} }{ \op{v} } \a $. Let  
\beq
\mc{F}_n(t) \, = \, \Int{ \R +\i \a_{\op{u}}  }{}    \hspace{-2mm}    \dd s  \; \chi(t,s) \, s^{n} \,  \ex{-\be_{t}s^2}
\enq
where $\be_t$ is as defined in \eqref{definition constante beta de t et suite mr} and $\chi(t,s)$ has been introduced in \eqref{definition chi et sigma ups}. 
Then, there exist  functions $\mc{F}_n^{(a)}(t)$, $a=1,2,3$ such that $\mc{F}_n=\mc{F}_n^{(1)} + \mc{F}_n^{(2)} + \mc{F}_n^{(3)}$, and enjoying the properties:
\begin{itemize}

 \item $\mc{F}_n^{(1)}$ and  $\mc{F}_n^{(2)}$ are holomorphic on the set 
\beq
\mc{S}_{\th_0;A} \, = \, \Big\{ t \in \Cx : |\Re(t)| \, > \, A \; \;  and  \; \; t=\rho \ex{\i \th} \quad  with  \quad  \rho \in \R \quad and \quad  |\th|<\th_0  \Big\} 
\enq
where $A$ is large enough while $\th_0$ is small enough;

\item when $ \Re(t) \tend + \infty $
\beq
\mc{F}_{n}^{(1)}(t) \; = \; \chi(t,0) \cdot \Big( \tfrac{1}{\be_t} \Big)^{ \f{1+n}{2} } \cdot  u_n \cdot \Big( \tfrac{1}{ t \sqrt{\be_{t}} }\Big)^{w_n}
\cdot \Big( 1+ \e{O}\Big( \f{1}{ t  }\Big) \Big)
\enq
 for some constant 
\beq
\left\{ \ba{ccc}   u_n =  \Ga\Big( \tfrac{n+1}{2} \Big)   & \e{if} & n \in  2\mathbb{N}  \vspace{1mm} \\ 
	    u_n  \not=0 & \e{if} & n \in  2\mathbb{N} +1		\ea \right.   \; , 
    \quad \; an \; integer \quad \left\{ \ba{ccc}   w_n = 0   & \e{if} & n \in  2\mathbb{N}  \vspace{1mm} \\ 
	    w_n  \geq 1 & \e{if} & n \in  2\mathbb{N} +1		\ea \right.  \;, 
\label{definition proprietes un et wn}
\enq
and with a remainder that is uniform and holomorphic on $\mc{S}_{\th_0;A}$;

\item there exists $C_1,C_2>0$ such that, for $\rho \tend  \pm \infty$ and $\th>0$,  
\beq
\Big|  \mc{F}_n^{(2)}\Big( \rho \ex{-\i \th \vsg \e{sgn}(\rho) } \Big) \Big| \, \leq \, C_1 \ex{ - C_2 |\rho| \th } \quad  and  \quad
\big|  \mc{F}_n^{(3)}( t ) \big| \, \leq \, C_1 \ex{ - C_2 |\Re(t)|  } \quad  for  \quad t\in \R+\i\a \; ,
\enq
where $\vsg  =    - \e{sgn} \Big( \sul{r=1}{\ell} \veps_r \xi_r^2  n_r \Big)$. 

\end{itemize}

\end{lemme}

 \Proof 
 
One has to distinguish between the two cases: $|\op{u}| \, > \, \op{v}$ or $|\op{u}| \, < \, \op{v}$. Also, recall the notation $\sg_{\ups} \, = \, 1-\ups \tfrac{ \op{u} }{ \op{v} }$

\subsection*{A)  The regime  $|\op{u}| \, > \, \op{v}$  }

 Let $\mf{s}_{\op{u}}= \e{sgn}(\op{u})$, so that, using that $\e{sgn}(\sg_{\ups}) \, = \,  -\ups  \mf{s}_{\op{u}} $ one can recast
\beq
\chi(t,s) \, = \, \pl{ \ups= \pm }{} \bigg\{  \f{   \i \ups }{ s \, - \, t  \mf{s}_{ \op{u} } |\sg_{\ups}|     }   \bigg\}^{ \de_{\ups} }
\enq
meaning that, for fixed $t$, the map $s \mapsto \chi(t,s)$ has cuts along $  \mf{s}_{\op{u}} t |\sg_{\ups}| -\i  \R^{\ups}$. 
Since the integration runs through $\R+\i \a_{\op{u}}$, and, for $t \in \R + \i \a $,  
\beq
\Im\Big( \i \a_{\op{u}} \, - \, \mf{s}_{\op{u}}|\sg_{\ups}| t  \Big) \; = \; \a \ups 
\enq
the cut along $  \mf{s}_{\op{u}} t |\sg_{+}| -\i \R^{+}$ lies below the integration line  $\R+\i \a_{\op{u}}$ and the one along 
$  \mf{s}_{\op{u}} t |\sg_{-}| - \i \R^{-}$ lies above the integration line  $\R+\i \a_{\op{u}}$. 
One can represent $\be_{t} \, = \, \ex{ \i \th_{\be_t} } |\be_{t}|$. Since, for $t$  large, 
\beq
\be_{t}\;  = \; \i \f{ \vsg  C_{\be} }{ t } \; + \; \e{O}\big( t^{-2} \big)
\quad \e{one} \, \e{has} \quad 
\th_{\be_t} \, \sim \, \f{\pi}{2} \mf{s}_t \vsg  \qquad \e{with} \qquad 
 \left\{ \ba{ccc}     \vsg  &  =  &  - \e{sgn} \Big( \sul{r=1}{\ell} \veps_r \xi_r^2  n_r \Big)   \\ 
		          C_{\be}  &  =    &  \Big| \sul{r=1}{\ell} \veps_{r} \xi_r^2 \op{v}^2 n_r \Big|   \ea \right.  \;. 
\label{definition de xiu et C beta}
\enq
Here, I introduced the shorthand notation $\mf{s}_{t}=\e{sgn}\Big( \Re(t) \Big)$. Also, for further convenience, it is useful to set 
\beq
\vsg_{\op{u}} \, = \, \mf{s}_{\op{u}} \cdot \vsg  \;. 
\enq

It is then convenient to deform the integration contour towards the curve depicted in Figure \ref{Figure contour deformes pour u bigger than v re(t) su negatif} in the case 
when $\Re(t) \mf{s}_{\op{u}}<0$ and the one depicted in Figure \ref{Figure contour deformes pour u bigger than v re(t) su positif} in the case when $\Re(t) \mf{s}_{\op{u}}>0$.

\begin{figure}[ht]
\begin{center}

\begin{pspicture}(12,5)

\psline[linestyle=dashed, dash=3pt 2pt](0,2.5)(11.8,2.5)
\psline[linewidth=2pt]{->}(11.8,2.5)(11.9,2.5)

\rput(11,2.2){$\R + \i \a_{\op{u}}$}

\rput(0.1,3.1){ $\mf{s}_{\op{u}} t |\sg_-|$ }
\psdots(1,3)
\psline(1,3)(1,4.6)
\psdot(1,4.6)
\pscurve(1,4.6)(1.2,3.5)(1.3,2.9)(1,2.8)(0.7,2.9)(0.8,3.5)(1,4.6)
\rput(1.3,4.7){$ r_- $}
\rput(1.7 , 3.1){$\Ga_-^{\be_t}$}

\rput(2.8,2){ $\mf{s}_{\op{u}} t |\sg_+|$ }
\psdots(2,2)
\psline(2,2)(2,0.8)
\psdots(2,0.85)
\pscurve(2,0.85)(2.2,1.6)(2,2.1)(1.8,1.6)(2,0.85)
\rput(2.35,0.7){$ r_+ $}
\rput(1.5,1.5){$\Ga_+^{\be_t}$}

\psline(0,0)(12,5)
\psline[linewidth=2pt]{->}(8,3.35)(8.1,3.39)
\psline[linewidth=2pt]{->}(1,2.8)(1.1,2.8)

\rput(11,0){$ \vsg_{\op{u}}<0$}
\rput(11.5,0.8){ $ \ex{ -\tfrac{\i}{2} \th_{\be_t} } \R  $ }

\psline(0,5)(12,0)
\psline[linewidth=2pt]{->}(8,1.65)(8.1,1.60)
\psline[linewidth=2pt]{->}(1.8,1.6)(1.8,1.65)

\rput(11,5){$ \vsg_{\op{u}}>0$}
\rput(11,4.2){ $\ex{-\tfrac{\i}{2} \th_{\be_t}  } \R$ }

\end{pspicture}
\caption{ Deformed contours in the case $|\op{u}|>\op{v}$ and for $\Re(t) \mf{s}_{\op{u}}<0$
\label{Figure contour deformes pour u bigger than v re(t) su negatif} }
\end{center}

\end{figure}
\begin{figure}[ht]
\begin{center}

\begin{pspicture}(12,5)

\psline[linestyle=dashed, dash=3pt 2pt](0,2.5)(11.8,2.5)
\psline[linewidth=2pt]{->}(11.8,2.5)(11.9,2.5)

\rput(11,2.2){$\R + \i \a_{\op{u}}$}

\rput(11.9,3.1){ $\mf{s}_{\op{u}} t |\sg_-|$ }
\psdots(11,3)
\psline(11,3)(11,4.6)
\psdot(11,4.6)
\pscurve(11,4.6)(10.8,3.5)(10.7,2.9)(11,2.8)(11.3,2.9)(11.2,3.5)(11,4.6)
\rput(10.7,4.7){$ r_- $}
\rput(10.3, 3.1){$\Ga_-^{\be_t}$}

\rput(9.2,2){ $\mf{s}_{\op{u}} t |\sg_+|$ }
\psdots(10,2)
\psline(10,2)(10,0.8)
\psdots(10,0.85)
\pscurve(10,0.85)(9.8,1.6)(10,2.1)(10.2,1.6)(10,0.85)
\rput(9.65,0.7){$ r_+ $}
\rput(10.5,1.5){$\Ga_+^{\be_t}$}

\psline(0,0)(12,5)
\psline[linewidth=2pt]{->}(8,3.35)(8.1,3.39)
\psline[linewidth=2pt]{->}(11,2.8)(11.1,2.8)

\rput(1,0){$ \vsg_{\op{u}}<0$}
\rput(0.5,0.8){ $ \ex{ -\tfrac{\i}{2} \th_{\be_t} } \R  $ }

\psline(0,5)(12,0)
\psline[linewidth=2pt]{->}(8,1.65)(8.1,1.60)
\psline[linewidth=2pt]{->}(10.2,1.65)(10.2,1.6)

\rput(1,5){$ \vsg_{\op{u}}>0$}
\rput(1,4.3){ $\ex{-\tfrac{\i}{2} \th_{\be_t}  } \R$ }

\end{pspicture}
\caption{ Deformed contours in the case $|\op{u}|>\op{v}$ and for $\Re(t) \mf{s}_{\op{u}} > 0$
\label{Figure contour deformes pour u bigger than v re(t) su positif} }
\end{center}

\end{figure}

Upon the change of variables $s=\be_t^{-\tf{1}{2}} s^{\prime}$ in the integration along $\ex{-\tfrac{\i}{2}\th_{\be_t} } \R$, one decomposes $\mc{F}_n$ as 
\beq
\mc{F}_n(t) \, = \,  \Big( \f{ 1 }{ \be_t } \Big)^{ \frac{1}{2}(n+1) } \Int{ \R }{}  \!  \dd s \, \ex{-s^2}\, s^{n} \, \chi\Big(t, \tfrac{ s }{ \sqrt{\be_t} } \Big) \; + \; 
\Int{ \Ga^{\be_t}_{ \vsg_{\op{u}}} }{}\,  \dd s\, s^n \, \chi(t,s) \ex{-\be_t s^2 } \;. 
\enq

\subsubsection*{ $\bullet$  The Gaussian integral }

The first integral appearing in this decomposition can be analysed by observing that 
\beq
 \chi\Big(t, \tfrac{ s }{ \sqrt{\be_t} } \Big) \; = \; \chi(t,0) \cdot \Bigg\{ 1\, - \, \f{ s }{ t \sqrt{\be_t} } \sul{\ups=\pm}{} \f{ \ups \de_{\ups} }{ \sg_{\ups} }   
\, + \, \e{O}\Big( \f{ s^2 }{ t^2 \cdot \be_t } \Big) \Bigg\}
\enq
uniformly in $|s| \leq |\Re(t)|^{\f{1}{4}}$. Thus, it is convenient to introduce the intervals
\beq
 I_{\e{in}} \, = \, \intoo{ - \tau }{ \tau } \;  \quad \e{and} \quad 
I_{ \e{out} } \, = \, \intoo{ - \infty }{ - \tau  } \cup \intoo{  \tau  }{ + \infty } \qquad \e{with} \quad \tau=|\Re(t)|^{\f{1}{4}} \;. 
\enq
Then, one has 
\beq
\Big( \f{ 1 }{ \be_t } \Big)^{ \frac{1}{2}(n+1) } \Int{ \R }{}  \!   \dd s \,   \ex{-s^2} s^n \chi\Big(t, \tfrac{ s }{ \sqrt{\be_t} } \Big) \; = \; \mc{H}_{ I_{\e{in}} }(t) + \mc{H}_{ I_{\e{out}} }(t)
\label{ecriture decomposition integrale Gaussienne sur H in et H out}
\enq
where, for even $n$, one has  
\beq
\mc{H}_{ I_{\e{in}} }(t) \; = \;    \chi(t,0) \cdot  \Big( \f{ 1 }{ \be_t } \Big)^{ \frac{1}{2}(n+1) }  \Bigg\{ \Int{ I_{\e{in}} }{} \ex{-s^{2}} s^n \cdot \dd s \; + \;  
\e{O}\bigg(  \Int{ I_{\e{in}} }{} \f{ s^{2+n} \ex{-s^{2}}   }{ \be_t \, t^2 }  \cdot \dd s  \bigg)  \Bigg\}
\enq
The integrations can be extended to infinity upon adding some $\e{O}\Big( \ex{-\frac{1}{2} \tau^2} \Big)$ corrections, so that, for even $n$, one has  
\beq
\mc{H}_{ I_{\e{in}} }(t) \; = \;  \chi(t,0) \cdot  \Big( \f{ 1 }{ \be_t } \Big)^{ \frac{1}{2}(n+1) }  \Bigg\{  \Ga\Big( \tfrac{n+1}{2} \Big)   
\; + \;   \e{O}\bigg( \f{ 1 }{ \be_t \, t^2 } \bigg) \, + \, \e{O}\bigg(   \ex{ - \tfrac{1}{2}\sqrt{|\Re(t)|} }     \bigg)   \Bigg\} \;. 
\enq
Similar handlings in the case of odd $n$ entail that in such a case 
\beq
\mc{H}_{ I_{\e{in}} }(t) \; = \;  \chi(t,0) \cdot  \Big( \f{ 1 }{ \be_t } \Big)^{ \frac{1}{2}(n+1) } \cdot \Big( \tfrac{1}{ t \sqrt{\be_{t}} }\Big)^{w_n}
\cdot \Bigg\{ u_n    \; + \;   \e{O}\bigg( \f{ 1 }{ \be_t \, t^2 } \bigg) \, + \, \e{O}\bigg(   \ex{ - \tfrac{1}{2}\sqrt{|\Re(t)|} }     \bigg)   \Bigg\}  
\enq
for some integer $w_n \geq 1$ and a coefficient $u_n \not=0$. 

Regarding to the second contribution in \eqref{ecriture decomposition integrale Gaussienne sur H in et H out}, it can be presented as 
\beq
\mc{H}_{ I_{\e{out}} }(t) \; = \;  \Big( \f{ 1 }{ \be_t } \Big)^{ \frac{1}{2}(n+1) } \Int{0}{+\infty} \dd s  \, \ex{-(s+\tau)^2}  \,  (s+\tau )^{n} \bigg\{  \chi\Big(t, \tfrac{ s+\tau }{ \sqrt{\be_t} } \Big) 
\; + \; (-1)^n    \chi\Big(t, - \tfrac{ s+\tau }{ \sqrt{\be_t} } \Big)  \bigg\} \;. 
\enq
Since, for $s\geq 0$,  
\beq
\Im\Big(  t  \sg_{\ups}  \pm \ups  \tfrac{ s+\tau }{ \sqrt{\be_t} }  \Big) \, = \, \a \sg_{\ups} \mp \ups  \frac{ s+\tau }{ \sqrt{|\be_t|} } \sin\big[ \tfrac{\th_{\be_t} }{2} \big]
\enq
which obviously does not vanish for $|t|$ large enough, with $t \in \R +\i\a $, and is dominated in this regime by the second term,  it follows that 
\beq
 \Big| \chi\Big(t, \pm  \tfrac{ s+\tau }{ \sqrt{\be_t} } \Big)  \Big| \, \leq \,  C  \cdot 
 \big| \be_t \big|^{ \tfrac{ \de_{+} + \de_{-} }{2} }  
\enq
so that, upon inserting the large-$t$ behaviour of $\be_t$, one has 
\beq
\Big| \mc{H}_{ I_{\e{out}} }(t)  \Big| \;  \leq  \; C^{\prime}  |t|^{ \tfrac{ 1}{ 2 }( n+1-\de_{+} - \de_{-}) } \cdot \ex{ - \sqrt{ |\Re(t)| } }  
\Int{0}{+\infty} \dd s  \ex{-s^2-2s\tau}  \Big( \,  |t|^{ \tfrac{ 1}{ 4 }( |\de_{+}| + |\de_{-}| +n)}  +  |s|^{ |\de_{+}| + |\de_{-}| + n }    \Big)  \bigg\} 
=\e{O}\bigg(   \ex{ - \tfrac{3}{4}\sqrt{|\Re(t)|} } \bigg)   \;. 
\enq
Thence, all in all, since $\chi(t,0)\cdot \be_t^{-\tfrac{1}{2}(n+1) }$ has at most an algebraic growth in $t$, 
\beq
\Int{ \R }{} \hspace{-2mm} \f{ \dd s }{ \sqrt{\be_t} } \ex{-s^2} \chi\Big(t, \tfrac{ s }{ \sqrt{\be_t} } \Big) \; = \; 
\chi(t,0) \cdot \Big( \f{ 1 }{ \be_t } \Big)^{ \frac{1}{2}(n+1) } \cdot   \Big( \tfrac{1}{ t \sqrt{\be_{t}} }\Big)^{w_n}\cdot  \Bigg\{   u_n    \; + \;   \e{O}\bigg( \f{ 1 }{  t  } \bigg) \Bigg\} 
\, + \, \e{O}\bigg(   \ex{ - \tfrac{1}{2}\sqrt{|\Re(t)|} }     \bigg) \;. 
\enq
Above, $u_n$ and $w_n$ are as appearing in \eqref{definition proprietes un et wn}. Finally, it is readily seen that both remainders are
holomorphic in $t \in \mc{S}_{\th_0,A}$ for some $\th_0$ small enough and $A$ large enough.

\subsubsection*{ $\bullet$  The loop integral contribution}

It now remains to focus on the integral along $\Ga_{  \vsg_{\op{u}} }^{ \be_t }$. The contour $\Ga_{\ups}^{\be_t}$ can be deformed as 
\beq
\Ga_{\ups}^{\be_t} \; \hookrightarrow \; \Big\{ \mf{s}_{\op{u}} t |\sg_{\ups}| \, - \, \ups \Dp{}\mc{D}_{0,\eps}  \Big\} \cup
\Big\{ \mf{s}_{\op{u}} t |\sg_{\ups}| \, +\, \i  \intff{ -\ups\eps }{ -\ups T_{\ups} }  \Big\} \;. 
\enq
Here, $ - \, \ups \Dp{}\mc{D}_{0,\eps}$ stands for the circle of radius $\eps$ centred at $0$ and oriented $-\ups$ counterclockwise. 

However, in doing so, one has to take into account the discontinuity of the integrand along the line $\mf{s}_{\op{u}} t |\sg_{\ups}| \, - \, \i \ups \intff{ \eps }{ T_{\ups} }$, where 
I have set 
\beq
 T_{\ups} \; = \; - \ups \Big( \Im\big( r_{\ups} \big) - \mf{s}_{\op{u}} |\sg_{\ups}| \a \Big) \; \sim \;   C \cdot |\Re(t)| 
\label{ecriture T ups et estimation de sa croissance à l'infini}
\enq
for some $C>0$ and when $ |\Re(t)|  \tend +\infty$. This yields that 
\beq
\Int{ \Ga^{\be_t}_{\ups} }{} \! \dd s\;  \chi(t,s) s^n\,  \ex{-\be_t s^2 }  \; = \; \sul{a=1}{3}\Psi_{a;\ups }(t) 
\enq
where, given $\eta>0$ small enough,  
\beq
\Psi_{1;\ups }(t) \; = \; - \ups \eps  \hspace{-1mm}  \Int{ \Dp{} \mc{D}_{0,1} }{}  \hspace{-2mm} \dd z \;  \f{ \ex{- \be_t \big(\eps z + \mf{s}_{\op{u}} t |\sg_{\ups}| \big)^2   } }{ \big[ -\i \eps \ups z \big]^{\de_{\ups}}  }
\cdot \big(\eps z + \mf{s}_{\op{u}} t |\sg_{\ups}| \big)^n \cdot \bigg\{   \f{ \i }{ 2t - \ups \eps z    }    \bigg\}^{ \de_{-\ups} }  \;, 
\enq
\beq
\Psi_{2;\ups }(t) \; = \; (-\i \ups)^n 2 \sin[\pi \de_{\ups} ]  \hspace{-1mm}  \Int{ \eps }{ \eta(t-\i\a) \mf{s}_{t} } \hspace{-3mm} \dd w \,  h_{\ups}(w,t)
\qquad \e{and} \qquad 
\Psi_{3;\ups }(t) \; = \; (-\i \ups)^n  2 \sin[\pi \de_{\ups} ]  \hspace{-1mm} \Int{ \eta |\Re(t)| }{T_{\ups} } \hspace{-3mm} \dd w \,  h_{\ups}(w,t) \;. 
\enq
Above, I have introduced
\beq
h_{\ups}(w,t) \; = \; \big(w + \i \ups \mf{s}_{\op{u}} |\sg_{\ups}| \, t \big)^n \cdot  \f{  \ex{ \, \be_t \big(w + \i \ups \mf{s}_{\op{u}} |\sg_{\ups}| \, t \big)^2   } }
{  w^{ \de_{\ups} } \cdot \big[ w - 2  \i t \big]^{ \de_{-\ups}  }  } \;. 
\enq

\subsubsection*{ $\bullet$  The bound on  $\Psi_{3; \vsg_{\op{u}} } (t) $}

  Using the expansion \eqref{definition de xiu et C beta}, one gets, uniformly in $w \in \Big[ \eta |\Re(t)| ;  T_{\ups}  \Big] $ and in respect to $|\Re(t)| \tend + \infty$
\bem
 \be_t \big(w + \i \ups \mf{s}_{\op{u}} |\sg_{\ups}| \, t \big)^2 \; = \; \f{ \i  \vsg  C_{\be} }{t}
\Big( w^2 \, + \,  2 w \i \ups \mf{s}_{\op{u}} |\sg_{\ups}| \, t  \, - \,  \sg_{\ups}^{2} \, t^2  \Big)  \; + \; \e{O}(1) \\
\; = \; \i  \vsg  C_{\be} \Big\{  \tfrac{ w^2 }{ t } \, - \, \sg_{\ups}^2 t \Big\} \, - \,  2 C_{\be} \ups  \vsg_{\op{u}} |\sg_{\ups}| w \,  + \, \e{O}\big( 1 \big) \;. 
\end{multline}
Also, given $t\in \R + \i \a$, one has that for $|\Re(t)| \tend +\infty$ and $w \in \Big[ \eta |\Re(t)| ;  T_{\ups}  \Big]$
\beq
\big| w + \i \ups \mf{s}_{\op{u}} |\sg_{\ups}| \, t \big|^n  \, \leq \, C |\Re(t) |^{n} \qquad \e{and} \qquad 
 \Big|w^{ -\de_{\ups} } \cdot \big[ w - 2  \i t \big]^{ -\de_{-\ups}  } \Big| \, \leq \, C |\Re(t) |^{ |\de_{+}| + |\de_{-}| }
\enq
for some constant $C>0$ and where one uses in the intermediate steps that both $|w|$ and $|w - 2  \i t|$ are bounded from below. 
Hence, by putting these bounds together, one gets for some constants $C, C^{\prime}, C^{\prime \prime}$, 
\beq
\Psi_{3;  \vsg_{\op{u}} }(t)  \; \leq  \;     C   |\Re(t)|^{ n-\de_+ - \de_- }   \Int{ \eta |\Re(t)| }{  T_{\ups} } \ex{-2 C_{\be} |\sg_{  \vsg_{\op{u}}} | w } \cdot \dd w 
\; \leq  \;  C^{\prime} \ex{ -C^{\prime \prime}|\Re(t)| }
\enq
where the last bound follows from \eqref{ecriture T ups et estimation de sa croissance à l'infini} and holds for $t\in \R+\i\a$ with $|t|$-large enough.

\subsubsection*{ $\bullet$  The bound on  $\Psi_{2; \vsg_{\op{u}} }(t) $}

Assume that $t = \rho \ex{\i \th} $ with $\rho \in \R$, $\rho \tend \pm \infty$ and $|\th|<\th_0$ with $\th_0$ small enough. 
Then, it is convenient to rescale the integration variable as 
\beq
w_{x}\, = \; (1-x) \eps \, + \, x \eta \mf{s}_{t} (t-\i\a)
\enq
so that
\beq
 \Psi_{2;\ups }(t) \; = \; 2 (-\i \ups)^{n} \sin[\pi \de_{\ups} ]\,  \Big(  \eta \mf{s}_{t} (t-\i\a) \, -\, \eps  \Big)  \Int{ 0 }{1 } \hspace{-1mm} \dd x \,  h_{\ups}(w_x,t) \;. 
\enq
Furthermore, one has that 
\beq
\be_t \big(w_x + \i \ups \mf{s}_{\op{u}} |\sg_{\ups}| \, t \big)^2 \; = \;  - \i   \vsg C_{\be}   t  \Big( |\sg_{\ups}|-\i \ups \mf{s}_{\op{u}} \mf{s}_{t}\eta x \Big)^2 \; + \; \e{O}\big( 1 \big)
\enq
and, for $\eta$ small enough, it holds 
\beq
|\sg_{\ups}|-\i \ups \mf{s}_{\op{u}} \mf{s}_{t}\eta x \, = \, g_{x,\ups} \ex{-\tfrac{\i}{2} \vp_{x} \ups \mf{s}_{\op{u}} \mf{s}_{t} }
\qquad \e{for} \, \e{some} \quad 
0\leq \vp_{x} < C \eta \quad \e{and} \quad g_{x,\ups} \, > \, C^{\prime}  \, > \, 0
\enq
uniformly in $x$. Thus, for $\ups =  \vsg_{\op{u}}$, 
\beq
\be_t \big(w_x + \i  \vsg_{\op{u}} \mf{s}_{\op{u}} |\sg_{  \vsg_{\op{u}} }| \, t \big)^2 \; = \;   |\rho| \,   C_{\be} \,     g_{x,\vsg_{\op{u}} }^2 \exp\Big\{ \i \big[\th-(\vp_{x}+\tfrac{\pi}{2})\mf{s}_{t}  \vsg \big] \Big\}  \; + \; \e{O}\big( 1 \big) \;. 
\enq
In particular, for $\th=-\mf{s}_t  \vsg \psi$ with $\psi>0$, one gets that  
\beq
\Re\Big[ \be_t \big(w_x + \i \ups \mf{s}_{\op{u}} |\sg_{\ups}| \, t \big)^2 \Big] \; \leq  \;  - C^{\prime\prime} \, |\rho|  \, \sin\Big( \psi +\vp_{x}  \Big) \, \leq \, - C^{(3)} \, |\rho|  \,  \psi
\enq
uniformly in $x \in \intff{0}{1}$. Furthermore, 
\beq
\Re\big( w_{x} \big) \, = \, (1-x) \eps + x \eta |\rho| \cos(\th) \quad \e{so}\; \e{that} \quad 
 \big| w_x^{-\de_{\ups} } \big|  \; \leq \; C 
\enq
since $|w_x|$ is bounded from below away from $0$. Finally, one also has that 
\beq
\Im\Big( w_x -2 \i  t  \Big) \; = \; -\a x \eta \mf{s}_{t} \, - \, 2 \rho \cos(\th) + x \eta \rho \mf{s}_{t} \sin(\th)
\enq
so that  uniformly in $x \in \intff{0}{1}$  and for $|\rho|$ large enough and $\th_0$ small enough $\big| \Im\big( w_x -2 \i  t  \big) \big|  \; \geq  \;  C |\rho|$
hence ensuring that $| w_x -2 \i  t |$ is bounded from below and thus 
\beq
\Big| \big( w_x -2 \i  t  \big)^{-\de_{-\ups} } \Big|  \; \leq \; C \;. 
\enq
 The 
numerator in $h_{\vsg_{\op{u}} }(w_x,t)$ generates a power-law  bound in $\rho$ so that, all-in-all, 
\beq
\big| h_{\vsg_{\op{u}} }(w_x,t) \big| \, \leq \,  C^{\prime} \ex{- C \psi |\rho|} \qquad \e{for} \qquad t =  |\rho| \ex{-\i  \vsg \mf{s}_t \psi } \;. 
\enq
This entails that 
\beq
 \Big| \Psi_{2;  \vsg_{\op{u}} }(t)  \Big|  \; \leq \;  C^{\prime}  \ex{- C \psi |\rho|}   \qquad \e{for} \qquad t = |\rho| \ex{-\i \psi  \vsg  \mf{s}_t} \;. 
\enq

Also, the above reasonings and estimates ensure that $\Psi_{2; \vsg_{\op{u}} }$ is holomorphic on $\mc{S}_{\th_0;A}$ with  $A$ large enough and $\th_0$ small enough.

\subsubsection*{ $\bullet$  The bound on  $\Psi_{1;\ups }(t) $}

Similar handlings to what has been exposed show that, for $z=\ex{\i\psi}$,  
\beq
-\Re\Big( \be_t \, (\eps z + \mf{s}_{\op{u}} t |\sg_{\ups} | \big)^2  \Big) \, = \, - \sg_{\ups}^2 C_{\be} |\rho| \sin(\th) \, + \, \e{O}(1) \qquad \e{with} \quad 
 t \, = \, \rho \ex{-\i  \vsg \mf{s}_{t} \th } \;. 
\enq
 Furthermore, for $z=\ex{\i \psi}$ one has 
\beq
\Big| \Im\Big( \eps z \,- \,2 \i  t    \Big)\Big| \, = \, \Big| \eps \sin(\psi) \, - \, 2   \rho \cos(\th)   \Big|  \geq C |\rho|
\enq
provided that $\th$ is small enough and $|\rho|$ large enough, one readily gets that, for some constants $C, C^{\prime}$,  
\beq
 \Big| \Psi_{1;\ups }(t)  \Big|  \; \leq \;  C^{\prime}  \ex{- C \th |\rho|}   \qquad \e{for} \qquad t=\rho \ex{-\i \th  \vsg  \mf{s}_t} \;. 
\enq
Again, the above  also ensures that $\Psi_{1;\ups }$ is holomorphic on $\mc{S}_{\th_0;A}$ with  $A$ large enough and $\th_0$ small enough.

Thus, upon putting all the intermediate bounds together, the claim follows.

 \subsubsection*{B) The regime $|\op{u}|< \op{v}$} 
 The analysis is quite similar to the previous regime. I thus only highlight the main steps. 
 
In the present case, since $\e{sgn}( \sg_{\pm} )=+$, it is convenient to represent
\beq
\chi(t,s) \, = \, \pl{ \ups= \pm }{} \bigg\{  \f{   \i \ups }{ s \,+ \, t  \ups \sg_{\ups}     }   \bigg\}^{ \de_{\ups} }
\enq
meaning that, for fixed $t$, the map $s \mapsto \chi(t,s)$ has cuts along the lines $  -\ups  t \sg_{\ups} - \ups \i  \R^{+}$. 
Since $|\a_{\op{u}}| <\a$, the cut along  $  -t \sg_{+} -  \i  \R^{+}$ is located below the original integration line $\R + \i \a_{ \op{u} }$
while the one along  $  t \sg_{-} +  \i  \R^{+}$ is located above $\R + \i \a_{ \op{u} }$.
\begin{figure}[ht]
\begin{center}

\begin{pspicture}(12,5)

\psline[linestyle=dashed, dash=3pt 2pt](0,2.5)(11.8,2.5)
\psline[linewidth=2pt]{->}(11.8,2.5)(11.9,2.5)

\rput(12,2.2){$\R + \i \a_{\op{u}}$}

\rput(0.4,2.9){ $  t \sg_-$ }
\psdots(1,3)
\psline(1,3)(1,4.6)
\psdot(1,4.6)
\pscurve(1,4.6)(1.2,3.5)(1.3,2.9)(1,2.8)(0.7,2.9)(0.8,3.5)(1,4.6)
\rput(1.3,4.7){$ r_- $}
\rput(1.7 , 3.1){$\Ga_-^{\be_t}$}

\rput(9.4,2){ $- t \sg_+ $ }
\psdots(10,2)
\psline(10,2)(10,0.8)
\psdots(10,0.85)
\pscurve(10,0.85)(9.8,1.6)(10,2.1)(10.2,1.6)(10,0.85)
\rput(9.65,0.7){$ r_+ $}
\rput(10.5,1.5){$\Ga_+^{\be_t}$}

\psline(0,0)(12,5)
\psline[linewidth=2pt]{->}(8,3.35)(8.1,3.39)
\psline[linewidth=2pt]{->}(1,2.8)(1.1,2.8)

\rput(11,0){$ \vsg <0$}
\rput(11.5,0.8){ $ \ex{ -\tfrac{\i}{2} \th_{\be_t} } \R  $ }

\psline(0,5)(12,0)
\psline[linewidth=2pt]{->}(8,1.65)(8.1,1.60)
\psline[linewidth=2pt]{->}(9.8,1.6)(9.8,1.65)

\rput(11,5){$ \vsg >0$}
\rput(11,4.2){ $\ex{-\tfrac{\i}{2} \th_{\be_t}  } \R$ }

\end{pspicture}
\caption{ Deformed contours in the case $|\op{u}|<\op{v}$ and for $\Re(t) < 0$
\label{Figure contour deformes pour u bigger than v re(t) negatif} }
\end{center}

\end{figure}
\begin{figure}[ht]
\begin{center}

\begin{pspicture}(12,5)

\psline[linestyle=dashed, dash=3pt 2pt](0,2.5)(11.8,2.5)
\psline[linewidth=2pt]{->}(11.8,2.5)(11.9,2.5)

\rput(11,2.2){$\R + \i \a_{\op{u}}$}

\rput(11.8,3.1){ $  t \sg_-$ }
\psdots(11,3)
\psline(11,3)(11,4.6)
\psdot(11,4.6)
\pscurve(11,4.6)(10.8,3.5)(10.7,2.9)(11,2.8)(11.3,2.9)(11.2,3.5)(11,4.6)
\rput(10.7,4.7){$ r_- $}
\rput(10.3, 3.1){$\Ga_-^{\be_t}$}

\rput(2.6,2){ $ -t \sg_+$ }
\psdots(2,2)
\psline(2,2)(2,0.8)
\psdots(2,0.85)
\pscurve(2,0.85)(2.2,1.6)(2,2.1)(1.8,1.6)(2,0.85)
\rput(2.35,0.7){$ r_+ $}
\rput(1.5,1.5){$\Ga_+^{\be_t}$}

\psline(0,0)(12,5)
\psline[linewidth=2pt]{->}(8,3.35)(8.1,3.39)
\psline[linewidth=2pt]{->}(11,2.8)(11.1,2.8)

\rput(1,0){$ \vsg < 0$}
\rput(0.5,0.8){ $ \ex{ -\tfrac{\i}{2} \th_{\be_t} } \R  $ }

\psline(0,5)(12,0)
\psline[linewidth=2pt]{->}(8,1.65)(8.1,1.60)
\psline[linewidth=2pt]{->}(2.2,1.65)(2.2,1.6)

\rput(1,5){$  \vsg > 0$}
\rput(1,4.3){ $\ex{-\tfrac{\i}{2} \th_{\be_t}  } \R$ }

\end{pspicture}
\caption{ Deformed contours in the case $|\op{u}|<\op{v}$ and for $\Re(t) >0$
\label{Figure contour deformes pour u bigger than v re(t) positif} }
\end{center}

\end{figure}

The analysis then depends on the sign of $ \vsg$

\subsubsection*{$\bullet$  $ \vsg >0$ }

In this case, one can deform the contour as in Figures \ref{Figure contour deformes pour u bigger than v re(t) negatif} or \ref{Figure contour deformes pour u bigger than v re(t) positif}, 
depending on the sign of $\Re(t)$, without having to deal with the cuts of $\chi(t,s)$. This yields
\beq
\mc{F}_n(t) \, = \, \Big( \f{ 1 }{ \be_t } \Big)^{ \frac{1}{2}(n+1) }   \Int{ \R }{}    \dd s \,  s^n  \ex{-s^2} \, \chi\Big(t, \tfrac{ s }{ \sqrt{\be_t} } \Big)   
\enq
and one can conclude by the previous analysis.

\subsubsection*{$\bullet$   $ \vsg < 0$ }

In this case, when deforming the contour as in Figures \ref{Figure contour deformes pour u bigger than v re(t) negatif} or \ref{Figure contour deformes pour u bigger than v re(t) positif}, 
according to the sign of $\Re(t)$, one observes that one has to take into account both cuts stemming from $\chi(t,s)$. This yields

\beq
\mc{F}_n(t) \, = \,  \Big( \f{ 1 }{ \be_t } \Big)^{ \frac{1}{2}(n+1) } \Int{ \R }{}    \dd s  \,   s^n  \ex{-s^2} \,  \chi\Big(t, \tfrac{ s }{ \sqrt{\be_t} } \Big) \; + \; 
\sul{\ups= \pm }{} \Int{ \Ga^{\be_t}_{ \ups } }{} \dd s\,  s^n \, \chi(t,s) \ex{-\be_t s^2 } \;. 
\enq
  The two cut-issued integrals are very similar in structure to those studied previously, and, eventually,
one ends up with the same conclusions. \qed


\begin{thebibliography}{10}

\bibitem{AbadaBougourziSiLakhalFourSpinonDSFXXXMultInt}
A.~Abada, A.H. Bougourzi, and B.~Si-Lakhal, \emph{{"Exact four-spinon dynamical
  correlation function of the Heisenberg model."}}, Nucl. Phys. B \textbf{497}
  (1997), 733--753.

\bibitem{AffleckPereiraWhiteEdgeSingInSpin1-2}
I.~Affleck, R.~G. Pereira, and S.~R. White, \emph{{"Exact edge singularities
  and dynamical correlation functions in spin-$1/2$ chains."}}, Phys. Rev.
  Lett. \textbf{100} (2008), 027206.

\bibitem{AffleckPereiraWhiteSpectralFunctionsfor1DLatticeFermionsBoundStatesContributions}
\bysame, \emph{{"Spectral function of spinless fermions on a one-dimensional
  lattice."}}, Phys. Rev. \textbf{B 79} (2009), 165113.

\bibitem{BarnesDoubleGaFctn1}
E.W. Barnes, \emph{{"Genesis of the double gamma function."}}, Proc. London
  Math. Soc. \textbf{\bf{31}} (1900), 358--381.

\bibitem{BarnesDoubleGaFctn2}
\bysame, \emph{{"The theory of the double gamma function."}}, Philos. Trans.
  Roy. Soc. London, Ser. A \textbf{196} (1901), 265--388.

\bibitem{BeckBonnerMullerFenomenologicalFormDSFinXXX}
H.~Beck, J.C. Bonner, and G.~M\"{u}ller, \emph{{"Zero-Temperature Dynamics of
  the $s= 1/2$ Linear Heisenberg Antiferromagnet."}}, Phys. Rev. Lettµ.
  \textbf{\bf 43} (1979), 75--78.

\bibitem{BeckBonnerMullerThomasSpectralFctsXXXGeneralFeatures}
H.~Beck, J.C. Bonner, G.~M\"{u}ller, and H.~Thomas, \emph{{"Quantum spin
  dynamics of the antiferromagnetic linear chain in zero and nonzero magnetic
  field."}}, Phys. Rev. B \textbf{\bf 24} (1981), 1429--1467.

\bibitem{BeliavinPolyakovZalmolodchikovCFTin2DQFT}
A.A. Beliavin, A.M. Polyakov, and A.B. Zalmolodchikov, \emph{{"Infinite
  conformal symmetry in two-dimensional quantum field theory"}}, Nucl. Phys. B
  \textbf{\bf 241} (1984), 333--380.

\bibitem{BiegelKarbachMullerDSFXXXNumericsFromFormCBAEigenvectors}
D.~Biegel, M.~Karbach, and G.~M\"{u}ller, \emph{{"Quasiparticles governing the
  zero-temperature dynamics of the one-dimensional spin-$1/2$ Heisenberg
  antiferromagnet in a magnetic field."}}, Phys. Rev. B \textbf{\bf 66} (2002),
  054405,12pp.

\bibitem{BiegelKarbachMullerDSFXXXLongItudandTransverseAtqPiPisur2VariousExcitationClasses}
\bysame, \emph{{"Transition rates via Bethe ansatz for the spin-1/2 Heisenberg
  chain."}}, Europhys. Lett. \textbf{\bf 59} (2002), 882--888.

\bibitem{BiegelKarbachMullerDSFXXZTransverseAtqPiPisur2VariousvaluesDelta}
\bysame, \emph{{"Transition rates via Bethe ansatz for the spin-1/2 planar XXZ
  antiferromagnet."}}, J. Phys. A: Math. Gen. \textbf{\bf 36} (2003),
  5361--5368.

\bibitem{BierstoneWhitneyExtensionTheoremAndOtherDiffFctsProperties}
E.~Bierstone, \emph{{"Differentiable functions."}}, Bol. Soc. Bras. Mat.
  \textbf{\bf 11} (1980), 139--190.

\bibitem{BinerBoehmCauxGudelHabichtKieferKramerLauchliMcMorrowMesotNormandRueggRonnowStahnThielemann(C5H12N)2CuBr4AsDeltaHalfXXZChain}
D.~Biner, M.~Boehm, J.-S. Caux, H.-U. G\"{u}del, K.~Habicht, K.~Kiefer,
  W.~Kr\"{a}mer, A.M. L\"{a}uchli, D.F. McMorrow, J.~Mesot, B.~Normand, Ch.
  R\"{u}egg, H.M. R\o{}nnow, J.~Stahn, and B.~Thielemann, \emph{{"Direct
  Observation of Magnon Fractionalization in the Quantum Spin Ladder"}}, Phys.
  Rev. Lett. \textbf{\bf 102} (2009), 107204, 5pp.

\bibitem{BloteCardyNightingalePredictionL-1correctionsEnergyAscentralcharge}
H.W.J. Bl\"{o}te, J.L. Cardy, and M.P. Nightingale, \emph{{"Conformal
  invariance, the central charge, and universal finite-size amplitudes at
  criticality."}}, Phys. Rev. Lett. \textbf{\bf 56} (1986), 742--745.

\bibitem{BougourziCoutureKacirAll2SpinonDSFforXXXFromVOpsColseFormulae}
A.H. Bougourzi, M.~Couture, and M.~Kacir, \emph{{"Exact two-spinon dynamic
  structure factor of the one-dimensional $s=1/2$ Heisenberg-Ising
  antiferromagnet."}}, Phys. Rev. B \textbf{54} (1996), 12669?12672.

\bibitem{BougourziFledderjohanKarbachMullerMutterDynamicTwoSpinonStructureFactor}
A.H. Bougourzi, A.~Fledderjohann, M.~Karbach, G.~M\"{u}ller, and
  K.-H.-M\"{u}tter, \emph{{"Two-spinon dynamic structure factor of the
  one-dimensional $S=1/2$ {H}eisenberg antiferromagnet."}}, Phys. Rev. B
  \textbf{55} (1997), 12510--12517.

\bibitem{BougourziKarbachMuller2SpinonDSFMassiveXXZFromVopExplicit}
A.H. Bougourzi, M.~Karbach, and G.~M\"{u}ller, \emph{{"Exact two-spinon dynamic
  structure factor of the one-dimensional $s=1/2$ Heisenberg-Ising
  antiferromagnet."}}, Phys. Rev. B \textbf{58} (1998), 11429--11438.

\bibitem{CampbellCarmeloMachadoSacramentoLowerTresholdsXXXLongAndTransDSFPseudoFermionDynTheor}
D.K. Campbell, J.M.P. Carmelo, J.D.P. Machado, and P.D. Sacramento,
  \emph{{"Singularities of the dynamical structure factors of the spin-1/2 XXX
  chain at finite magnetic field"}}, J. Phys.: Condens. Matter \textbf{\bf 27}
  (2015), 406001, 16pp.

\bibitem{CardyConformalDimensionsFromLowLSpectrum}
J.L. Cardy, \emph{{"Operator content of two-dimensional conformally invariant
  theories."}}, Nucl. Phys. B \textbf{\bf 270} (1986), 186--204.

\bibitem{CauxCalabreseDynamicalStructureFactoBoseGas}
J.-S. Caux and P.~Calabrese, \emph{{"Dynamical density-density correlations in
  the one-dimensional Bose gas."}}, Phys. Rev. A \textbf{\bf 74} (2006),
  031605.

\bibitem{CauxCalabreseSlavnovSpectralFunctionBoseGas}
J.-S. Caux, P.~Calabrese, and N.A. Slavnov, \emph{{"One-particle dynamical
  correlations in the one-dimensional Bose gas."}}, J. Stat. Mech. \textbf{\bf
  74} (2007), P01008.

\bibitem{CauxHagemansDSFXXXFourSpinonContributionDeeperAnalysis}
J.-S. Caux and R.~Hagemans, \emph{{"The four-spinon dynamical structure factor
  of the Heisenberg chain."}}, J. Stat. Mech: Th. and Exp. (2006), P12013.

\bibitem{CauxHagemansMailletDynamicalCorrFunctXXZinFieldPlots}
J.-S. Caux, R.~Hagemans, and J.-M. Maillet, \emph{{"Computation of dynamical
  correlation functions of Heisenberg chains: the gapless anisotropic
  regime."}}, J. Stat. Mech. \textbf{\bf 95} (2005), P09003.

\bibitem{CauxImambekovPanfilShashiHeuristicsPrefactorsForEdgeExpInIntModels}
J.-S. Caux, A.~Imambekov, M.~Panfil, and A.~Shashi, \emph{{"Exact prefactors in
  static and dynamic correlation functions of 1D quantum integrable models:
  applications to the Calogero-Sutherland, Lieb-Liniger and XXZ models."}},
  hys. Rev. B \textbf{85} (2012), 155136, 95pp.

\bibitem{CauxKonnoSorrelWestonFFofMasslessXXZfromXYZResults}
J.-S. Caux, H.~Konno, M.~Sorrell, and R.~Weston, \emph{{"Exact form-factor
  results for the longitudinal structure factor of the massless XXZ model in
  zero field."}}, J. Stat. Mech. (2012), P01007.

\bibitem{CauxMailletDynamicalCorrFunctXXZinFieldPlots}
J.-S. Caux and J.-M. Maillet, \emph{{"Computation of dynamical correlation
  functions of Heisenberg chains in a field."}}, Phys. Rev. Lett. \textbf{\bf
  95} (2005), 077201.

\bibitem{CheianovPustilnikXXZLoweredgeNLLLEdgeExp}
V.V. Cheianov and M.~Pustilnik, \emph{{"Threshold singularities in the dynamic
  response of gapless integrable models"}}, Phys. Rev. Lett. \textbf{100}
  (2008), 126403.

\bibitem{DavisFodaJimboMiwaNakayashikiDiagonalizationXXZinfiniteDelta>1}
B.~Davies, O.~Foda, M.~Jimbo, T.~Miwa, and A.~Nakayashiki,
  \emph{{"Diagonalization of the XXZ Hamiltonian by vertex operators."}}, Comm.
  Math. Phys. \textbf{\bf 151} (1993), 83--153.

\bibitem{KozDugaveGohmannThermoFunctionsZeroTXXZMassless}
M.~Dugave, F.~G\"{o}hmann, and K.K. Kozlowski, \emph{{"Functions characterizing
  the ground state of the XXZ spin-1/2 chain in the thermodynamic limit."}},
  SIGMA \textbf{10} (2014), 043, 18 pages.

\bibitem{FaddeevSklyaninTakhtajanSineGordonFieldModel}
L.D. Faddeev, E.K. Sklyanin, and L.A. Takhtadzhan, \emph{{"The quantum inverse
  scattering method."}}, Teor. Math. Phys. \textbf{\bf{40}} (1979), 194.

\bibitem{GlazmanImambekovComputationEdgeExpExact1DBose}
L.I. Glazman and A.~Imambekov, \emph{{"Exact exponents of edge singularities in
  dynamic correlation functions of 1D Bose gas."}}, Phys. Rev. Lett.
  \textbf{100} (2008), 206805.

\bibitem{GlazmanImambekovDvPMTCompletTheoryNNLL}
\bysame, \emph{{"Universal theory of non-linear Luttinger liquids."}}, Science
  \textbf{323} (2009), 228--231.

\bibitem{GlazmanImambekovSchmidtReviewOnNLLuttingerTheory}
L.I. Glazman, A.~Imambekov, and T.L. Schmidt, \emph{{"One-dimensional quantum
  liquids: beyond the Luttinger liquid paradigm."}}, Rev. Mod. Phys.
  \textbf{84} (2012), 1253--1306.

\bibitem{GolubitskyGuilleminStableMappingAndSings}
M.~Golubitsky and V.~Guillemin, \emph{{"Stable mappings and their
  singularities"}}, Graduate texts in mathematics, Springer-Verlag, 1973.

\bibitem{GoffTennantNaglerCsCoCl3IndetifiedasMassiveXXZChain}
D.~A.~Tennant J.~P.~Goff and S.~E. Nagler, \emph{{"Exchange mixing and soliton
  dynamics in the quantum spin chain $\e{CsCoCl}_3$"}}, Phys. Rev. B
  \textbf{\bf 52} (1995), 15992--16000.

\bibitem{CauxMosselPerezCastillo2SpinonDSFMassiveXXZReloaded}
J.~Mossel J.-S.~Caux and I.~Perez Castillo, \emph{{"The two-spinon transverse
  structure factor of the gapped Heisenberg antiferromagnetic chain."}}, J.
  Stat. Mech.: Th. and Exp. (2008), P08006.

\bibitem{JimboMiwaFormFactorsInMassiveXXZ}
M.~Jimbo and T.~Miwa, \emph{{"Algebraic analysis of solvable lattice models"}},
  Conference Board of the Mathematical Sciences, American Mathematical Society,
  1995.

\bibitem{KarbachMullerDSFXXXSelectionRuleAndPlotsFromCBA}
M.~Karbach and G.~M\"{u}ller, \emph{{"Line-shape predictions via Bethe ansatz
  for the one-dimensional spin- $1/2$ Heisenberg antiferromagnet in a magnetic
  field."}}, Phys. rev. B \textbf{\bf 62} (2000), 14871, 9pp.

\bibitem{KozKitMailSlaTerEffectiveFormFactorsForXXZ}
N.~Kitanine, K.K. Kozlowski, J.-M. Maillet, N.A. Slavnov, and V.~Terras,
  \emph{{"On the thermodynamic limit of form factors in the massless XXZ
  Heisenberg chain."}}, J. Math. Phys. \textbf{50} (2009), 095209.

\bibitem{KozKitMailSlaTerRestrictedSums}
\bysame, \emph{{"A form factor approach to the asymptotic behavior of
  correlation functions in critical models."}}, J. Stat. Mech. : Th. and Exp.
  \textbf{1112} (2011), P12010.

\bibitem{KozKitMailSlaTerThermoLimPartHoleFormFactorsForXXZ}
\bysame, \emph{{"Thermodynamic limit of particle-hole form factors in the
  massless XXZ Heisenberg chain."}}, J. Stat. Mech. : Th. and Exp.
  \textbf{1105} (2011), P05028.

\bibitem{KozKitMailSlaTerRestrictedSumsEdgeAndLongTime}
\bysame, \emph{{"Form factor approach to dynamical correlation functions in
  critical models."}}, J. Stat. Mech. \textbf{1209} (2012), P09001.

\bibitem{KitanineMailletTerrasFormfactorsperiodicXXZ}
N.~Kitanine, J.-M. Maillet, and V.~Terras, \emph{{"Form factors of the XXZ
  Heisenberg spin-$1/2$ finite chain."}}, Nucl. Phys. B \textbf{554} (1999),
  647--678.

\bibitem{KitanineMailletTerrasElementaryBlocksPeriodicXXZ}
\bysame, \emph{{"Correlation functions of the XXZ Heisenberg spin-$1/2$ chain
  in a magnetic field."}}, Nucl. Phys. B \textbf{567} (2000), 554--582.

\bibitem{KohnoDynamDominantStringExcitationXXZChainVariousDSFXXXFiniteandZeroh}
M.~Kohno, \emph{{"Dynamically dominant excitations of string solutions in the
  spin-$1/2$ antiferromagnetic Heisenberg chain in a magnetic field."}}, Phys.
  Rev. Lett. \textbf{\bf 102} (2009), 037203.

\bibitem{KorepinSlavnovNonlinearIdentityScattPhase}
V.E. Korepin and N.A. Slavnov, \emph{{"The new identity for the scattering
  matrix of exactly solvable models."}}, Eur. Phys. J. \textbf{\bf B 5} (1998),
  555--557.

\bibitem{KozLongDistanceLargeTimeXXZ}
K.K. Kozlowski, \emph{{"Large-distance and long-time asymptotics of two-point
  dynamical correlation functions in the massless regime of the XXZ chain."}},
  to appear.

\bibitem{KozProofOfStringSolutionsBetheeqnsXXZ}
\bysame, \emph{{"On string solutions to the Bethe equations for the XXZ chain:
  a rigorous approach."}}, to appear.

\bibitem{KozProofOfAsymptoticsofFormFactorsXXZBoundStates}
\bysame, \emph{{"Form factors of bound states in the XXZ chain."}}, J. Phys. A:
  Math. Theor. Special Issue "Emerging talents" \textbf{50} (2017), 184002.

\bibitem{KozProofOfDensityOfBetheRoots}
\bysame, \emph{{"On condensation properties of Bethe roots associated with the
  XXZ chain."}}, Comm. Math. Phys. \textbf{357} (2018), 1009--1069.

\bibitem{KozMasslessFFSeriesXXZ}
\bysame, \emph{{"On the thermodynamic limit of form factor expansions of
  dynamical correlation functions in the massless regime of the XXZ spin $1/2$
  chain."}}, J. Math. Phys. "Ludwig Faddeev memorial volume" (2018),
  https://doi.org/10.1063/1.5021892.

\bibitem{LakeTennantCauxBarthelSchollwockNaglerFrostKCUF3DSFComparisionFromExpMEasureAndABA}
B.~Lake, D.A. Tennant, J.-S. Caux, T.~Barthel, U.~Schollw\"{o}ck, S.E. Nagler,
  and C.D. Frost, \emph{{"Multispinon Continua at Zero and Finite Temperature
  in a Near-Ideal Heisenberg Chain"}}, Phys. Rev. Lett. \textbf{111} (2013),
  137205, 5pp.

\bibitem{LashkevichPugaiFormFactorsEightVertex}
M.~Lashkevich and Y.~Pugai, \emph{{"Free field construction for correlation
  functions of the eight-vertex model."}}, Nucl. Phys. \textbf{B \bf{516}}
  (1998), 623--651.

\bibitem{LutherPeschelCriticalExponentsXXZZeroFieldLuttLiquid}
A.~Luther and I.~Peschel, \emph{{"Calculation of critical exponents in two
  dimensions from quantum field theory in one dimension."}}, Phys. Rev. B
  \textbf{\bf{12}(9)} (1975), 3908--3917.

\bibitem{StoneReichBroholmLefmannRischelLandeeTurnbullVeryClearDSFMeasureForXXXMagnetCu(C4H4N2)(NO3)2}
C.~Broholm K. Lefmann C. Rischel C. P.~Landee M.~B.~Stone, D. H.~Reich and
  M.~M. Turnbull, \emph{{"Extended Quantum Critical Phase in a Magnetized Spin-
  1/2 Antiferromagnetic Chain"}}, Phys. Rev. Lett. \textbf{\bf 91} (2003),
  037205, 4pp.

\bibitem{GlazmanKamenevKhodasPustilnikNLLLTheoryAndSpectralFunctionsFremionsFirstAnalysis}
A.~Kamenev L.I.~Glazman M.~Khodas, M.~Pustilnik, \emph{{"Dynamic response of
  one-dimensional interacting fermions."}}, Phys. Rev. Lett. \textbf{\bf 96}
  (2006), 196405.

\bibitem{GlazmanKamenevKhodasPustilnikDSFfor1DBosons}
\bysame, \emph{{"Dynamics of excitations in a one-dimensional Bose liquid."}},
  Phys. Rev. Lett. \textbf{\bf 99} (2007), 110405.

\bibitem{GlazmanKamenevKhodasPustilnikNLLLTheoryAndSpectralFunctionsFremionsBetterStudy}
\bysame, \emph{{"Fermi-Luttinger liquid: Spectral function of interacting
  one-dimensional fermions."}}, Phys. Rev. B \textbf{\bf 76} (2007), 155402.

\bibitem{MullerShrockDynamicCorrFnctsTIandXXAsymptTimeAndFourier}
G.~M\"{u}ller and R.E. Shrock, \emph{{"Dynamic correlation functions for
  one-dimensional quantum-spin systems: new results based on a rigorous
  approach."}}, Phys. Rev. B \textbf{\bf{29}} (1984), 288--301.

\bibitem{NaglerTennantCowleyPerringSatijaTestOfXXXDispersionRelationForKCuF3}
S.E. Nagler, D.A. Tennant, R.A. Cowley, T.G. Perring, and S.K. Satija,
  \emph{{"Spin dynamics in the quantum antiferromagnetic chain compound
  $\e{KCuF}_3$"}}, Phys. Rev. B \textbf{\bf{ 44}} (1991), 12361--12368.

\bibitem{HammarStoneReichBroholmGibsonLandeeOshikawaCu(C4H4N2)(NO3)2AsXXXMagnetIdentified}
D.~H. Reich C. Broholm P. J. Gibson M.~T. P.~R.~Hammar, M. B.~Stone, C.~P.
  Landee, and M.~Oshikawa, \emph{{"Characterization of a quasi-one-dimensional
  spin-1/2 magnet which is gapless and paramagnetic for $g \mu_B H \ll J$ and
  $k_B T \ll J$"}}, Phys. Rev. B \textbf{\bf 59} (1999), 1008--1015.

\bibitem{PustilnikDSFForCalogero}
M.~Pustilnik, \emph{{"Dynamic structure factor of the Calogero-Sutherland
  model."}}, Phys. Rev. Lett. \textbf{\bf 97} (2006), 036404.

\bibitem{RuelleRigorousResultsForStatisticalMechanics}
D.~Ruelle, \emph{{"Statistical mechanics: rigorous results"}}, W.A. Benjamin,
  Inc., 1969.

\bibitem{SatoShiroishiTakahashiDSFLongitudXXZABA}
J.~Sato, M.~Shiroishi, and M.~Takahashi, \emph{{"Evaluation of Dynamic spin
  structure factor for the spin-1/2 XXZ chain in a magnetic field."}}, J. Phys.
  Soc. Jpn \textbf{\bf 79} (2004), 3008--3014.

\bibitem{CauxGlazmanImambekovShasiNonUniversalPrefactorsFromFormFactors}
A.~Shashi, L.I. Glazman, J.-S. Caux, and A.~Imambekov, \emph{{"Non-universal
  prefactors in correlation functions of 1D quantum liquids."}}, Phys. Rev. B.
  \textbf{84} (2011), 045408.

\bibitem{SlavnovFormFactorsNLSE}
N.A. Slavnov, \emph{{"Non-equal time current correlation function in a
  one-dimensional Bose gas."}}, Theor. Math. Phys. \textbf{\bf 82} (1990),
  273--282.

\bibitem{StamperetalMeasureStructureFactorbyBraggSpectroscopyonBEC}
D.M. Stamper-Kurn, A.P. Chikkatur, A.~G\"{o}rlitz, S.~Inouye, S.~Gupta, D.E.
  Pritchard, and W.~Ketterle, \emph{{"Excitation of phonons in a Bose-Einstein
  condensate by light scattering."}}, Phys. Rev. Lett. \textbf{\bf 83} (1999),
  2876--2879.

\bibitem{TakahashiThermodynamics1DSolvModels}
M.~Takahashi, \emph{{"Thermodynamics of one dimensional solvable models."}},
  Cambridge university press, 1999.

\bibitem{YangYangXXZproofofBetheHypothesis}
C.N. Yang and C.P. Yang, \emph{{"One dimensional chain of anisotropic spin-spin
  interactions: I proof of Bethe's hypothesis."}}, Phys. Rev. \textbf{\bf 150}
  (1966), 321--327.

\end{thebibliography}
\end{document}